\documentclass{article}
\usepackage{asherspector}

\usepackage{csquotes}
\usepackage[utf8]{inputenc}
\usepackage{soul}
\usepackage{hyperref}
\usepackage{float}
\usepackage{algorithm}
\usepackage{algpseudocode}
\usepackage{url}
\usepackage{caption}
\usepackage{subcaption}
\usepackage{xcolor}
\usepackage{multirow}
\usepackage{apptools}
\usepackage[margin=0.9in]{geometry}
\usepackage{natbib}
\usepackage{enumitem}
\usepackage{booktabs}
\usepackage{tablefootnote}
\usepackage[normalem]{ulem}
\usepackage{booktabs,tabularx,array}
\makeatletter
\let\thmt@original@suspendcounter\thmt@suspendcounter
\renewcommand{\thmt@suspendcounter}[2]{\thmt@original@suspendcounter{#1}{#2}\expandafter\def\csname theH#1\endcsname{continued.#1.\arabic{thmt@dummyctr}}}
\makeatother

\title{Model-Agnostic Covariate-Assisted Inference on Partially Identified Causal Effects}
\author{Wenlong Ji \thanks{Department of Statistics, Stanford University} \and Lihua Lei \thanks{Graduate School of Business and Department of Statistics (by courtesy), Stanford University} \and Asher Spector \thanks{Department of Statistics, Stanford University}}
\date{\today}

\newcommand\blfootnote[1]{\begingroup
  \renewcommand\thefootnote{}\begin{NoHyper}\footnotetext{#1}\end{NoHyper}\endgroup
}

\newcommand{\lcb}{_\mathrm{LCB}}
\newcommand{\ucb}{_\mathrm{UCB}}
\newcommand{\aug}{^\mathrm{aug}}
\newcommand{\given}{|}
\newcommand{\error}{\mathrm{error}}

\newcommand{\mrswap}{\mathrm{swap}}
\newcommand{\swap}{^\mrswap}
\newcommand{\crossfit}{^\mathrm{crossfit}}
\newcommand{\errpi}{\error_n(\hat \pi)}
\newcommand{\errc}{\error_n(\hat c)}

\newcommand{\conv}{^{\dagger}}
\newcommand{\nan}{}
\newcommand{\kdagger}{^{(k, \dagger)}}
\newcommand{\kadagger}{^{(k, a_k, \dagger)}}

\newcommand{\1}{\mathbf 1}
\begin{document}

\maketitle
\blfootnote{\hspace{-0.55cm}Authors are ordered alphabetically. We thank Alberto Abadie, Donald Andrews, 
Andres Aradillas-Lopez, P. M. Aronow, Jushan Bai, Stephen Bates, Stéphane Bonhomme, Yong Cai, Denis Chetverikov, Guilherme Duarte, Bulat Garafov, Isaac Gibbs, Patrik Guggenberger, Kevin Guo, Jinyong Hahn, Marc Henry, Keisuke Hirano, Guido Imbens, Sung Jae Jun, Vishal Kamat, Samir Khan, Yuichi Kitamura, Dean Knox, Sokbae Lee, Zhipeng Liao, Elena Manresa, Rosa Matzkin, Serena Ng, Joris Pinkse, Kirill Ponomarev, Guillaume Pouliot, Dominik Rothenh{\"a}usler, Fredrik S\"{a}vje, Vira Semenova, Azeem Shaikh, Shuyang Shen, Ruoyao Shi, Jann Spiess, Max Tabord-Meehan, Alexander Torgovitsky, Edward Vytlacil, and Martin Wainwright for helpful discussion. The authors used OpenAI's ChatGPT for assistance with proof exploration, exposition, and editing. L.L. is grateful for the support of National Science Foundation grant DMS-2338464. A.S. was partially supported by the Two Sigma Graduate Fellowship Fund,
the Citadel GQS PhD Fellowship, and a Graduate Research Fellowship from the National Science Foundation.}
\begin{abstract}
    Many causal estimands are only partially identifiable since they depend on the unobservable joint distribution between potential outcomes. 
    Stratification on pretreatment covariates can yield sharper bounds; however, unless the covariates are discrete with relatively small support, this approach typically requires binning covariates or estimating the conditional distributions of potential outcomes given the covariates. Binning can result in substantial efficiency loss and become challenging to implement, even with a moderate number of covariates. Estimating conditional distributions, on the other hand, may yield invalid inference if the distributions are inaccurately estimated, such as when a misspecified model is used or when the covariates are high-dimensional.
    In this paper, we propose a unified and model-agnostic inferential approach for a wide class of partially identified estimands. Our method, based on duality theory for optimal transport problems, has four key properties. First, in randomized experiments, our approach can wrap around any estimates of the conditional distributions and provide uniformly valid inference, even if the initial estimates are arbitrarily inaccurate. A simple extension of our method to observational studies is valid under the usual product-rate condition for the treatment modeling and outcome modeling errors, or under a separate conservative-bias condition. Second, if nuisance parameters are estimated at semiparametric rates, our estimator is asymptotically unbiased for the sharp bound. Third, we can apply the multiplier bootstrap to select covariates and models without sacrificing validity even if the true model is not selected. Finally, it is computationally efficient. In particular, our method and theory work for discrete and continuous potential outcomes. Overall, in three empirical applications, our method consistently reduces the width of estimated identified sets and confidence intervals without making additional structural assumptions.

\end{abstract}

\section{Introduction}\label{sec::intro}

\subsection{Motivation and problem statement}\label{subsec::motivation}

Many parameters of interest in econometrics and causal inference are only \textit{partially identifiable} \citep{manski2003partial, tamer2010partial, molinari2020microeconometrics}. Even in randomized experiments, we cannot observe the joint distribution of the potential outcomes $(Y_i(1), Y_i(0))$ since we observe at most one outcome per subject; thus, the distribution of the individual treatment effect $Y_i(1) - Y_i(0)$ is unidentifiable. However, most causal parameters of interest can be \textit{bounded} using the marginal distributions of $Y_i(1)$ and $Y_i(0)$. Furthermore, incorporating information from covariates $X_i \in \R^p$ can substantially reduce the width of the partially identified set. 

However, partial identification bounds involving covariates can depend delicately on the relationship between the outcome and the covariates, making inference challenging. For illustration, we now give three motivating examples, although we will state a general problem formulation in Section \ref{sec::method}. As notation, assume that we observe $n$ i.i.d. observations $\{(X_i, W_i, Y_i)\}_{i=1}^n$ for covariates $X_i \in \mcX$, a binary treatment $W_i \in \{0,1\}$ and an outcome $Y_i \in \mcY$ with potential outcomes $Y_i(1), Y_i(0)$.  This paper focuses on randomized experiments (see Assumption \ref{assump::rand_experiment}) where the marginal distributions of $(Y_i(1), X_i)$ and $(Y_i(0), X_i)$ are identified, though the extension to observational data is rather straightforward (see Section \ref{subsec::aipw} for details). Thus, we say that a parameter is \textit{identified} if it depends only on these marginal distributions.

\begin{example}[\fh\, bounds]\label{ex::fh} For fixed $y_1, y_0 \in \R$, let $\theta = \P(Y_i(1) \le y_1, Y_i(0) \le y_0)$ denote the joint CDF of the potential outcomes. $\theta$ is not identified but can be bounded. Indeed, without covariates, \cite{hoeffding1940, frechet1951} showed that the sharp lower bound on $\theta$ is
\begin{equation}\label{eq::fhnocov}
\theta \ge \theta_L \defeq \max(0, \P(Y_i(1) \le y_1) + \P(Y_i(0) \le y_0) - 1).
\end{equation}
With covariates, applying Eq. (\ref{eq::fhnocov}) conditional on $X_i$ and integrating yields the sharp lower bound:
\begin{equation}\label{eq::fhwithcov}
\theta \ge \theta_L \defeq \E\left[\max(0, \P(Y_i(1) \le y_1 \mid X_i) + \P(Y_i(0) \le y_0 \mid X_i) - 1) \right].
\end{equation}
\end{example}

\begin{example}[Variance of the Individual Treatment Effect]\label{ex::varite} A natural measure of treatment effect heterogeneity is the variance of the individual treatment effect $\theta = \var(Y_i(1) - Y_i(0))$. If $\theta$ is large relative to the average treatment effect (ATE), the treatment may harm many individuals, and it is unclear if it should be given to the general population. 
The sharp lower bound on $\theta$ can be written as
\begin{equation}\label{eq::ex1bnd}
    \theta \ge \theta_L \defeq \var\left(\E[Y_i(1) - Y_i(0) \mid X_i]\right) + \E[\var_{U \sim \Unif(0,1)}(P_{Y(1)\mid X}^{\star\,-1}(U \mid X_i) - P_{Y(0)\mid X}^{\star\,-1}(U \mid X_i))],
\end{equation}
where $P_{Y(k) \mid X}\opt$ denotes the true conditional CDF of $Y_i(k) \mid X_i$ for $k \in \{0,1\}$. 
\end{example}

\begin{example}[ATE with selection bias]\label{ex::lee} Suppose we only observe outcomes for a set of ``selected'' individuals, where selection may depend on treatment status. E.g., we only observe wages for individuals who are employed \citep{lee2009training}, but treatment may affect employment. Formally, let $S_i \in \{0,1\}$ be the indicator for the selection event, with $S_i(1), S_i(0)$ its potential outcomes. A natural estimand is the average treatment effect (ATE) for the individuals who would be selected with or without the treatment: 
\begin{equation}
    \theta \defeq \E[Y_i(1) - Y_i(0) \mid S_i(1) = S_i(0) = 1].
\end{equation}
$\theta$ is only partially identifiable, but as in Example \ref{ex::varite}, if we can learn the relationship between $Y_i, S_i$ and $X_i$, then we can give sharp bounds on $\theta$. In particular, \cite{semenova2021generalized} showed that if one assumes that selection is ``monotone'' in the treatment, meaning $S_i(1) \ge S_i(0)$ a.s., and the conditional law $Y_i(1)\mid S_i(1)=1,X_i$ is continuous, then the sharp lower bound is
\begin{equation}\label{eq::lee_formula}
    \theta \ge \theta_L \defeq
    \E_{X_i\mid S_i(0)=1}
    \left[
        \E\left[
            Y_i(1)
            \mid S_i(1)=1,X_i,
            Y_i(1)\le Q_{\eta(X_i)}(X_i)
        \right]
    \right]
    -\E[Y_i(0)\mid S_i(0)=1].
\end{equation}
where
$\eta(x)\defeq
\frac{\P(S_i(0)=1\mid X_i=x)}
     {\P(S_i(1)=1\mid X_i=x)}$
and $Q_{\alpha}(x)$ denotes the $\alpha$ conditional quantile of
$Y_i(1)\mid S_i(1)=1,X_i=x$. These bounds are colloquially known as ``Lee bounds'' \citep{zhang2003estimation, lee2009training}.
\end{example}

Given a partially identified parameter $\theta$, this paper aims to estimate sharp bounds $[\theta_L, \theta_U]$ which incorporate information from covariates. This problem is challenging because the bounds typically depend delicately on the conditional distribution of $Y \mid X, W$, as exemplified by Equations (\ref{eq::fhwithcov})-(\ref{eq::lee_formula}).  Thus, most existing approaches to estimate $\theta_L, \theta_U$ make assumptions allowing uniformly consistent estimation of such nuisance parameters (see Section \ref{subsec::literature} for a review and Section \ref{sec: refined theory for discrete outcomes} for a technical description). This assumption is often implausible when $X_i$ is continuous or high-dimensional, unless the researcher is willing to impose further assumptions on the conditional distributions (e.g., a parametric model, smoothness, sparsity), which may not hold in applications.

Thus, in this work, we ask the question: can we convert 
a working estimate of the conditional distribution of $Y \mid X, W$ into inferential bounds on the sharp identified set $[\theta_L, \theta_U]$ which are (i) sharp when the working estimate is consistent and (ii) conservative but valid when the working estimate is arbitrarily inaccurate?

We end this section by noting that this question is motivated by the core philosophy of partial identification. Indeed, why not simply make enough assumptions so that the parameter $\theta$ is identified? In his seminal book, \cite{manski2003partial} answers this question by formulating the principle of decreasing credibility:

\begin{displayquote}
The credibility of inference decreases with the strength of the assumptions maintained. 
\end{displayquote}

Our objective is to enhance \textit{credibility} by removing any assumptions about the accuracy of the researcher's working model  of nuisance parameters without sacrificing \textit{power} when the researcher's model matches the ground truth.

\subsection{Overview of results and main contributions}\label{subsec::contrib}

Our work introduces a framework for inference on sharp, covariate-assisted partial identification bounds on causal parameters. Let $P\opt$ denote the true joint distribution of $(Y(1),Y(0),X)$, and suppose that $\{(Y_i(1),Y_i(0),X_i)\}_{i=1}^n$ are i.i.d. draws from $P\opt$. We consider estimands of the form
\begin{equation}\label{eq::estimand_def}
    \theta(P\opt) \defeq \E_{P\opt}\left[f(Y(0), Y(1), X)\right]
\end{equation}
for some known function $f : \mcY^2 \times \mcX \to \R$. Many estimands can be reduced to this case, including Examples \ref{ex::fh}-\ref{ex::lee}, certain conditional expectations, quantiles of treatment effects, and more (see Section \ref{subsec::examples}). We let $\theta_L \le \theta(P\opt) \le \theta_U$ denote the sharp (population) lower and upper partial identification bounds on $\theta(P\opt)$; these quantities are defined formally in Section \ref{subsec::setting}.

Our method outputs estimates $\hat\theta_L, \hat\theta_U$ of the sharp bounds $\theta_L, \theta_U$ as well as lower and upper confidence bounds $\hat\theta\lcb, \hat\theta\ucb$. The main idea is to leverage duality theory for optimal transport problems (reviewed in Section \ref{subsec::setting}) to convert any estimate $\hat P_{Y \mid X, W}$ of the conditional distribution of the outcome into robust partial identification bounds $\hat\theta_L, \hat\theta_U$. We emphasize that this method works automatically for any estimand defined above---in our software, the analyst can specify any function $f$ and does not need to do any additional calculations to obtain the results. This eliminates the need for a closed-form representation of $\theta_L, \theta_U$.

Our main results and contributions are as follows.

\begin{enumerate}[label=\textbf{\arabic*.},leftmargin=*,itemsep=0.5em]
\item \textbf{Uniform validity.} Our method allows analysts to estimate the distribution of $Y \mid X,W$ using any statistical or machine learning technique, e.g., quantile regression, boosting, neural networks, etc. However, in randomized experiments with known propensity scores, the resulting confidence bounds are valid \textit{even if} the estimate $\hat P_{Y \mid X, W}$ is arbitrarily inaccurate relative to the ground truth $P\opt_{Y \mid X, W}$. In this sense, our method is ``model-agnostic'': it can leverage models for power without relying on them for validity.

Formally, $\hat\theta_L$ and $\hat\theta_U$ are always \textit{conservatively} biased in the sense that $\E[\hat\theta_L] \le \theta_L$ and $\E[\hat\theta_U] \ge \theta_U$. Furthermore, the confidence bounds have uniform asymptotic coverage 
without any assumptions on the accuracy of $\hat P_{Y \mid X, W}$ 
(see Theorem \ref{thm::alwaysvalid}). Finally we extend our method to observational studies where the propensity scores are not known (see Theorem \ref{thm::alwaysvalid_aug}).

\item \textbf{Tightness.} Our estimators $\hat\theta_L,\hat\theta_U$ can be asymptotically unbiased and $\sqrt n$-consistent for the sharp bounds $\theta_L,\theta_U$, even though the relevant nuisance parameters are estimated at slower-than-$\sqrt n$ rates. The actual rate requirement depends on the margin condition. We establish the conditions for both continuous and discrete potential outcomes in Section \ref{subsec::tightness}.

For continuous potential outcomes, we leverage algebraic properties of the Kantorovich duals for a broad class of estimands and develop a unified theory of higher-order accuracy, rather than deriving separate bias arguments on an estimand-by-estimand basis as in much of the existing literature \citep[e.g.][]{semenova2020better,fava2024predicting,kaji2023subgrouptreat}. For discrete potential outcomes, we develop a stronger tightness theory based on the local geometry of the conditional dual linear program. This theory substantially relaxes the requirement in \citet{semenova2023aggregated} that nuisance estimators converge uniformly over the entire covariate space. It replaces this uniform requirement with a more realistic average-error condition.

\item \textbf{Easy model selection.} A major question in empirical applications is (i) how to select the subset of the covariates used in the analysis and (ii) how to estimate the outcome model $Y_i \mid X_i, W_i$. Our method permits the analyst to use the multiplier bootstrap \citep{chernozhukov2013multbootstrap} to select the tightest bound based on different models or subsets of the covariates. Bootstrap validity follows from both uniform validity, which ensures that no candidate model is invalid, and the estimator's simple linear structure, as required by multiplier bootstrap theory.

\item \textbf{Computational efficiency.} To compute our bounds, we propose an algorithm that is computationally efficient even when $X_i$ is high-dimensional. For partially identified causal estimands without additional constraints, we can leverage tailored optimal transport solvers, which are generally much faster than generic linear programming solvers. The python package \texttt{dualbounds} implements this algorithm: \url{https://dualbounds.readthedocs.io/en/latest/}.
\end{enumerate}

It is noteworthy that the same construction delivers uniform validity without
requiring an accurate outcome model and tightness when the outcome model is
consistently estimated on average over the covariate space. If only validity is required, one
can simply throw away all covariates and use covariate-independent bounds,
which sacrifice tightness. A common remedy is to stratify on a few discrete
variables or to bin covariates. Unless the covariates have a small joint
support, however, stratification can lose substantial information, while
data-driven binning must balance approximation quality against the number of
observations per bin. Conversely, most provably tight covariate-assisted
procedures rely on consistent estimation of features of the conditional
outcome distributions, making robustness to misspecification difficult
\citep[e.g.][]{semenova2023classification, levis2023covariate}.

\begin{figure}
    \centering
    \includegraphics[width=\linewidth]{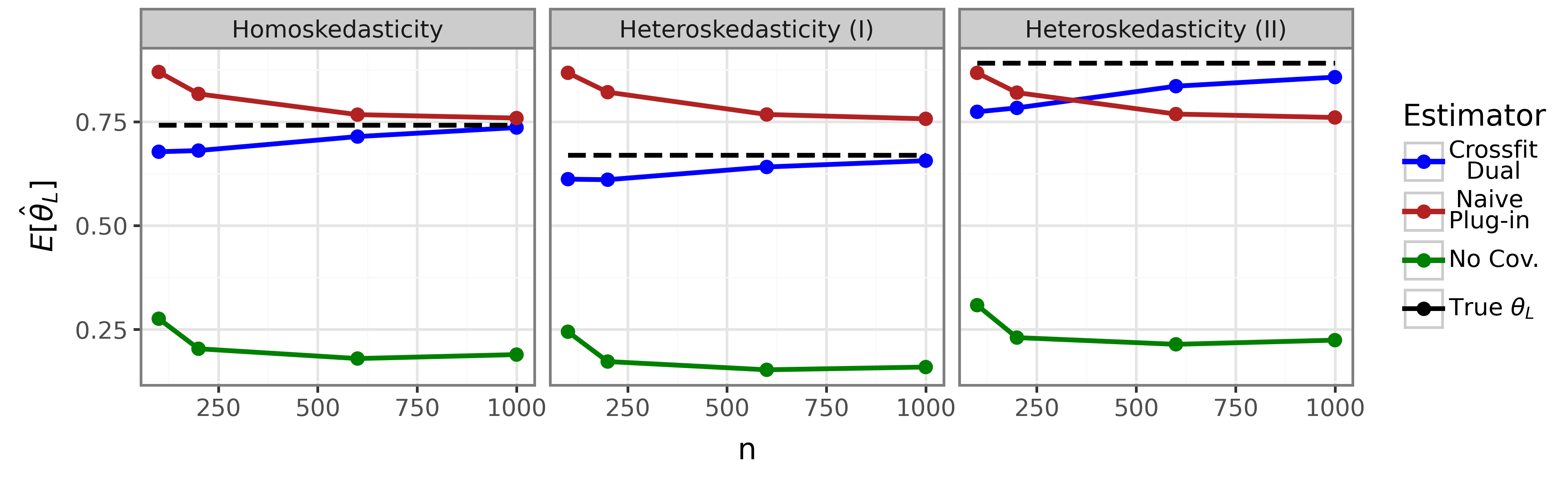}
    \caption{This figure illustrates the robustness gain from dual bounds in a
    simulation with lower Lee bounds (Example~\ref{ex::lee}). It reports the
    average estimate $\widehat\theta_L$; the dotted black line is the true sharp
    lower bound $\theta_L$. The covariate-free estimator is conservative, while
    a covariate-assisted plug-in estimator can be conservative or
    anticonservative under two forms of heteroskedastic misspecification. In
    these designs, the cross-fit dual estimator remains conservative under
    misspecification and becomes sharp as the first stage improves. See
    Section~\ref{sec::sims} for details.}
    \label{fig::contribution}
\end{figure}

To see how the plug-in approach is vulnerable to misspecification or inconsistent nuisance estimation, Figure~\ref{fig::contribution} illustrates the robustness gain in a numerical
experiment with lower Lee bounds. We fit a homoskedastic Gaussian working
model for $Y_i(k)\mid X_i$. A naive estimator that plugs this model into
\eqref{eq::lee_formula} performs well under correct specification but can be
conservatively or anticonservatively biased under heteroskedasticity. The
cross-fit dual estimator uses the same working model and remains conservative
in these designs. Its formal validity is covered by
Theorem~\ref{thm::crossfit_validity} under the conditions stated there. See
Section~\ref{sec::sims} for the simulation design and coverage results.

Although this paper focuses on partially identified causal effects, strong duality provides a general principle for simultaneously achieving robustness to inconsistent nuisance estimation and tightness to sufficiently accurate nuisance estimation for estimands defined by mathematical programs, including infinite-dimensional problems. Inference for estimands characterized by finite-dimensional linear programs or systems of linear moment inequalities is now well developed \citep[e.g.][]{semenova2023classification,semenova2023aggregated,hsieh2022lp,andrews2023,coxshi2023,fang2023lp}. Our framework opens the door to covariate-assisted inference with continuous outcomes and nonlinear mathematical programs.

\subsection{Related literature}\label{subsec::literature}

Partial identification has a long history in econometrics and causal inference, and a great deal of work has been done to characterize and estimate sharp bounds in various settings
\citep[e.g.][]{manski1990treateffects, manski1997monotone, balke1997bounds, heckman1997, manski2002intervaldata, imbensmanski2004, firpo2008, molinari2008partial, beresteanu2008asymptotic, lee2009training, stoye2009more, fan2010, chiburis2010semiparametric, romano2010inference, beresteanu2011sharp,
fan2012confidence, tetenov2012positivetreatment, andrewsshi2013, aronow2014, fan2017partial, firpo2019, kaido2019confidence, kline2021moment,
russell2021sharp, jun2023persuasion, oberreynolds2023estimating,fava2024predicting, byun24a}; see \cite{manski2003partial, tamer2010partial, molinari2020microeconometrics} for a review. When covariates are available, the bounds can be improved by conditioning on the covariates and aggregating covariate-specific sharp bounds \citep{chernozhukov2007econometrica, chandrasekhar2012, chernozhukov2013intersection, semenova2020cate, semenova2021generalized, 
 semenova2023classification,  lee2023partial, levis2023covariate}. 
However, unless the covariates are discrete with a few values, these methods generally either (a) make assumptions that allow the conditional distributions of the potential outcomes to be consistently estimated at semiparametric rates or (b) have to discretize covariates in a non-disciplined way at the cost of efficiency loss. In contrast, our method can handle any type of covariates without making any assumptions that enable consistent estimates of the conditional distributions, while still achieving sharp inference when those estimates are consistent. For Lee bounds, the first arXiv version of \citet{semenova2020better} discusses a result with a similar flavor for covariate selection, which was removed from later versions. In particular, the bounds are sharp when all relevant covariates are selected and valid even if only a subset is selected. Nevertheless, that approach still requires the nuisance parameters to be consistently estimated.

Our key technical tool is the theory of duality in optimization. This tool is of course not new, although we use it in a novel way. In particular, many existing works use duality theory as part of an inference strategy, for example in analysis of certain linear programming problems (e.g., \cite{hsieh2022lp, andrews2023, fang2023lp}) and in sensitivity analysis \citep{dornguo2021, dorn2022doublyvaliddoublysharp}. Most recently, \cite{semenova2023classification} independently developed a dual-based estimator for a class of intersection bounds \citep{chernozhukov2013intersection}. However, unlike our work, they require discrete potential outcomes to obtain a conditional finite-dimensional linear program and consistent estimates of the conditional distributions at semiparametric rates uniformly over the covariate space.

\section{Core Methodology}\label{sec::method}

To aid comprehension, we mostly defer measure-theoretic details to Appendix \ref{sec: comp proof_sec2} and computational details to Section \ref{sec::comp}. For brevity, we focus on the sharp lower bound $\theta_L$, but the same method can be used to estimate upper bounds by simply multiplying $\theta(P\opt)$ by negative one.

\subsection{Assumptions and background on Kantorovich duality}\label{subsec::setting}

We assume the setting of a randomized experiment, although Section \ref{subsec::aipw} relaxes the assumption that the propensity scores are known.

\begin{assumption}\label{assump::rand_experiment} The propensity scores $\pi(X_i) \defeq \P(W_i = 1 \mid X_i)$ are known and bounded away from zero and one, and the potential outcomes $(Y_i(1), Y_i(0))$ are conditionally independent of the treatment $W_i$ given the covariates $X_i$.
\end{assumption}

We also allow the analyst to \textit{optionally} specify additional assumptions about $P\opt$, the joint distribution of $(Y(1), Y(0), X)$, via conditional moment inequalities, as defined below.

\begin{assumption}\label{assumption::condmoment} For each $x \in \mcX$, let $\mcW_x = \{w_{x,1}, \dots, w_{x,L}\}$ denote a finite collection of user-specified functions mapping $\mcY^2 \to \R$ for $L \in \N$.\footnote{Our theory allows $|\mcW_x| = L$ to vary with $x$ but for simplicity our notation suppresses this dependence.} Let $\mcP$ be the following set of distributions:
\begin{equation*}
\mcP=\left\{P\text{ on }\mcY^2\times\mcX:
\E_P[w(Y(0),Y(1))\mid X=x]\le0
\quad\forall w\in\mcW_x,\ x\in\mcX
\right\}.
\end{equation*}
Then we assume $P\opt \in \mcP$.
\end{assumption}
For example, $\mcP$ is the unrestricted set of all joint distributions over $\mcY^2 \times \mcX$ if $\mcW_x$ is empty for each $x \in \mcX$, in which case Assumption \ref{assumption::condmoment} is always satisfied. For Lee bounds (Example \ref{ex::lee}), define, with the convention that $Y(w)S(w)=0$ when $S(w)=0$,
\[
Z(w)=\bigl(Y(w)S(w),S(w)\bigr),
\qquad
\mcZ=\{(0,0)\}\cup\bigl(\R\times\{1\}\bigr),
\]
and apply the framework with $Z(w)$ as the potential outcome and $\mcY=\mcZ$. Monotone selection is represented by the conditional moment function
\[
w_x^{\mathrm{Lee}}(z_0,z_1)=\I\{s_0>s_1\},
\qquad z_w=(r_w,s_w),
\]
included in $\mcW_x$. Because this function is nonnegative, the restriction
$\E[w_x^{\mathrm{Lee}}\{Z(0),Z(1)\}\mid X=x]\le0$ is equivalent to
$S(0)\le S(1)$ almost surely conditional on $X=x$. The first coordinate of an
unselected augmented outcome is fixed at zero, so this formulation does not assign an
unobserved value to $Y(w)$ when $S(w)=0$.

Given $\mcP$, the lower bound $\theta_L$ is the minimum value of $\theta(P)$ for all $P \in \mcP$ which are consistent with the true marginal distributions $P_{Y(1), X}\opt$ and $P_{Y(0),X}\opt$:
\begin{equation}
\label{eq::init_thetaL_def}
    \theta_L = \inf_{P \in \mcP} \E_P[f(Y(0), Y(1), X)] \suchthat P_{Y(1), X} = P_{Y(1), X}\opt \text{ and } P_{Y(0), X} = P_{Y(0), X}\opt.
\end{equation}

Now, we introduce the dual to this optimization problem. We refer to a collection of functions $\nu_{0,x}, \nu_{1,x} : \mcY \to \R$ indexed by $x \in \mcX$ as \textit{dual variables}; we use the notation $\nu = (\nu_{0,x}, \nu_{1,x})_{x \in \mcX}$ to denote the collection of these functions. Given dual variables $\nu$,  the Kantorovich dual function is
\begin{equation}\label{eq:g_nu}
    g(\nu) \defeq \E_{P\opt}[\nu_{0,X}(Y(0)) + \nu_{1,X}(Y(1))].
\end{equation}
Intuitively, $g(\nu)$ is an ``average treatment effect'' of the transformed potential outcomes $Y'(1) \defeq \nu_{1,X}(Y(1))$ and $Y'(0) \defeq - \nu_{0,X}(Y(0))$. Thus, $g(\nu)$ is easy to estimate for any fixed $\nu$.

We aim to use $g(\nu)$ as a lower bound on $\theta_L$. To ensure $g(\nu) \le \theta_L$ holds, we will enforce a collection of known constraints on $\nu$. In the simplest case where $\mcW_x$ is empty, we require that $\nu_{0,x}(y_0) + \nu_{1,x}(y_1) \le f(y_0, y_1, x)$ for all $(y_0,y_1)\in\mcY^2$. In the general case where $\mcW_x = \{w_{x,1}, \dots, w_{x,L}\}$ is nonempty, we can slightly loosen these constraints to take advantage of additional assumptions on $\mcP$. Namely, for $x \in \mcX$, we say that $\nu_{0,x}, \nu_{1,x}$ are \textit{conditionally valid} at $x$ if there are a collection of nonnegative constants $\{\lambda_{x,\ell}\}_{\ell=1}^{L}$ such that the following holds:
\begin{equation}\label{eq::cond_dual_const}
    \nu_{0,x}(y_0) + \nu_{1,x}(y_1) 
    \le f(y_0, y_1, x) + \sum_{\ell=1}^{L} \lambda_{x,\ell} \cdot w_{x, \ell}(y_0, y_1) 
    \,\,\,\,\,\,\,\,\,\,\,\,\,\, \text{ for all } (y_0,y_1)\in\mcY^2
\end{equation}
and we let $\mcV_x \subset \{\mcY \to \R^2\}$ denote the set of all pairs of functions satisfying this condition. Finally, we say that the full set of dual variables $\nu = (\nu_{0,x}, \nu_{1,x})_{x \in \mcX}$ are fully valid or ``dual-feasible'' if $(\nu_{0,x},\nu_{1,x})\in\mcV_x$ for every $x \in \mcX$, and we let $\mcV \subset \{\mcY \times \mcX \to \R^2\}$ denote the set of all valid dual variables.

To see where \eqref{eq::cond_dual_const} comes from, consider the special case
in which, for every \(x\), both \(Y(0)\mid X=x\) and \(Y(1)\mid X=x\)
have finite support \(\{y_1,\ldots,y_m\}\). At a fixed \(x\), any joint
distribution of \((Y(0),Y(1))\mid X=x\) is represented by an \(m\times m\)
coupling matrix \(\Gamma=(\Gamma_{jh})\) with row sums
\(\P\{Y(0)=y_j\mid X=x\}\) and column sums
\(\P\{Y(1)=y_h\mid X=x\}\). The conditional sharp lower bound minimizes
\(\sum_{j,h}\Gamma_{jh}f(y_j,y_h,x)\) over these coupling matrices, subject
also to the restrictions generated by \(\mcW_x\). Its standard Lagrangian
dual is
\[
\begin{aligned}
\max_{\nu_0,\nu_1,\lambda\ge 0}\quad
&
\sum_{j=1}^{m}\P\{Y(0)=y_j\mid X=x\}\nu_{0,x}(y_j)
+
\sum_{h=1}^{m}\P\{Y(1)=y_h\mid X=x\}\nu_{1,x}(y_h)
\\
\mathrm{s.t.}\quad
&
\nu_{0,x}(y_j)+\nu_{1,x}(y_h)
\le
f(y_j,y_h,x)
+
\sum_{a=1}^L
\lambda_a w_{x,a}(y_j,y_h),
\quad j,h\in[m].
\end{aligned}
\]
Thus, \eqref{eq::cond_dual_const} is the standard Lagrangian-dual constraint,
written for general outcome spaces. When \(\mcW_x\) is empty, this is the
usual Kantorovich dual of the discrete optimal transport problem with cost
\(f(y_0,y_1,x)\).

Computational issues aside (see Section \ref{sec::comp}), we emphasize that $\mcV$ is a \textit{known} set which does not depend on $P\opt$. Satisfying this known constraint ensures that weak duality holds, i.e., $g(\nu) \le \theta_L$. The theorem below states this formally; it also states a strong duality result and gives a useful characterization of the optimal dual variables $\nu\opt \in \argmax_{\nu \in \mcV} g(\nu)$. \footnote{We indeed need normalizations of the duals. For continuous potential outcomes, we consider plug-in estimates of the Kantorovich duals normalized in a certain way, as described in Examples~\ref{ex::fh}--\ref{ex:QTE} in Section~\ref{subsec:structured-margins}. For discrete potential outcomes, we require $\nu_{0,1}=0$, as in the definition of $\mathcal V_{f,x}^0$ in Section~\ref{sec: refined theory for discrete outcomes}.} To ease readability, we defer technical regularity conditions regarding measurability and the proof of Theorem \ref{thm::kantorovich} to Appendix \ref{sec: comp proof_sec2}.

\begin{theorem}[Kantorovich duality]\label{thm::kantorovich} Under Assumption \ref{assumption::condmoment}, the following holds:
\begin{enumerate}[leftmargin=*, topsep=0.5pt, itemsep=0.5pt]
    \item \emph{Weak duality}: For any valid dual variables $\nu \in \mcV$, $g(\nu) \le \theta_L$.
    \item \emph{Strong duality}: under mild measurability and regularity conditions on $f$ and $\mcW_x$ stated in Appendix \ref{sec: comp proof_sec2} and \ref{sec: general measurability}, there exist  $\nu\opt = (\nu_{0,x}\opt, \nu_{1,x}\opt)_{x \in \mcX} \in \mcV$ such that $g(\nu\opt) = \theta_L$.\footnote{For Lee bounds, we use the augmented potential outcomes $Z_i(w)=\bigl(Y_i(w)S_i(w),S_i(w)\bigr)$; see Appendix~\ref{sec: comp proof_sec2} for details.} Furthermore, for each $x \in \mcX$, $\nu\opt$ satisfies
    \begin{equation}\label{eq::separability}
        (\nu_{0,x}\opt,\nu_{1,x}\opt) \in \argmax_{(\nu_{0,x},\nu_{1,x}) \in \mcV_x} \E_{P\opt_{Y(0) \mid X = x}}[\nu_{0,x}(Y(0))] + \E_{P\opt_{Y(1) \mid X = x}}[\nu_{1,x}(Y(1))].
    \end{equation}
\end{enumerate}
\end{theorem}

Theorem \ref{thm::kantorovich} has two statistical implications. First, to estimate $\theta_L$, we need only (i) estimate $\nu\opt$ and (ii) estimate $g(\nu\opt)$. Second, $\nu\opt$ is only a functional of $\PCopt$ and does not depend on $P_X\opt$. We will use these insights to estimate $\theta_L$ in the next section.

It is useful to distinguish the two roles played by duality in our framework. Weak
duality is the source of validity. In the unrestricted case, feasibility means
\(\nu_{0,x}(y_0)+\nu_{1,x}(y_1)\le f(y_0,y_1,x)\) for all
\((y_0,y_1,x)\).
Thus, no matter what the unobserved joint distribution of \((Y(0),Y(1))\mid X=x\) is,
the average transformed value \(g(\nu)\) cannot exceed the sharp lower bound \(\theta_L\).
Hence, weak duality allows us to use a first-stage
working model to construct valid bounds without trusting the model.
Strong duality plays a different role. It says that among all feasible duals, there
exists an optimal one whose dual value equals the sharp lower bound. Therefore, strong
duality is needed to ensure our method does not create a constant identification gap by moving from primals to duals. In short, weak duality ensures robust validity (Section~\ref{subsec::validity}),
while strong duality is necessary for tightness (Section~\ref{subsec::tightness}).

\subsection{Inference via dual bounds}\label{subsec::dualbnds}

To motivate our method, recall that the dual function $g(\nu)$ is easy to estimate for any fixed choice of $\nu \in \mcV$ using an inverse probability weighting (IPW) estimator. Thus, if only we knew the value of $\nu\opt$, we could easily estimate $\theta_L = g(\nu\opt)$. The main idea is to use the first split of the data to estimate $\hat \nu \approx \nu\opt$, and the second split of the data to estimate $g(\hat \nu)$. Sample splitting and cross-fitting have also been used for inference on partially identified parameters \citep[e.g.][]{andrewsshi2013,shi2015simple,kedagni2020generalized,semenova2020better,semenova2023setidentifiedlms,semenova2023classification}. Our primary motivation is different: sample splitting makes the second-stage target a fixed dual value, so any first-stage dual yields valid inference, while strong duality governs tightness when the first stage is accurate. Crucially, even if our first-stage estimate $\hat \nu$ is poor, our inference will be conservative but valid, since weak duality ensures $g(\hat \nu) \le \theta_L$. And as we will see in Section \ref{sec::theory}, if $\hat \nu$ is close to $\nu\opt$, then our confidence interval will be tight.

\begin{definition}[Dual lower bounds]\label{def::dualbnds} Given data $\{(Y_i, W_i, X_i)\}_{i=1}^n$, we first randomly split the data into two disjoint subsets $\mcD_1$ and $\mcD_2$. Then we perform the following steps:

\begin{enumerate}[label=\underline{Step \arabic*:},ref=\arabic*,leftmargin=*]
\item \label{step:dual-first-stage}On $\mcD_1$, compute any estimator $\hat \nu \in \mcV$ for $\nu\opt \in \argmax_{\nu \in \mcV} g(\nu)$. There are many reasonable ways to do this, but we suggest the following method:
\begin{enumerate}[label=\textnormal{Step \arabic{enumi}\alph*:},ref=\arabic{enumi}\alph*,topsep=0pt,leftmargin=*]
    \item \label{step:dual-estimate-distributions}Compute an estimate $\hatPZC,\hatPOC$ of the conditional distributions $\PZC\opt,\POC\opt$. To do this, one can use any machine-learning or regression algorithm, such as lasso-based techniques, regularized quantile regression, or distributional regression --- see Section \ref{subsec::hatpobs} for more details. 
    
    \item \label{step:dual-optimize}Let $\hat \nu$ maximize the ``empirical dual'' $\hat g$ which plugs in $\hatPZC,\hatPOC$ for $P\opt$. Formally, we use the characterization from Theorem \ref{thm::kantorovich}. For each $x \in \mcX$, define $\hat\nu_{0,x}, \hat\nu_{1,x} : \mcY \to \R$ as the solution to
    \begin{equation}\label{eq::hatnu_def}
        (\hat\nu_{0,x},\hat\nu_{1,x}) \in \argmax_{(\nu_{0,x},\nu_{1,x}) \in \mcV_x} \E_{\hat{P}_{Y(0)\mid X = x}}[\nu_{0,x}(Y(0))] + \E_{\hat{P}_{Y(1)\mid X = x}}[\nu_{1,x}(Y(1))].\end{equation}
When Eq. (\ref{eq::hatnu_def}) does not have a unique solution, we apply the specific normalization discussed in Sections \ref{subsec:structured-margins} and \ref{sec: refined theory for discrete outcomes}. Computing $\hat \nu$ may seem challenging, but we will discuss simple methods to do this in Section \ref{sec::comp}. For now, we merely note that the final estimator depends only on $\hat\nu_{0,x},\hat\nu_{1,x}$ for $x \in \{X_i : i \in \mcD_2\}$ and thus we do not need to solve Eq. (\ref{eq::hatnu_def}) for all $x \in \mcX$.

\end{enumerate}

\item \label{step:dual-second-stage}Define $\tilde{\theta}_L \defeq g(\hat \nu)$, and note by weak duality that $\tilde{\theta}_L \le \theta_L$ holds deterministically. On $\mcD_2$, we will define a conservative estimator of $\theta_L$ by using an IPW estimator that is unbiased for $\tilde{\theta}_L$. Formally:
\begin{equation}\label{eq::hattheta_def}
    \hat\theta_L=\frac{1}{|\mcD_2|}\sum_{i\in\mcD_2}\left\{\frac{W_i\hat\nu_{1,X_i}(Y_i)}{\pi(X_i)}+\frac{(1-W_i)\hat\nu_{0,X_i}(Y_i)}{1-\pi(X_i)}\right\}.
\end{equation}
Conditional on $\mcD_1$, $\hat \theta_L$ is a sample mean of i.i.d. terms, and $\hat \theta_L$ is conservatively biased for $\theta_L$. Thus, we can compute a lower confidence bound on $\theta_L$ via the univariate central limit theorem. In particular, let $\hat \sigma_{S}$ denote the sample standard deviation of the summands $\left\{\frac{\hat \nu_{1,X_i}(Y_i) W_i}{\pi(X_i)} + \frac{\hat \nu_{0,X_i}(Y_i) (1-W_i)}{1 - \pi(X_i)}\right\}_{i \in \mcD_2}$. Then a $1-\alpha$ lower confidence bound (LCB) for $\theta_L$ is
\begin{equation}\label{eq::hatthetalcb_def}
    \hat \theta\lcb = \hat \theta_L - \Phi^{-1}(1-\alpha) \frac{\hat \sigma_{S}}{\sqrt{|\mcD_2|}}
\end{equation}
where $\Phi$ is the standard Gaussian CDF. 
\end{enumerate}
\end{definition}

\begin{remark}
The IPW estimator in \eqref{eq::hattheta_def} can be replaced by an augmented IPW estimator. Given the
first-fold dual variables \(\widehat\nu\), define
\[
c_k(x)=\mathbb E\{\widehat\nu_{k,x}(Y(k))\mid X=x,\mathcal D_1\},\qquad k=0,1.
\]
Then one may replace the IPW summand in \eqref{eq::hattheta_def} by
\[
\frac{W_i}{\pi(X_i)}
\{\widehat\nu_{1,X_i}(Y_i)-\widehat c_1(X_i)\}
+
\frac{1-W_i}{1-\pi(X_i)}
\{\widehat\nu_{0,X_i}(Y_i)-\widehat c_0(X_i)\}
+
\widehat c_1(X_i)+\widehat c_0(X_i).
\]
When \(\pi\) is known, this summand remains conditionally unbiased for
\(g(\widehat\nu)\) for any choice of \(\widehat c_0,\widehat c_1\), so the validity
argument in Theorem~\ref{thm::alwaysvalid} is unchanged. If \(\widehat c_k\) consistently estimates
\(c_k\), the augmented estimator can reduce the variance and improve estimation efficiency. We use the non-augmented IPW form in the main algorithm for simplicity. Later, for observational studies with unknown propensity scores, the augmentation removes first-order sensitivity of raw IPW and leaves a remainder proportional to the product of the propensity and regression errors. Root-$n$ validity at a sharp pseudo-target therefore requires the product-rate condition in Theorem \ref{thm::alwaysvalid_aug}, unless one nuisance is exactly correct.
\end{remark}

\begin{remark}[Quasi-linear targets]
\label{rmk:quasilinear}
The procedure above directly applies to targets that are linear in the joint distribution,
\(\theta(P)=\E_P[f(Y(0),Y(1),X)]\). Some targets are not themselves linear, but can be
reduced to a family of linear problems through one-dimensional inversion. Suppose that
there exists a known family of functions \(\{f_c:c\in\mathbb R\}\) such that
\[
    \theta(P)\le c
    \quad\Longleftrightarrow\quad
    \E_P[f_c(Y(0),Y(1),X)]\le 0.
\]
Define the sharp lower bound for the linearized problem
\[
    \phi_L(c)
    =
    \inf_{P\in\mcP}
    \E_P[f_c(Y(0),Y(1),X)]
\]
subject to the same marginal constraints as in \eqref{eq::init_thetaL_def}.
Assume that the above infimum is attained. Then
$
    \theta_L\le c
    \Longleftrightarrow
    \phi_L(c)\le 0.
$
For each fixed \(c\), \(\phi_L(c)\) is exactly a lower bound for a linear functional, so
Definition~\ref{def::dualbnds} can be applied without modification with \(f_c\)
in place of \(f\). Let \(\widehat\phi_{\mathrm{LCB}}(c)\) denote the resulting lower
confidence bound for \(\phi_L(c)\). A lower confidence bound for \(\theta_L\) is then
obtained by inversion:
\[
    \widehat\theta_{\mathrm{LCB}}^{\mathrm{ql}}
    =
    \inf\{c:\widehat\phi_{\mathrm{LCB}}(c)\le 0\}.
\]
We can solve this inversion by grid search. On a fixed bounded search interval,
attaining grid resolution \(\epsilon\) requires \(O(1/\epsilon)\) evaluations
of the dual bounds problem.
\end{remark}
We will see in Section \ref{sec::theory} that this procedure is uniformly valid (in randomized experiments) and that it provides an asymptotically exact and sharp lower confidence bound if we can estimate $\hatPC$ consistently at certain rate. The main drawbacks of this procedure are that it requires splitting the data and that Eq. (\ref{eq::hattheta_def}) assumes the propensity scores are known. In Section \ref{sec::theory}, we overcome these drawbacks by employing cross-fitting and by plugging in estimates $\hat \pi$ of the propensity scores in observational data. 
Before presenting these additional results, however, we first give a few guidelines and examples of how to apply this procedure.

\subsection{Model selection via the multiplier bootstrap}\label{subsec::multbootstrap}

To compute dual bounds, one must estimate the dual variables $\nu\opt$---in practice, we recommend first estimating $\hatPC$ and then computing $\hat \nu$ as per Eq. (\ref{eq::hatnu_def}). However, there are many ways to estimate $\hatPC$. E.g., analysts may prefer to use only a subset of the covariates to predict $Y$, but even after observing $\mcD_1$, it is not clear which subset of the covariates to choose. And even after making this decision, as discussed in Section \ref{subsec::hatpobs}, there are still countless existing methods to estimate $\hatPC$. This raises the question: in practice, how should analysts choose between $K$ candidate estimates $\hat \nu^{(1)}, \dots, \hat \nu^{(K)} \in \mcV$ of the dual variables $\nu\opt$? Or more colloquially, how should we perform model selection?

One solution is to perform cross-validation within the first fold ($\mcD_1$) and pick the best-performing model that gives the highest empirical dual value, i.e., the tightest estimated lower bound. This approach is clearly valid since the final estimated dual variables $\hat \nu$ still depend only on $\mcD_1$, satisfying Definition \ref{def::dualbnds}. In Section \ref{subsec::aipw}, we recommend this approach for observational studies, where the validity of the final bounds may depend on the accuracy of the outcome model. However, in randomized experiments, we can improve upon this method.

In particular, let $\tilde{\theta}_L^{(k)} = g(\hat \nu^{(k)})$ denote the dual lower bound on $\theta_L$ implied by the estimate $\hat \nu^{(k)}$, for $k=1, \dots, K$. We will estimate $\max_{k\in [K]} \tilde{\theta}_L^{(k)}$, the tightest possible lower bound on $\theta_L$ based on $\{\hat \nu^{(k)}\}_{k=1}^K$, using the Gaussian multiplier bootstrap \citep{chernozhukov2013multbootstrap}, as defined below.

\begin{definition}[Dual bounds with the multiplier-bootstrap]\label{def::multidualbounds} Given dual variables $\hat \nu^{(1)}, \dots, \hat \nu^{(K)} \in \mcV$, for $i \in \mcD_2$, define the IPW summands as:
\begin{equation}
    S_i^{(k)} \defeq \frac{\hat \nu_{1,X_i}^{(k)}(Y_i) W_i}{\pi(X_i)} + \frac{\hat \nu_{0,X_i}^{(k)}(Y_i) (1-W_i)}{1 - \pi(X_i)} \text{ for } k \in [K]. \end{equation}
Define $\hat\theta_L^{(k)} = \frac{1}{|\mcD_2|} \sum_{i\in \mcD_2} S_i^{(k)}$ and $\hat \sigma_k^2 = \frac{1}{|\mcD_2|} \sum_{i \in \mcD_2} (S_i^{(k)} - \hat\theta_L^{(k)})^2$ to be the dual estimators and associated sample variances for each $k \in [K]$. The main idea is to use $T \defeq \max_{k\in [K]} \frac{\sqrt{|\mathcal D_2|} \hat\theta_L^{(k)}}{\hat \sigma_k}$ as a test statistic and compute its quantile using the Gaussian multiplier bootstrap. Precisely:
\begin{enumerate}[noitemsep, topsep=0pt]
    \item Sample $\epsilon_i \iid \mcN(0,1)$ for each $i \in \mcD_2$.
    \item Let $T^{(b)} = \max_{k\in [K]} \hat \sigma_k^{-1} \left[\frac{1}{\sqrt{|\mcD_2|}} \sum_{i \in \mcD_2} \epsilon_i (S_i^{(k)} - \hat\theta_L^{(k)}) \right]$ be the bootstrapped test statistic.
    \item Let $\hat q_{1-\alpha} \defeq Q_{1-\alpha}(T^{(b)} \mid \mcD)$ be the $1-\alpha$ quantile of $T^{(b)}$ conditional on the data. This can be computed by simulating many bootstrap samples.
\end{enumerate}
Then, return the following multiplier bootstrap (MB) lower confidence bound:
\begin{equation}\label{eq::lcb_mbs}
    \hat\theta\lcb\mb \defeq \max_{k \in [K]} \left\{\hat\theta_L^{(k)} - \hat q_{1-\alpha} \frac{\hat \sigma_k}{\sqrt{|\mcD_2|}}\right\}.
\end{equation}
\end{definition}

The multiplier bootstrap is well-suited to this problem for two reasons. First, our bounds are valid no matter which model we select, i.e., $\tilde{\theta}^{(k)}_L \le \theta_L$ always holds. This may not be true in other problems---for example, when estimating regression coefficients, selecting different subsets of covariates may lead to anticonservative bias, but in our setting, any bias from misspecification is conservative. Second, after estimating $\{\hat \nu^{(k)}\}_{k=1}^K$, the dual bounds $\{\tilde{\theta}_L^{(k)}\}_{k=1}^K$ can be expressed as marginal moments and estimating them does not require (e.g.) any complicated M-estimation. As a result, in Section \ref{subsec::validity}, we conclude that the multiplier bootstrap quantile $\hat q_{1-\alpha}$ is consistent even if $K$ grows exponentially with a power of $n$ \citep{chernozhukov2018multibootstrap}.

\begin{remark} In some of our empirical applications (Section \ref{sec::applications}), the estimands can only be expressed as the \textit{ratio} of two marginal moments. We can extend the multiplier bootstrap methodology to that setting under the restriction that $K$ cannot grow with $n$. For brevity, we present this extension in Appendix \ref{appendix::delta_method}.
\end{remark}

\subsection{Guidelines on estimating the conditional distributions $\hatPC$}\label{subsec::hatpobs}

The first step in computing a dual bound $\hat\theta\lcb$ is to estimate $\hatPC$, or equivalently, to estimate the conditional distribution of $Y_i \mid X_i, W_i$. An immense literature exists on this modeling problem \citep[e.g.][]{koenker1978regression, chernozhukov2010quantile, chernozhukov2013inference, friedman2020contrast}, and any choice will yield valid inferences. However, we make a few recommendations here.

To start, note that it is usually insufficient to model the \textit{conditional mean} $\E_{P\opt}[Y_i \mid X_i, W_i]$, since the sharp lower bound $\theta_L$ may depend on the whole conditional distribution (e.g., Examples \ref{ex::fh}--\ref{ex::lee}). Instead, we can apply \textit{distributional regression}, which is devoted to estimating the distribution of $Y_i \mid X_i, W_i$ (see \cite{kneib2023distreg} for a review). One way to do this is to fit many quantile regressions. Another simple method is to assume a Gaussian linear model, i.e., 
\begin{equation}
    Y_i = \phi(X_i, W_i)^T \beta + \epsilon_i
\end{equation}
where $\phi(X_i, W_i) \in \R^d$ is some feature transformation of $X_i, W_i$ and $\epsilon_i \iid \mcN(0, \sigma^2)$. To fit this model, one can (i) adaptively fit the feature representation $\phi$ using the first fold $\mcD_1$, (ii) fit a regularized estimate $\hat \beta$ of $\beta$ using (e.g.) a cross-validated lasso on $\mcD_1$, and (iii) estimate $\sigma^2$ using the usual OLS estimator of the residual variance. 
Of course, the Gaussian assumption may not always be realistic. Instead, our default implementation in \texttt{dualbounds} fits the same coefficients $\hat \beta$ and uses the empirical residuals $\hat \epsilon_i \defeq Y_i - \phi(X_i, W_i)^T \hat \beta$ to nonparametrically estimate the distribution of $\epsilon_i$. Similarly, in the presence of heteroskedasticity, we can estimate $\var(Y_i \mid X_i, W_i)$ using a nonparametric estimator like a random forest; clearly, the possibilities are endless. The main point is that misspecification of these models will not affect the validity of $\hat\theta\lcb$, although better models will yield tighter estimates and confidence intervals.  

For the examples in the next subsection, we can often simplify this first-stage step further by exploiting the explicit functional forms of the conditional duals derived in Appendix~\ref{sec: dual form}. Rather than estimating the full conditional distributions of \(Y(0)\mid X\) and \(Y(1)\mid X\), these formulas show that it can be enough to estimate selected features of those distributions, such as conditional probabilities, CDFs, quantiles, or other problem-specific summaries. Section~\ref{sec::comp} discusses these computational simplifications in more detail.

\subsection{Examples}\label{subsec::examples}

In this section, we give a few examples of estimands that fit into the framework from Section \ref{subsec::dualbnds}. We start by revisiting Examples \ref{ex::fh}--\ref{ex::lee} in Section \ref{subsec::motivation} and then discuss four other examples.

\begin{example}[name={\fh\, bounds},continues=ex::fh] The joint CDF of the potential outcomes evaluated at a fixed point $(y_1, y_0) \in \mcY^2$ is clearly an expectation under $P^\star$, i.e., $\theta(P^\star) \defeq \E[\I(Y_i(1) \le y_1, Y_i(0) \le y_0)]$.
\end{example}

\begin{example}[name={Variance of ITE},continues=ex::varite] If $\theta(P^\star) = \var(Y_i(1) - Y_i(0))$, we can write
\begin{equation}\label{eq:varite-decomposition}
    \theta(P^\star) = \E[(Y_i(1) - Y_i(0))^2] - \left(\E[Y_i(1) - Y_i(0)]\right)^2.
\end{equation}
Note that the first term on the right-hand side is an expectation under $P^\star$, and the second term is identifiable: it is just the ATE squared. Thus, we can apply our  methodology to the first term, and we can estimate the second term by squaring an (e.g.) IPW estimator of the ATE. The only adjustment from Definition \ref{def::dualbnds} is that we use the bivariate delta method to compute standard errors (see Appendix \ref{appendix::delta_method} for a full derivation).
\end{example}

We now return to the case of Lee bounds (Example \ref{ex::lee}) from Section \ref{subsec::motivation}. 

\begin{example}[name={Lee bounds},continues=ex::lee] Suppose $\theta(P^\star) \defeq \E[Y_i(1) - Y_i(0) \mid S_i(1) = S_i(0) = 1]$ is the ATE for the ``always takers,'' i.e., the subset of individuals who would be selected under treatment or control. Formally, use the observable augmented potential outcomes $Z_i(w)=\bigl(Y_i(w)S_i(w),S_i(w)\bigr)$, with $Y_i(w)S_i(w)$ defined as zero when $S_i(w)=0$; this avoids assigning an outcome value when $S_i(w)=0$. For the Lee construction on $\mcD_1$, it is enough to estimate the selection probability $\P(S_i=1\mid X_i,W_i)$ and the selected-outcome law $Y_i\mid S_i=1,X_i,W_i$. No model for $Y_i\mid S_i=0,X_i,W_i$ is required; in that state the outcome may be unobserved or undefined. More specifically, Appendix \ref{sec:lee} shows that the Lee cutoff and duals require only $q_0(x)=\P\{S_i(0)=1\mid X_i=x\}$ and the selected subdensities.

Although $\theta(P^\star)$ is not an expectation under $P^\star$, it can be reduced to this case. In particular, note
\begin{equation}\label{eq:lee-ratio-decomposition}
    \theta(P^\star) = \frac{\E[\{Y_i(1)S_i(1)-Y_i(0)S_i(0)\}S_i(0)S_i(1)]}{\E[S_i(0)S_i(1)]}.
\end{equation}
We assume the denominator
\(\E_P[S_i(0)S_i(1)]>0\) for all \(P\) under consideration.
To analyze this, there are two cases. First, analysts often make assumptions (e.g., monotonicity) which ensure that the denominator is identifiable \citep{lee2009training, semenova2021generalized}. In this case, we can first apply the standard dual bound methodology to the numerator, which is linear in $P^\star$. Then, on the second fold $\mcD_2$, we also estimate the (identifiable) denominator. Finally, we combine estimates for the numerator and denominator using the bivariate delta method, as in Example \ref{ex::varite} (see Appendix \ref{appendix::delta_method} for an explicit calculation).

Second, even when the denominator is unidentifiable, $\theta(P^\star)$ is still quasilinear in $P^\star$ because $\theta(P^\star) \le c$ if and only if
\begin{equation*}
    \theta^{(c)}(P^\star) \defeq \E[\{Y_i(1)S_i(1)-Y_i(0)S_i(0)\}S_i(0)S_i(1)] - c\E[S_i(0)S_i(1)] \le 0.
\end{equation*}
Therefore we can apply the strategy discussed in Remark \ref{rmk:quasilinear} to compute valid dual bounds. 
\end{example}

\begin{example}[Makarov bounds]\label{ex::makarov}  Define $\theta(P^\star) \defeq \E[\I(Y_i(1) - Y_i(0) \le t)]$ to be the CDF of the ITE at a fixed point $t \in \R$. Then $\theta(P^\star)$ is clearly an expectation under $P^\star$.
\end{example}

\begin{example}[Positive treatment effect]
\label{ex::positive_ite}
For \(u\in\mathbb R\), let \((u)_+\defeq\max\{u,0\}\).
Define
\[
    \theta(P^\star) = \E[(Y_i(1)-Y_i(0))_+].
\]
This is clearly an expectation under \(P^\star\), with
\(f(y_0,y_1,x)=(y_1-y_0)_+\).
\end{example}

The ideas in Example \ref{ex::lee} apply to any quasilinear function of $P^\star$. Two examples are given below.

\begin{example}[Baseline-disadvantaged subgroup treatment effect]
\label{ex:CTE}
Following \citet{kaji2023subgrouptreat}, consider the baseline-disadvantaged subgroup treatment effect
\[
    \theta(P^\star) = \E[Y_i(1) - Y_i(0) \mid Y_i(0) \le c],
\]
where $c \in \R$ is a constant and $P^\star\{Y_i(0) \le c\}>0$. This estimand is quasilinear in $P^\star$, so we can compute valid dual bounds as in Remark \ref{rmk:quasilinear}. When $Y_i(1)$ and $Y_i(0)$ measure income, it is the treatment effect for disadvantaged individuals whose income would be below $c$ without treatment.
\end{example}

\begin{example}[Quantile of ITE]\label{ex:QTE} Suppose $\theta(P^\star) = Q_{\alpha}(Y_i(1) - Y_i(0))$, where $Q_{\alpha}(\cdot)$ denotes the $\alpha$-quantile function. Then $\theta(P^\star)$ is quasilinear in $P^\star$ \citep{boyd2004}, and we can again compute valid dual bounds as in Remark \ref{rmk:quasilinear}.
Specifically, fix \(\alpha\in(0,1)\). For every
\(c\in\mathbb R\), define
\[
f_c(y_0,y_1)=\alpha-\I\{y_1-y_0\le c\}.
\]
Then
\[
Q_\alpha\{Y(1)-Y(0)\}\le c
\quad\Longleftrightarrow\quad
\E_P[f_c\{Y(0),Y(1)\}]\le0.
\]
Let
\[
\theta_{L,c}
=\inf_{P\in\mathcal P}\E_P[f_c\{Y(0),Y(1)\}]
\]
and let \(\widetilde\theta_{L,c}\) denote its plug-in dual lower bound.
Inverting these lower bounds gives
\[
  q_L=\inf\{c\in\mathbb R:\theta_{L,c}\le0\},
  \qquad
  \widetilde q_L
  =\inf\{c\in\mathbb R:\widetilde\theta_{L,c}\le0\}.
\]
\end{example}
\section{Theory}\label{sec::theory}

\subsection{Uniform validity}\label{subsec::validity}

For expositional convenience, we assume $|\mcD_2| \ge cn$ for some constant $c > 0$ throughout the section. Throughout, $\E[\cdot \mid \mcD_1]$ denotes an expectation conditional on the first fold of data. All proofs will be presented in Appendix \ref{appendix::proofs}. 

Our first main theoretical result is that in randomized experiments, $\hat\theta\lcb$ is a valid $1-\alpha$ lower confidence bound on $\theta_L$ under arbitrary model misspecification. Note that the following result allows for the analyst to use any method to estimate the optimal dual variables $\hat \nu$ as long as $\hat \nu \in \mcV$ are dual-feasible.
It also places no restrictions on the relationship between the potential outcomes and $X_i$, although we do require the following moment condition on $\hat \nu$.

\begin{assumption}\label{assumption::nu_moment} For $k \in \{0,1\}$, we assume the fourth moment $\E_P[\hat \nu_{k,X}(Y(k))^4 \mid \mcD_1] \le B < \infty$ is bounded  conditional on $\mcD_1$ and the conditional variance of  $S_i = \frac{\hat \nu_{1,X_i}(Y_i) W_i}{\pi(X_i)} + \frac{\hat \nu_{0,X_i}(Y_i) (1-W_i)}{1-\pi(X_i)}$ is bounded away from zero, i.e., $\var_P(S_i \mid \mcD_1) \ge \frac{1}{B}$.
\end{assumption}

Assumption \ref{assumption::nu_moment} can be substantially relaxed at the cost of a more technical statement (see Appendix \ref{appendix::validity_proofs}, Remark \ref{remark::weaken_assump_nu_moment}). All we need is for the moments of $S_i$ to be sufficiently regular such that we can apply a univariate central limit theorem (CLT) to $\{S_i\}_{i \in \mcD_2}$ conditional on $\mcD_1$.

\begin{theorem}\label{thm::alwaysvalid} 
Assume Assumption \ref{assump::rand_experiment}. For any $B > 0$, let $\mcP_B \subset \mcP$ denote the set of all distributions $P \in \mcP$ such that $\hat\nu$ satisfies Assumption \ref{assumption::nu_moment} under $P$. Then
\begin{equation*}
    \liminf_{n \to \infty} \inf_{P \in \mcP_B} \P(\hat\theta\lcb \le \theta_L) \ge 1 - \alpha.
\end{equation*}
\begin{proofsketch} Let $\tilde{\theta}_L = g(\hat \nu)$ denote the effective estimand, as in Definition \ref{def::dualbnds}. Then
\begin{equation*}
    \theta_L - \hat \theta\lcb = \underbrace{\theta_L - \tilde{\theta}_L}_{\text{Term A}} + \underbrace{\tilde{\theta}_L - \hat\theta\lcb}_{\text{Term B}}.
\end{equation*}
Term A is positive deterministically by weak duality. Term B is positive with probability equal to $1-\alpha$ asymptotically by the standard CLT.
\end{proofsketch}
\end{theorem}

Of course, by multiplying $\theta(P\opt)$ by negative one, these theorems prove that we can get a $1-\alpha$ upper confidence bound $\hat \theta_{\mathrm{UCB}}$ on the sharp upper bound $\theta_U$. These bounds can be combined to cover either the partially identified set or the parameter $\theta(P\opt)$ \citep{imbensmanski2004, stoye2009more} (see Section \ref{subsec::twosided}).

Theorem \ref{thm::alwaysvalid} has two key ingredients---(i) weak duality plus (ii) the fact that $\tilde{\theta}_L$ has a representation as a marginal moment, which allows us to apply the CLT. As discussed in Section \ref{subsec::multbootstrap}, these properties also allow us to use the multiplier bootstrap to select a ``good'' choice of $\hat \nu$. In particular, the multiplier bootstrap is asymptotically valid as long as the central moments of the IPW summands $S_i^{(k)}$ do not grow too quickly with $n$ and $K$, as stated formally below. 

\begin{assumption}[\cite{chernozhukov2018multibootstrap}]\label{assump::mbs} For $K$ estimates $\hat \nu^{(1)}, \dots, \hat \nu^{(K)}$ of $\nu\opt$, for $i \in \mcD_2$, define the IPW summands \begin{equation*}\label{eq::tilde_si_eq}
    S_{i}^{(k)} \defeq  \frac{\hat \nu_{1,X_i}^{(k)}(Y_i) W_i}{\pi(X_i)} + \frac{\hat \nu_{0,X_i}^{(k)}(Y_i) (1-W_i)}{1 - \pi(X_i)} \text{ and } Z_{ik} = \frac{S_i^{(k)} - \E[S_i^{(k)} \mid \mcD_1]}{\sqrt{\var(S_i^{(k)}\mid\mcD_1)}}.
\end{equation*}
We assume that, almost surely, the conditional variances are uniformly nondegenerate, i.e., $\min_{k\in[K]}\var(S_i^{(k)}\mid\mcD_1)\ge 1/B$ for some fixed $B<\infty$, and that there exist $\epsilon \in (0,1/4), c > 0$ such that
\begin{equation*}\label{eq::Bn_def}
    B_n \defeq
    \left\{
    \max_{k \in [K]}
    \left(
    \E[|Z_{ik}|^4 \mid \mcD_1]^{1/2}
    \vee
    \E[|Z_{ik}|^3 \mid \mcD_1]
    \right)
    \right\}
    \vee
    \E\left[\max_{k \in [K]} |Z_{ik}|^4 \mid \mcD_1 \right]^{1/4}
    \le c \frac{n^{1/4 - \epsilon}}{\log(K n)^{7/4}}.
\end{equation*}
\end{assumption}

This assumption is standard in the literature. If, in addition, the standardized
third and fourth moments and the maximal fourth-moment envelope entering
\(B_n\) are uniformly bounded, then it is satisfied when
$\log K = O(n^{1/7 - \epsilon})$ for some $\epsilon > 0$. Thus, under this
uniform moment-envelope condition, we can select from many different models
without sacrificing validity.

\begin{corollary}\label{cor::multiboot_validity} Suppose the analyst computes $K$ estimates $\hat \nu^{(1)}, \dots, \hat \nu^{(K)}$ of $\nu\opt$ on $\mcD_1$ and uses the multiplier bootstrap to compute a lower bound $\hat\theta\lcb\mb$ as defined in Def. \ref{def::multidualbounds}. Fix $c > 0, \epsilon \in(0,1/4)$ and let $\mcP_{c,\epsilon}$ denote the set of distributions $P \in \mcP$ such that Assumption \ref{assump::mbs} holds. Then under Assumption \ref{assump::rand_experiment},
\begin{equation*}
    \liminf_{n \to \infty} \inf_{P \in \mcP_{c,\epsilon}} \P(\hat\theta\lcb\mb \le \theta_L) \ge 1 - \alpha.
\end{equation*}
\end{corollary}
\begin{remark}
\label{rmk:pseudo_target}
Importantly, the uniform validity results in Theorem \ref{thm::alwaysvalid}
and Corollary \ref{cor::multiboot_validity} allow for conservative inference,
and therefore do not contradict the conclusion that many intersection bounds do
not admit regular estimators. For example,
consider the lower bound on \(P\{S(1)=S(0)=1\}\) in the
no-covariate case, where
\[
\theta_L=\max\{p_0+p_1-1,0\},
\qquad
p_w=P(S=1\mid W=w),\quad w=0,1.
\]
This functional is not pathwise differentiable when \(p_0+p_1=1\).
Consequently, \citet{hirano2012impossibility} show that there is no regular
estimator for \(\theta_L\). Nevertheless, uniformly valid conservative
inference is possible if we shift the estimand based on our nuisance parameter
estimates. In particular, on the second fold, the dual bounds method targets
\[
\widetilde\theta_L=g(\widehat\nu),
\]
which is guaranteed to satisfy \(\widetilde\theta_L\le\theta_L\) by weak duality.
Therefore, margin-type assumptions such as those used by
\citet{semenova2023classification} and \citet{levis2023covariate} are
unnecessary for uniform validity in our setting, despite the nonregularity
highlighted by \citet{hirano2012impossibility}; however, similar conditions
are needed for tightness, as we will see in Section~\ref{subsec::tightness}.
\end{remark}

\subsection{Tightness}\label{subsec::tightness}

Each of the previous results relies on the weak duality result that $\tilde{\theta}_L \le \theta_L$ holds deterministically. Although this ensures that $\hat\theta\lcb$ and $\hat\theta\lcb\mb$ are valid lower confidence bounds, one might worry that it will make inference too conservative. We investigate this question in this subsection.

\subsubsection{General analysis}

We now give high-level conditions under which $\hat\theta\lcb$ converges to $\theta_L$ at oracle rates. The main intuition follows from the decomposition
\begin{equation*}
    \theta_L - \hat \theta\lcb = \underbrace{\theta_L - \tilde{\theta}_L}_{\text{first-stage bias}} + \underbrace{\tilde{\theta}_L - \hat\theta\lcb}_{\text{variance from the CLT}}.
\end{equation*}
The univariate CLT suggests that the second term is asymptotically exact. Thus, the main question is how large the first-stage bias is. 

The following theorem tells us that the first stage bias is bounded by the product of the errors in estimating $(\PZC\opt,\POC\opt)$ and $\nu\opt$. Thus, if the product of the errors decays at an $o(n^{-1/2})$ rate, the first stage bias will be negligible compared to the variance from the univariate CLT. As notation, let $p_0\opt(y_0 \mid x),  p_1\opt(y_1 \mid x)$ denote the conditional densities of $Y(0) \mid X$ and $Y(1) \mid X$ with respect to some base measure $\psi$ on $\mcY$ \footnote{We choose $\psi$ to be the Lebesgue measure for continuous outcomes and the counting measure for discrete outcomes.}; similarly, let $\hat p_1(y_1 \mid x), \hat p_0(y_0 \mid x)$ denote the estimated densities under $\hatPC$.

Fix any \(r,s\in[1,\infty]\) satisfying \(1/r+1/s=1\). For each
\(x\in\mcX\), define
\begin{equation}
\error_{P,r}(x) \defeq
\begin{cases}
\left(\displaystyle\sum_{k\in\{0,1\}}\int_{y\in\mcY}
\left|p_k\opt(y\mid x)-\hat p_k(y\mid x)\right|^r\psi(dy)\right)^{1/r},
& r<\infty,\\[6pt]
\displaystyle\max_{k\in\{0,1\}}\operatorname*{ess\,sup}_{y\in\mcY}
\left|p_k\opt(y\mid x)-\hat p_k(y\mid x)\right|,
& r=\infty.
\end{cases}
\end{equation}
Similarly, define
\begin{equation}
\error_{\nu,s}(x) \defeq
\begin{cases}
\left(\displaystyle\sum_{k\in\{0,1\}}\int_{y\in\mcY}
\left|\hat\nu_{k,x}(y)-\nu_{k,x}\opt(y)\right|^s\psi(dy)\right)^{1/s},
& s<\infty,\\[6pt]
\displaystyle\max_{k\in\{0,1\}}\operatorname*{ess\,sup}_{y\in\mcY}
\left|\hat\nu_{k,x}(y)-\nu_{k,x}\opt(y)\right|,
& s=\infty.
\end{cases}
\end{equation}
The \(L^2\) specialization used below is written explicitly as
\(\error_{P,2}(x)\) and \(\error_{\nu,2}(x)\).

\begin{theorem}\label{thm::inprodbound} Suppose strong duality holds, i.e., $g(\nu\opt) = \theta_L$. Let $\hat \nu$ be the empirical dual optimizer and $\tilde \theta_L =  g(\hat \nu)$. For any \(r,s\in[1,\infty]\) satisfying \(1/r+1/s=1\), the first-stage bias is bounded by the product of the \(L^r\) error in estimating the distributions of $Y(k) \mid X$, \(k \in \{0,1\}\), and the \(L^s\) error in estimating $\nu\opt$. Formally,
\begin{align}\label{eq::inprodbound}
        0 \le \theta_L - \tilde{\theta}_L
    &\le 
        \E\left[\error_{P,r}(X) \cdot \error_{\nu,s}(X)\mid \mcD_1\right].
\end{align}
\end{theorem}

Overall, Theorem \ref{thm::inprodbound} gives intuition that if strong duality holds, the first-stage bias should decay at a faster rate than $\E_{X \sim P_X\opt}[\error_{P,r}(X) \mid \mcD_1]$, which represents the $L^r$ error in estimating the outcome model. Intuitively, this is because if $\hatPC$ is close to $\PCopt$, then $\hat \nu$ should be close to $\nu\opt$, since $\hat \nu$ maximizes the empirical dual based on $\hatPC$, and $\nu\opt$ solves the population dual based on $\PCopt$.

We emphasize that as long as strong duality holds, Theorem \ref{thm::inprodbound} makes no assumptions whatsoever about the form of $\theta(P)$, the dimension of the covariates $X$, or the model class $\mcP$---furthermore, it is a finite-sample result with no ``hidden'' constants.

\subsubsection{Refined theory for continuous potential outcomes}
\label{subsec:structured-margins}

The product-bias inequality in Theorem~\ref{thm::inprodbound} reduces
tightness to investigating the dependence of the dual error to the first-stage
primal error. We begin with continuous potential outcomes, for which the conditional
transport problem need not be a finite-dimensional linear program and a generic
LP stability argument is therefore unavailable
\citep[e.g.,][]{semenova2023aggregated,khandamiryan2026adaptive,hsieh2022lp,fang2023lp}.
The general theory would
require careful analysis of infinite-dimensional linear programs. In this
paper, we focus on the seven examples in Section~\ref{subsec::examples}.
Appendix~\ref{sec: dual form} shows that the true Kantorovich duals
are either known functions of \((p_1^\star,p_0^\star)\) or belong to a class of
such functions parametrized by an \(x\)-dependent scalar. This structure reduces
dual error to lower-dimensional objects.

Here, we describe the estimated duals, relevant margin
condition, and the bias guarantee for each example. To simplify the exposition,
we consider stronger margin conditions than those in Appendix~\ref{sec:proofs-structured-margins}. We refer to
Appendix~\ref{sec: dual form} for the derivation
of Kantorovich duals and Sections~\ref{app:fh-proof}--\ref{sec:qte-proof} of Appendix~\ref{sec:proofs-structured-margins} for the formal assumptions, statements, and
proofs for the bias guarantees. Throughout this subsection, \(A\preceq B\)
means that \(A\le C B\) for a finite constant \(C\) that does not depend on
the sample size or the first-stage estimates; the appendices report the
corresponding explicit constants. For all but Examples~\ref{ex::fh} and~\ref{ex::makarov}, we assume
that \(Y(1)\) and \(Y(0)\) have bounded support
\([\underline y,\overline y]\) for some
\(-\infty<\underline y<\overline y<\infty\). Throughout,
\(F_{w,x}(y)=P\{Y(w)\le y\mid X=x\}\), \(\widehat F_{w,x}\) denotes its fitted
counterpart, and
\(\widehat Q_{w,x}(u)=\inf\{y:\widehat F_{w,x}(y)\ge u\}\) denotes its generalized inverse.
\begin{example}[name={\fh\, bounds},continues=ex::fh]
Define
\[
  \Gamma_x
  =
  F_{1,x}(y_1)+F_{0,x}(y_0)-1.
\]
\begin{itemize}[leftmargin=*]
\item \textit{Choice of estimated duals.}
Set
\[
  \widehat\Gamma_x
  =\widehat F_{1,x}(y_1)+\widehat F_{0,x}(y_0)-1.
\]
The estimated duals in our dual bounds method are given by
\[
  \widehat\nu_{0,x}(z)
  =\I\{\widehat\Gamma_x>0\}\I\{z\le y_0\},
  \qquad
  \widehat\nu_{1,x}(z)
  =\I\{\widehat\Gamma_x>0\}\bigl(\I\{z\le y_1\}-1\bigr).
\]
\item \textit{Margin condition.}
Assume
\[
  \P\{0<|\Gamma_X|\le s\}\le C s^b,\qquad s\downarrow0.
\]
\item \textit{Bias guarantee.}
When the corresponding first-stage errors are sufficiently
small, Appendix~\ref{app:fh-proof} gives the following \(r=2\) and
\(r=\infty\) bounds, respectively:
\[
\begin{aligned}
  0\le \theta_L-\widetilde\theta_L
  &\preceq
  \left\{
  \E\!\left[(\widehat\Gamma_X-\Gamma_X)^2\mid\mathcal D_1\right]
  \right\}^{(b+1)/(b+2)},
  \\
  0\le \theta_L-\widetilde\theta_L
  &\preceq
  \left\{
  \operatorname*{ess\,sup}_{x\sim P_X\opt}
  |\widehat\Gamma_x-\Gamma_x|
  \right\}^{b+1}.
\end{aligned}
\]
\end{itemize}
\end{example}

\begin{example}[name={Variance of ITE},continues=ex::varite]
As noted in \eqref{eq:varite-decomposition}, the variance of the individual
treatment effect equals the second moment of the individual treatment effect
minus the squared ATE. Since the ATE is point identified, we focus on
\[
  \theta(P)=\E_P[(Y_i(1)-Y_i(0))^2].
\]
\begin{itemize}[leftmargin=*]
\item \textit{Choice of estimated duals.}
Define
\[
  \widehat T_x(y)=\widehat Q_{1,x}\!\left(\widehat F_{0,x}(y)\right),
  \qquad
  \widehat\varphi_x(y)=\int_{\underline y}^{y}\widehat T_x(s)\,ds,
\]
and
\[
  \widehat\varphi_x^\star(z)
  =\sup_{y\in[\underline y,\overline y]}
  \{yz-\widehat\varphi_x(y)\}.
\]
The estimated duals in our dual bounds method are given by
\[
  \widehat\nu_{0,x}(y_0)=y_0^2-2\widehat\varphi_x(y_0),
  \qquad
  \widehat\nu_{1,x}(y_1)=y_1^2-2\widehat\varphi_x^\star(y_1).
\]

\item \textit{Margin condition.}
No separate margin condition is needed.

\item \textit{Bias guarantee.}
Appendix~\ref{sec:smoothmonge} gives
\[
  0\le \theta_L-\widetilde\theta_L
  \preceq
  \E\!\left[ H_f(X)
    \operatorname{error}_{P,2}(X)^2
    \mid \mathcal D_1
  \right],
\] for the function $H_f(X)$ defined in Appendix~\ref{sec:smoothmonge}.
\end{itemize}
\end{example}

\begin{example}[name={Lee bounds under monotone selection},continues=ex::lee]
By \eqref{eq:lee-ratio-decomposition}, we only need to work with the numerator,
denoted by \(\beta(P^\star)\), and its sharp lower bound \(\beta_L=\inf_P\beta(P)\),
because the denominator is point identified under monotone selection.
Define
\[
  q_0(x)=P\{S(0)=1\mid X=x\},
  \qquad
  p_{1,1}(y\mid x)
  =\frac{d\,P\{Y(1)\le y,\;S(1)=1\mid X=x\}}{dy}.
\]
\begin{itemize}[leftmargin=*]
\item \textit{Choice of estimated duals.}
Let \(\widehat q_0(x)\) and \(\widehat p_{1,1}(\cdot\mid x)\) be
the corresponding estimates. These are the only first-stage quantities needed
for this example. We constrain or project the first-stage estimates so that
\[
0\le\widehat q_0(x)\le
\int_{\underline y}^{\overline y}\widehat p_{1,1}(y\mid x)\,dy\le1;
\]
this guarantees that $\widehat a_x$ below is well-defined. Define
\[
  \operatorname{error}_{P,2}(x)^2
  =|\widehat q_0(x)-q_0(x)|^2
  +\int_{\underline y}^{\overline y}
    |\widehat p_{1,1}(y\mid x)-p_{1,1}(y\mid x)|^2\,dy,
\]
and set
\[
  \widehat a_x
  =\inf\left\{a\in[\underline y,\overline y]:
    \int_{\underline y}^{a}\widehat p_{1,1}(y\mid x)\,dy
    \ge\widehat q_0(x)
  \right\}.
\]
With \((u)_-=\min\{u,0\}\), write the augmented potential outcomes as
$z_w=(r_w,s_w)=\bigl(Y(w)S(w),S(w)\bigr)$ and choose
\[
  \widehat\nu_{0,x}(r_0,s_0)=\widehat a_xs_0-r_0,
  \qquad
  \widehat\nu_{1,x}(r_1,s_1)=(r_1-\widehat a_xs_1)_-,
  \qquad
  \widehat\lambda_x=(\widehat a_x)_+.
\]
Here $\widehat\lambda_x$ is the multiplier on
$w_x^{\mathrm{Lee}}(z_0,z_1)=\I\{s_0>s_1\}$ in the generic dual constraint.
\item \textit{Margin condition.}
The score function is
\[
  \Psi_x^{\mathrm{Lee}}(c)
  =\int_{\underline y}^{c}p_{1,1}(y\mid x)\,dy-q_0(x).
\]
The margin condition is
for every \(s\ge 0\),
\[
  \operatorname{Leb}\{c\in[\underline y,\overline y]:
  0\le|\Psi_x^{\mathrm{Lee}}(c)|\le s\}
  \le
  m_b(x) s^b.
\]
\item \textit{Bias guarantee.}
Appendix~\ref{sec:lee} gives
\[
  0\le \beta_L-\widetilde\beta_L
  \preceq
  \E\!\left[
    m_b(X)\operatorname{error}_{P,2}(X)^{b+1}
    \mid \mathcal D_1
  \right].
\]
In particular, \citet{semenova2021generalized} considers a different margin
condition and effectively proves the same bias rate with \(b=1\). Unlike our
guarantee, which controls an average over \(X\) of
\(\operatorname{error}_{P,2}(X)\), \citet{semenova2021generalized} requires \(L^\infty\) control
over \(x\) for the estimate of \(q_0(x)\).
\end{itemize}
\end{example}

\begin{example}[name={Distribution treatment effect},continues=ex::makarov]
We consider the case $t=0$ without loss of generality:
\[
  \theta(P)=\P\{Y(1)\le Y(0)\}.
\]
Write
\[
  \Delta_x(s)=F_{1,x}(s)-F_{0,x}(s),
  \qquad
  s_x^\star=\inf\argmax_{s\in\mathbb R}\Delta_x(s).
\]
\begin{itemize}[leftmargin=*]
\item \textit{Choice of estimated duals.}
Define
\[
  \widehat\Delta_x(s)=\widehat F_{1,x}(s)-\widehat F_{0,x}(s),
  \qquad
  \widehat s_x
  =\inf\argmax_{s\in\mathbb R}\widehat\Delta_x(s).
\]
Then choose
\[
  \widehat\nu_{0,x}(y_0)=-\I\{y_0<\widehat s_x\},
  \qquad
  \widehat\nu_{1,x}(y_1)=\I\{y_1\le\widehat s_x\}.
\]
\item \textit{Margin condition.}
When \(\Delta_x\) is differentiable, the score function is
\[
  \Psi_x^{\mathrm{DTE}}(t)=\Delta_x'(t)=p_{1,x}(t)-p_{0,x}(t).
\]
Assume the true and fitted conditional distributions admit densities on
\(\mathbb R\), \(s_x^\star\) and \(\widehat s_x\) are finite,
and \(\Delta_x\) and \(\widehat\Delta_x\) are strictly single-peaked at
\(s_x^\star\) and \(\widehat s_x\), respectively. Define
\[
M_x^{\mathrm{DTE}}
=
(s_x^\star\wedge\widehat s_x,\,
 s_x^\star\vee\widehat s_x].
\]
For every \(u>0\), assume
\[
  \operatorname{Leb}\{t\in M_x^{\mathrm{DTE}}:
  0<|\Psi_x^{\mathrm{DTE}}(t)|\le u\}
  \le
  m_b(x) u^b.
\]
\item \textit{Bias guarantee.}
Appendix~\ref{sec:makarov} gives the following \(r=2\)
and \(r=\infty\) bounds, respectively:
\begin{align*}
  0\le \theta_L-\widetilde\theta_L
  &\preceq
  \E\!\left[
    m_b(X)^{1/(b+2)}
    \operatorname{error}_{P,2}(X)^{(2b+2)/(b+2)}
    \mid \mathcal D_1
  \right],
  \\
  0\le \theta_L-\widetilde\theta_L
  &\preceq
  \E\!\left[
    m_b(X)\operatorname{error}_{P,\infty}(X)^{b+1}
    \mid \mathcal D_1
  \right].
\end{align*}
\end{itemize}
\end{example}

\begin{example}[name={Positive treatment effect},continues=ex::positive_ite]
The population CDF gap is
\[
  \Delta_x(t)=F_{0,x}(t)-F_{1,x}(t),
  \qquad
  t\in[\underline y,\overline y].
\]
\begin{itemize}[leftmargin=*]
\item \textit{Choice of estimated duals.}
Let
\[
  \widehat\Delta_x(t)=\widehat F_{0,x}(t)-\widehat F_{1,x}(t),
  \qquad
  \widehat\alpha_x(t)=\I\{\widehat\Delta_x(t)>0\}.
\]
The estimated duals in our dual bounds method are given by
\[
  \widehat\nu_{0,x}(y_0)
  =\int_{y_0}^{\overline y}\widehat\alpha_x(t)\,dt,
  \qquad
  \widehat\nu_{1,x}(y_1)
  =-\int_{y_1}^{\overline y}\widehat\alpha_x(t)\,dt.
\]
\item \textit{Margin condition.}
The score function is \(\Psi_x^{\mathrm{PTE}}(t)=\Delta_x(t)\).
The margin condition is for every \(s\ge 0\),
\[
  \operatorname{Leb}\{t\in[\underline y,\overline y]:
  0\le|\Psi_x^{\mathrm{PTE}}(t)|\le s\}
  \le
  m_b(x) s^b.
\]
\item \textit{Bias guarantee.}
Appendix~\ref{sec:pte} shows
\[
  0\le \theta_L-\widetilde\theta_L
  \preceq
  \E\!\left[
    m_b(X)\operatorname{error}_{P,2}(X)^{b+1}
    \mid \mathcal D_1
  \right].
\]
\end{itemize}
\end{example}

\begin{example}[name={Baseline-disadvantaged subgroup treatment effect},continues=ex:CTE]
Only the numerator
\[
  \beta(P^\star)=\E[Y(1)\I\{Y(0)\le c\}]
\]
is partially identified.
\begin{itemize}[leftmargin=*]
\item \textit{Choice of estimated duals.}
Set
\[
  \widehat a_x
  =\inf\{a:\widehat F_{1,x}(a)\ge\widehat F_{0,x}(c)\}.
\]
The estimated duals in our dual bounds method are given by
\[
  \widehat\nu_{0,x}(y_0)=-\widehat a_x\I\{y_0>c\},
  \qquad
  \widehat\nu_{1,x}(y_1)=\min\{y_1,\widehat a_x\}.
\]
\item \textit{Margin condition.}
The score function is
\[
  \Psi_x^{\mathrm{sub}}(\tau)=F_{1,x}(\tau)-F_{0,x}(c).
\]
The margin condition is for every \(s\ge 0\),
\[
  \operatorname{Leb}\{\tau\in[\underline y,\overline y]:
  0\le|\Psi_x^{\mathrm{sub}}(\tau)|\le s\}
  \le
  m_b^{\mathrm{sub}}(x) s^b.
\]
\item \textit{Bias guarantee.}
Appendix~\ref{app:subgroup-proof} gives
\[
  0\le \beta_L-\widetilde\beta_L
  \preceq
  \E\!\left[
    m_b^{\mathrm{sub}}(X)\operatorname{error}_{P,2}(X)^{b+1}
    \mid \mathcal D_1
  \right].
\]
\end{itemize}
\end{example}

\begin{example}[name={Quantile of ITE},continues=ex:QTE]
Recall the definitions in Example~\ref{ex:QTE}.
\begin{itemize}[leftmargin=*]
\item \textit{Choice of estimated duals.}
Define
\[
\begin{aligned}
\Delta_{x,c}(a)&=F_{1,x}(a+c)-F_{0,x}(a),
&a_{x,c}^\star
&=\inf\argmin_{a\in[\underline y,\overline y]}\Delta_{x,c}(a),\\
\widehat\Delta_{x,c}(a)&=\widehat F_{1,x}(a+c)-\widehat F_{0,x}(a),
&\widehat a_{x,c}
&=\inf\argmin_{a\in[\underline y,\overline y]}
\widehat\Delta_{x,c}(a).
\end{aligned}
\]
The estimated duals are
\[
\widehat\nu_{0,x}^{c}(y_0)
=\alpha-\I\{y_0\ge\widehat a_{x,c}\},
\qquad
\widehat\nu_{1,x}^{c}(y_1)
=-\I\{y_1\le\widehat a_{x,c}+c\}.
\]
\item \textit{Margin condition.}
For every \(c\in\mathbb R\), the score function is
\[
  \Psi_{x,c}^{\mathrm{QITE}}(a)
  =\partial_a\Delta_{x,c}(a)
  =p_{1,x}(a+c)-p_{0,x}(a),
  \qquad
  \Delta_{x,c}(a)=F_{1,x}(a+c)-F_{0,x}(a).
\]
Assume \(-\Delta_{x,c}\) and
\(-\widehat\Delta_{x,c}\) are strictly single-peaked at
\(a_{x,c}^\star\) and \(\widehat a_{x,c}\), respectively. Also assume, for every \(s>0\),
\[
  \operatorname{Leb}\{a\in[\underline y,\overline y]:
  0<|\Psi_{x,c}^{\mathrm{QITE}}(a)|\le s\}
  \le
  m_b(x) s^b.
\]
In addition, assume that the infimum defining \(q_L\) is
attained and that, for some \(\ell_q>0\) and \(\eta>0\),
\[
  \theta_{L,c}\ge\ell_q(q_L-c)
  \qquad
  \text{for all }c\in[q_L-\eta,q_L].
\]
\item \textit{Bias guarantee.}
Appendix~\ref{sec:qte-proof} gives the following \(r=2\)
and \(r=\infty\) bounds, respectively:
\begin{align*}
  0\le q_L-\widetilde q_L
  &\preceq
  \E\!\left[
    m_b(X)^{1/(b+2)}
    \operatorname{error}_{P,2}(X)^{(2b+2)/(b+2)}
    \mid \mathcal D_1
  \right],
  \\
  0\le q_L-\widetilde q_L
  &\preceq
  \E\!\left[
    m_b(X)\operatorname{error}_{P,\infty}(X)^{b+1}
    \mid \mathcal D_1
  \right].
\end{align*}
\end{itemize}
\end{example}

\subsubsection{Refined theory for discrete potential outcomes}
\label{sec: refined theory for discrete outcomes}
We next turn to discrete potential outcomes with finite support. Unlike
Section~\ref{subsec:structured-margins}, we allow an arbitrary cost \(f\) whose conditional range is
uniformly bounded. We impose no conditional coupling restrictions beyond the
two marginal distributions, so the problem has the finite-dimensional
transportation-LP formulation below. Write the finite outcome support as
\(\{y_1,\ldots,y_m\}\), let
\(p_w(y_j\mid x)=P\{Y(w)=y_j\mid X=x\}\). For
\(w\in\{0,1\}\), define the conditional probability vectors and their
stacking by
\begin{equation}\label{eq:discrete-conditional-marginal-vector}
\begin{aligned}
\mu_w(x)
&=\big(p_w(y_1\mid x),\ldots,p_w(y_m\mid x)\big)\in\R^m,
\qquad w\in\{0,1\},\\
\mu(x)
&=\big(\mu_0(x),\mu_1(x)\big)\in\R^{2m}.
\end{aligned}
\end{equation}
Let \(\widehat\mu(x)\) be the analogous vector formed from the estimated
conditional probabilities. Under counting measure, the Euclidean first-stage
error is
\begin{equation}\label{eq:discrete-primal-l2-error}
\error_{P,2}(x)
=\|\widehat\mu(x)-\mu(x)\|_2
=
\left\{
\sum_{w=0}^1\sum_{j=1}^m
\left|\widehat p_w(y_j\mid x)-p_w(y_j\mid x)\right|^2
\right\}^{1/2}.
\end{equation}

For \(j_0,j_1\in\{1,\ldots,m\}\), define the normalized dual feasible
set
\begin{equation*}
\mathcal V_{f,x}^0
=
\left\{
(v_0,v_1)\in\R^{2m}:
v_{0,j_0}+v_{1,j_1}\le f(y_{j_0},y_{j_1},x)
\ \text{for all }j_0,j_1,
\quad v_{0,1}=0
\right\}
\end{equation*}
and assume that a fixed finite set
\(\mathcal T\) and Borel maps
\(x\mapsto(\nu_{0,x}^t,\nu_{1,x}^t)\), \(t\in\mathcal T\), enumerate its
distinct vertices for almost every \(x\). Fix an ordering of \(\mathcal T\)
to resolve all ties measurably. This condition is automatic when \(f\) does
not depend on \(x\). For
\(\eta=(\eta_0,\eta_1)\in\R^{2m}\), define
\begin{equation}\label{eq:discrete-vertex-objective}
\phi_x(t;\eta)
=-\eta_0^\top\nu_{0,x}^t-\eta_1^\top\nu_{1,x}^t,
\qquad t\in\mathcal T.
\end{equation}
Strong duality gives
\begin{equation*}
\theta_L(x)
=-\min_{t\in\mathcal T}\phi_x(t;\mu(x)),
\qquad
\theta_L
=-\E\!\left[\min_{t\in\mathcal T}\phi_X(t;\mu(X))\right].
\end{equation*}

To obtain our dual bound estimator, we choose the estimated duals as
\begin{equation*}
\widehat t_x\in\argmin_{t\in\mathcal T}\phi_x(t;\widehat\mu(x)),
\qquad
(\widehat\nu_{0,x},\widehat\nu_{1,x})
=(\nu_{0,x}^{\widehat t_x},\nu_{1,x}^{\widehat t_x}).
\end{equation*}
The corresponding conditional pseudo-target is
\begin{equation*}
\widetilde\theta_L(x)
=\mu_0(x)^\top\widehat\nu_{0,x}
+\mu_1(x)^\top\widehat\nu_{1,x},
\qquad
\widetilde\theta_L
=\E\{\widetilde\theta_L(X)\mid\mcD_1\}.
\end{equation*}
Define the set of optimal duals as
\begin{equation*}
\mathcal T_x^\star
=\argmin_{t\in\mathcal T}\phi_x(t;\mu(x))
\end{equation*}
and the Euclidean distance from the estimated dual to this set by
\begin{equation*}
\error_{\nu,2}(x)
=\inf_{t\in\mathcal T_x^\star}
\left\{
\|\widehat\nu_{0,x}-\nu_{0,x}^t\|_2^2
+\|\widehat\nu_{1,x}-\nu_{1,x}^t\|_2^2
\right\}^{1/2}.
\end{equation*}

Let
\begin{equation*}
\mathcal E_x
=\left\{
(t,t')\in\mathcal T^2:t\ne t',\ 
\operatorname{conv}\left\{
(\nu_{0,x}^t,\nu_{1,x}^t),
(\nu_{0,x}^{t'},\nu_{1,x}^{t'})
\right\}
\text{ is an edge of }\mathcal V_{f,x}^0
\right\}
\end{equation*}
be the set of ordered adjacent-vertex pairs. Define the Euclidean diameter
of the vertex set by
\begin{equation}\label{eq:discrete-l2-vertex-diameter}
D_2(x)
=\max_{t,t'\in\mathcal T}
\left\{
\|\nu_{0,x}^t-\nu_{0,x}^{t'}\|_2^2
+\|\nu_{1,x}^t-\nu_{1,x}^{t'}\|_2^2
\right\}^{1/2}.
\end{equation}
Finally, for \(z\ge0\), define
\begin{equation}\label{eq:causal-vertex-margin-modulus}
\omega_x(z)
=
\1\!\left\{
\exists (t,t')\in\mathcal E_x:
0<
\left|\phi_x(t';\mu(x))-\phi_x(t;\mu(x))\right|
\le z
\right\}.
\end{equation}
Thus, \(D_2(x)\) is the largest Euclidean distance between two normalized
dual vertices, whereas \(\omega_x(z)=1\) exactly when some adjacent pair has
a nonzero population objective gap no larger than \(z\).

The next result bounds the dual error and the resulting first-stage bias for
finite-support potential outcomes.
\begin{theorem}
\label{prop:discrete-lower-level}
The dual error $\error_{\nu,2}(x)$ can be bounded by
\begin{equation}\label{eq:causal-vertex-direct-bound}
\error_{\nu,2}(x)
\le D_2(x)\,
\omega_x\!\left(D_2(x)\error_{P,2}(x)\right).
\end{equation}
Suppose \(E|f(y_1,y_1,X)|<\infty\) and that, for some constant
$B<\infty$ and every $x$,
\begin{equation}\label{eq:discrete-cost-oscillation-bound}
\max_{1\le j_0,j_1\le m}f(y_{j_0},y_{j_1},x)
-
\min_{1\le j_0,j_1\le m}f(y_{j_0},y_{j_1},x)
\le B.
\end{equation}
Define
\begin{equation*}
\overline{\error}_{P,2}(x)
=2B\sqrt{2m-1}\,\error_{P,2}(x).
\end{equation*}
Then the first-stage bias satisfies
\begin{equation}\label{eq:causal-vertex-envelope-bias}
0\le \theta_L-\tilde\theta_L
\le
\E\!\left[
\overline{\error}_{P,2}(X)
\omega_X\!\left(\overline{\error}_{P,2}(X)\right)
\mid \mcD_1
\right].
\end{equation}
\end{theorem}

We next show that, for the class of causal estimands considered
in this paper, Theorem~\ref{prop:discrete-lower-level} nests
\citet{semenova2023aggregated}'s bias bound as a special case that imposes a
uniform convergence requirement on $\mu(x)$. In
Appendix~\ref{sec:semenova-margin-comparison}, we elaborate on
\citet{semenova2023aggregated}'s setting, assumptions, and results and explain
how they map to our setting.

\begin{corollary}
\label{thm:discrete-bias-under-semenova}
Consider the setting of Theorem~\ref{prop:discrete-lower-level}.
Identify the fixed
index set \(T\) in \citet{semenova2023aggregated} with \(\mathcal T\), and
identify the nuisance function called \(\nu_0\) there with \(\mu\) in
\eqref{eq:discrete-conditional-marginal-vector}. Assume the following
counterparts of the stated assumptions in
\citet{semenova2023aggregated}:
\begin{enumerate}[label=\textnormal{(\roman*)},ref=\textnormal{(\roman*)},leftmargin=2.8em]
\item \label{cond:semenova-uniform}\emph{Assumption~4.2.}
There is a deterministic sequence \(\nu_n^\infty=o(n^{-1/4})\) such that
the cross-fitted estimate satisfies
\[
\Pr\!\left\{
\sup_{x\in\mathcal X}\|\widehat\mu(x)-\mu(x)\|_2
\le\nu_n^\infty
\right\}\to1.
\]
\item\label{cond:semenova-gradient} \emph{Assumption~4.3(2).}
For a constant \(B_\phi<\infty\), the gradient, which is constant in
\(v\), satisfies
\[
\sup_{x\in\mathcal X,\,t\in\mathcal T}
\left\|
\nabla_v\phi_x(t;v)
\right\|_2
\le B_\phi.
\]
\item\label{cond:semenova-margin} \emph{Assumption~4.4.}
For constants \(\bar B,\delta\in(0,\infty)\),
\[
\sup_{\substack{t,t'\in\mathcal T\\t\ne t'}}
\Pr\!\left\{
0\le
\phi_X(t;\mu(X))-\phi_X(t';\mu(X))
\le s
\right\}
\le\bar B s,
\qquad s\in(0,\delta).
\]
\end{enumerate}
Fix one fold and suppress its index, writing its first-stage estimate as
\(\widehat\mu\). Whenever
\[
\sup_{x\in\mathcal X}
\|\widehat\mu(x)-\mu(x)\|_2
\le \nu_n^\infty,
\]
the bias bound in Theorem~\ref{prop:discrete-lower-level} satisfies
\begin{equation}
\label{eq:prop-d2-bias-under-semenova}
0\le \theta_L-\widetilde\theta_L
\le C(\nu_n^\infty)^2,
\end{equation}
for some constant \(C\) that only depends on \(m\), \(B_\phi\), and
\(\bar B\).
Consequently, by condition~\ref{cond:semenova-uniform} and
\(\nu_n^\infty=o(n^{-1/4})\),
\begin{equation}\label{eq:prop-d2-root-n-bias-under-semenova}
\sqrt n(\theta_L-\widetilde\theta_L)=o_p(1).
\end{equation}
\end{corollary}

As a result of Corollary~\ref{thm:discrete-bias-under-semenova}, our bound can be viewed as relying on weaker conditions than
Assumption~4.2 of \citet{semenova2023aggregated}. In particular, it does not
require uniform control of the primal error and accommodates any exponent
\(b>0\).

Examples~\ref{ex:logistic-unbounded-covariate}
and~\ref{ex:honest-tree-integrated-loss} in
Appendix~\ref{sec:discrete-margin-comparison} show that this uniform control can
be overly restrictive even for standard estimators of $\mu(x)$, such as
logistic regression or decision trees with low-dimensional covariates. In
contrast, Theorem~\ref{prop:discrete-lower-level} may continue to produce an
$o_P(n^{-1/2})$ bias bound in those cases.

\subsection{Cross fitting}\label{subsec::crossfit}

This subsection shows that cross-fitting can recover the factor of two lost by sample splitting without sacrificing validity under the conditions below. Tightness holds when the fold-specific bias is $o_p(n^{-1/2})$. As notation, let $\hat\theta_L\swap$ denote the same estimator as $\hat \theta_L$ but with the roles of $\mcD_1$ and $\mcD_2$ swapped. The cross-fit estimator is then
\begin{equation}
\label{eqn: cross-fit estimator}
    \hat\theta_L\crossfit \defeq \frac{\hat\theta_L + \hat\theta_L\swap}{2}.
\end{equation}
A cross-fit lower confidence bound can be computed as follows. Let $\hat \nu$ and $\hat \nu\swap$ denote the estimated dual variables from $\mcD_1$ and $\mcD_2$, respectively. For ease of exposition, we assume $n$ is even and $|\mcD_1| = |\mcD_2| = n/2$. \footnote{The results in this section can be easily extended to $M$-fold cross-fitting for $M > 2$.} Let $S_i = \frac{\hat\nu\swap_{1,X_i}(Y_i) W_i}{\pi(X_i)} + \frac{\hat\nu\swap_{0,X_i}(Y_i) (1-W_i)}{1 - \pi(X_i)}$ if $i \in \mcD_1$. If $i \in \mcD_2$, let $S_i$ be defined analogously but with $\hat \nu\swap$ replaced with $\hat \nu$. Then if $\hat \sigma_s\crossfit$ is the empirical standard deviation of $\{S_i\}_{i=1}^n$, the cross-fit lower confidence bound is
\begin{equation}\label{eq::lcb_crossfit}
    \hat\theta\lcb\crossfit = \hat\theta_L\crossfit - \Phi^{-1}(1-\alpha)\frac{\hat\sigma_s\crossfit}{\sqrt{n}}.
\end{equation}

\begin{remark}
Here our cross-fit estimator \eqref{eqn: cross-fit estimator} is written as the average of $\hat \theta_L$ obtained on two folds following the convention known as DML1 \cite{chernozhukov2018dml}. Several papers have advocated the use of DML2 \cite{kallus2022s, levis2023covariate, semenova2023aggregated, velez2024asymptotic}, which pools the estimating equations across all folds instead of solving an estimating equation separately for each fold. Here, our dual bound estimator is a simple average corresponding to a linear moment in $\theta$. Therefore, the DML1 and DML2 estimators are numerically identical in our setting with balanced folds. The DML1/DML2 distinction matters for nonlinear estimating equations.
\end{remark}

We first establish validity when $\hat{\nu}$ is potentially inconsistent. Due to the dependence introduced by cross-fitting, we need more regularity conditions to show an analogue of Theorem \ref{thm::alwaysvalid}. Interestingly, we show that $\hat\theta\lcb\crossfit$ is valid under two separate and non-nested conditions.

\begin{theorem}\label{thm::crossfit_validity} 
Assume that $\hat\nu\swap$ is computed using the same procedure as $\hat \nu$ (but applied to $\mcD_2$ instead of $\mcD_1)$, so that Assumption \ref{assumption::nu_moment} holds for $\hat \nu\swap$.
Under Assumption \ref{assump::rand_experiment},
\begin{equation*}
    \liminf_{n \to \infty} \P(\hat\theta\lcb\crossfit \le \theta_L) \ge 1 - \alpha,
\end{equation*}
if one of the following holds:
\begin{enumerate}[label=\textnormal{Condition \arabic*:},ref=\arabic*,leftmargin=*]
    \item \label{cond:crossfit-stable}There exist deterministic dual variables $\nu\conv \in \mcV$, which are not necessarily optimal, satisfying the moment conditions in Assumption \ref{assumption::nu_moment} such that $\E\left[\left(\hat \nu_{k,X}(Y(k)) - \nu\conv_{k,X}(Y(k))\right)^2\right] \to 0$ holds at any rate for $k \in \{0,1\}$. Note that we allow $\{\nu\conv_k\}_{k\in \{0,1\}}$ to change with $n$.
    \item \label{cond:crossfit-bias}The outcome model is sufficiently misspecified such that the first-stage bias is strictly larger than $n^{-1/2}$ in order, i.e., $\sqrt{n}( \theta_L -\tilde\theta_L) \toprob \infty.$ 
\end{enumerate}

\end{theorem}

Condition~\ref{cond:crossfit-stable} of Theorem \ref{thm::crossfit_validity} shows that if the estimated dual functions $\hat \nu_0, \hat \nu_1$ are asymptotically deterministic, although their limits may differ from $(\nu_0\opt, \nu_1\opt)$, $\hat\theta\lcb\crossfit$ is a valid lower confidence bound. Similar conditions on estimated nuisance parameters have been studied in other contexts \citep{chernozhukov2020adversarial, arkhangelsky2021double}. A strength of this result is that it allows $\hat \nu_0, \hat \nu_1$ to converge at arbitrarily slow rates. Indeed, the proof technique for this result is based on a novel argument leveraging weak duality; it is not necessarily true that under Condition~\ref{cond:crossfit-stable}, $\hat\theta\lcb\crossfit$ is equivalent to an ``oracle'' confidence bound of any form. Condition~\ref{cond:crossfit-bias} suggests that even if this is not true and the fluctuations of $\hat \nu_0, \hat \nu_1$ do not vanish asymptotically, cross-fitting can be valid if the first-stage bias is sufficiently large, making $\hat\theta\lcb\crossfit$ conservative but valid. Except in pathological examples, we expect Condition~\ref{cond:crossfit-bias} to hold whenever Condition~\ref{cond:crossfit-stable} does not. Intuitively, if $\hat\nu$ has non-vanishing fluctuations, this suggests that $\hat\nu$ is not consistently estimating $\nu\opt$, in which case we should expect a substantial conservative bias, satisfying Condition~\ref{cond:crossfit-bias}. Thus, in practice, we recommend using cross-fitting.

\begin{remark} Under the conditions of Theorem \ref{thm::crossfit_validity}, one can use cross-fitting in combination with a multiplier-bootstrap-like procedure to perform model selection as long as one chooses among a finite number of (fit) outcome models. We present this result in Appendix \ref{appendix::delta_method} for brevity.
\end{remark}

Now we turn to tightness of cross-fitting under the refined tightness theories
developed in Section~\ref{subsec:structured-margins} and
Theorem~\ref{prop:discrete-lower-level}. Analogous to
$\tilde{\theta}_L$, we define the effective estimand of
$\hat\theta\lcb\crossfit$ as 
\begin{equation}\label{eq::tilde_lcb_crossfit}
\tilde{\theta}_L\crossfit= (g(\hat\nu) + g(\hat\nu\swap))/2.
\end{equation}
The next result shows that $\tilde{\theta}_L\crossfit$ is statistically
indistinguishable from the sharp lower bound $\theta_L$ whenever the
fold-specific bias bounds from either theory are $o_p(n^{-1/2})$.

\begin{corollary}\label{corr::crossfit_tightness}
Suppose that the duals estimated on each fold satisfy one of the following
fold-specific tightness conditions:
\begin{enumerate}[label=(\roman*),leftmargin=*]
\item For one of Examples~\ref{ex::fh}--\ref{ex:QTE}, the corresponding
lower-level conditions in Section~\ref{subsec:structured-margins} and
Appendix~\ref{sec:proofs-structured-margins} hold on each fold, and the
resulting example-specific bias is \(o_p(n^{-1/2})\) on each fold.
\item The finite-support conditions of
Theorem~\ref{prop:discrete-lower-level} hold and
\begin{equation*}
\E\!\left[
\overline{\error}_{P,2}(X)
\omega_X\!\left(\overline{\error}_{P,2}(X)\right)
\mid \mcD_1
\right]
=o_p(n^{-1/2}),
\end{equation*}
with the analogous condition holding when the roles of $\mcD_1$ and
$\mcD_2$ are reversed.
\end{enumerate}
Then
\begin{equation}
    \sqrt{n}(\tilde{\theta}_L\crossfit - \theta_L) = o_p(1).\end{equation}
\end{corollary}

\subsection{Dual bounds for observational studies}
\label{subsec::aipw}

So far, our theory has assumed that the propensity scores are known. However, when $\pi(X_i)$ is unknown, we can replace the IPW estimator with an augmented IPW (AIPW) estimator to increase robustness. In particular, define the conditional mean of the estimated dual variables $\hat\nu$ as

\[c_0(x) \defeq \E_{P\opt_{Y(0)\mid X = x}}[\hat \nu_{0,x}(Y(0))] \text{ and } c_1(x) \defeq \E_{P\opt_{Y(1)\mid X = x}}[\hat \nu_{1,x}(Y(1))]\]

so $c_k(X_i)$ is the conditional mean of $\hat \nu_{k,X_i}(Y(k))$ given $X_i$ and $\mcD_1$. Also, let $\hat c_0(x), \hat c_1(x)$ denote estimators of $c_0(x), c_1(x)$ fit on $\mcD_1$; for example, one can automatically compute $\hat c_0(x), \hat c_1(x)$ by plugging in $\hatPC$. Lastly, for any $i \in \mcD_2$, define the AIPW summand 
\begin{equation}\label{eq::si_aug}
    S_i \defeq W_i \frac{\hat \nu_{1,X_i}(Y_i) - \hat c_1(X_i)}{\hat \pi(X_i)} + (1-W_i) \frac{\hat \nu_{0,X_i}(Y_i) - \hat c_0(X_i)}{1 - \hat \pi(X_i)} + \hat c_1(X_i) + \hat c_0(X_i),
\end{equation}
where $\hat \pi$ are propensity scores estimated on $\mcD_1$. Then, if $\hat \sigma_s\aug$ is the sample standard deviation of $\{S_i\}_{i \in \mcD_2}$ on $\mcD_2$, the ``augmented'' version of $\hat\theta\lcb$ is
\begin{equation}\label{eq::auglcb}
    \hat\theta\lcb\aug \defeq \frac{1}{|\mcD_2|} \sum_{i \in \mcD_2} S_i - \Phi^{-1}(1-\alpha) \frac{\hat \sigma_s\aug}{\sqrt{|\mcD_2|}}.
\end{equation}

We can now prove a validity result for $\hat\theta\lcb\aug$. In the first case, we impose the product-rate condition
\[
\left\{\E[\errpi^2]\E[\errc^2]\right\}^{1/2}=o(n^{-1/2})
\]
and a nondegeneracy condition for the actual AIPW summand. This includes the exact propensity-correct case $\hat\pi=\pi$ and the exact regression-correct case $\hat c_k=c_k$, while also allowing both nuisances to be estimated. In the second case, validity follows from conservativeness when the dual gap $\theta_L-\tilde\theta_L$ dominates the nuisance-induced error.

\begin{theorem}\label{thm::alwaysvalid_aug} Suppose Assumption \ref{assump::rand_experiment} holds except that the propensity scores are not known. 
For $k \in \{0,1\}$, assume the fourth moments $\E[|\hat\nu_{k,X}(Y(k))|^4 \mid \mcD_1] \le B < \infty$ and $\E[|\hat c_k(X)|^4 \mid \mcD_1] \le B < \infty$ are uniformly bounded. Finally, assume that the estimated propensity scores $\hat \pi(X_i)$ are uniformly bounded away from zero and one.
 
Let $\errpi \defeq \E[(\hat \pi(X) - \pi(X))^2 \mid \mcD_1]^{1/2}$ denote the $\ell_2$ error in estimating the propensity scores and let $\errc = \max_{k \in \{0,1\}} \E[(\hat c_k(X) - c_k(X))^2 \mid \mcD_1]^{1/2}$ denote the $\ell_2$ error in estimating the conditional mean of $\hat \nu$, where $X$ is an independent draw from the distribution of $X_i$.  Consider the two conditions below:
\begin{enumerate}[label=\textnormal{Condition \arabic*:},ref=\arabic*,topsep=0pt,leftmargin=*] 
    \item \label{cond:aipw-product}Assume the product-rate condition
    \begin{equation}\label{eq::riskdecay}
        \E[\errpi^2] \E[\errc^2] = o(1/n).
    \end{equation}
    Furthermore, for the actual AIPW summand $S_i$ defined in \eqref{eq::si_aug}, assume $\var(S_i\mid\mcD_1)\ge 1/B$.
    \item \label{cond:aipw-bias}The outcome model is sufficiently misspecified such that the first-stage bias $\theta_L - \tilde{\theta}_L$ dominates either $\errpi$ or $\errc$. More precisely, assume
    \begin{equation*}
        \frac{n^{-1/2}+\min(\errc, \errpi)}{\theta_L - \tilde{\theta}_L } \toprob 0.
    \end{equation*}

\end{enumerate}
If either Condition~\ref{cond:aipw-product} or Condition~\ref{cond:aipw-bias} holds, then $\hat\theta\lcb\aug$ is asymptotically valid:
\begin{equation*}
    \liminf_{n \to \infty} \P(\hat\theta\lcb\aug \le \theta_L) \ge 1 - \alpha.
\end{equation*}
\end{theorem}

See Appendix \ref{appendix::validity_proofs} for a proof.

\begin{remark} In observational studies, the multiplier bootstrap method for model selection from Section \ref{subsec::multbootstrap} is not appropriate because the validity of the final bounds may depend on the accuracy of the outcome model. For example, the multiplier bootstrap might select a highly inaccurate outcome model that yields (misleadingly) tight bounds. Thus, in observational studies, we recommend that the analyst perform cross-validation on $\mcD_1$, as discussed in Section \ref{subsec::multbootstrap}, to select the best-performing outcome model.
\end{remark}

\section{Computation}\label{sec::comp}

\subsection{General strategy and ensuring validity}\label{subsec::compstrat}\label{subsec::condcomp}

In this section, we discuss how to compute the dual bounds in Definition \ref{def::dualbnds}. Computation is straightforward except for two questions:
\begin{itemize}[topsep=0pt, itemsep=0.5pt]
    \item Dual bounds will yield valid results for any estimated dual variables as long as $\hat\nu \in \mcV$ is dual-feasible. However, it is not obvious how to ensure that dual-feasibility holds. 

    \item Our recommended approach to estimating the dual variables requires solving the optimization problem
    \begin{equation}\label{eq::hatnu_def_comp_sec}
    (\hat\nu_{0,x},\hat\nu_{1,x}) \in \argmax_{(\nu_{0,x}, \nu_{1,x}) \in \mcV_x} \E_{\hatPZC}[\nu_{0,x}(Y(0)) \mid X = x] + \E_{\hatPOC}[\nu_{1,x}(Y(1)) \mid X = x].
    \end{equation}
    If $Y$ is continuous, this is an infinite-dimensional program, so it is unclear how to solve it unless the functional forms of the optimizers can be derived, as in the examples in Section~\ref{subsec::examples}.
\end{itemize}

We now outline a general strategy to answer these questions based on two key observations. Note that for simplicity, in this section, we assume the response $Y$ is real-valued.

\begin{enumerate}[label=\textbf{Observation \arabic*:},ref=\arabic*,leftmargin=*]
\item \label{obs:conditional-separability}\textbf{The problem separates in $\mcX$.} Theorem \ref{thm::kantorovich} makes clear that to compute $\hat\nu \approx \argmax_{\nu \in \mcV} g(\nu)$, it suffices to repeatedly solve the problem conditional on $x$. The solutions to these problems are independent in the sense that the value of $\hat\nu_{0,x},\hat\nu_{1,x}$ does not affect the value of $\hat\nu_{0,x'},\hat\nu_{1,x'}$ for some $x' \ne x$. Similarly, by definition we have that $\hat \nu \in \mcV$ is dual-feasible if and only if $(\hat\nu_{0,x},\hat\nu_{1,x})\in\mcV_x$ for all $x \in \mcX$.

\item \label{obs:compute-needed-only}\textbf{Only compute what we need.} To apply dual bounds, we need to only compute $\{\hat\nu_{0,x}, \hat\nu_{1,x}\}_{x \in \{X_i : i \in \mcD_2\}}$ to compute the IPW estimator $\hat\theta_L$ and lower confidence bound $\hat\theta\lcb$---i.e., we do not need to solve Eq. (\ref{eq::hatnu_def_comp_sec}) for all $x \in \mcX$. 
\end{enumerate}

These observations have two implications.

\begin{enumerate}[label=\textbf{Implication \arabic*:},ref=\arabic*,leftmargin=*]
\item \label{imp:ensure-validity}\textbf{Ensuring validity.} Given \textit{any} initial estimate $\hat\nu\init$ which may or may not be dual-feasible, we can convert $\hat\nu\init$ into dual-feasible estimators as follows:
\begin{itemize}[noitemsep, topsep=0.5pt]
    \item For $x \in \mcX$, we define $c_x$ to be half of the maximum violation of the conditional feasibility constraint. Namely, for any estimated $\hat \lambda_{x,1}, \dots, \hat \lambda_{x,L} \ge 0$, we define:
    \begin{equation*}
        2c_x \defeq \sup_{(y_0,y_1)\in\mcY^2}\left[\hat\nu_{0,x}^{\rm init}(y_0)+\hat\nu_{1,x}^{\rm init}(y_1)-\sum_{\ell=1}^L\hat\lambda_{x,\ell}w_{x,\ell}(y_0,y_1)-f(y_0,y_1,x)\right].
    \end{equation*}

    \item Then we define the final estimators
    \begin{equation}\label{eq::hatnu_adjust}
        \hat\nu_{0, x}(y_0) \defeq \hat\nu_{0, x}\init(y_0) - c_x \text{ and } \hat\nu_{1, x}(y_1) \defeq \hat\nu_{1, x}\init(y_1) - c_x
    \end{equation}
    which are guaranteed to be dual-feasible by definition of $\mcV$.
\end{itemize}

For each $x \in \mcX$, $c_x$ can be computed or approximated using a two-dimensional grid search---crucially, because this grid search is low-dimensional, we can accurately compute $c_x$. Furthermore, Observation~\ref{obs:compute-needed-only} implies that we only need to compute $c_x$ for $\{X_i : i \in \mcD_2\}$. As a result, the steps above represent a generic algorithm to convert \textit{any} initial estimates $\hat\nu\init$ into valid dual estimates $\hat\nu\in\mcV$ via $|\mcD_2|$ grid searches.
Here the correction in \eqref{eq::hatnu_adjust} is a deterministic step for enforcing dual feasibility. If the conditional dual problem is solved exactly, which implies the dual variables must be feasible, the correction does not reduce tightness. For discrete outcomes, there are many LP solvers that can reliably find the optimal solution. For continuous outcomes, the generic optimal transport problem is often difficult to solve exactly, thus it is more practical to solve a penalized or discretized conditional problem to approximate the optimal solution and it is necessary to enforce the validity of approximate solution.

\item \label{imp:compute-duals}\textbf{A generic strategy for computing optimal dual variables.} Similarly, to compute $\hat\nu \in \argmax_{\nu \in \mcV} g(\nu)$, we have the following general strategy:
\begin{enumerate}[label=\textnormal{Step \arabic*:},ref=\arabic*,noitemsep,topsep=0pt,leftmargin=*]
    \item \label{step:compute-estimate-distribution}Estimate $\hatPC$ on $\mcD_1$.
    \item \label{step:compute-conditional-duals}For $i \in \mcD_2$, solve the ``conditional problem'' Eq. (\ref{eq::hatnu_def_comp_sec}) for $x = X_i$ and use the outputs $\hat\nu_{0,X_i}, \hat\nu_{1,X_i}$ to compute the IPW summands in the definition of $\hat\theta_L$ and $\hat\theta\lcb$.
\end{enumerate}
In other words, we need to only solve the conditional problem $|\mcD_2|$ times to compute the dual bounds. The pseudo-code is given in Algorithm~\ref{alg:crossfit-dual-lcb}. We discuss how to do this in the next section.

\end{enumerate}

\begin{remark}\label{rem::gridsearch} For finite support outcomes, we emphasize that no matter how poorly we solve Eq. (\ref{eq::hatnu_def_comp_sec}), as long as we adjust our final dual variables using Eq. (\ref{eq::hatnu_adjust}), we will get valid lower confidence bounds on $\theta_L$.
\end{remark}
\begin{remark}
When the discretization is sufficiently fine and the estimand function $f$ is not too irregular between neighboring grid points, Algorithm~\ref{alg:conditional-dual} typically produces only small off-grid feasibility violations, so the feasibility adjustment has a limited effect on validity and sharpness. Assume that the number of grid points $J_n$ satisfies $J_n/\sqrt{n}\to\infty$.
When $f$ is Lipschitz and the discretization is sufficiently fine, the feasibility-adjustment bias is $o_P(n^{-1/2})$, so the resulting bound retains root-$n$ validity. If the initial duals are optimal, first-order sharpness is unaffected as well because the adjustment changes the objective by only $o_P(n^{-1/2})$.
\end{remark}
{
\begin{algorithm}[t]
\caption{Cross-fit dual lower confidence bound}
\label{alg:crossfit-dual-lcb}
\begin{algorithmic}[1]
\Require Data $\{(X_i,W_i,Y_i)\}_{i=1}^n$, propensity score $\pi(\cdot)$,
estimand function $f$, conditional feasibility sets $\{\mcV_x:x\in\mathcal X\}$,
conditional-distribution learner $\mathcal A$, number of folds $M$,
confidence level $1-\alpha$.
\Ensure A lower confidence bound $\widehat\theta_{\mathrm{LCB}}$ for the sharp lower bound $\theta_L$.

\State Randomly partition $\{1,\ldots,n\}$ into folds $I_1,\ldots,I_M$.

\For{$r=1,\ldots,M$}
    \State Let $I_{-r}=\{1,\ldots,n\}\setminus I_r$.
    \State Fit $\mathcal A$ on $I_{-r}$ to estimate the conditional distributions
    $
    \widehat P^{(-r)}_{Y(0)\mid X},
    \widehat P^{(-r)}_{Y(1)\mid X}.
    $

    \For{$i\in I_r$}
        \State Set $x=X_i$.
        \State Solve the (approximate) conditional dual problem at $x$ using
        $\widehat P^{(-r)}_{Y(0)\mid X=x}$ and
        $\widehat P^{(-r)}_{Y(1)\mid X=x}$ to obtain the conditional dual variables
        $
        \widehat\nu^{(-r),\init}_{0,x},\widehat\nu^{(-r),\init}_{1,x}
        $
        \State If an approximate algorithm is used, project $
        \widehat\nu^{(-r),\init}_{0,x},\widehat\nu^{(-r),\init}_{1,x}
        $ into the feasible set $\mcV_x$ according to \eqref{eq::hatnu_adjust} and obtain feasible dual variables 
        $\widehat\nu^{(-r)}_{0,x},\widehat\nu^{(-r)}_{1,x}$.
        \State Define the held-out IPW summand
        \[
        S_i
        =
        \frac{W_i\widehat\nu^{(-r)}_{1,X_i}(Y_i)}{\pi(X_i)}
        +
        \frac{(1-W_i)\widehat\nu^{(-r)}_{0,X_i}(Y_i)}{1-\pi(X_i)}.
        \]
    \EndFor
\EndFor

\State Compute
\[
\widehat\theta_L
=
\frac1n\sum_{i=1}^n S_i,
\qquad
\widehat\sigma^2
=
\frac1n\sum_{i=1}^n (S_i-\widehat\theta_L)^2 .
\]

\State Return
\[
\widehat\theta_{\mathrm{LCB}}
=
\widehat\theta_L
-
\Phi^{-1}(1-\alpha)\frac{\widehat\sigma}{\sqrt n}.
\]
\end{algorithmic}
\end{algorithm}

When the functional form of the true conditional dual variables is known up to a
finite-dimensional parameter, the generic conditional solver in Algorithm
\ref{alg:crossfit-dual-lcb} can be replaced by a smaller optimization problem. Suppose
that, for each \(x\), the relevant optimal dual variables are known to lie in a family
\[
    \left\{
    \left(\nu_{0,x}^{(a)},\nu_{1,x}^{(a)}\right):
    a\in\mathcal A_x\subseteq\mathbb R^d
    \right\},
\]
where the unknown parameter \(a=a(x)\) may depend on \(x\). If this family contains a
conditional dual optimizer, restricting the search to \(a\in\mathcal A_x\) does not
sacrifice strong duality. This is the structure used in Section
\ref{subsec:structured-margins} for Examples \ref{ex::lee}--\ref{ex:QTE}. After
estimating the conditional distributions on the training folds, solve, for each held-out
covariate value \(x\),
\[
\widehat a_x
\in
\argmax_{a\in\mathcal A_x}
\left\{
\E_{\widehat P_{Y(0)\mid X=x}}\!\left[\nu_{0,x}^{(a)}(Y(0))\right]
+
\E_{\widehat P_{Y(1)\mid X=x}}\!\left[\nu_{1,x}^{(a)}(Y(1))\right]
\right\}.
\]
Then set the initial duals to
\[
\widehat\nu_{0,x}^{\init}=\nu_{0,x}^{(\widehat a_x)},
\qquad
\widehat\nu_{1,x}^{\init}=\nu_{1,x}^{(\widehat a_x)}.
\]
If the structured search or numerical implementation is approximate, apply the
feasibility correction in \eqref{eq::hatnu_adjust}; otherwise use these duals directly in
the held-out IPW summands of Algorithm \ref{alg:crossfit-dual-lcb}.

\paragraph{Alternative generic solver: discretization.}
The structured calculation above should be used when the finite-dimensional
structure from Section \ref{subsec:structured-margins} is available. When that structure
is unavailable, or when the analyst wants a fully generic implementation, the
discretization-based conditional solver below replaces the structured parameter search
with a linear program over discretized fitted conditional distributions and can be plugged into the
generic cross-fit routine in Algorithm \ref{alg:crossfit-dual-lcb}.

We suggest a discretization-based method to approximately solve this conditional problem \eqref{eq::hatnu_def_comp_sec} and obtain initial estimates $\hat\nu_{0,x}\init, \hat\nu_{1,x}\init$.\footnote{Our software implements this method by default, although Appendix \ref{appendix::series} discusses an alternative approach based on series estimators.} 
The idea is to approximate $\hat P_{Y(k) \mid X = x}$ as a discrete distribution with support $\{y_{k,1,x}, \dots, y_{k,\nvals,x}\}$ and probability mass function (PMF) $p_{k,1,x}, \dots, p_{k,\nvals,x} \in (0,1)$ so that
\begin{equation*}
    \hat P_{Y(0) \mid X = x} \approx \sum_{j=1}^{\nvals} p_{0,j,x}\delta_{y_{0,j,x}}  \text{ and } \hat P_{Y(1) \mid X = x} \approx \sum_{i=1}^{\nvals} p_{1,i,x}\delta_{y_{1,i,x}} ,
\end{equation*}
where $\delta_z$ denotes the point mass on $z \in \R$. In particular, we suggest taking $y_{k,j,x}$ as the $\frac{j}{\nvals+1}$th quantile of $\hat P_{Y(k) \mid X = x}$ and setting $p_{k,j,x} = \frac{1}{\nvals}$ for $k \in \{0,1\}, j\in \{1, \ldots, \nvals\}$. The conditional optimization problem then becomes a discrete linear program with $2\nvals+L$ variables and $\nvals^2+L$ constraints. We summarize the procedure in Algorithm \ref{alg:conditional-dual} below. 

\begin{algorithm}[t]
\caption{\texorpdfstring{Alternative discretization-based conditional dual computation}{Alternative discretization-based conditional dual computation}}
\label{alg:conditional-dual}
\begin{algorithmic}[1]
\Require Covariate value $x$, fitted conditional distributions
$\widehat P_{Y(0)\mid X=x}$ and $\widehat P_{Y(1)\mid X=x}$,
feasibility functions $\{w_{x,\ell}\}_{\ell=1}^L$, estimand function $f$,
discretization size $\nvals$, discretization rule \(\mathsf{Disc}_{\nvals}\)
\Ensure Conditionally feasible dual pair
$(\widehat\nu_{0,x},\widehat\nu_{1,x})\in\mcV_x$.
\State Apply the discretization rule to the fitted conditional distributions:
\[
\{(y_{k,j,x},p_{k,j,x})\}_{j=1}^{\nvals}
=
\mathsf{Disc}_{\nvals}\!\left(\widehat P_{Y(k)\mid X=x}\right),
\qquad k\in\{0,1\}.
\]
By default, we take
\[
y_{k,j,x}
=
\widehat F^{-1}_{k,x}\!\left(\frac{j}{{\nvals}+1}\right),
\qquad
p_{k,j,x}=\frac{1}{{\nvals}},
\qquad j=1,\ldots,{\nvals}.
\]

\State Solve the finite-dimensional linear program
\[
\begin{aligned}
\max_{\nu_0,\nu_1,\lambda}\quad
&
\sum_{j=1}^{\nvals} p_{0,j,x}\nu_{0,j}
+
\sum_{j=1}^{\nvals} p_{1,j,x}\nu_{1,j}
\\
\mathrm{s.t.}\quad
&
\nu_{0,j}+\nu_{1,\ell}
-
\sum_{a=1}^L \lambda_a w_{x,a}(y_{0,j,x},y_{1,\ell,x})
\le
f(y_{0,j,x},y_{1,\ell,x},x),
\qquad j,\ell\in[{\nvals}],
\\
&
\lambda_1,\ldots,\lambda_L\ge 0 .
\end{aligned}
\]
Let the solution define initial dual functions
$\widehat\nu^{\mathrm{init}}_{0,x}$ and
$\widehat\nu^{\mathrm{init}}_{1,x}$.
\State Compute the piecewise-linear interpolation of $\widehat\nu^{\mathrm{init}}_{0,x}$ and
$\widehat\nu^{\mathrm{init}}_{1,x}$ to obtain the continuous dual variables $\widehat\nu_{0,x},\widehat\nu_{1,x}$.

\State Return $\widehat\nu_{0,x},\widehat\nu_{1,x}$.
\end{algorithmic}
\end{algorithm}
}
Here the linear programming problem can be solved efficiently using off-the-shelf LP solvers if, e.g., $\nvals \le 100$.  After solving this problem, we obtain initial values $\{\hat\nu_{0,x}\init(y_{0,j,x})\}_{j=1}^{\nvals}, \{\hat\nu\init_{1,x}(y_{1,i,x})\}_{i=1}^{\nvals}$ and we define the full functions $\hat\nu_{0,x}, \hat\nu_{1,x} : \R \to \R$ via linear interpolation.  
In the absence of additional constraints, the discretized program is a standard
optimal transport problem, allowing us to replace generic LP routines with
tailored optimal transport solvers. These include entropy-regularized methods
built on the iterative proportional fitting procedure of
\citet{deming1940least} and Sinkhorn matrix scaling
\citep{sinkhorn1964relationship,cuturi2013sinkhorn}, near-linear-time Sinkhorn
variants \citep{altschuler2017near}, and primal-dual extragradient methods
\citep{jambulapati2019direct}. Such methods can be substantially faster than
generic LP solvers on large grids. See Appendix~\ref{sec: complexity} for a
detailed computational complexity analysis. In
Appendix~\ref{sec: computational-comparison-semenova}, we compare the
computational performance of our formulation with the approaches of
\citet{semenova2020better} and \citet{semenova2023aggregated} for Lee bounds.
In particular, when the one-dimensional reduction is applied as in
\citet{semenova2020better} and Section~\ref{subsec:structured-margins}, the two
methods have similar computational efficiency. When off-the-shelf solvers are
applied as in \citet{semenova2023aggregated} and our
Algorithm~\ref{alg:conditional-dual}, tailored optimal transport solvers are
substantially more efficient than generic LP solvers.

\section{Empirical applications}\label{sec::applications}

We now illustrate our method in applications to two randomized experiments and one observational study. For all applications, we implement the dual bound estimators based on Algorithm~\ref{alg:conditional-dual} without leveraging the structures of the duals. Code and data are publicly available at \url{https://github.com/amspector100/dual_bounds_paper}.

\subsection{Persuasion effects of political news}
\label{subsec::persuasion}
We first analyze data from \cite{gerber2009}, who in 2005 randomly assigned a set of individuals in Prince William County, Virginia, to receive a free subscription offer for the Washington Post.\footnote{The original experiment had a third treatment condition, namely to receive a free subscription offer for the Washington Times. For simplicity, we follow \cite{jun2023persuasion} and only analyze subjects in the Washington Post or control treatment groups.} Using administrative data, they also determined whether each subject voted in the November 2006 elections. Thus, for $n=2400$ individuals, $W_i \in \{0,1\}$ denotes whether individual $i$ received a free subscription to the Washington Post, and $Y_i \in \{0,1\}$ denotes whether individual $i$ voted in the 2006 elections.

In this context, \citet{jun2023persuasion} studied the ``persuasion effect'' of the treatment, defined as the probability that the treatment causes an individual who would not otherwise have voted:
\begin{equation}\label{eq::persuasion_effect}
    \theta(P\opt) \defeq P\opt(Y(1) = 1 \mid Y(0) = 0) = \frac{P\opt(Y(1) = 1, Y(0) = 0)}{P\opt(Y(0) = 0)}.
\end{equation}
This estimand is also known as the Probability of Sufficiency \citep{pearl1999probabilities}. As noted by \citet{jun2023persuasion}, without covariates, the sharp bounds on $\theta(P\opt)$ are rescaled \fh\, bounds:
\begin{equation*}
    \theta_L^{\mathrm{no-covariates}} \defeq \max\left(\frac{\E_{P\opt}[Y(1) - Y(0)]}{1-\E_{P\opt}[Y(0)]}, 0\right) \le \theta(P\opt) \le \min\left(\frac{\E_{P\opt}[Y(1)]}{1-\E_{P\opt}[Y(0)]}, 1\right) \defeq \theta_U^{\mathrm{no-covariates}}.
\end{equation*}
However, \cite{gerber2009} also collected a rich set of covariate information, including demographic information, political preferences, and previous voter turnout data. Furthermore, $\theta(P\opt)$ takes the form of an unidentifiable expectation divided by an identifiable expectation (since $P\opt(Y(0) = 0)$ is identified in Eq. (\ref{eq::persuasion_effect})). Thus, we can use our methodology to form covariate-assisted estimates of the numerator and apply the bivariate delta method to perform inference on $\theta(P\opt)$, as described in Appendix \ref{appendix::delta_method}.

To form the dual bounds, we estimate the conditional distributions of $Y(1) \mid X$ and $Y(0) \mid X$ using three outcome models: a cross-validated logistic ridge regression, a random forest, and a k-nearest neighbors (KNN) classifier, where the covariates are the $43$ baseline covariates from \cite{gerber2009} plus interaction terms with the treatment. For each outcome model, we form dual bounds following the methodology from Sections \ref{subsec::crossfit} and \ref{sec::comp} using $10$-fold cross-fitting. We also compute non-robust plug-in bounds, which plug in the estimated conditional distributions and the empirical distribution of $X$ into Eq. (\ref{eq::init_thetaL_def}); unlike dual bounds, these bounds can be anti-conservatively biased. We also aggregate the results across all dual bounds using the multiplier bootstrap-like procedure detailed in Appendix \ref{appendix::delta_method}.

Table \ref{table::persuasion} shows the results, from which we report three main findings. First, the covariate-assisted dual bounds are more than twice as narrow as the covariate-free bounds. Second, the dual bounds appear to be more reliable than the covariate-assisted plug-in bounds. For example, the KNN and random forest outcome models produce plug-in lower bounds larger than $15\%$. This is implausible because the ATE point estimate is $0.029$ and not significant; indeed, we do not even have power to reject the sharp null that $Y(1) = Y(0)$ with probability one. In contrast, dual bounds can leverage each outcome model to provide provably valid confidence bounds without assuming that the outcome model is accurate. Third, the multiplier bootstrap method successfully selects the tightest lower and upper bounds while providing rigorous uncertainty quantification.
\begin{table}[h!]
\centering

\begin{tabular}{llllll}
\toprule
Outcome model & $R^2$ & Dual LB & Dual UB & Plug-in LB & Plug-in UB \\
\midrule
No covariates & 0.0 & 0.056 & 0.966 & 0.057 & 0.966 \\
 &  & (0.04) [0.0] & (0.039) [1.0] & (0.034) [0.0] & (0.031) [1.0] \\
Ridge & 0.49 & 0.038 & 0.365 & 0.046 & 0.376 \\
 &  & (0.027) [0.0] & (0.019) [0.403] &   &   \\
RF & 0.365 & 0.003 & 0.41 & 0.158 & 0.348 \\
 &  & (0.022) [0.0] & (0.021) [0.451] &   &   \\
KNN & 0.358 & 0.0 & 0.409 & 0.192 & 0.366 \\
 &  & (0.019) [0.0] & (0.021) [0.45] &   &   \\
Multiplier bootstrap$\opt$ & - & 0.056 & 0.365 &  &  \\
 &  &  [0.0] &  [0.412] &   &   \\
\bottomrule
\end{tabular}

\caption{This table shows the lower and upper bounds on $P\opt(Y(1) = 1 \mid Y(0) = 0)$ for the experimental data from \cite{gerber2009}, with standard errors shown in parentheses and confidence bounds shown in brackets. We do not know how to compute standard errors for the covariate-assisted plug-in bounds, so we do not list them. $\opt$Note that we do not use the exact multiplier bootstrap methodology from Section \ref{subsec::multbootstrap}. Rather, we use the variant from Appendix \ref{appendix::delta_method}, which permits the use of cross-fitting. Solving the dual problem takes 0.44-0.52 seconds in each setting, while fitting the nuisance takes 0.23-1.66 seconds depending on the model choice. 
\label{table::persuasion}}
\end{table}

\begin{remark} Our analysis is inspired by Jun and Lee (2022), but it differs from theirs in three ways. First, they do not leverage covariates in their main empirical results. Second, they consider the monotone treatment response assumption that $Y(1) \ge Y(0)$ almost surely \citep{manski1997monotone}. We chose to avoid this assumption, since prior work has shown that media exposure can sometimes depress turnout \citep[e.g.,][]{gentzkow2006}, suggesting the treatment effect may be heterogeneous even if it is positive on average. Lastly, they perform an instrumental variables (IV) analysis where the exposure is whether an individual \textit{read} the Washington Post and the outcome is whether an individual voted for a Democrat. However, their exposure and outcome were only collected for $\approx 30\%$ of the sample who responded to a follow-up survey; thus, by performing an ITT analysis with voter turnout as the outcome, we avoid any missing data problems. It is possible to extend our methodology to IV analyses, but it requires new methodological ideas which we defer to a separate work \citep{dualivnote2024}.
\end{remark}

\subsection{Estimating intensive margins}\label{subsec::carranza_chenroth}

\cite{carranza2022} conducted a randomized experiment in South Africa where treated individuals received assessment results that they could share with potential employers. They found that treated individuals had higher employment rates and higher earnings, suggesting that the tests provided useful information about workers' skills. However, we might wonder: is the treatment effect driven by increases in employment (extensive margin), or does the treatment increase hours worked for individuals who would have been employed with or without the treatment (intensive margin)?

To estimate the intensive margin, \citet{chenroth2023} analyzed the following quantities:
\begin{equation}\label{eq:intensive-margin-targets}
    \E[Y(1) - Y(0) \mid Y(1) > 0, Y(0) > 0] \text{ and } \E\left[\log(Y(1)) - \log(Y(0)) \mid Y(1) > 0, Y(0) > 0\right],
\end{equation}
In \eqref{eq:intensive-margin-targets}, the outcome $Y$ measures the average hours worked per week post-treatment, and the logs in the latter estimand ensure that it is scale-invariant and can roughly be interpreted as a ``percentage'' effect. \citet{chenroth2023} bounded these quantities using the methodology from \cite{lee2009training}, which assumes that $Y(1) > 0$ holds whenever $Y(0) > 0$, i.e., the treatment does not cause any individual to be unemployed. To defend this assumption, \citet{chenroth2023} noted that individuals with poor test results likely did not share them with their employers, and we agree that this assumption seems plausible in this setting.

However, the dataset from \cite{carranza2022} contains a rich set of pre-treatment covariates, including baseline earnings, demographic information, and educational history. Thus, we produce covariate-assisted variants of the bounds from \citet{chenroth2023}. To fit the outcome model, we use the default settings in the \texttt{dualbounds} package, which employs a linear model with interactions:
\begin{equation}
    Y_i = X_i^T \beta + W_i X_i^T \gamma + \epsilon_i.
\end{equation}
To estimate $\beta$ and $\gamma$, we use a cross-validated ridge regression. We estimate the distribution of $\epsilon_i \mid X_i, W_i$ as the empirical distribution of the estimated residuals $\{Y_i - X_i^T\hat\beta - W_i X_i^T \hat\gamma : i \in [n], W_i = w\}$.\footnote{This estimate severely restricts the heteroskedasticity pattern, since it asserts that the residuals are independent of the covariates given the treatment, i.e., $\epsilon_i \Perp X_i \mid W_i$. That said, we emphasize that the final dual bounds are valid even if the model for the distribution of $\epsilon_i \mid X_i, W_i$ is completely inaccurate.} We then convert this outcome model into a cross-fit dual bound using the methodology from Sections \ref{sec::method} and \ref{sec::comp}.

Table \ref{table::carranza2022} shows the results: for both the logged and non-logged outcome, the covariate-assisted bounds are only $\approx 60\%$ as wide as the covariate-free bounds. Although the bounds are still quite wide, this analysis nonetheless shows that covariate adjustment can substantially sharpen partial identification bounds without requiring additional assumptions.

\begin{table}[h!]
\centering
\begin{tabular}{|c|c|c|c|c|}
\hline
& \multicolumn{2}{|c|}{Log-hours} & \multicolumn{2}{|c|}{Hours} \\ \hline
Outcome model & Lower bound & Upper bound & Lower bound & Upper bound \\ \hline
No covariates (plug-in) & -0.193 & 0.281 & -6.64 & 2.69 \\ 
  &  (0.062) & (0.111) & (1.36) & (2.06) \\\hline
Ridge (dual) & -0.115 & 0.185 & -4.74 & 1.18 \\ 
  & (0.060) & (0.130) & (1.46) & (2.00) \\
  \hline
\hline
\end{tabular}
\caption{This table shows the lower and upper bounds on $\E[\log(Y(1)) - \log(Y(0)) \mid Y(1) > 0, Y(0) > 0]$ and $\E[Y(1) - Y(0) \mid Y(1) > 0, Y(0) > 0]$ for the dataset from \cite{carranza2022}. Standard errors are shown in parentheses and clustered at the assessment date level (following \cite{chenroth2023}). Solving the dual problem takes 1.83-2.11 seconds in each setting, while fitting the nuisance takes 30.62 seconds for ridge regression. }\label{table::carranza2022}
\end{table}

\subsection{401k eligibility}
\label{subsec::401k}
We now study how 401(k) eligibility impacts wealth. 
An extensive literature argues that 401(k) eligibility is essentially exogenous conditional on covariates \citep[e.g.,][]{poterba1995, poterba1998, poterba2000, chernhansen2004}, since workers likely choose employers based on job characteristics besides 401(k) eligibility, e.g., income. We adopt this assumption; thus, the outcome $Y \in \R$ measures total household wealth and the treatment $W \in \{0,1\}$ indicates 401(k) eligibility.

We obtain data from \cite{chernozhukov2018dml}, who estimated average treatment effects (ATEs) using a sample of households from the 4th wave of the 1990 Survey of Income and Program Participation. Yet the literature emphasizes that treatment effects may be highly heterogeneous. For instance, 401(k) eligibility may reduce wealth for households who otherwise would participate in a different retirement plan or whose 401(k) contributions adversely reduce their liquidity.
To study whether the ATE reflects broad wealth gains or masks offsetting negative gains for some households, we now bound the \textit{positive treatment effect}:
\begin{equation}
    \theta(P\opt) = \E_{P\opt}\left[\max(Y(1) - Y(0), 0)\right].
\end{equation}

Our analysis uses the same covariates as \cite{chernozhukov2018dml}, including income, demographics, and financial indicators such as homeownership status. Following \cite{chernozhukov2018dml}, we use the raw covariates except when fitting regularized GLMs, where we include polynomial transformations and pairwise interactions
We estimate cross-fit propensity scores using a cross-validated logistic elastic net. For this observational application, the reported AIPW confidence guarantee relies on one of the conditions in Theorem \ref{thm::alwaysvalid_aug}.

The outcome model takes the form $Y_i = \E[Y_i \mid X_i, W_i] + \epsilon_i$. To estimate $\E[Y_i \mid X_i, W_i]$, we use  
(i) a cross-validated elastic net,
(ii) a KNN regressor, (iii) an \texttt{sklearn} histogram gradient boosting (HGBoost) regressor  \citep{sklearn, lightgbm2017} with the constraint that $\E[Y_i \mid X_i, W_i]$ is increasing in $W_i$,\footnote{Note that while the ``HGBoost monotone'' model asserts that the conditional average treatment effect $\E[Y(1) - Y(0) \mid X]$ is nonnegative, it nonetheless allows the positive treatment effect $\theta(P\opt) = \E[\max(Y(1) - Y(0), 0)]$ to differ from the ATE, for example, because there may be unobserved covariates $U$ such that $\E[Y(1) - Y(0) \mid X, U]$ may be negative.} and (iv) an intercept-only model as a baseline.
We estimate the distribution of $\epsilon_i \mid X_i, W_i$ as in Section \ref{subsec::carranza_chenroth}. For each model, we report cross-fit AIPW dual bounds (as described in Sections \ref{subsec::aipw}-\ref{sec::comp}) as well as non-robust ``plug-in'' bounds, which plug $\hat P_{Y \mid X, W}$ and the empirical distribution of the covariates into Eq. (\ref{eq::init_thetaL_def}).

\begin{table}[h!]
\centering
\begin{tabular}{lll|ll|ll}
\toprule
 & $R^2$ & ATE & Dual LB & Dual UB & Plug-in LB & Plug-in UB \\
Method &  &  &  &  &  &  \\
\midrule
HGBoost & 0.4255 & 6381 & 5564 & 47286 & 5834 & 51075 \\
 &  & (1882) & (1201) & (1258) & \nan & \nan \\
KNN & 0.3776 & 6792 & 4476 & 47424 & 10135 & 45800 \\
 &  & (1966) & (1640) & (1333) & \nan & \nan \\
Elastic net & 0.1672 & 10637 & 6938 & 60940 & 9078 & 55005 \\
 &  & (2284) & (2091) & (1579) & \nan & \nan \\
Intercept only & 0.0 & 11851 & 11326 & 66622 & 11446 & 64696 \\
 &  & (2579) & (2484) & (1763) & (1806) & (1481) \\
\bottomrule
\end{tabular}
\caption{This table shows estimated average treatment effects as well as lower and upper bounds on $\E[\max(Y(1) - Y(0), 0)]$ for the 401(k) eligibility dataset. Standard errors are shown in parentheses. We do not know how to compute standard errors for the covariate-assisted plug-in bounds, so we do not list them. Solving the dual problem takes 14.57-18.42 seconds in each setting, while fitting the nuisance takes 2.01-17.18 seconds depending on the model choice. }\label{table::401k}
\end{table}

Table \ref{table::401k} shows the out-of-sample $R^2$, cross-fit AIPW ATE estimates, dual bounds, and plug-in bounds for each outcome model. We report two main conclusions.

\begin{enumerate}[label=\underline{\arabic*.},leftmargin=*]
\item \textbf{Incorporating covariates improves robustness.} It is known that in observational studies, accurate outcome models can reduce the bias of ATE estimates. E.g., in Table \ref{table::401k}, more accurate outcome models yield smaller ATE estimates, ranging from $\approx \$11$K to $\approx \$6$K.  Table \ref{table::401k} suggests that the same logic applies to partially identified estimands (see Theorem \ref{thm::alwaysvalid_aug}), since the dual lower bounds decrease with the ATE estimates. 

\item \textbf{Plug-in bounds can be anticonservative.}  The KNN plug-in lower bound is $\approx 
\$10$K. 
This value seems implausible, since it is twice as large as the corresponding dual bound and $50\%$ larger than the ATE estimate from the best-performing model.
Indeed, covariate-assisted plug-in bounds rely entirely on the outcome model, whereas the AIPW dual bounds reduce the requirement on the convergence rate of the nuisance parameters. That said, it is reassuring that the best-performing model (HGBoost) yields similar plug-in and dual bounds.\end{enumerate}

\begin{remark} Although the ATE is a lower bound on $\theta(P\opt)$, the estimated dual lower bounds are $0$--$2$ standard errors below the corresponding ATE estimates. This discrepancy is a consequence of fitting an imperfect outcome model, which leads to conservative bounds.
\end{remark}

\subsection{A Monte-Carlo simulation}\label{sec::sims}

In this section, we run simulations to demonstrate the power, validity, and computational efficiency of dual bounds. Throughout, we consider randomized experiments where the propensity scores $\pi(x) = \frac{1}{2}$ are known. 

\label{sec::two_stage_sim}

We perform simulations where we estimate lower Lee bounds (Example \ref{ex::lee}). We sample covariates $X_{i} \iid \mcN(0,I_p)$ for $p=20$ covariates and draw $Y_i \mid X_i$ from a homoskedastic Gaussian linear model:
\begin{equation}\label{eq::y_dgp}
    Y_i(1) \mid X_i \sim \mcN(X_i^T \beta + \tau, \sigma^2) \text{ and } Y_i(0) \mid X_i \sim \mcN(X_i^T \beta, \sigma^2)
\end{equation}
for variance $\sigma^2 = 1$, coefficients $\beta \in \R^p$ chosen such that $\var(Y_i(1)) = \var(Y_i(0)) = 10$, and average treatment effect $\tau = 2$. We sample the selection events $S_i \mid X_i$ from a logistic regression model:
\begin{equation}\label{eq::s_dgp}
    \P(S_i(0) = 1 \mid X_i) = \mathrm{logit}^{-1}\left(X_i^T \beta_S + \tau_{S,0} \right) \text{ and } \P(S_i(1) = 1 \mid X_i) = \mathrm{logit}^{-1}\left(X_i^T \beta_S + \tau_{S,1} \right) \text{ for } \beta_S \in \R^p
\end{equation}
with $\|\beta_S\|_2 = 1$ and $\tau_{S,0} = 0, \tau_{S,1} = 1$. Following general practice in the literature, our simulations enforce the monotonicity condition $S(1) \ge S(0)$ a.s., and we assume that the practitioner knows this a-priori.
Specifically, conditional on $X_i$, we draw a common $U_i\sim\operatorname{Unif}(0,1)$ and set $S_i(w)=\I\{U_i\le p_w(X_i)\}$, where $p_w$ is given by \eqref{eq::s_dgp}; because $p_1(x)\ge p_0(x)$, this enforces monotonicity. The Gaussian outcome shocks are drawn independently of $U_i$ conditional on $X_i$. We compare three methods for estimating the sharp bound $\theta_L$ in this problem:
\begin{enumerate}[leftmargin=*, topsep=0pt, itemsep=0.5pt]
\item The ``naive plug-in'' method estimates the selection probabilities $\P(S_i=1\mid X_i,W_i)$ by logistic ridge regression and estimates the selected-outcome law $Y_i\mid S_i=1,X_i,W_i$ by fitting the Gaussian ridge model only to selected observations. In particular, $\hat\sigma$ is the sample standard deviation of the fitted residuals among observations with $S_i=1$; no outcome is imputed or modeled for $S_i=0$. We plug these estimates into \eqref{eq::lee_formula}, yielding $\hat\theta_L\plugin$. It is a natural model-based extension to the estimator proposed in \cite{lee2009training}. In general, it is not clear how to compute standard errors for $\hat\theta_L\plugin$; to be as generous as possible, we compute oracle lower confidence bounds using the true variance 
\begin{equation}
    \hat\theta\lcb\plugin = \hat\theta_L\plugin - \Phi^{-1}(1-\alpha) \sqrt{\var\left(\hat\theta_L\plugin\right)}.
\end{equation}
We compute the true value of $\var(\hat\theta_L\plugin)$ numerically by sampling many datasets from the true data-generating process.

\item The ``dual crossfit'' approach estimates the same selection probabilities and selected-outcome laws, now using cross-fitting. After computing these estimates on $K=5$ folds of the data, we apply the cross-fit dual bounds methodology from Section \ref{subsec::crossfit}. For each arm $w$, these estimates induce the fitted augmented outcome marginal
\[
\widehat P_{Z(w)\mid X=x}
=\{1-\widehat p_w(x)\}\delta_{(0,0)}
+\widehat p_w(x)\,\widehat P_{(Y(w),1)\mid S(w)=1,X=x}.
\]
We retain the atom $(0,0)$ exactly and discretize the first continuous component using the approach in Section \ref{subsec::condcomp}, with $\nvals=50$ selected-outcome grid points. \footnote{We remind the reader that the final dual lower confidence bound will be valid no matter how small $\nvals$ is, although increasing $\nvals$ may yield higher power.} Computing dual bounds using this method takes less than $5$ seconds with $n=1000$ observations in our simulations.

\item The ``no covariates'' method is identical to the naive plug-in approach except that it estimates only the marginal selection probabilities and the selected-outcome distributions $Y_i\mid S_i=1,W_i=w$, without conditioning on covariates.
\end{enumerate}

\begin{figure}
    \centering
    \includegraphics[width=\linewidth]{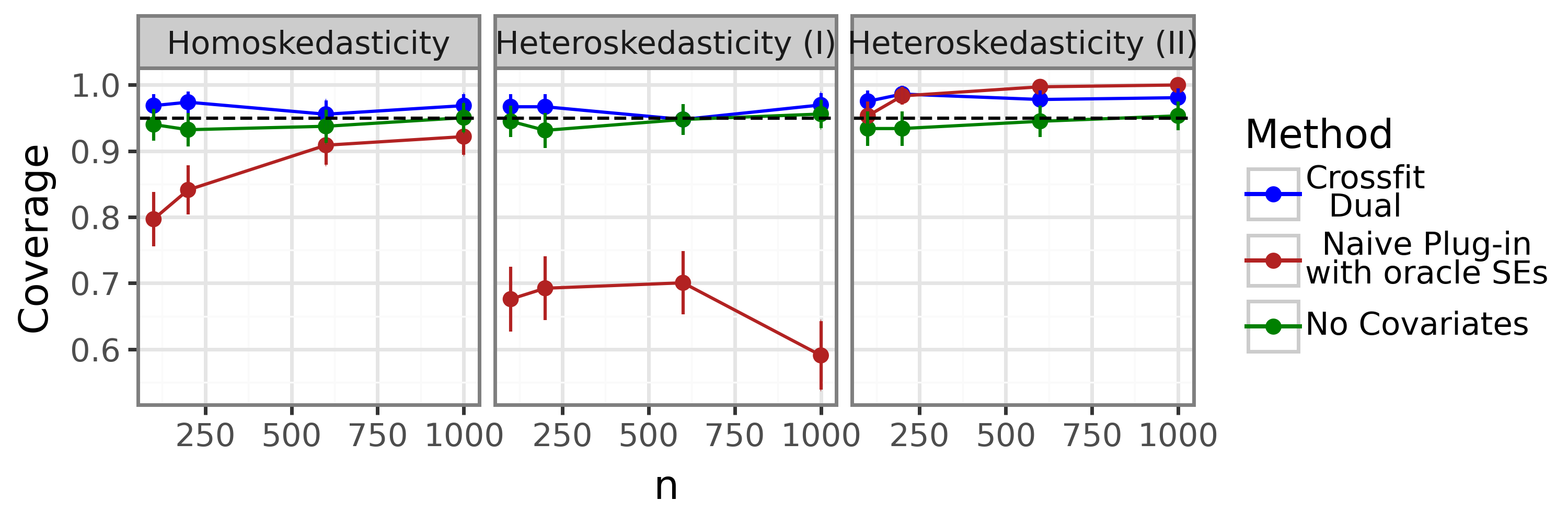}
    \caption{This figure shows the coverage of the lower Lee confidence bounds from Figure \ref{fig::contribution}. The nominal level is $95\%$, shown by the dotted black line.}
    \label{fig::lee_coverage}
\end{figure}

We also consider the performance of each method in two misspecified settings where $Y_i \mid X_i$ is actually heteroskedastic:
\begin{equation}\label{eq::y_dgp_heterosked}
    Y_i(1) \mid X_i \sim \mcN(X_i^T \beta + \tau, \sigma_1^2(X_i)) \text{ and } Y_i(0) \mid X_i \sim \mcN(X_i^T \beta, \sigma_0^2(X_i))
\end{equation}
for the functions $\sigma_1^2(X) = \sigma_1^2 \|X\|_2^2, \sigma_0^2(X) = \sigma_0^2 \|X\|_2^2$ for constants $\sigma_1, \sigma_0 \ge 0$. In this case, for both the naive plug-in and dual crossfit methods, the estimated outcome model is misspecified, since it incorrectly assumes homoskedasticity. In the first setting (labelled as ``Heteroskedasticity (I)''), we set $\sigma_1 / \sigma_0 = 3$; in the other setting (``Heteroskedasticity (II)''), we set $\sigma_0 / \sigma_1 = 0.3$.

Figures \ref{fig::contribution} and \ref{fig::lee_coverage} show the results with $n \in \{100, 200, 600, 1000\}$. Figure \ref{fig::contribution} shows the average value of the estimate $\hat\theta_L$ and the lower confidence bound $\hat\theta\lcb$; it shows that the naive plug-in estimator is biased when $n$ is small (due to the effect of regularization) and when the model is misspecified. In contrast, the cross-fit dual bounds are (i) guaranteed to be conservatively biased at worst and (ii) less sensitive to errors in estimating the outcome model, yielding valid and reasonably sharp inference in all three settings. Figure \ref{fig::lee_coverage} confirms that dual bounds provide $\ge 95\%$ coverage in all settings, whereas the naive plug-in method can be quite conservative or anticonservative, depending on the form of heteroskedasticity. Overall, in this setting, cross-fit dual bounds perform well even in small samples.

\section{Discussion}

This paper introduces a dual bound method to estimate and perform inference on a class of partially identified causal parameters. The method can leverage any statistical and machine learning techniques to learn the conditional distribution of the outcome $Y$ given the covariates $X$. In randomized experiments, the resulting bounds are always valid regardless of whether the estimates are consistent and asymptotically sharp even when the nuisance parameters are estimated at slower-than-$\sqrt n$ rates. The theory works for both continuous and discrete potential outcomes. In addition, one can apply the multiplier bootstrap to perform model selection. For observational studies, the method has an AIPW extension that is valid under the standard product-rate condition or a less standard conservativeness condition in Theorem \ref{thm::alwaysvalid_aug}. In all settings, the dual bounds can be computed efficiently. 

Our analysis leaves open many questions.
Perhaps the most pressing question is whether the techniques developed in this paper can be applied more generally. In particular, we use a duality argument to guarantee the robustness of our method. Does this same argument apply to settings beyond causal inference? In the next two sections, we begin to address this question. Then, Section \ref{subsec::deep_dual} discusses an alternative computational strategy, and Section \ref{subsec::twosided} discusses two-sided intervals.

\subsection{Extensions beyond causal effects}

In this section, we discuss whether our method can be extended to settings that cannot be reduced to estimating an expectation of the form $\E[f(Y(0), Y(1), X)]$. Indeed, the ideas in Section \ref{subsec::dualbnds} apply to many estimands in economics which can be written as the optimal value of an optimization problem. We describe two classes of problems below.

\begin{example}[Inference for linear programs] Suppose that $\theta$ is the optimal objective value of a linear program of the form
$$\theta \defeq \min_{z \in \R^d} c^T z \suchthat A z \le b(P),$$
where $A \in \R^{m \times d}$ is a known matrix and $b(P) \in \R^m$ is a vector of moments or conditional moments of a probability distribution $P$. Many estimands can be written this way \citep[e.g.][]{gafarov2019inference, fang2023lp}, including those arising in models of demand \citep{tebaldi2019demandlp, nevo2016demandlp} and income mobility \citep{chetty2016mobility}. If we observe i.i.d. samples from $P$, the exact same method from Section \ref{subsec::dualbnds} can be applied to obtain a $1-\alpha$ lower confidence bound on $\theta$.
\end{example}

\begin{example}[Variance of the CATE] In Example \ref{ex::varite}, we noted that $\var(Y(1) - Y(0))$ is a natural measure of treatment effect heterogeneity. Another interesting estimand is the variance of the conditional average treatment effect (CATE) $\tau(X) \defeq \E[Y(1) - Y(0) \mid X]$. Using Fenchel conjugacy or Cauchy--Schwarz, we can derive a dual representation:
\begin{align*}
        \var(\tau(X))
    &=
        \max_{h : \mcX \to \R} 2 \cov(h(X), \tau(X)) - \var(h(X)) \\
    &=
        \max_{h : \mcX \to \R} 2 \cov(h(X), Y(1) - Y(0)) - \var(h(X)).
\end{align*}
Crucially, the bound $B(h) \defeq 2 \cov(h(X), Y(1) - Y(0)) - \var(h(X))$ is easy to estimate (in randomized experiments) for any fixed $h : \mcX \to \R$; thus, we can obtain a robust lower confidence bound on $\var(\tau(X))$ by selecting a function $\hat h \approx \argmax_{h} B(h)$ on the first split of data and estimating $B(\hat h)$ on the second split. This idea is connected to \cite{floodgate2020,tvfloodgate2023}, who also select a lower bound (albeit a different one) for nonparametric variance estimation using a different variational representation. \end{example}

\subsection{Cost of robustness when the propensity scores are unknown}

However, dual bounds as defined in Section \ref{subsec::dualbnds} are not appropriate for every problem. For example, one might hope that a simple modification of Definition \ref{def::dualbnds} can produce always valid bounds on the average treatment effect when the propensity scores $\pi(X_i)$ are not known. Unfortunately,  our method yields valid yet trivial lower and upper bounds due to the lack of strong duality. This is consistent with \cite{aronow2021nonparametric}, who prove that no uniformly consistent estimator of ATE exists under strong ignorability and strict overlap without further assumptions if one of the covariates is continuous.

Suppose the vectors $(X_i, W_i, Y_i(0), Y_i(1))$ are sampled i.i.d. from some population distribution $P\opt \in \mcP$, where $\mcP$ is the set of distributions on $\mcX \times \{0,1\} \times \mcY^2$ satisfying unconfoundedness, i.e., $\{Y_i(1), Y_i(0)\} \Perp W_i \mid X_i$, and strict overlap, i.e., $0 < \pi_P(x) < 1$ for all $x\in \mcX$ and $P \in \mcP$, where $\pi_P(X) \defeq \E_P[W \mid X]$.

For the result below, assume additionally that $\mcY$ contains its minimum and
maximum. For every $x\in\mcX$ and $(y_0,y_1)\in\mcY^2$, assume that $\mcP$
contains sequences of distributions under which $X=x$ almost surely,
$(Y(0),Y(1))=(y_0,y_1)$ almost surely, and the treatment propensity converges
to zero or to one. Thus strict overlap is imposed separately for each
distribution, without a uniform overlap constant over $\mcP$.

Given i.i.d. observations $(X_i, W_i, Y_i)$, we seek to form a lower bound on the average treatment effect $\theta(P\opt) \defeq \E_{P\opt}[Y_i(1) - Y_i(0)]$ which is valid even under arbitrary misspecification of $\pi_P(X_i)$ and the outcome model. Although $\theta$ is identifiable, it can still be written as the solution to the optimization problem
\begin{align}
    \theta(P\opt) = \min_{P \in \mcP} \E_P[Y(1) - Y(0)] \suchthat P_{X,W,Y} = P\opt_{X,W,Y}
\end{align}
where $P_{X,W,Y}$ is the distribution of $(X, W, Y)$ under $P$ and $P\opt_{X,W,Y}$ is the true distribution of $(X, W, Y)$. Note that the optimization variable is $P$, which is a joint distribution over $(X,W,Y(0),Y(1))$, and $P_{X,W,Y}$ is a functional of $P$. For any $h : \mcX \times \{0,1\} \times \mcY \to \R$, the Lagrange dual to this problem is
\begin{equation*}
    g(h) \defeq \E_{P\opt}\left[h(X, W, Y)\right] + \kappa(h),
\end{equation*}
where $\kappa(h) \defeq \inf_{P \in \mcP} \E_P[Y(1) - Y(0) - h(X, W, Y)]$ is a known constant depending on $h$. For any $h$, $g(h)$ is a valid lower bound on $\theta(P\opt)$ by weak duality, but unfortunately, strong duality does not hold. In particular, for any $h$, we have that
\begin{align}\label{cex::ate}
    h(X_i, W_i, Y_i) + \kappa(h) \le \begin{cases} Y_i - \max(\mcY) & W_i = 1 \\ \min(\mcY) - Y_i & W_i = 0. \end{cases}
\end{align}
 See Appendix \ref{appendix::cex_ate} for a proof.

This result tells us that any dual bound (in the sense of Def. \ref{def::dualbnds}) on the ATE which is valid under arbitrary misspecification must also be trivial, since it must impute $Y_i(1)$ to have the minimum possible value whenever it is not observed, and it must impute $Y_i(0)$ to have the maximum possible value when it is not observed. Thus, applied in this way, our method reduces to the nonparametric bounds from \cite{manski1989}, even though we have made the extra assumptions of unconfoundedness and strict overlap, which are not made in \cite{manski1989}.

\subsection{An alternative computational strategy}\label{subsec::deep_dual}

To compute our recommended estimator $\hat \nu$ of the optimal dual variables, one must first model the conditional distributions $\PCopt$. This procedure may not be feasible when conditional distributions are hard to model. For example, when $X$ includes unstructured data such as images (e.g., profile pictures as in \cite{athey2022smiles}), texts (e.g., resumes as in \cite{vafa2022career}), and embeddings \citep{vafa2024estimating}, existing machine learning algorithms may not be able to provide distribution estimates. In Appendix \ref{Appendix: deep dual}, we present Deep Dual Bounds, an alternative approach that directly learns the optimal dual variables by solving the dual problem. 
This approach parametrizes $\nu_{0,X}(Y(0)), \nu_{1,X}(Y(1))$ by neural networks, which, unlike the two-stage approach, can exploit the smoothness of the dual functions in $X$. 

Although this end-to-end formulation is conceptually clear, standard gradient-based algorithms cannot be directly applied since potential outcomes cannot be observed simultaneously. We resolve this issue by matching treated and control units and optimizing an approximate objective function. We emphasize that the approximation error does not affect the validity of Dual Bounds, as only dual feasibility is required. The details and experimental results of the algorithm are discussed in Appendix \ref{Appendix: deep dual} and Table \ref{table::deepdual}. 

\subsection{Two-sided confidence intervals}\label{subsec::twosided}
In previous sections we focused on one-sided confidence bounds on the sharp population bounds $\theta_L, \theta_U$. To cover the full identified set, we can simply construct $(1-\alpha/2)$ lower/upper confidence bounds on the lower/upper bounds \citep{horowitz2000nonparametric}. However, in many applications, it suffices to cover the true parameter. It is well-known \citep{imbensmanski2004, stoye2009more} that tighter uniform confidence intervals can be constructed by estimating the gap between upper and lower bounds and/or the correlation between two estimators. 

With data splitting, the upper and lower dual bounds are both empirical moments:
\[\hat{\theta}_L = \frac{1}{|\mathcal{D}_2|}\sum_{i\in \mathcal{D}_2}S_i^{L}, \quad \hat{\theta}_U = \frac{1}{|\mathcal{D}_2|}\sum_{i\in \mathcal{D}_2}S_i^{U}.\]
Under mild regularity assumptions on the marginal moments of $S_i^L$ and $S_i^U$ discussed in Section \ref{sec::theory}, $(\hat{\theta}_L, \hat{\theta}_U)$ is asymptotically bivariate Gaussian and the empirical covariance matrix of $(S_i^L, S_i^U)_{i\in \mathcal{D}_2}$ is a consistent estimate of the true asymptotic covariance matrix. While the superefficiency assumption in \cite{imbensmanski2004} does not necessarily hold in our case, we can apply the construction studied in Proposition 3 of \cite{stoye2009more} to guarantee the uniform coverage of the true parameter.

\bibliography{ref}
\bibliographystyle{apalike}

\appendix
\counterwithin{example}{section}
\renewcommand{\theHexample}{\theHsection.\arabic{example}}
\section{Proofs from Section \ref{sec::theory}}\label{appendix::proofs}

In this section we present the proofs of all theorems, lemmas, and corollaries in Section \ref{sec::theory}, some of which are followed by remarks. Some technical lemmas will be deferred to the end of each subsection. For notational convenience, we will denote by $n_1$ and $n_2$ the size of $\mcD_1$ and $\mcD_2$ and assume $n_2 / n > c$ for some constant $c > 0$ throughout as mentioned at the beginning of Section \ref{sec::theory}. For cross-fitting (Appendix \ref{appendix::crossfit_proofs}), we will assume $n_1 = n_2 = n/2$. Finally, we assume $\pi(X) \in [\Gamma, 1- \Gamma]$ for some $\Gamma > 0$ as implied by Assumption \ref{assump::rand_experiment}.

\subsection{Main proofs from Section \ref{subsec::validity}}\label{appendix::validity_proofs}

Although we give a proof sketch of Theorems \ref{thm::alwaysvalid} and \ref{thm::alwaysvalid_aug} in Section \ref{subsec::validity}, we give a few more details here for the sake of completeness. We also prove Corollary \ref{cor::multiboot_validity}.

\subsubsection{Proof of Theorem \ref{thm::alwaysvalid}}
As in Section \ref{subsec::validity}, we begin with the decomposition
\begin{equation*}
    \theta_L - \hat \theta\lcb = \underbrace{\theta_L - \tilde{\theta}_L}_{\text{Term A}} + \underbrace{\tilde{\theta}_L - \hat\theta\lcb}_{\text{Term B}}.
\end{equation*}
Term A is positive deterministically by weak duality. To analyze Term B, let $S_i = \frac{\hat \nu_{1,X_i}(Y_i) W_i}{\pi(X_i)} + \frac{\hat \nu_{0,X_i}(Y_i) (1-W_i)}{1 - \pi(X_i)}$ for $i \in \mcD_2$ and let $\hat\sigma_s$ be the sample standard deviation of $\{S_i\}_{i \in \mcD_2}$. Now, by construction, 
\begin{align*}
        \tilde{\theta}_L - \hat\theta\lcb 
    &= 
        \E\left[\hat \nu_{0,X_i}(Y_i(0)) + \hat \nu_{1,X_i}(Y_i(1)) \mid \mcD_1 \right] - \frac{1}{n_2} \sum_{i\in \mcD_2} S_i + \Phi^{-1}(1-\alpha) \frac{\hat \sigma_s}{\sqrt{n_2}} \\
    &=
        \E[S_i \mid \mcD_1] - \frac{1}{n_2} \sum_{i\in\mcD_2} S_i + \Phi^{-1}(1-\alpha) \frac{\hat \sigma_s}{\sqrt{n_2}}.
\end{align*}
One approach to analyze this sum would be to apply the standard univariate CLT conditional on $\mcD_1$ and to let $n_2$ grow to $\infty$. However, we must be slightly careful, because the rate of the convergence of the CLT depends on (e.g.) the higher moments of $S_i$, which depend on $\hat \nu$, and $\hat \nu$ changes with $n$ (since $\mcD_1$ changes with $n$). Instead, we will apply the Lyapunov CLT for triangular arrays.

Indeed, Assumption \ref{assumption::nu_moment} specifies that the conditional variance of $S_i$ is bounded away from zero and its fourth conditional moment is uniformly bounded. In particular, the latter follows because the fourth conditional moment of $\hat\nu(Y(k), X)$ is uniformly bounded and because of strict overlap. This moment condition, in combination with the fact that $\{S_i\}_{i \in \mcD_2}$ are i.i.d. conditional on $\mcD_1$, allows us to apply the Lyapunov CLT conditionally on $\mcD_1$: 
\begin{equation*}
    \frac{\sqrt{n_2}}{\sqrt{\var(S_i \mid \mcD_1)}} \left(\frac{1}{n_2} \sum_{i\in \mcD_2} (S_i - \E[S_i \mid \mcD_1])\right) \tod \mcN(0, 1).
\end{equation*}
Note that the Lyapunov CLT holds for any triangular array of random variables as long as the moment condition from Assumption \ref{assumption::nu_moment} holds; therefore, this convergence is uniform over $\mcP_B$. A similar argument based on the LLN for triangular arrays implies that $\hat \sigma_s \toprob \sqrt{\var(S_i \mid \mcD_1)}$ as $n \to \infty$, and furthermore that this convergence is uniform over $\mcP_B$.\footnote{In particular, since the fourth moment of $S_i$ given $\mcD_1$ is bounded, Chebyshev's inequality implies the uniform convergence of $\hat\sigma_s^2 \toprob \var(S_i \mid \mcD_1)$.} Then, Slutsky's theorem implies that
\begin{equation*}
    \frac{\sqrt{n_2}}{\hat \sigma_s} \left(\frac{1}{n_2} \sum_{i\in \mcD_2} (S_i - \E[S_i \mid \mcD_1])\right) \tod \mcN(0, 1).
\end{equation*}
This proves that $\liminf_{n \to \infty} \inf_{P \in \mcP_B} \P(\tilde{\theta}_L - \hat\theta\lcb \ge 0) = 1 - \alpha$, completing the proof.

\begin{remark}\label{remark::weaken_assump_nu_moment} We can substantially relax Assumption \ref{assumption::nu_moment} without changing the proof of Theorem \ref{thm::alwaysvalid}. The required condition need not hold with probability one. Let
\(\mu_n=\E[S_i\mid\mcD_1]\) and
\(\sigma_n^2=\var(S_i\mid\mcD_1)\).
It suffices that, for some \(\delta,\epsilon,B>0\),
\begin{equation}\label{eq:relaxed-lyapunov-condition}
\frac{\E[|S_i-\mu_n|^{2+\delta}\mid\mcD_1]}
{\sigma_n^{2+\delta}}
\le B n_2^{\delta/2-\epsilon}
\end{equation}
with probability approaching one, together with
\(\hat\sigma_s\toprob\sigma_n\). Then, asymptotically, we can apply the Lyapunov CLT conditional on $\mcD_1$ with probability uniformly approaching one for any distribution satisfying \eqref{eq:relaxed-lyapunov-condition}.

The only other time we use Assumption \ref{assumption::nu_moment} is to show \(\hat\sigma_s\toprob\sigma_n\) holds uniformly over $\mcP_B$. Here, we can again replace Assumption \ref{assumption::nu_moment} with any assumption guaranteeing uniform convergence (in probability) of $\hat\sigma_s$.
\end{remark}

\subsubsection{Proof of Corollary \ref{cor::multiboot_validity}}
We first review a result from \cite{chernozhukov2018multibootstrap} (labeled as Theorem 4.3 in the original paper).

\begin{proposition}[\cite{chernozhukov2018multibootstrap}]\label{prop::chernozhukov_mbs} Let $X_1, \dots, X_n \in \R^p$ be i.i.d. variables satisfying the following:
\begin{enumerate}[noitemsep, topsep=0pt]
    \item $\mu \defeq \E[X_1] \le 0$ holds elementwise, $\sigma_j^2 = \var(X_{1j})>0$. 
    \item Let $Z_{ij} \defeq \frac{X_{ij} - \mu_j}{\sigma_j}$ be the centered normalized variables and define
    $$B_n \defeq
    \left\{\max_{j \in [p]}
    \max\!\left(\E[|Z_{1j}|^4]^{1/2},\E[|Z_{1j}|^3]\right)\right\}
    \vee
    \E\!\left[\max_{j \in [p]} |Z_{1j}|^4\right]^{1/4}<\infty.$$
\end{enumerate}
Let $\hat \mu_j = \frac{1}{n} \sum_{i=1}^n X_{ij}$ and $\hat \sigma_j^2 = \frac{1}{n} \sum_{i=1}^n (X_{ij} - \hat \mu_j)^2$. Also, define the statistic
$$T = \max_{j \in [p]} \frac{\sqrt{n} \hat \mu_j}{\hat \sigma_j}$$ and let
$$T^{(b)}=\max_{j\in[p]}
\frac{n^{-1/2}\sum_{i=1}^n W_i^{(b)}(X_{ij}-\hat\mu_j)}
{\hat\sigma_j} $$
for $W_i \iid \mcN(0,1)$. Let $\hat q_{1-\alpha}$ denote the conditional $1-\alpha$ quantile of $T^{(b)}$ given $X_1, \dots, X_n$. Then if there exist constants $c_1 \in (0,1/2), C_1 \ge 0$ such that
\begin{equation}
    B_n^2 \log(p n)^{7/2} \le C_1 n^{1/2 - c_1},
\end{equation}
we have that there exist constants $c, C > 0$ depending only on $c_1, C_1$ such that
\begin{equation*}
    \P(T > \hat q_{1-\alpha}) \le \alpha + C n^{-c} \to \alpha.
\end{equation*}
\end{proposition}

Now we prove Corollary \ref{cor::multiboot_validity}. The proof is a straightforward application of Proposition \ref{prop::chernozhukov_mbs} and weak duality, but for completeness, we will state it here. Recall the notation from Section \ref{subsec::multbootstrap}: we let $\tilde{\theta}_L^{(k)} \defeq g(\hat \nu^{(k)})$, define $\tilde{\theta}_L = \max_{k\in [K]} \tilde{\theta}_L^{(k)}$, and define the summands
\begin{equation*}
    S_i^{(k)} \defeq \frac{\hat \nu_{1,X_i}^{(k)}(Y_i) W_i}{\pi(X_i)} + \frac{\hat \nu_{0,X_i}^{(k)}(Y_i) (1-W_i)}{1 - \pi(X_i)} \text{ for } k \in [K]. \end{equation*}
We also set $\bar S_k$ and $\hat \sigma_k^2$ to be the empirical mean and variance of $\{S_i^{(k)} : i \in \mcD_2\}$:
\begin{equation*}
    \bar S_k = \frac{1}{|\mcD_2|} \sum_{i \in \mcD_2} S_i^{(k)} \text{ and } \hat \sigma_k^2 = \frac{1}{|\mcD_2|}\sum_{i \in \mcD_2} (S_i^{(k)} - \bar S_k)^2.
\end{equation*}
For any $a \in \R$, define the test statistic
\begin{equation*}
    T(a) \defeq \max_{k \in [K]}\frac{\sqrt{|\mcD_2|}(\bar S_k - a)}{\hat \sigma_k}.
\end{equation*}
We also define the \textit{multiplier bootstrap variant}: for $\xi_i \iid \mcN(0,1)$, we define
\begin{align*}
        T^{(b)}(a) 
    &\defeq 
        \max_{k \in [K]} \frac{|\mcD_2|^{-1/2} \sum_{i \in \mcD_2} \xi_i (S_i^{(k)} - a - (\bar S_k - a))}{\hat \sigma_k} = T^{(b)}
\end{align*}
where we note that $T^{(b)}(a)$ does not depend on $a$, so we abbreviate it by $T^{(b)}$. Let $\hat q_{1-\alpha} \defeq Q_{1-\alpha}(T^{(b)} \mid \mcD)$ denote the conditional quantile of $T^{(b)}$ given all the data. 

Note that for any $a \ge \tilde{\theta}_L$, $\E[\bar S - a \mathbf{1}_K \mid \mcD_1] \le 0$ holds elementwise. This, combined with Assumption \ref{assump::mbs}, allows us to apply Proposition \ref{prop::chernozhukov_mbs} conditional on $\mcD_1$. Thus, there exist universal constants $c, C > 0$ such that
\begin{equation}
    \sup_{P \in \mcP_{c,\epsilon}} \P_P\left(T(\tilde{\theta}_L) > \hat q_{1-\alpha} \mid \mcD_1 \right) \le \alpha + C n^{-c}.
\end{equation}
Applying the tower property and taking limits yields
\begin{equation}\label{eq::chern_cor}
    \limsup_{n \to \infty} \sup_{P \in \mcP_{c,\epsilon}} \P_P(T(\tilde{\theta}_L) > \hat q_{1-\alpha}) \le \alpha.
\end{equation}

Recall also the definition of $\hat\theta\lcb\mb$ from Eq. (\ref{eq::lcb_mbs}):
\begin{align*}
        \hat\theta\lcb\mb 
    &\defeq 
        \max_{k \in [K]} \left\{\bar S_k - \hat q_{1-\alpha}
        \frac{\hat \sigma_k}{\sqrt{|\mcD_2|}}\right\} \\
    &=
        \sup\left\{a \in \R : \max_{k \in [K]} \frac{\sqrt{|\mcD_2|} (\bar S_k - a)}{\hat \sigma_k} > \hat q_{1-\alpha} \right\} \\
    &= 
        \sup\left\{a \in \R : T(a) > Q_{1-\alpha}(T^{(b)} \mid \mcD)\right\}.
\end{align*}
Since $T(a)$ is decreasing in $a$, we have that
\begin{equation*}
    \hat\theta\lcb\mb > \tilde{\theta}_L \Leftrightarrow T(\tilde{\theta}_L) > Q_{1-\alpha}(T^{(b)} \mid \mcD).
\end{equation*}
Therefore, by Eq. (\ref{eq::chern_cor}), we conclude
\begin{equation*}
    \limsup_{n \to \infty} \sup_{P \in \mcP_{c,\epsilon}} \P_P(\hat\theta\lcb\mb > \tilde{\theta}_L) \le \limsup_{n \to \infty} \sup_{P \in \mcP_{c,\epsilon}} \P_P(T(\tilde{\theta}_L) > \hat q_{1-\alpha}) \le \alpha.
\end{equation*}
Since $\tilde{\theta}_L \le \theta_L$ by weak duality, this completes the proof.

\subsection{Proof of Theorem \ref{thm::inprodbound}}

As notation, for any functions $f_0, f_1, h_0, h_1 : \mcY \times \mcX \to \R$ and any $x \in \mcX$, define the inner product
\begin{equation}\label{eq:proof-inner-product}
    \langle (f_0, f_1), (h_0, h_1) \rangle_x = \sum_{k \in \{0,1\}} \int_{y \in \mcY} f_k(y, x) h_k(y,x) \psi(dy).
\end{equation}
For \(q\in[1,\infty]\), let \(\|(f_0,f_1)\|_{q,x}\) denote the \(L^q\) norm over
\(\{0,1\}\times\mcY\) induced by counting measure on \(\{0,1\}\) and \(\psi\) on \(\mcY\).
The definitions in Section \ref{subsec::tightness} imply that
\begin{equation}
    \error_{P,r}(x) = \|\hat p-p\opt\|_{r,x}
    \quad\text{and}\quad
    \error_{\nu,s}(x) = \|\hat\nu-\nu\opt\|_{s,x}.
\end{equation}
where $\hat p = (\hat{p}_0, \hat{p}_1) : \mcY \times \mcX \to \R^2$ denotes the estimated conditional densities of $Y(k) \mid X, k \in \{0,1\}$ and $p\opt = (p_0\opt, p_1\opt) : \mcY \times \mcX \to \R^2$ denotes the true conditional densities. With this notation, let $\hat g : \mcV \to \R$ denote the estimate of the dual which plugs in $\hat p$ for $p\opt$:
\begin{align*}
        \hat g(\nu) 
    &\defeq
        \E_{\hatPZC \times P_X\opt}[\nu_{0,X}(Y(0))] + \E_{\hatPOC \times P_X\opt}[\nu_{1,X}(Y(1))] \\
    &=
        \E_{X \sim P_X\opt}\left[\langle \hat p, \nu \rangle_X \right],
\end{align*}
where in the inner product in \eqref{eq:proof-inner-product}, we think of $\nu = (\nu_{0,x}, \nu_{1,x})_{x \in \mcX}$ as a function $\nu : \mcY \times \mcX \to \R^2$ defined by $\nu_{k,x}(y) = \nu_{k,x}(y)$. With this notation, we observe that
\begin{align}
        \theta_L - \tilde{\theta}_L 
    &=
        g(\nu\opt) - g(\hat \nu) & \text{ by strong duality and defn. of } \tilde{\theta}_L \notag\\
    &\le 
        g(\nu\opt) - \hat g(\nu\opt) + \hat g(\hat \nu) - g(\hat \nu) & \text{ since }  \hat\nu \defeq \argmax_{\nu} \hat g(\nu) \notag\\
    &=
        \E_{X \sim P_X\opt}\left[
        \langle p\opt - \hat p, \nu\opt - \hat\nu \rangle_X
        \mid \mcD_1 \right] & \text{ using linearity of inner products } \label{eq:inprod-linearity-step}\\
    &\le 
        \E_{X \sim P_X\opt}\left[\|p\opt - \hat p\|_{r,X} \|\nu\opt - \hat \nu\|_{s,X}\mid \mcD_1 \right] & \text{by H\"older's inequality} \notag\\
    &=
        \E_{X \sim P_X\opt}\left[\error_{P,r}(X) \cdot \error_{\nu,s}(X) \mid \mcD_1\right] & \text{by definition.} \notag
\end{align}
The equality in \eqref{eq:inprod-linearity-step} uses the fact that for any fixed $\nu$, $\hat g(\nu) - g(\nu) = \E_{X \sim P_X\opt}[\langle \hat p -p\opt, \nu \rangle_X\mid \mcD_1].$ Throughout, we condition on $\mcD_1$ since $\tilde\theta_L$ and $\hat \nu$ are random and $\mcD_1$-measurable. This completes the proof.

\subsection{Main proofs from Section \ref{subsec::crossfit}}\label{appendix::crossfit_proofs}

\subsubsection{Proof of Theorem \ref{thm::crossfit_validity}}

We handle the two conditions separately. 

\textbf{Proof under Condition~\ref{cond:crossfit-stable}:} We first introduce some notation. Define the summand
\begin{equation*}
    S_i = \begin{cases} \frac{\hat \nu\swap_{1,X_i}(Y_i) W_i}{\pi(X_i)} + \frac{\hat\nu\swap_{0,X_i}(Y_i) (1-W_i)}{1 - \pi(X_i)} & i \in \mcD_1 \\ \frac{\hat \nu_{1,X_i}(Y_i) W_i}{\pi(X_i)} + \frac{\hat\nu_{0,X_i}(Y_i) (1-W_i)}{1 - \pi(X_i)} & i \in \mcD_2 \end{cases}
\end{equation*}
so that by definition, 
\begin{equation*}
    \hat\theta_L = \frac{1}{|\mcD_2|} \sum_{i \in \mcD_2} S_i \text{ and } \hat \theta_L\swap = \frac{1}{|\mcD_1|} \sum_{i \in \mcD_1} S_i \text{ and } \hat\theta_L\crossfit = \bar S = \frac{\hat\theta_L + \hat\theta_L\swap}{2}.
\end{equation*}
and $\hat\theta\lcb\crossfit = \hat\theta_L\crossfit - \Phi^{-1}(1-\alpha) \frac{\hat\sigma_s}{\sqrt{n}}$, where throughout this proof,  $\hat\sigma_s$ is the sample standard deviation of $\{S_i\}_{i=1}^n$.

Now, we will compare $\hat \theta\lcb$ to an oracle estimator. Define $S_i\conv$ to be the analogue of $S_i$, but replacing $\hat \nu$ and $\hat\nu\swap$ with $\nu\conv$:
\begin{equation*}
    S_i\conv \defeq \frac{\nu\conv_{1,X_i}(Y_i) W_i}{\pi(X_i)} + \frac{\nu\conv_{0,X_i}(Y_i) (1-W_i)}{1 - \pi(X_i)}
\end{equation*}
and the oracle estimator and lower confidence bound are defined as 
\begin{equation*}
    \hat\theta\conv_L \defeq \frac{1}{n} \sum_{i=1}^n S_i\conv \text{ and } \hat\theta\conv\lcb \defeq \hat\theta\conv_L - \Phi^{-1}(1-\alpha) \frac{\hat\sigma_s\conv}{\sqrt{n}}
\end{equation*}
where $\hat\sigma_s\conv$ is the sample standard deviation of $\{S_i\conv\}_{i = 1}^n$.

$\hat\theta\lcb\conv$ is clearly a valid $1-\alpha$ lower confidence bound on $\E[S_i\conv]$ by the univariate central limit theorem; thus, by weak duality, $\hat\theta\lcb\conv$ is a valid $1-\alpha$ lower confidence bound on $\theta_L$ (see Theorem \ref{thm::alwaysvalid}). Thus, a standard proof technique in the literature is to show that $\hat\theta\lcb\crossfit$ is asymptotically equivalent to $\hat\theta\lcb\conv$. However, this is \textit{not} true in this setting: in general, $\sqrt{n}(\hat\theta\lcb\conv - \hat\theta\lcb\crossfit) \not \toprob 0$. Nonetheless, a careful application of weak duality will allow us to show the desired result. In particular, $\hat\theta\lcb\crossfit$ may have more fluctuations than $\hat\theta\lcb\conv$, but weak duality will guarantee that $\hat\theta\lcb\crossfit$ will not fluctuate above $\theta_L$ with probability more than $\alpha$ asymptotically.

In particular, Lemma \ref{lem::crossfit_validity} proves the standard result that
\begin{align}\label{eq::crossfit_magic}
        \hat\theta\lcb\crossfit - \hat\theta\lcb\conv 
    &= 
        \frac{\E[\hat\theta_L \mid \mcD_1] + \E[\hat\theta_L\swap \mid \mcD_2]}{2} - \E[S_i\conv] + o_p(n^{-1/2}) \\
    &=
        \frac{g(\hat\nu) + g(\hat\nu\swap)}{2} - g(\nu\conv) + o_p(n^{-1/2}). \nonumber
\end{align}
where $g$ is the Kantorovich dual function defined in Section
\ref{sec::method}. By weak duality,
$$
\frac{g(\hat\nu)+g(\hat\nu\swap)}{2}\le \theta_L.
$$
Combining this inequality with \eqref{eq::crossfit_magic}, there exists
$r_n=o_p(n^{-1/2})$ such that
$$
\hat\theta\lcb\crossfit-\theta_L
\le
\hat\theta\lcb\conv-g(\nu\conv)+r_n.
$$
Consequently,
\begin{align}\limsup_{n\to\infty}\P(\hat\theta\lcb\crossfit>\theta_L)
&\le\limsup_{n\to\infty}
\P\!\left(\hat\theta\lcb\conv-g(\nu\conv)+r_n>0
\right)\le \alpha.
\label{eq:crossfit-tail-probability}
\end{align}
The second inequality in \eqref{eq:crossfit-tail-probability} follows from the Lyapunov central limit theorem for
the deterministic oracle summands $S_i\conv$, the consistency of
$\hat\sigma_s\conv$, and the variance lower bound in Assumption
\ref{assumption::nu_moment}. Therefore
$$
\liminf_{n\to\infty}
\P(\hat\theta\lcb\crossfit\le\theta_L)
\ge 1-\alpha.
$$ 

\textbf{Proof under Condition~\ref{cond:crossfit-bias}:} Under this condition, a similar proof as Lemma \ref{lem::validity_cond2} shows that
\begin{equation*}
    \hat\theta_L = g(\hat \nu) + O_p(n^{-1/2}), \hat\theta_L\swap = g(\hat\nu\swap) + O_p(n^{-1/2}) \text{ and } \hat\sigma_s = O_p(1).
\end{equation*}
As a result, we have that
\begin{equation*}
    \hat\theta\lcb\crossfit = \frac{g(\hat \nu) + g(\hat \nu\swap)}{2} + O_p(n^{-1/2})  = \widetilde\theta_L^{\mathrm{crossfit}} +  O_p(n^{-1/2}).
\end{equation*}
Condition~\ref{cond:crossfit-bias} tells us that $\theta_L - \tilde{\theta}_L \defeq \theta_L - g(\hat \nu) \gg n^{-1/2}$, and the same result holds with $g(\hat\nu\swap)$ replacing $g(\hat\nu)$ by symmetry. Therefore
\begin{equation*}
    \hat\theta\lcb\crossfit = \theta_L + \widetilde\theta_L^{\mathrm{crossfit}} - \theta_L + O_p(n^{-1/2}).
\end{equation*}
Since $\sqrt{n}(\theta_L - \widetilde\theta_L^{\mathrm{crossfit}})\toprob \infty$, the second term is deterministically nonpositive by weak duality and dominates the $O_p(n^{-1/2})$ term, we conclude
\begin{equation*}
    \liminf_n \P(\hat\theta\lcb\crossfit \le \theta_L) = 1 \ge 1 - \alpha
\end{equation*}
which proves the desired result.

\subsubsection{Proof of Corollary \ref{corr::crossfit_tightness}}

Under the finite-support alternative, Theorem~\ref{prop:discrete-lower-level}
gives
\begin{equation*}
0\le \theta_L-g(\hat\nu)
\le
\E\!\left[
\overline{\error}_{P,2}(X)
\omega_X\!\left(\overline{\error}_{P,2}(X)\right)
\mid \mcD_1
\right]
=o_p(n^{-1/2}).
\end{equation*}
Under the example-specific alternative, the corresponding result in
Appendix~\ref{sec:proofs-structured-margins} gives the same fold-specific
conclusion \(0\le \theta_L-g(\hat\nu)=o_p(n^{-1/2})\). Applying the same
argument with the roles of $\mcD_1$ and $\mcD_2$ reversed gives
$\theta_L-g(\hat\nu\swap)=o_p(n^{-1/2})$. Therefore
\begin{equation}
0\le
\theta_L-\tilde\theta_L\crossfit
=
\frac{\theta_L-g(\hat\nu)}{2}
+\frac{\theta_L-g(\hat\nu\swap)}{2}
=o_p(n^{-1/2}),
\end{equation}
which implies
\(\sqrt{n}(\tilde\theta_L\crossfit-\theta_L)=o_p(1)\).

\subsubsection{Technical lemmas}
\begin{lemma}\label{lem::crossfit_validity} Assume the conditions and notation of Theorem \ref{thm::crossfit_validity}. Then
\begin{equation*}
        \hat\theta\lcb\crossfit - \hat\theta\lcb\conv 
    = 
        \frac{g(\hat\nu) + g(\hat\nu\swap)}{2} - g(\nu\conv) + o_p(n^{-1/2}).
\end{equation*}
\end{lemma}
\begin{proof} First, we observe that
\begin{align}
        \hat\theta\lcb\crossfit - \hat\theta\lcb\conv 
    &= 
        \hat\theta_L\crossfit - \hat\theta_L\conv - \frac{\Phi^{-1}(1-\alpha)}{\sqrt{n}} \left[\hat\sigma_s - \hat\sigma_s\conv \right] \\
    &=
        \hat\theta_L\crossfit - \hat\theta_L\conv + o_p(n^{-1/2})
        \label{eq:crossfit-lcb-reduction}
\end{align}
where the last line follows from Condition~\ref{cond:crossfit-stable}, since
\[
|\hat\sigma_s-\hat\sigma_s\conv|
\le
\left\{\frac1n\sum_{i=1}^n\bigl[(S_i-\bar S)-(S_i\conv-\bar S\conv)\bigr]^2\right\}^{1/2}
\le
\left\{\frac1n\sum_{i=1}^n(S_i-S_i\conv)^2\right\}^{1/2}
=o_p(1).
\]

Next, we observe
\begin{align}
        \hat\theta_L\crossfit - \hat\theta_L\conv
    &=
        \frac{1}{2 |\mcD_1|} \sum_{i \in \mcD_1} (S_i-S_i\conv) + \frac{1}{2 |\mcD_2|} \sum_{i \in \mcD_2} (S_i-S_i\conv).
        \label{eq:crossfit-two-sums}
\end{align}
For simplicity, we focus on the first sum in \eqref{eq:crossfit-two-sums}. Note that $\{S_i - S_i\conv\}_{i \in \mcD_1}$ are i.i.d. conditional on $\mcD_2$. We will apply Chebyshev's inequality to this sum; to do this, we analyze its mean and variance.\begin{enumerate}
    \item \textit{Mean}: Since $\pi(X_i)$ are known propensity scores, we have the exact result that
    \begin{equation*}
        \E[S_i \mid \mcD_2] - \E[S_i\conv \mid \mcD_2] = g(\hat\nu\swap) - g(\nu\conv).
    \end{equation*}
    \item \textit{Variance}: Observe that
    \begin{align*}
            \var(S_i - S_i\conv \mid \mcD_2)
        &\le 
            \E[(S_i - S_i\conv)^2 \mid \mcD_2] \\
        &=
            \E\left[\left(\frac{W_i (\hat \nu\swap_{1,X_i}(Y_i) - \nu\conv_{1,X_i}(Y_i))}{\pi(X_i)}\right)^2 \mid \mcD_2 \right] \\
        &+
            \E\left[\left(\frac{(1-W_i) (\hat \nu\swap_{0,X_i}(Y_i) - \nu\conv_{0,X_i}(Y_i))}{1-\pi(X_i)}\right)^2 \mid \mcD_2 \right] \\
        &\le 
            \Gamma^{-2} \sum_{k \in \{0,1\}} \E[(\hat \nu\swap_{k,X_i}(Y_i) - \nu\conv_{k,X_i}(Y_i))^2 \mid \mcD_2].
    \end{align*}
    However, we assume that $\E[(\hat \nu\swap_{k,X_i}(Y_i) - \nu\conv_{k,X_i}(Y_i))^2] \to 0$ for $k \in \{0,1\}$. Thus, $\E[ \var(S_i - S_i\conv \mid \mcD_2)] \to 0$ and thus $ \var(S_i - S_i\conv \mid \mcD_2) = o_p(1)$.
\end{enumerate}
From this analysis, we conclude by Chebyshev's inequality that
\begin{equation}\label{eq:crossfit-first-sum}
    \frac{1}{2 |\mcD_1|} \sum_{i \in \mcD_1} (S_i-S_i\conv)  = \frac{g(\hat \nu\swap) - g(\nu\conv)}{2} + o_p(n^{-1/2}).
\end{equation}
The same analysis applied to the other sum yields that
\begin{equation}\label{eq:crossfit-second-sum}
    \frac{1}{2 |\mcD_2|} \sum_{i \in \mcD_2} (S_i-S_i\conv)  = \frac{g(\hat \nu) - g(\nu\conv)}{2} + o_p(n^{-1/2}).
\end{equation}
Combining \eqref{eq:crossfit-lcb-reduction}, \eqref{eq:crossfit-first-sum}, and \eqref{eq:crossfit-second-sum} yields the desired result
\begin{align*}
        \hat\theta\lcb\crossfit - \hat\theta\lcb\conv 
    &=
        \hat\theta_L\crossfit - \hat\theta_L\conv + o_p(n^{-1/2}) \\
    &=
        \frac{g(\hat \nu\swap) + g(\hat \nu)}{2} - g(\nu\conv) + o_p(n^{-1/2}).
\end{align*}
\end{proof}

\subsection{Main proofs from Section \ref{subsec::aipw}}\label{appendix::aipw_proofs}

\subsubsection{Proof of Theorem \ref{thm::alwaysvalid_aug}}

Unlike the standard analysis of AIPW estimators for average treatment effects, the proof uses conditional studentization of the actual AIPW summands rather than asymptotic equivalence to an oracle summand.

\textbf{Analysis of Condition~\ref{cond:aipw-product}:}
Lemma \ref{lem::validity_cond1} shows directly that
\[
\liminf_{n\to\infty}\P\!\left(\hat\theta\lcb\aug\le g(\hat\nu)\right)\ge 1-\alpha.
\]
Weak duality gives $g(\hat\nu)\le\theta_L$ deterministically, proving the desired coverage under Condition~\ref{cond:aipw-product}.

\textbf{Analysis of Condition~\ref{cond:aipw-bias}:} Lemma \ref{lem::validity_cond2} shows that

    \begin{equation}\label{eq:aipw-condition-two-rate}
    \hat\theta\aug\lcb - \tilde{\theta}_L = O_p(n^{-1/2} + \min(\errpi, \errc)).
    \end{equation}

Combining \eqref{eq:aipw-condition-two-rate} directly with Condition~\ref{cond:aipw-bias} gives
\[
\hat\theta\aug\lcb-\tilde\theta_L
=O_p\!\left(n^{-1/2}+\min(\errpi,\errc)\right)
=o_p(\theta_L-\tilde\theta_L).
\]
Thus, $\theta_L - \tilde{\theta}_L$ dominates $\hat\theta\aug\lcb - \tilde{\theta}_L$. Furthermore, weak duality implies that $\theta_L \ge \tilde{\theta}_L$ deterministically. Thus, using the decomposition
\begin{equation*}
    \theta_L - \hat \theta\lcb\aug = \underbrace{\theta_L - \tilde{\theta}_L}_{\text{strictly positive}} + \underbrace{\tilde{\theta}_L - \hat\theta\lcb\aug}_{\text{negligible}},
\end{equation*}
we conclude that $\liminf_{n \to \infty} \P(\hat\theta\lcb\aug \le \theta_L) = 1$. As a result, under this form of misspecification, the dual bound $\hat\theta\lcb\aug$ is valid and in fact very conservative.

\subsubsection{Technical lemmas}\label{subsec::validity_technical}

\begin{lemma}\label{lem::validity_cond1}
Under Condition~\ref{cond:aipw-product},
\[
\liminf_{n\to\infty}\P\!\left(\hat\theta\lcb\aug\le g(\hat\nu)\right)\ge 1-\alpha.
\]
\end{lemma}
\begin{proof}
Let
\[
\mu_S\defeq\E[S_i\mid\mcD_1],
\qquad
\sigma_S^2\defeq\var(S_i\mid\mcD_1),
\qquad
\hat\theta_L\aug\defeq\frac1{|\mcD_2|}\sum_{i\in\mcD_2}S_i.
\]
Using the definitions of $c_0$ and $c_1$ and conditioning first on $(X_i,\mcD_1)$ gives
\begin{align*}
\mu_S-g(\hat\nu)
={}&\E\!\left[
(\hat\pi(X)-\pi(X))
\left\{
\frac{\hat c_1(X)-c_1(X)}{\hat\pi(X)}
-\frac{\hat c_0(X)-c_0(X)}{1-\hat\pi(X)}
\right\}
\mathrel{\Big|}\mcD_1
\right].
\end{align*}
Thus, by overlap and Cauchy--Schwarz,
\[
|\mu_S-g(\hat\nu)|\le C\errpi\errc.
\]
Moreover,
\[
\E[\errpi\errc]
\le\{\E[\errpi^2]\E[\errc^2]\}^{1/2}
=o(n^{-1/2}),
\]
so Markov's inequality implies $\mu_S-g(\hat\nu)=o_p(n^{-1/2})$.

Conditional on $\mcD_1$, the variables $\{S_i:i\in\mcD_2\}$ are i.i.d. The fourth-moment and overlap assumptions give a uniform conditional fourth-moment bound, while Condition~\ref{cond:aipw-product} gives $\sigma_S^2\ge1/B$. Hence the conditional Lyapunov central limit theorem, the conditional law of large numbers, and Slutsky's theorem yield
\[
\sup_{t\in\mathbb R}
\left|
\P\!\left(
\frac{\sqrt{|\mcD_2|}(\hat\theta_L\aug-\mu_S)}{\hat\sigma_s\aug}
\le t
\mathrel{\Big|}\mcD_1
\right)-\Phi(t)
\right|\toprob0.
\]
Since
\[
\frac{\sqrt{|\mcD_2|}\{\mu_S-g(\hat\nu)\}}{\hat\sigma_s\aug}=o_p(1),
\]
the definition of $\hat\theta\lcb\aug$ implies the stated result.
\end{proof}

\begin{lemma}\label{lem::validity_cond2} Assume the conditions of Theorem \ref{thm::alwaysvalid_aug} except for Condition~\ref{cond:aipw-product}. Then
\begin{equation*}
    \tilde{\theta}_L - \hat\theta\lcb\aug = O_p(n^{-1/2}) + O_p(\min(\errpi, \errc)).
\end{equation*}
\end{lemma}
\begin{proof} Throughout the proof, let $(Y(1),Y(0),X)$ be a draw from the distribution of $(Y_i(1),Y_i(0),X_i)$ that is independent of the data. Recall also that $n_2 \ge cn$ by assumption.

The proof has two steps. In \ref{step:aipw-error-rate}, we show that $\tilde\theta_L - \hat\theta\aug_L = O_p(n^{-1/2}) + O_p(\min(\errpi, \errc))$. Then, \ref{step:aipw-lcb} gives the final result by replacing $\hat\theta\aug_L$ with $\hat\theta\aug\lcb$.

\begin{enumerate}[label=\underline{Step \arabic*:},ref=Step~\arabic*,leftmargin=*]
\item\label{step:aipw-error-rate} We show that $\tilde\theta_L - \hat\theta\aug_L = O_p(n^{-1/2}) + O_p(\min(\errpi, \errc))$. We establish the two component rates in \ref{part:aipw-outcome-rate} and \ref{part:aipw-propensity-rate}. Specifically, \ref{part:aipw-outcome-rate} proves the outcome-regression rate
\[
\hat\theta_L\aug-\tilde\theta_L=O_p(n^{-1/2}+\errc),
\]
whereas \ref{part:aipw-propensity-rate} proves the propensity-score rate
\[
\hat\theta_L\aug-\tilde\theta_L=O_p(n^{-1/2}+\errpi).
\]
Taking the smaller of these two remainder bounds yields the claimed rate.

\begin{enumerate}[label=\textit{Part (\alph*):},ref=Part~(\alph*),leftmargin=*]
\item\label{part:aipw-outcome-rate} Define
$$
e_k(x)=\hat c_k(x)-c_k(x),\qquad k\in\{0,1\}.
$$
For each $i\in\mcD_2$, the augmented summand admits the exact
decomposition
\begin{align*}
S_i
={}&c_1(X_i)+c_0(X_i)\\
&+\frac{W_i}{\hat\pi(X_i)}
  \{\hat\nu_{1,X_i}(Y_i)-c_1(X_i)\}\\
&+\frac{1-W_i}{1-\hat\pi(X_i)}
  \{\hat\nu_{0,X_i}(Y_i)-c_0(X_i)\}\\
&+e_1(X_i)\left\{1-\frac{W_i}{\hat\pi(X_i)}\right\}\\
&+e_0(X_i)\left\{1-\frac{1-W_i}{1-\hat\pi(X_i)}\right\}.
\end{align*}
Conditional on $\mcD_1$, the empirical average of
$c_1(X_i)+c_0(X_i)$ equals
$\widetilde\theta_L+O_p(n^{-1/2})$. Moreover, by the definitions of
$c_1$ and $c_0$,
$$
\E[\hat\nu_{k,X}(Y(k))-c_k(X)\mid\mcD_1,X]=0,
\qquad k\in\{0,1\}.
$$
The next two weighted residual averages are therefore
$O_p(n^{-1/2})$ under the stated moment and overlap assumptions.

The last two terms are not generally conditionally mean zero. However,
uniform overlap and Cauchy--Schwarz imply
\begin{align*}
\E\!\left[
\left|e_1(X)\left\{1-\frac{W}{\hat\pi(X)}\right\}\right|
\middle|\mcD_1\right]
&\le C\,\E[e_1(X)^2\mid\mcD_1]^{1/2},\\
\E\!\left[
\left|e_0(X)\left\{1-\frac{1-W}{1-\hat\pi(X)}\right\}\right|
\middle|\mcD_1\right]
&\le C\,\E[e_0(X)^2\mid\mcD_1]^{1/2}.
\end{align*}
Their empirical averages are consequently
$O_p(\errc+n^{-1/2})$. Combining the terms yields$$\hat\theta_L\aug-\widetilde\theta_L=O_p(\errc+n^{-1/2}),$$
which proves \ref{part:aipw-outcome-rate}.

\item\label{part:aipw-propensity-rate} In this case, we can decompose
\begin{align*}
        \hat\theta_L\aug
    &\defeq
        \frac{1}{n_2} \sum_{i \in \mcD_2} \left\{\hat c_1(X_i) + \hat c_0(X_i) + \frac{W_i}{\hat \pi(X_i)}(\hat \nu_{1,X_i}(Y_i) - \hat c_1(X_i)) + \frac{1-W_i}{1 - \hat \pi(X_i)} (\hat \nu_{0,X_i}(Y_i) - \hat c_0(X_i))\right\} \\
    &=
        \underbrace{\frac{1}{n_2} \sum_{i \in \mcD_2} \left\{\hat c_1(X_i) \left[1 - \frac{W_i}{\pi(X_i)} \right] + \hat c_0(X_i)\left[1 - \frac{1-W_i}{1-\pi(X_i)} \right]\right\}}_{\text{term~\ref{term:aipw-outcome-centering}}}  \\
    &+
        \underbrace{\frac{1}{n_2} \sum_{i \in \mcD_2} \left\{\frac{W_i \hat \nu_{1,X_i}(Y_i)}{\pi(X_i)} + \frac{(1-W_i) \hat \nu_{0,X_i}(Y_i)}{1-\pi(X_i)}\right\}}_{\text{term~\ref{term:aipw-ipw}}}  \\
    &+
        \underbrace{\frac{1}{n_2} \sum_{i \in \mcD_2} \left(\hat \pi(X_i)^{-1} - \pi(X_i)^{-1}\right) W_i (\hat \nu_{1,X_i}(Y_i) - \hat c_1(X_i))}_{\text{term~\ref{term:aipw-treated-propensity}}} \\
    &+
        \underbrace{\frac{1}{n_2} \sum_{i \in \mcD_2} \left\{(1-\hat \pi(X_i))^{-1} - (1 - \pi(X_i))^{-1}\right\} (1 - W_i) (\hat \nu_{0,X_i}(Y_i) - \hat c_0(X_i))}_{\text{term~\ref{term:aipw-control-propensity}}}.
\end{align*}
To analyze these terms, observe that
\begin{enumerate}[label=\arabic*.,ref=\arabic*]
    \item\label{term:aipw-outcome-centering} The summands in term~\ref{term:aipw-outcome-centering} are mean zero and i.i.d. conditional on $\mcD_1$. Furthermore, their $2+\delta$ moment is uniformly bounded conditional on $\mcD_1$ since the $2+\delta$ moments of $\hat c_1, \hat c_0$ are uniformly bounded and $\pi(X_i)$ is bounded away from zero and one. Chebyshev's inequality thus implies that term~\ref{term:aipw-outcome-centering} is $O_p(n^{-1/2})$.
    \item\label{term:aipw-ipw} Term~\ref{term:aipw-ipw} is simply an IPW estimator of $\tilde{\theta}_L$, so under the fourth-moment and overlap assumptions of Theorem \ref{thm::alwaysvalid_aug} it converges to $\tilde{\theta}_L$ plus $O_p(n^{-1/2})$.
    \item\label{term:aipw-treated-propensity} Term~\ref{term:aipw-treated-propensity} is an i.i.d. sum conditional on $\mcD_1$. Its summands have uniformly bounded $2+\delta$ moment because the fourth-moment conditions in Theorem \ref{thm::alwaysvalid_aug} imply that $\hat \nu_{1,X}(Y(1)) - \hat c_1(X)$ has a uniformly bounded $2 + \delta$ moment and $\hat \pi(X_i)^{-1}, \pi(X_i)^{-1}$ are uniformly bounded. Furthermore, H\"older's inequality yields that
    \begin{align*}
        &
            \E\left[\left|\left(\hat \pi(X_i)^{-1} - \pi(X_i)^{-1}\right) \left(W_i (\hat \nu_{1,X_i}(Y_i) - \hat c_1(X_i))\right)\right| \mid \mcD_1 \right] \\
        \le&
            \sqrt{\E[\left(\hat \pi(X_i)^{-1} - \pi(X_i)^{-1}\right)^2 \mid \mcD_1] \E\left[\left(W_i (\hat \nu_{1,X_i}(Y_i) - \hat c_1(X_i)\right)^2 \mid \mcD_1 \right]} \\
        =&
            O\left(\sqrt{\E[\left(\hat \pi(X_i) - \pi(X_i)\right)^2 \mid \mcD_1]} \right) \\
        \defeq&
            O(\errpi).
    \end{align*}
    where the penultimate line follows because (a) $\E\left[\left(W_i (\hat \nu_{1,X_i}(Y_i) - \hat c_1(X_i))\right)^2 \mid \mcD_1 \right]$ is uniformly bounded and (b) $\hat \pi, \pi$ are uniformly bounded away from zero, and the function $x \mapsto x^{-1}$ is Lipschitz on $[\Gamma, \infty)$. Thus, $|\hat \pi(X_i)^{-1} - \pi(X_i)^{-1}| \le L |\hat \pi(X_i) - \pi(X_i)|$ where $L$ is some Lipschitz constant not depending on $n$. 

    Applying Chebyshev's inequality to term~\ref{term:aipw-treated-propensity} yields that it is $O_p(\errpi) + O_p(n^{-1/2})$.
    \item\label{term:aipw-control-propensity} Term~\ref{term:aipw-control-propensity} is $O_p(\errpi) + O_p(n^{-1/2})$ for the same reason that term~\ref{term:aipw-treated-propensity} is.
\end{enumerate}

\end{enumerate}

The analysis in \ref{part:aipw-propensity-rate} yields $\hat\theta\aug_L = \tilde{\theta}_L + O_p(n^{-1/2}) + O_p(\errpi)$. Combining \ref{part:aipw-outcome-rate} and \ref{part:aipw-propensity-rate} implies that
\begin{equation*}
    \hat\theta\aug_L - \tilde{\theta}_L = O_p(n^{-1/2}) + O_p(\min(\errpi, \errc)).
\end{equation*}

\item\label{step:aipw-lcb} To complete the proof of the lemma, we note that
\begin{equation*}
    \hat\theta\aug\lcb - \tilde{\theta}_L = \hat\theta\aug_L - \tilde{\theta}_L - \Phi^{-1}(1 - \alpha) \frac{\hat \sigma_s\aug}{\sqrt{n_2}}.
\end{equation*}
Thus, to prove the lemma, it suffices to prove that $\hat \sigma_s\aug = O_p(1)$. To do this, we observe that $\hat \sigma_s\aug$ is defined as
\begin{equation*}
    (\hat\sigma_s\aug)^2 = \frac{1}{n_2} \sum_{i \in \mcD_2} (S_i - \bar S)^2.
\end{equation*}
Since the $S_i$ are i.i.d. and have a uniformly bounded $2+\delta$ moment conditional on $\mcD_1$, a uniform LLN applied to $S_i$ and then $S_i^2$ yields that $\hat\sigma_s\aug = \sqrt{\var(S_i \mid \mcD_1)} + o_p(1)$ where $\var(S_i \mid \mcD_1)$ is uniformly bounded by the theorem assumptions. This proves that $\hat\sigma_s\aug = O_p(1)$. 
\end{enumerate}
\end{proof}

\section{Deriving Kantorovich Duals for Examples in Section~\ref{subsec::examples}}
\label{sec: dual form}
For concrete estimands, the dual problems can often be written in explicit form or significantly simplified. Here, we derive the dual function for each of our examples explicitly. 
\subsection{Example \ref{ex::fh}}\label{app:dual-fh} For the Fr\'echet-Hoeffding bound, fix $(y_1,y_0)\in\mathbb R^2$. The cost function is $f(y_0',y_1',x)=\mathbf 1\{y_0'\le y_0,\ y_1'\le y_1\}.$ Define the conditional marginal probabilities
\[
p_0(x)=\mathbb P(Y(0)\le y_0\mid X=x),\qquad
p_1(x)=\mathbb P(Y(1)\le y_1\mid X=x).
\]
By the Fr\'echet--Hoeffding lower bound \citep[see, e.g.,][]{nelsen2006introduction}, the optimal value of the conditional primal problem is 
\[\theta_L(x)=\inf_{\gamma\in\Pi(P_{Y(0)\mid x},P_{Y(1)\mid x})}\int \I\{y_0'\le y_0,\ y_1'\le y_1\}\,d\gamma(y_0',y_1')
=\max\{0,p_0(x)+p_1(x)-1\}.\]

The conditional dual problem is defined as 
\[
\sup_{\nu_{0,x},\nu_{1,x}}
\Big\{\mathbb E[\nu_{0,x}(Y(0))\mid x]+\mathbb E[\nu_{1,x}(Y(1))\mid x]\Big\}
\quad \text{s.t.}\quad
\nu_{0,x}(y_0')+\nu_{1,x}(y_1')\le \mathbf \I\{y_0'\le y_0,\ y_1'\le y_1\}.
\]
If $p_0(x)+p_1(x)\le 1$, then $\theta_L(x)=0$ and we can take the trivial optimizer
\[
\nu^\star_{0,x}(y_0')\equiv 0,\qquad \nu^\star_{1,x}(y_1')\equiv 0.
\] If $p_0(x)+p_1(x)> 1$, then $\theta_L(x)=p_0(x)+p_1(x)-1$ and one can choose
\begin{align*}
\nu^\star_{0,x}(y_0') = \mathbf \I\{y_0'\le y_0\}, \quad \nu^\star_{1,x}(y_1') &= \mathbf \I\{y_1'\le y_1\}-1.
\end{align*}
It is easy to check that for any $(y_0',y_1')$,
\[
\nu^\star_{0,x}(y_0')+\nu^\star_{1,x}(y_1')
=
\mathbf \I\{y_0'\le y_0\}+\mathbf \I\{y_1'\le y_1\}-1
\;\le\;
\mathbf \I\{y_0'\le y_0,\ y_1'\le y_1\}=f(y_0',y_1',x),
\]
thus this dual pair is feasible, and it attains the optimal objective
\[
\mathbb E[\nu^\star_{0,x}(Y(0))\mid x]+\mathbb E[\nu^\star_{1,x}(Y(1))\mid x]
=
p_0(x)+(p_1(x)-1)=\theta_L(x).
\]
\subsection{Example \ref{ex::varite}}\label{app:dual-varite} For variance of ITE, since the ATE is identifiable, we can focus on the second moment $\E_P[(Y_i(1)-Y_i(0))^2]$, which corresponds to the quadratic cost function $f(y_0,y_1,x) = (y_1 - y_0)^2$. This problem has been extensively studied in the optimal transport literature, and an explicit formula for the optimal dual variables is available. Let $F_{1,x}(y)$ and $F_{0,x}(y)$ denote the conditional CDFs of $Y(1)$ and $Y(0)$, respectively. Assume that both CDFs are continuous. Define $T_x(y_0) = F_{1,x}^{-1}(F_{0,x}(y_0))$ (the one-dimensional monotone transport map, or monotone rearrangement; see \citealp[Chapter~2, Sections~2.1--2.2]{santambrogio2015optimal}), $u_x(y_0) = \int_0^{y_0} T_x(s)ds$, and its Legendre-Fenchel conjugate $u_x^\star(y_1) = \sup_{y_0}\{y_0y_1-u_x(y_0)\}$. Then the optimal dual variables are given by
\[
\nu^\star_{0,x}(y_0) = y_0^2 - 2u_x(y_0),\quad  \nu^\star_{1,x}(y_1) = y_1^2 - 2u_x^\star(y_1).
\]

This result is often known as Brenier's theorem \citep{brenier1991polar}; we refer interested readers to \cite{villani2021topics} for a detailed derivation. 

\subsection{Example \ref{ex::lee}}
\label{sec: lee reduction}
\label{app:dual-lee}
Here we derive the dual variables for the Lee bound under the
monotonicity condition \(S_i(1)\ge S_i(0)\). Under monotonicity,
the always-selected population is \(\{S_i(0)=1\}\), and hence
\[
\theta(P)
=
\frac{\beta(P)}{\P_P(S_i(0)=1)},
\qquad
\beta(P)=\E_P[\{Y_i(1)S_i(1)-Y_i(0)S_i(0)\}S_i(0)].
\]
Since \(\P_P(S_i(0)=1)\) is point identified, we study \(\beta_L\).

\begin{samepage}
Define the augmented potential outcome as
\[
Z_i(w)=\bigl(Y_i(w)S_i(w),S_i(w)\bigr),
\qquad
\mcZ=\{(0,0)\}\cup\bigl(\R\times\{1\}\bigr).
\]
Thus the augmented potential outcomes are $(0,0)$ for unselected units.
\end{samepage}
Write
$z_w=(r_w,s_w)$. On the full product space $\mcZ^2$, the estimand is given by
\[
f^{\mathrm{Lee}}(z_0,z_1,x)=s_0(r_1-r_0),
\]
and the monotonicity constraint is characterized by
\[
w_x^{\mathrm{Lee}}(z_0,z_1)=\I\{s_0>s_1\}.
\]

Define
\[
p_w(x)=\P(S_i(w)=1\mid X_i=x),
\qquad
\eta(x)=
\begin{cases}
p_0(x)/p_1(x),&p_1(x)>0,\\
0,&p_1(x)=0,
\end{cases}
\qquad
\mu_0(x)=\E[Y_i(0)S_i(0)\mid X_i=x].
\]
The term \(\mu_0(x)\) is point identified from the \(w=0\) marginal distribution and only shifts the \(\beta_L\) objective.

For \(a\in\R\), define
\begin{equation}\label{eq:lee-dual-family}
\nu^{(a)}_{0,x}(r_0,s_0)=a s_0-r_0,
\qquad
\nu^{(a)}_{1,x}(r_1,s_1)=(r_1-a s_1)_{-},
\qquad
\lambda_x^{(a)}=(a)_{+}.
\end{equation}
Now, we prove $(\nu_{0,x}^{(a)}, \nu_{1,x}^{(a)}, \lambda_x^{(a)})$ is dual feasible. Indeed, if \(s_0=0\), then
\[
\nu^{(a)}_{0,x}(r_0,s_0)+\nu^{(a)}_{1,x}(r_1,s_1)
\le 0=f^{\mathrm{Lee}}(z_0,z_1,x)
+\lambda_x^{(a)}w_x^{\mathrm{Lee}}(z_0,z_1),
\]
whereas if \(s_0=s_1=1\), then
\[
\nu^{(a)}_{0,x}(r_0,s_0)+\nu^{(a)}_{1,x}(r_1,s_1)
=a-r_0+(r_1-a)_-\le r_1-r_0=f^{\mathrm{Lee}}(z_0,z_1,x)
+\lambda_x^{(a)}w_x^{\mathrm{Lee}}(z_0,z_1).
\]
Finally, if $s_0=1$ and $s_1=0$, then $r_1=0$ and
\[
\nu^{(a)}_{0,x}(r_0,1)+\nu^{(a)}_{1,x}(0,0)
=a-r_0
\le -r_0+(a)_+
=f^{\mathrm{Lee}}(z_0,z_1,x)
+\lambda_x^{(a)}w_x^{\mathrm{Lee}}(z_0,z_1).
\]

Using \((u)_{-}=\min\{u,0\}\), its conditional dual objective is
\begin{align*}
g_x(a)
&=
\E[\nu^{(a)}_{0,x}\{Z_i(0)\}\mid X_i=x]
+
\E[\nu^{(a)}_{1,x}\{Z_i(1)\}\mid X_i=x] \\
&=
p_1(x)
\E[(Y_i(1)-a)_{-}\mid S_i(1)=1,X_i=x]
+a\,p_0(x)-\mu_0(x).
\end{align*}

Next, we prove strong duality.

\textbf{(Case 1)} If $0<p_0(x)<p_1(x)$, then $\eta(x)\in(0,1)$ and the conditional quantile $Q_{\eta(x)}(x)$ defined in Example~\ref{ex::lee} is finite.
By the Rockafellar--Uryasev representation \citep{rockafellar2000optimization,sahoo2022learningbiased,lei2023policybiased,chen2026adversarialcvar}, any \(a\) satisfying
\begin{equation}\label{eq:lee-quantile-condition}
\P(Y_i(1)<a\mid S_i(1)=1,X_i=x)
\le \eta(x)
\le
\P(Y_i(1)\le a\mid S_i(1)=1,X_i=x)
\end{equation}
maximizes \(g_x(a)\).

\textbf{(Case 2)} If $p_0(x)=0$, then $\mu_0(x)=\beta_L(x)=0$.
If $p_1(x)=0$, every triple in \eqref{eq:lee-dual-family} has conditional
objective zero and is therefore optimal. Suppose instead that $p_1(x)>0$.
If the conditional distribution of $Y_i(1)$ given $S_i(1)=1$ and $X_i=x$
has a finite lower endpoint $\underline y_1(x)$, then any
$a\le\underline y_1(x)$ gives $g_x(a)=0$ and hence yields an optimal feasible
triple. If the conditional lower support is unbounded and
$\E[Y_i(1)_{-}\mid S_i(1)=1,X_i=x]<\infty$, then
$g_x(a)\uparrow0$ as $a\downarrow-\infty$. Thus
$\sup_{a\in\R}g_x(a)=\beta_L(x)=0$, although the supremum need not be attained
at a finite cutoff. The zero dual is not used because it need not be feasible
on the full product space $\mcZ^2$.

\textbf{(Case 3)} If $p_0(x)=p_1(x)=p>0$, monotonicity forces $S_i(0)=S_i(1)$ almost surely conditional on $X_i=x$, whence
\[
\beta_L(x)=\E[Y_i(1)S_i(1)\mid X_i=x]-\mu_0(x)
\]
and
\[
g_x(a)=p\E[\min\{Y_i(1),a\}\mid S_i(1)=1,X_i=x]-\mu_0(x),
\quad
\beta_L(x)-g_x(a)=p\E[(Y_i(1)-a)_+\mid S_i(1)=1,X_i=x].
\]
If $Q_1(x)$ is finite, \eqref{eq:lee-quantile-condition} has a finite solution. For unbounded upper support, the last term is positive for every $a\in\R$ and decreases to zero as $a\uparrow\infty$. Thus the supremum equals $\beta_L(x)$ but is unattained; a coupling with $S_i(0)=S_i(1)$ attains the primal value \citep{lee2009training,semenova2021generalized}. Hence
\[
\sup_{a\in\R}g_x(a)=\beta_L(x).
\]
Then, the conditional dual problem reduces to a one-dimensional optimization problem
\begin{equation}\label{eq:lee-dual-reduction}
\sup_{(\nu_{0,x},\nu_{1,x})\in\mcV_x}
\left\{
\E[\nu_{0,x}\{Z_i(0)\}\mid X_i=x]
+
\E[\nu_{1,x}\{Z_i(1)\}\mid X_i=x]
\right\}
=
\sup_{a\in\R}g_x(a),
\end{equation}
Whenever the supremum is attained at a finite $a$, an optimizer is
\begin{equation}\label{eq:lee-dual-optimizer}
(\nu^\star_{0,x},\nu^\star_{1,x})
=
(\nu^{(a_x^\star)}_{0,x},\nu^{(a_x^\star)}_{1,x}),
\qquad
a_x^\star\in\argmax_{a\in\R}g_x(a),
\qquad
\lambda_x^\star=(a_x^\star)_+.
\end{equation}
\subsection{Example \ref{ex::makarov}}\label{app:dual-makarov} For the Makarov bound, fix $t\in\R$. The cost function is $f(y_0,y_1,x) = \I(y_1 - y_0 \le t)$. Let $F_{1,x}(y)$ and $F_{0,x}(y)$ denote the conditional CDFs of $Y_1$ and $Y_0$, respectively. The sharp lower bound is \citep{makarov1982, frank1987}:
\begin{equation}
\label{eq:makarov-sharp-lower-bound}
\theta_L(x)\;=\;\max\left\{\sup_{a\in\mathbb R}\big[ F_{1,x}(a+t)-F_{0,x}(a-)\big], 0\right\},
\end{equation}
where $F_{0,x}(a-)=\lim_{u\uparrow a}F_{0,x}(u)$ is the left limit. For every $a\in\mathbb R$, define the dual variables
\begin{equation}
\label{eq:makarov-dual-family}
\nu_{0,x}^{(a)}(y_0)=-\I\{y_0< a\},\quad
\nu_{1,x}^{(a)}(y_1)=\I\{y_1\le a+t\}.
\end{equation}
Then for all $(y_0,y_1)$,
\[
\nu_{0,x}^{(a)}(y_0)+\nu_{1,x}^{(a)}(y_1)
=
\I\{y_1\le a+t\}-\I\{y_0< a\}
\;\le\;\I\{y_1-y_0\leq t\},
\]
because the left-hand side can only be $1$ when $y_1\le a+t$ and $y_0\geq a$,
which implies $y_1-y_0\leq t$. Hence $(\nu_{0,x}^{(a)},\nu_{1,x}^{(a)})$ is dual-feasible, and the corresponding dual objective value is
\[
g_x(a)=\mathbb E[\nu_{0,x}^{(a)}(Y(0))\mid x]+\mathbb E[\nu_{1,x}^{(a)}(Y(1))\mid x]
=
F_{1,x}(a+t)-F_{0,x}(a-).
\]
Taking the supremum over $a\in\mathbb R$ and also allowing the zero dual pair
$(\nu_{0,x},\nu_{1,x})=(0,0)$, whose objective value is zero, gives
\[
\max\{0,\sup_{a\in\mathbb R}g_x(a)\}=\theta_L(x)
\]
by \eqref{eq:makarov-sharp-lower-bound}. Hence the primal formula
\eqref{eq:makarov-sharp-lower-bound} and the feasible dual family
\eqref{eq:makarov-dual-family} establish strong duality directly. Therefore, it suffices to search over the family
\[
\{(0,0)\}\cup \left\{\left(\nu_{0,x}^{(a)}(y_0), \nu_{1,x}^{(a)}(y_1)\right): a\in \R \right\},
\]
which reduces to a one-dimensional optimization problem. 

\subsection{Example \ref{ex::positive_ite}}\label{app:dual-positive-ite}

Consider
\[
    \theta(P) = \E_P[(Y(1)-Y(0))_+],
\]
so that \(f(y_0,y_1,x)=(y_1-y_0)_+\). For simplicity, suppose
\(Y(0),Y(1)\in [\underline y,\overline y]\). For \(k\in\{0,1\}\), write
\[
    F_{k,x}(t) = P^\star(Y(k)\le t\mid X=x).
\]
We show that the conditional sharp lower bound is
\[
    \theta_L(x)
    =
    \int_{\underline y}^{\overline y}
    \{F_{0,x}(t)-F_{1,x}(t)\}_+\,dt .
\]

We can write the dual variables using a scalar-indexed
parametrization. For each \(t\in[\underline y,\overline y]\), define
\[
\phi_{0}^{(t)}(y_0)=\mathbf 1\{y_0\le t\},
\qquad
\phi_{1}^{(t)}(y_1)=-\mathbf 1\{y_1\le t\}.
\]
Let \(\mathcal A\) be the set of measurable functions
\(\alpha:[\underline y,\overline y]\to[0,1]\). For
\(\alpha\in\mathcal A\), set
\begin{equation}\label{eq:positive-ite-alpha-duals}
\nu_{0,x}^{(\alpha)}(y_0)
=\int_{\underline y}^{\overline y}\alpha(t)\phi_{0}^{(t)}(y_0)\,dt
=\int_{y_0}^{\overline y}\alpha(t)\,dt,
\qquad
\nu_{1,x}^{(\alpha)}(y_1)
=\int_{\underline y}^{\overline y}\alpha(t)\phi_{1}^{(t)}(y_1)\,dt
=-\int_{y_1}^{\overline y}\alpha(t)\,dt .
\end{equation}
For any \(y_0,y_1\),
\[
\nu_{0,x}^{(\alpha)}(y_0)+\nu_{1,x}^{(\alpha)}(y_1)
=
\int_{\underline y}^{\overline y}
\alpha(t)\{\mathbf 1(y_0\le t)-\mathbf 1(y_1\le t)\}\,dt
\le
\int_{\underline y}^{\overline y}\mathbf 1\{y_0\le t<y_1\}\,dt
=(y_1-y_0)_+,
\]
so \((\nu_{0,x}^{(\alpha)},\nu_{1,x}^{(\alpha)})\) is dual feasible.
Its conditional dual objective is
\[
\begin{aligned}
G_x(\alpha)
&=
\E[\nu_{0,x}^{(\alpha)}(Y(0))+\nu_{1,x}^{(\alpha)}(Y(1))\mid X=x] \\
&=
\int_{\underline y}^{\overline y}
\alpha(t)\{F_{0,x}(t)-F_{1,x}(t)\}\,dt .
\end{aligned}
\]
Thus the dual maximization over this scalar-indexed family is pointwise in
\(t\):
\[
\sup_{\alpha\in\mathcal A}G_x(\alpha)
=
\int_{\underline y}^{\overline y}
\sup_{r\in[0,1]} r\{F_{0,x}(t)-F_{1,x}(t)\}\,dt
=
\int_{\underline y}^{\overline y}
\{F_{0,x}(t)-F_{1,x}(t)\}_+\,dt .
\]
One optimal choice is
\begin{equation}\label{eq:positive-ite-alpha-star}
\alpha_x^\star(t)
=\mathbf 1\{F_{0,x}(t)>F_{1,x}(t)\},
\qquad t\in[\underline y,\overline y],
\end{equation}
with arbitrary values on ties. With this choice,
\((\nu_{0,x}^{(\alpha_x^\star)},\nu_{1,x}^{(\alpha_x^\star)})\) is an optimal
dual pair, and the conditional lower bound is
\[
    \theta_L(x)
    =
    \int_{\underline y}^{\overline y}
    \{F_{0,x}(t)-F_{1,x}(t)\}_+\,dt .
\]
Let
\[
Q_{k,x}(u)=\inf\{y:F_{k,x}(y)\ge u\},\qquad k\in\{0,1\},
\]
and let \(U\sim\mathrm{Unif}(0,1)\).  The comonotone coupling
\(Y(k)=Q_{k,x}(U)\) has the  conditional marginals.  Moreover,
by Tonelli's theorem,
\begin{align*}
\E[\{Q_{1,x}(U)-Q_{0,x}(U)\}_+]
&=\int_{\underline y}^{\overline y}
\P\{Q_{0,x}(U)\le t<Q_{1,x}(U)\}\,dt\\
&=\int_{\underline y}^{\overline y}
\{F_{0,x}(t)-F_{1,x}(t)\}_+\,dt.
\end{align*}
Thus the sharp lower bound is $\int_{\underline y}^{\overline y}
\{F_{0,x}(t)-F_{1,x}(t)\}_+\,dt$.

\subsection{Example \ref{ex:CTE}}\label{app:dual-cte} Here we study the special case of \cite{kaji2023subgrouptreat}, where $$\theta(P)=\E_P[Y_i(1) - Y_i(0) \mid Y_i(0) \le c] = \frac{\E_P[Y_i(1)\I\{Y_i(0)\leq c\}] - \E_P[Y_i(0)\I\{Y_i(0)\leq c\}]}{\P(Y_i(0)\leq c)}.$$ 
Since the denominator and the term $\E_P[Y_i(0)\I\{Y_i(0)\leq c\}]$ are identifiable, it remains to find the lower bound of $\tilde{\theta}(P)=\E_P[Y_i(1)\I\{Y_i(0)\leq c\}]$, which corresponds to the cost function $f(y_0, y_1,x)=y_1\I\{y_0\leq c\}$. Denote $p(x) = \P(Y(0)\leq c|X=x)$ and $F_{1,x}(y)$ and $F_{0,x}(y)$ the conditional cdf of $Y_1$ and $Y_0$, \cite{kaji2023subgrouptreat} showed that the sharp lower bound is 
$$
\tilde\theta_L(x) = \int_0^{p(x)} F^{-1}_{1,x}(u)du. 
$$
We construct the optimal dual variables as follows: fix a scalar $a\in \R$, define 
\begin{align*}
\nu_{0,x}^{(a)}(y_0)
=-a\,\mathbf 1\{y_0>c\},\quad \nu_{1,x}^{(a)}(y_1)
= \min\{y_1,a\}.
\end{align*}
These dual variables are feasible, because if $y_0\le c$, then
$\nu_{0,x}^{(a)}(y_0)+\nu_{1,x}^{(a)}(y_1)=\min\{y_1,a\}\le y_1$;
if $y_0>c$, then $f(y_0,y_1,x)=0$ and
$\nu_{0,x}^{(a)}(y_0)+\nu_{1,x}^{(a)}(y_1)=-a+\min\{y_1,a\}\le 0$.
Thus $\nu_{0,x}^{(a)}+\nu_{1,x}^{(a)}\le y_1\mathbf 1\{y_0\le c\}$ for all $(y_0,y_1)$.

The dual objective equals
\begin{align*}
g_x(a)
&=-a\{1-p(x)\}+\E[\min\{Y(1),a\}\mid X=x]\\
&=\int_{(-\infty,a]}u\,dF_{1,x}(u)
+a\{p(x)-F_{1,x}(a)\}\\
&=\int_0^{F_{1,x}(a)}F_{1,x}^{-1}(u)\,du
+a\{p(x)-F_{1,x}(a)\}.
\end{align*}
Let
\(a^\star=F_{1,x}^{-1}(p(x))=\inf\{a:F_{1,x}(a)\ge p(x)\}\).
Write
\(p=p(x)\), \(F=F_{1,x}\), and \(q=a^\star=F^{-1}(p)\). Then
\(F(q-)\le p\le F(q)\), and \(F^{-1}(u)=q\) for
\(u\in(F(q-),F(q)]\).
\[
\int_0^p F^{-1}(u)\,du
=
\int_{(-\infty,q)}u\,dF(u)+q\{p-F(q-)\}
=
\int_{(-\infty,q]}u\,dF(u)+q\{p-F(q)\}.
\]
Applying this identity with \(F=F_{1,x}\) and \(q=a^\star\) gives
\[
g_x(a^\star)=\int_0^{p(x)}F_{1,x}^{-1}(u)\,du =\tilde{\theta}_L(x).
\]
Thus $\nu_{0,x}^{(a^\star)}(y_0), \nu_{1,x}^{(a^\star)}(y_1)$ form an optimal pair of dual variables, and searching for the optimal dual variables reduces to a one-dimensional optimization problem that can be easily solved.
\subsection{Example \ref{ex:QTE}}
\label{app:dual-qte}
Let
\[
q_\alpha(P)=Q_\alpha\{Y(1)-Y(0)\}.
\]
For every \(c\in\R\),
\[
q_\alpha(P)\le c
\quad\Longleftrightarrow\quad
\E_P[f_c\{Y(0),Y(1)\}]\le0,
\qquad
f_c(y_0,y_1)=\alpha-\I\{y_1-y_0\le c\}.
\]
Thus the sharp lower bound on the \(\alpha\)-quantile is obtained by inverting
the sharp lower bounds for the linearized costs \(f_c\). Define
\[
\theta_{L,c}
=
\inf_{P\in\mathcal P}\E_P[f_c\{Y(0),Y(1)\}],
\qquad
q_L=\inf\{c\in\R:\theta_{L,c}\le0\}.
\]

For each \(c\),
\[
\theta_{L,c}(x)
=
\alpha-\sup_P P\{Y(1)-Y(0)\le c\mid X=x\}.
\]
This is the complementary Makarov problem. Applying the result in
Appendix~\ref{app:dual-makarov} gives
\[
\theta_{L,c}(x)
=
\alpha-1-\inf_{a\in\R}
\{F_{1,x}(a+c)-F_{0,x}(a-)\}.
\]
Let
\[
a_{x,c}^\star
\in
\argmin_{a\in\R}
\{F_{1,x}(a+c)-F_{0,x}(a-)\}.
\]
If \(a_{x,c}^\star\) exists (e.g., when the potential outcomes are bounded),
the corresponding optimal lower-bound duals are
\[
\nu_{0,x}^{\star,c}(y_0)
=\alpha-\I\{y_0\ge a_{x,c}^\star\},
\qquad
\nu_{1,x}^{\star,c}(y_1)
=-\I\{y_1\le a_{x,c}^\star+c\}.
\]
\subsection{Takeaway from the examples}
The examples above show that, for the causal estimands that are commonly used in practice, the duals have specific structure that allows us to narrow the space of \((\nu_{0,x}, \nu_{1,x})\) in our Dual Bound method without sacrificing the strong duality. Specifically, for Examples \ref{ex::fh}, \ref{ex::varite}, \ref{ex::positive_ite}, we know the exact closed-form expressions of the duals; for Examples \ref{ex::lee}--\ref{ex:QTE}, the duals are known up to a scalar \(a\). This observation will be the key to establish the finer theory in Section \ref{subsec:structured-margins}.
\section{Theory of continuous potential outcomes}
\label{sec:proofs-structured-margins}
This section develops the refined sharpness results summarized in
Section~\ref{subsec:structured-margins}. Appendices~\ref{app:fh-proof}
and~\ref{sec:smoothmonge} first handle two cases for which the Kantorovich duals
are known functions of the primals, so the dual bound estimators can be obtained
without explicitly solving an optimization problem: the \fh\ bound
(Example~\ref{ex::fh}) and the variance of individual treatment effects
(Example~\ref{ex::varite}) and, more generally, smooth estimands that have
monotone differences. Appendix~\ref{sec:general-margin}
then establishes a generic bound on the dual errors that leverages the scalar
parametrization of the Kantorovich duals derived in Appendix~\ref{sec: dual form}.
Appendices~\ref{sec:lee}--\ref{sec:qte-proof} apply the generic result in
Appendix~\ref{sec:general-margin} to Lee bounds (Example~\ref{ex::lee}),
distribution treatment effects (Example~\ref{ex::makarov}), positive-part
treatment effects (Example~\ref{ex::positive_ite}), and
baseline-disadvantaged subgroup treatment effects (Example~\ref{ex:CTE}), and
quantiles of the ITE (Example~\ref{ex:QTE}).
Finally, in Appendix~\ref{sec:discrete-margin-comparison}, we develop the
finite-support counterpart and compare our margin condition with the margin
condition in \cite{semenova2023aggregated}.

The results for the \fh\ bound and the distribution
treatment effect allow outcomes supported on all of \(\mathbb R\). The results
for the variance of the ITE, the Lee bound, the positive-part treatment effect,
the baseline-disadvantaged subgroup effect, and quantiles of the ITE assume
compact outcome support.

\subsection{\fh\, bound}
\label{app:fh-proof}

For the estimand in Example~\ref{ex::fh}, take the cost function in the generic conditional primal problem to be
\[
  f_{y_1,y_0}(y_0',y_1',x)
  =
  \I\{y_1'\le y_1,\ y_0'\le y_0\}.
\]
Fix the threshold pair \((y_1,y_0)\in\R^2\) and let
\[
  \Gamma_x=F_{1,x}(y_1)+F_{0,x}(y_0)-1.
\]
By Appendix~\ref{app:dual-fh}, the conditional sharp lower bound is
\[
  \theta_L(x)=\{\Gamma_x\}_+.
\]
Now let \(\widehat F_{w,x}\) be first-stage estimates of the conditional CDFs and define
\[
  \widehat\Gamma_x
  =
  \widehat F_{1,x}(y_1)+\widehat F_{0,x}(y_0)-1.
\]
Using the duals from Appendix~\ref{app:dual-fh}, define the population and estimated duals directly by
\begin{equation}\label{eq:fh-population-estimated-duals}
\begin{array}{ll}
\nu_{0,X}^{\star}(y)=\I\{\Gamma_X>0\}\I\{y\le y_0\},
&
\nu_{1,X}^{\star}(y)=\I\{\Gamma_X>0\}\big(\I\{y\le y_1\}-1\big),
\\[2pt]
\widehat\nu_{0,X}(y)=\I\{\widehat\Gamma_X>0\}\I\{y\le y_0\},
&
\widehat\nu_{1,X}(y)=\I\{\widehat\Gamma_X>0\}\big(\I\{y\le y_1\}-1\big).
\end{array}
\end{equation}
By Appendix~\ref{app:dual-fh}, \((\nu_{0,X}^{\star},\nu_{1,X}^{\star})\) is a population optimal dual and \((\widehat\nu_{0,X},\widehat\nu_{1,X})\) is dual-feasible for every first-stage estimate.

Let
\[
  \widetilde\theta_L
  =
  \E\{\widehat\nu_{0,X}(Y(0))+\widehat\nu_{1,X}(Y(1))\given\mathcal D_1\}
  =
  \E[\I\{\widehat\Gamma_X>0\}\Gamma_X\given\mathcal D_1].
\]
The sharp lower bound is \(\theta_L=\E[\{\Gamma_X\}_+]\).

The following assumption controls the probability that the \fh\, gap is close
to zero.
\begin{assumption}
\label{asm:fh-margin}
For some \(b>0\) and finite \(C\),
\[
P\{0<|\Gamma_X|\le s\}
\le
C s^{b}
\qquad\text{for all sufficiently small }s>0.
\]
\end{assumption}

For this subsection and any \(r\in(1,\infty]\), define
\begin{equation}\label{eq:fh-primal-error}
\error_{P,r}
=
\begin{cases}
\left\{
\E\!\left[|\widehat\Gamma_X-\Gamma_X|^r\mid\mathcal D_1\right]
\right\}^{1/r}, & r<\infty,\\[6pt]
\displaystyle
\operatorname*{ess\,sup}_{x\sim P_X\opt}
|\widehat\Gamma_x-\Gamma_x|, & r=\infty,
\end{cases}
\end{equation}

\begin{theorem}
\label{thm:fh-bias}
Suppose Assumption~\ref{asm:fh-margin} holds and fix
\(r\in(1,\infty]\). Whenever \(\error_{P,r}\) is sufficiently small,
\[
0\le \theta_L-\widetilde\theta_L
\lesssim
\error_{P,r}^{(b+1)/(b/r+1)}.
\]
Here and below, \(b/\infty=0\).
\end{theorem}

\begin{proof}
For $P_X\opt$-almost every $x$, let
\(\delta_x=|\widehat\Gamma_x-\Gamma_x|\). Then
\[
\{\Gamma_x\}_+-\I\{\widehat\Gamma_x>0\}\Gamma_x
=
|\Gamma_x|\,
\I\{\I\{\widehat\Gamma_x>0\}\ne \I\{\Gamma_x>0\},\ \Gamma_x\ne0\}.
\]
On the event in the indicator, \(0<|\Gamma_x|\le\delta_x\). Therefore, for
any \(t>0\),
\[
\{\Gamma_x\}_+-\I\{\widehat\Gamma_x>0\}\Gamma_x
\le
|\Gamma_x|\I\{0<|\Gamma_x|\le t\}
+\delta_x\I\{\delta_x>t\}.
\]
Taking expectation over a fresh draw of \(X\), conditional on
\(\mathcal D_1\), and applying Assumption~\ref{asm:fh-margin} give
\[
0\le \theta_L-\widetilde\theta_L
\le
C t^{b+1}
+\E[\delta_X\I\{\delta_X>t\}\mid\mathcal D_1]
\]
for all sufficiently small \(t>0\). First suppose \(r<\infty\). Since
\(a\I\{a>t\}\le a^r/t^{r-1}\),
\[
0\le \theta_L-\widetilde\theta_L
\le
C t^{b+1}
+\frac{\error_{P,r}^r}{t^{r-1}}.
\]
If \(\error_{P,r}>0\), setting
\(t=\error_{P,r}^{r/(b+r)}\) yields
\[
0\le \theta_L-\widetilde\theta_L
\le
(C+1)\error_{P,r}^{(b+1)/(b/r+1)};
\]
the case \(\error_{P,r}=0\) is immediate. If \(r=\infty\), then
\(\delta_X\le\error_{P,\infty}\) almost surely, so a sign error can occur only
when \(0<|\Gamma_X|\le\error_{P,\infty}\). Hence
\[
0\le \theta_L-\widetilde\theta_L
\le
C\error_{P,\infty}^{b+1}.
\]
\end{proof}

\subsection{Variance of ITE}\label{sec:smoothmonge}
\begingroup
\newcommand{\norm}[1]{\left\lVert #1 \right\rVert}
\newcommand{\abs}[1]{\left\lvert #1 \right\rvert}

We first state the bounded support assumption that will be
used in Appendices~\ref{sec:smoothmonge}, \ref{sec:pte},
\ref{app:subgroup-proof}, and~\ref{sec:qte-proof}. In
Appendix~\ref{sec:lee}, we will introduce an analogous assumption tailored to
the setting of Lee bounds.
\begin{assumption}\label{asm:common-density-setup}
There exist constants \(-\infty<\underline y<\overline y<\infty\) such that, for
almost every \(x\), the following hold.
\begin{enumerate}[label=\textnormal{(\arabic*)},ref=\textnormal{(\arabic*)},leftmargin=2.8em]
\item\label{cond:common-density-true} The supports of \(Y(0)\mid X=x\) and \(Y(1)\mid X=x\) are contained in
\([\underline y,\overline y]\), and each \(Y(w)\mid X=x\) admits a jointly
measurable density \(p_w^\star(\cdot\mid x)\) on
\([\underline y,\overline y]\).
\item\label{cond:common-density-estimated} For each \(w\in\{0,1\}\), the first-stage estimator returns a jointly
measurable density \(\widehat p_w(\cdot\mid x)\) on
\([\underline y,\overline y]\).
\end{enumerate}
\end{assumption}

Appendix~\ref{app:dual-varite} derives the optimal Kantorovich duals for the variance
of individual treatment effects. Because the ATE is identified, the only partially
identified component is the second moment
\[
  \E\big[(Y(1)-Y(0))^2\big],
\]
which corresponds to the quadratic cost \(f_x(y_0,y_1)=(y_1-y_0)^2\). Let
\[
  F_{w,x}(y)=P\{Y(w)\le y\mid X=x\},
  \qquad
  Q_{w,x}(u)=\inf\{y\in[\underline y,\overline y]:F_{w,x}(y)\ge u\}.
\]
Define the transport map
\[
  T_x^\star(y_0)=Q_{1,x}\!\big(F_{0,x}(y_0)\big),
  \qquad y_0\in[\underline y,\overline y].
\]
Equivalently, if \(Y_{0,x}\sim P_{Y(0)\mid X=x}\), then
\[
  T_x^\star(Y_{0,x})\sim P_{Y(1)\mid X=x},
  \qquad
  T_x^\star{}_{\#}P_{Y(0)\mid X=x}=P_{Y(1)\mid X=x}.
\]
With this map, define
\[
  \varphi_x(y_0)=\int_{\underline y}^{y_0}T_x^\star(s)\,ds,
  \qquad
  \varphi_x^\star(y_1)=\sup_{z\in[\underline y,\overline y]}\{zy_1-\varphi_x(z)\},
\]
then one convenient normalization of the optimal duals is
\[
  \nu_{0,x}^\star(y_0)=y_0^2-2\varphi_x(y_0),
  \qquad
  \nu_{1,x}^\star(y_1)=y_1^2-2\varphi_x^\star(y_1).
\]
The plug-in estimate replaces the conditional CDFs and quantiles by their first-stage
estimates. Thus
\[
  \widehat T_x(y_0)=\widehat Q_{1,x}\!\big(\widehat F_{0,x}(y_0)\big),
  \qquad
  \widehat\varphi_x(y_0)=\int_{\underline y}^{y_0}\widehat T_x(s)\,ds,
\]
\[
  \widehat\varphi_x^\star(y_1)=
  \sup_{z\in[\underline y,\overline y]}\{zy_1-\widehat\varphi_x(z)\},
  \qquad
  \widehat\nu_{0,x}(y_0)=y_0^2-2\widehat\varphi_x(y_0),
  \qquad
  \widehat\nu_{1,x}(y_1)=y_1^2-2\widehat\varphi_x^\star(y_1).
\]

\noindent\hspace*{2em}In the following, we consider a broader class of estimands with smooth \(f_x\):
\[
  \E[f_x(Y(0),Y(1))\mid X=x],
  \qquad f_x(y_0,y_1)=f(y_0,y_1,x).
\]
The key requirement is that \(f_x\) is smooth and its cross partial
\(\partial_{12}f_x\) has a fixed sign on the domain. For example,
\(\partial_{12} f_x(y_0,y_1)=-2\le 0\) for the second moment of individual
treatment effect.

\begin{assumption}
\label{asm: monge assumption}
Assume the following holds:
\begin{enumerate}[label=\textnormal{(\arabic*)},ref=\textnormal{(\arabic*)},leftmargin=3em]
\item\label{cond:monge-density-bounds} The densities \(p_k^\star(\cdot\mid x)\), \(k\in\{0,1\}\), are bounded away from zero and infinity: there exist measurable functions
$0<\underline p(x)\le \overline p(x)<\infty$ such that for almost every $x$ and every $y\in[\underline y,\overline y]$,
\[
\underline p(x)\le p_k^\star(y\mid x)\le \overline p(x),\qquad k\in\{0,1\}.
\]
\item\label{cond:monge-smooth-cost} For almost every $x$, the conditional cost $f_x(y_0,y_1)=f(y_0,y_1,x)$ is in $C^2([\underline y,\overline y]^2)$.
\item\label{cond:monge-cross-partial} For almost every $x$, one of the following two cases holds on all of $[\underline y,\overline y]^2$:
\begin{enumerate}[label=\textnormal{(\alph*)},ref=\textnormal{(\arabic{enumi}\alph*)},leftmargin=2.5em]
\item\label{case:monge-submodular} $\partial_{12}f_x(y_0,y_1)\le 0$ for all $(y_0,y_1)\in[\underline y,\overline y]^2$;
\item\label{case:monge-supermodular} $\partial_{12}f_x(y_0,y_1)\ge 0$ for all $(y_0,y_1)\in[\underline y,\overline y]^2$.
\end{enumerate}
\item\label{cond:monge-integrability} Define
\[
L_{12}(x) = \sup_{(y_0,y_1)\in[\underline y,\overline y]^2}\abs{\partial_{12}f_x(y_0,y_1)}.
\]
Assume
\[
\E\left[\left(\frac{L_{12}(X)}{\underline p(X)}\right)^2\right] < \infty.
\]
\end{enumerate}
    
\end{assumption}
For each $x$ and $k\in\{0,1\}$, define the true and estimated conditional CDFs
\[
  F_{k,x}(y) = \int_{\underline y}^y p_k^\star(t\mid x)\,dt,
  \qquad
  \hat F_{k,x}(y) = \int_{\underline y}^y \hat p_k(t\mid x)\,dt,
\]
and the corresponding quantile functions
\[
  Q_{k,x}(u) = F_{k,x}^{-1}(u),
  \qquad
  \hat Q_{k,x}(u) = \inf\{ y\in[\underline y,\overline y] : \hat F_{k,x}(y) \ge u\},
  \qquad u\in[0,1].
\]
In case~\ref{case:monge-submodular}, define
\begin{equation}
\label{eq:monge-transport-increasing}
  T_x^\star(y) = Q_{1,x}\bigl(F_{0,x}(y)\bigr),
  \qquad
  \hat T_x(y) = \hat Q_{1,x}\bigl(\hat F_{0,x}(y)\bigr).
\end{equation}
In case~\ref{case:monge-supermodular}, define
\begin{equation}
\label{eq:monge-transport-decreasing}
  T_x^\star(y) = Q_{1,x}\bigl(1-F_{0,x}(y)\bigr),
  \qquad
  \hat T_x(y) = \hat Q_{1,x}\bigl(1-\hat F_{0,x}(y)\bigr).
\end{equation}
Set $\nu_{0,x}(\underline y)=0$, and define
\begin{equation}
\label{eq:smoothmonge-duals}
\begin{aligned}
  \nu_{0,x}^\star(y)
  &= \int_{\underline y}^y \partial_1 f_x\bigl(s,T_x^\star(s)\bigr)\,ds,
  &
  \hat\nu_{0,x}(y)
  &= \int_{\underline y}^y \partial_1 f_x\bigl(s,\hat T_x(s)\bigr)\,ds,
  \\
  \nu_{1,x}^\star(y)
  &= \inf_{z\in[\underline y,\overline y]} \bigl\{ f_x(z,y)-\nu_{0,x}^\star(z) \bigr\},
  &
  \hat\nu_{1,x}(y)
  &= \inf_{z\in[\underline y,\overline y]} \bigl\{ f_x(z,y)-\hat\nu_{0,x}(z) \bigr\}.
\end{aligned}
\end{equation}
Finally, define
\[
  \mathrm{error}_{P,2}(x)^2 = \sum_{k\in\{0,1\}} \int_{\underline y}^{\overline y} \bigl(\hat p_k(y\mid x)-p_k^\star(y\mid x)\bigr)^2\,dy,
\]
\[
  \mathrm{error}_{\nu,2}(x)^2 = \sum_{k\in\{0,1\}} \int_{\underline y}^{\overline y} \bigl(\hat\nu_{k,x}(y)-\nu_{k,x}^\star(y)\bigr)^2\,dy,
\]
\[
  H_f(x) = \frac{2(\overline y-\underline y)^2L_{12}(x)}{\underline p(x)},
  \qquad
  \tilde\theta_L = g(\hat\nu).
\]

\begin{theorem}\label{thm:smooth-monge-dual-rate}
If Assumptions~\ref{asm:common-density-setup} and~\ref{asm: monge assumption} hold, then:
\begin{enumerate}[label=\textnormal{(\arabic*)},ref=\textnormal{(\arabic*)},leftmargin=2.8em]
\item\label{part:monge-dual-optimality} For almost every $x$, $(\nu_{0,x}^\star,\nu_{1,x}^\star)$ is an optimal dual pair for the true conditional problem, and $(\hat\nu_{0,x},\hat\nu_{1,x})$ is an optimal dual pair for the estimated conditional problem.
\item\label{part:monge-dual-rate} For almost every $x$,
\[
\mathrm{error}_{\nu,2}(x) \le H_f(x)\,\mathrm{error}_{P,2}(x).
\]
\item\label{part:monge-bias} If
\[
\E\bigl[\mathrm{error}_{P,2}(X)^4\bigr] = o(n^{-1}),
\]
then
\[
\sqrt n\,\bigl(\tilde\theta_L-\theta_L\bigr)=o_p(1).
\]
\end{enumerate}
\end{theorem}
\begin{proof}
    
We write the proof in the following steps.
Throughout \ref{step:monge-preliminaries}--\ref{step:monge-dual-rate}, fix an \(x\) satisfying Assumptions~\ref{asm:common-density-setup} and~\ref{asm: monge assumption}; the qualification ``for almost every \(x\)'' is suppressed. The final step, \ref{step:monge-bias}, integrates the pointwise bounds over \(X\).

\begin{enumerate}[label=\textbf{Step \arabic*.},ref=Step~\arabic*,leftmargin=*]
\item\label{step:monge-preliminaries}
We first establish a few preliminary properties.
By Assumption~\ref{asm: monge assumption}, condition~\ref{cond:monge-density-bounds}, each true conditional CDF $F_{k,x}$ is continuous, strictly increasing, and has range $[0,1]$. Hence the true conditional quantile $Q_{k,x}=F_{k,x}^{-1}$ is defined.
Moreover, for any $u<v$,
\[
  F_{1,x}(v)-F_{1,x}(u)
  = \int_u^v p_1^\star(t\mid x)\,dt
  \ge \underline p(x)(v-u).
\]
Therefore $Q_{1,x}$ is $1/\underline p(x)$-Lipschitz:
\begin{equation}
  \abs{Q_{1,x}(r)-Q_{1,x}(s)}
  \le \frac{\abs{r-s}}{\underline p(x)}
  \qquad \forall r,s\in[0,1].
  \label{eq:true-quantile-lipschitz-explicit-x}
\end{equation}

By Assumption~\ref{asm:common-density-setup}, condition~\ref{cond:common-density-estimated}, each estimated conditional CDF $\hat F_{k,x}$ is absolutely continuous and nondecreasing, and satisfies
\[
  \hat F_{k,x}(\underline y)=0,
  \qquad
  \hat F_{k,x}(\overline y)=1.
\]
Therefore the generalized inverse $\hat Q_{k,x}$ is defined on $[0,1]$.

Define the functions $T_x^\star$ and $\hat T_x$ by \eqref{eq:monge-transport-increasing} in case~\ref{case:monge-submodular} and by \eqref{eq:monge-transport-decreasing} in case~\ref{case:monge-supermodular}. In case~\ref{case:monge-submodular} they are increasing; in case~\ref{case:monge-supermodular} they are decreasing. In either case they are measurable and have values in $[\underline y,\overline y]$.

\item\label{step:monge-couplings}
This step verifies that the transport maps generate valid couplings for the estimated and true conditional distributions.
For the estimated pair, $\hat T_x$ pushes the estimated distribution $\hat p_0(\cdot\mid x)$ forward to $\hat p_1(\cdot\mid x)$, so that the pushforward measure $(\mathrm{Id},\hat T_x)_\#(\hat p_0(y\mid x)\,dy)$ is a coupling of the estimated pair.
Here, for any Borel set $B\subset [\underline y,\overline y]^2$, this pushforward measure is defined by
\[
  \bigl((\mathrm{Id},\hat T_x)_\#(\hat p_0(y\mid x)\,dy)\bigr)(B)
  =
  \int_{\underline y}^{\overline y}
  \I\{(y,\hat T_x(y))\in B\}\hat p_0(y\mid x)\,dy .
\]

If $Y_{0,x}\sim \hat p_0(\cdot\mid x)$, then $U_x=\hat F_{0,x}(Y_{0,x})\sim \mathrm{Unif}[0,1]$ because $\hat F_{0,x}$ is continuous. In case~\ref{case:monge-submodular},
\[
  \hat T_x(Y_{0,x})=\hat Q_{1,x}\bigl(\hat F_{0,x}(Y_{0,x})\bigr)=\hat Q_{1,x}(U_x),
\]
and in case~\ref{case:monge-supermodular},
\[
  \hat T_x(Y_{0,x})=\hat Q_{1,x}\bigl(1-\hat F_{0,x}(Y_{0,x})\bigr)=\hat Q_{1,x}(1-U_x).
\]
Since both $U_x$ and $1-U_x$ are uniform on $[0,1]$, the definition of the generalized inverse gives
\[
  \hat T_x(Y_{0,x}) \sim \hat p_1(\cdot\mid x).
\]
Thus
\[
  \hat\pi_x = (\mathrm{Id},\hat T_x)_\#\bigl(\hat p_0(y\mid x)\,dy\bigr)
\]
is a coupling of $\hat p_0(\cdot\mid x)$ and $\hat p_1(\cdot\mid x)$.

Exactly the same argument applies to the true pair $(p_0^\star(\cdot\mid x),p_1^\star(\cdot\mid x))$ and yields the coupling
\[
  \pi_x^\star = (\mathrm{Id},T_x^\star)_\#\bigl(p_0^\star(y\mid x)\,dy\bigr).
\]

\item\label{step:monge-optimality}
In this step, we prove part~\ref{part:monge-dual-optimality}.
Consider any pair of densities $(p_{0,x},p_{1,x})$ on $[\underline y,\overline y]$, let $T_x$ be the function defined above, and let
\[
  \nu_{0,x}(y) = \int_{\underline y}^y \partial_1 f_x\bigl(s,T_x(s)\bigr)\,ds,
  \qquad
  \nu_{1,x}(y) = \inf_{z\in[\underline y,\overline y]} \Bigl\{ f_x(z,y)-\nu_{0,x}(z) \Bigr\}.
\]
By definition of the infimum,
\begin{equation}
  \nu_{0,x}(z)+\nu_{1,x}(y) \le f_x(z,y)
  \qquad \forall z,y\in[\underline y,\overline y].
  \label{eq:dual-feasible-general-explicit-x}
\end{equation}
So $(\nu_{0,x},\nu_{1,x})$ is dual-feasible.

Fix $z_0\in[\underline y,\overline y]$, and define
\[
  H_x(z) = f_x\bigl(z,T_x(z_0)\bigr)-\nu_{0,x}(z).
\]
Since $\nu_{0,x}$ is absolutely continuous,
\[
  H_x'(z)
  = \partial_1 f_x\bigl(z,T_x(z_0)\bigr) - \partial_1 f_x\bigl(z,T_x(z)\bigr)
  \qquad \text{for a.e. } z.
\]
We claim that $H_x$ is nonincreasing on $[\underline y,z_0]$ and nondecreasing on $[z_0,\overline y]$.

Indeed, in case~\ref{case:monge-submodular}, $T_x$ is increasing and $\partial_{12}f_x\le 0$, so $y\mapsto \partial_1 f_x(z,y)$ is nonincreasing for each fixed $z$. Hence:
\begin{itemize}[leftmargin=2em]
  \item if $z\le z_0$, then $T_x(z)\le T_x(z_0)$, so $\partial_1 f_x\bigl(z,T_x(z_0)\bigr)\le \partial_1 f_x\bigl(z,T_x(z)\bigr)$, and therefore $H_x'(z)\le 0$;
  \item if $z\ge z_0$, then $T_x(z)\ge T_x(z_0)$, so $\partial_1 f_x\bigl(z,T_x(z_0)\bigr)\ge \partial_1 f_x\bigl(z,T_x(z)\bigr)$, and therefore $H_x'(z)\ge 0$.
\end{itemize}
In case~\ref{case:monge-supermodular}, $T_x$ is decreasing and $\partial_{12}f_x\ge 0$, so $y\mapsto \partial_1 f_x(z,y)$ is nondecreasing for each fixed $z$. Hence:
\begin{itemize}[leftmargin=2em]
  \item if $z\le z_0$, then $T_x(z)\ge T_x(z_0)$, so $\partial_1 f_x\bigl(z,T_x(z_0)\bigr)\le \partial_1 f_x\bigl(z,T_x(z)\bigr)$, and therefore $H_x'(z)\le 0$;
  \item if $z\ge z_0$, then $T_x(z)\le T_x(z_0)$, so $\partial_1 f_x\bigl(z,T_x(z_0)\bigr)\ge \partial_1 f_x\bigl(z,T_x(z)\bigr)$, and therefore $H_x'(z)\ge 0$.
\end{itemize}
Thus, in either case, $z_0$ minimizes $H_x$, and therefore
\begin{equation}
  \nu_{1,x}\bigl(T_x(z_0)\bigr)
  = f_x\bigl(z_0,T_x(z_0)\bigr)-\nu_{0,x}(z_0).
  \label{eq:contact-general-explicit-x}
\end{equation}
Equivalently,
\[
  \nu_{0,x}(z_0)+\nu_{1,x}\bigl(T_x(z_0)\bigr)=f_x\bigl(z_0,T_x(z_0)\bigr).
\]
Now let
\[
  \pi_x = (\mathrm{Id},T_x)_\#\bigl(p_{0,x}(y)\,dy\bigr).
\]
By \ref{step:monge-couplings}, $\pi_x$ is a coupling of $p_{0,x}$ and $p_{1,x}$. Integrating \eqref{eq:contact-general-explicit-x} with respect to $p_{0,x}$ yields
\[
  \int f_x(y_0,y_1)\,d\pi_x
  = \int \nu_{0,x}(y)\,p_{0,x}(y)\,dy + \int \nu_{1,x}(y)\,p_{1,x}(y)\,dy.
\]
But \eqref{eq:dual-feasible-general-explicit-x} implies that the right-hand side is a lower bound on the transport cost for every coupling of $p_{0,x}$ and $p_{1,x}$. Hence $\pi_x$ is an optimal coupling and $(\nu_{0,x},\nu_{1,x})$ are optimal dual variables.

Applying this once to the true pair $(p_0^\star(\cdot\mid x),p_1^\star(\cdot\mid x))$ and once to the estimated pair $(\hat p_0(\cdot\mid x),\hat p_1(\cdot\mid x))$ proves part~\ref{part:monge-dual-optimality} of the theorem.

\item\label{step:monge-cdf-rate}
We bound the Kolmogorov distance by \(\mathrm{error}_{P,2}(x)\).
Define
\[
  \Delta p_{k,x}(y) = \hat p_k(y\mid x)-p_k^\star(y\mid x),
  \qquad k\in\{0,1\}.
\]
Then for any $y\in[\underline y,\overline y]$,
\[
  \abs{\hat F_{k,x}(y)-F_{k,x}(y)}
  = \abs{\int_{\underline y}^y \Delta p_{k,x}(t)\,dt}
  \le \int_{\underline y}^{\overline y} \abs{\Delta p_{k,x}(t)}\,dt
  \le \sqrt{\overline y-\underline y}\,\norm{\Delta p_{k,x}}_{L^2([\underline y,\overline y])}.
\]
Hence
\begin{equation}
  \norm{\hat F_{k,x}-F_{k,x}}_{\infty}
  \le \sqrt{\overline y-\underline y}\,\norm{\Delta p_{k,x}}_{L^2([\underline y,\overline y])}.
  \label{eq:cdf-error-bound-explicit-x}
\end{equation}
Then by Cauchy--Schwarz,
\begin{equation}
  \norm{\hat F_{0,x}-F_{0,x}}_{\infty} + \norm{\hat F_{1,x}-F_{1,x}}_{\infty}
  \le \sqrt{2(\overline y-\underline y)}\,\mathrm{error}_{P,2}(x).
  \label{eq:two-cdf-error-bound-explicit-x}
\end{equation}

\item\label{step:monge-map-rate}
We convert the bound on the Kolmogorov distance to an \(\ell_{\infty}\) bound on the estimated transport map.
For every $u\in[0,1]$, the generalized inverse satisfies
\[
  Q_{1,x}\bigl((u-\norm{\hat F_{1,x}-F_{1,x}}_{\infty})_+\bigr)
  \le \hat Q_{1,x}(u)
  \le Q_{1,x}\bigl((u+\norm{\hat F_{1,x}-F_{1,x}}_{\infty})\wedge 1\bigr).
\]
Using \eqref{eq:true-quantile-lipschitz-explicit-x}, it follows that
\begin{equation}
  \abs{\hat Q_{1,x}(u)-Q_{1,x}(u)}
  \le \frac{\norm{\hat F_{1,x}-F_{1,x}}_{\infty}}{\underline p(x)}
  \qquad \forall u\in[0,1].
  \label{eq:quantile-error-bound-explicit-x}
\end{equation}
Using a similar argument,
\begin{equation}
  \abs{\hat Q_{0,x}(u)-Q_{0,x}(u)}
  \le \frac{\norm{\hat F_{0,x}-F_{0,x}}_{\infty}}{\underline p(x)}
  \qquad \forall u\in[0,1].
  \label{eq:quantile-error-bound-control-explicit-x}
\end{equation}
In case~\ref{case:monge-submodular},
\[
  \hat T_x(y)=\hat Q_{1,x}\bigl(\hat F_{0,x}(y)\bigr),
  \qquad
  T_x^\star(y)=Q_{1,x}\bigl(F_{0,x}(y)\bigr).
\]
In case~\ref{case:monge-supermodular},
\[
  \hat T_x(y)=\hat Q_{1,x}\bigl(1-\hat F_{0,x}(y)\bigr),
  \qquad
  T_x^\star(y)=Q_{1,x}\bigl(1-F_{0,x}(y)\bigr).
\]
In either case,
\[
  \abs{\hat T_x(y)-T_x^\star(y)}
  \le \abs{\hat Q_{1,x}(r_{x,y})-Q_{1,x}(r_{x,y})}
  + \abs{Q_{1,x}(r_{x,y})-Q_{1,x}(r_{x,y}^\star)},
\]
where $r_{x,y}$ and $r_{x,y}^\star$ are either $(\hat F_{0,x}(y),F_{0,x}(y))$ or $(1-\hat F_{0,x}(y),1-F_{0,x}(y))$. Thus, by \eqref{eq:true-quantile-lipschitz-explicit-x} and \eqref{eq:quantile-error-bound-explicit-x},
\[
  \abs{\hat T_x(y)-T_x^\star(y)}
  \le \frac{\norm{\hat F_{1,x}-F_{1,x}}_{\infty} + \norm{\hat F_{0,x}-F_{0,x}}_{\infty}}{\underline p(x)}.
\]
Taking the supremum and using \eqref{eq:two-cdf-error-bound-explicit-x},
\begin{equation}
  \norm{\hat T_x-T_x^\star}_{\infty}
  \le \frac{\sqrt{2(\overline y-\underline y)}}{\underline p(x)}\, \mathrm{error}_{P,2}(x).
  \label{eq:transport-error-bound-explicit-x}
\end{equation}

\item\label{step:monge-control-dual-rate}
In this step, we derive the error of estimated duals corresponding to the control group.
For any $y\in[\underline y,\overline y]$,
\[
  \hat\nu_{0,x}(y)-\nu_{0,x}^\star(y)
  = \int_{\underline y}^y \Bigl[ \partial_1 f_x\bigl(s,\hat T_x(s)\bigr) - \partial_1 f_x\bigl(s,T_x^\star(s)\bigr) \Bigr] ds.
\]
By the mean value theorem in the second argument and the definition of $L_{12}(x)$,
\[
  \abs{\partial_1 f_x\bigl(s,\hat T_x(s)\bigr) - \partial_1 f_x\bigl(s,T_x^\star(s)\bigr)}
  \le L_{12}(x)\, \norm{\hat T_x-T_x^\star}_{\infty}.
\]
Hence
\[
  \norm{\hat\nu_{0,x}-\nu_{0,x}^\star}_{\infty}
  \le (\overline y-\underline y)L_{12}(x)\,\norm{\hat T_x-T_x^\star}_{\infty},
\]
and therefore
\begin{equation}
  \norm{\hat\nu_{0,x}-\nu_{0,x}^\star}_{L^2([\underline y,\overline y])}
  \le (\overline y-\underline y)^{3/2}L_{12}(x)\,\norm{\hat T_x-T_x^\star}_{\infty}.
  \label{eq:nu0-bound-explicit-x}
\end{equation}

\item\label{step:monge-treated-dual-rate}
In this step, we derive the error of estimated duals corresponding to the treatment group.
Define the $c$-transform
\[
  (\mathcal C_x \phi)(y) = \inf_{z\in[\underline y,\overline y]} \Bigl\{ f_x(z,y)-\phi(z) \Bigr\}.
\]
We claim that $\mathcal C_x$ is $1$-Lipschitz from $L^\infty([\underline y,\overline y])$ to $L^\infty([\underline y,\overline y])$. Indeed, if $\phi,\psi\in L^\infty([\underline y,\overline y])$, then for every $y$,
\[
  (\mathcal C_x \phi)(y)
  = \inf_{z\in[\underline y,\overline y]} \{ f_x(z,y)-\phi(z) \}
  \le \inf_{z\in[\underline y,\overline y]} \{ f_x(z,y)-\psi(z) + \norm{\phi-\psi}_{\infty} \}
  = (\mathcal C_x \psi)(y) + \norm{\phi-\psi}_{\infty}.
\]
The same argument gives the reverse inequality and hence
\begin{equation}
  \norm{\mathcal C_x \phi-\mathcal C_x \psi}_{\infty}
  \le \norm{\phi-\psi}_{\infty}.
  \label{eq:c-transform-contraction-explicit-x}
\end{equation}
Since $\nu_{1,x}^\star = \mathcal C_x \nu_{0,x}^\star$ and $\hat\nu_{1,x} = \mathcal C_x \hat\nu_{0,x}$, \eqref{eq:c-transform-contraction-explicit-x} implies
\[
  \norm{\hat\nu_{1,x}-\nu_{1,x}^\star}_{\infty}
  \le \norm{\hat\nu_{0,x}-\nu_{0,x}^\star}_{\infty}.
\]
Therefore
\begin{equation}
  \norm{\hat\nu_{1,x}-\nu_{1,x}^\star}_{L^2([\underline y,\overline y])}
  \le (\overline y-\underline y)^{3/2}L_{12}(x)\,\norm{\hat T_x-T_x^\star}_{\infty}.
  \label{eq:nu1-bound-explicit-x}
\end{equation}

\item\label{step:monge-dual-rate}
Now we prove part~\ref{part:monge-dual-rate}.
By \eqref{eq:nu0-bound-explicit-x} and \eqref{eq:nu1-bound-explicit-x},
\[
  \mathrm{error}_{\nu,2}(x)
  = \Bigl( \norm{\hat\nu_{0,x}-\nu_{0,x}^\star}_{L^2([\underline y,\overline y])}^2
    + \norm{\hat\nu_{1,x}-\nu_{1,x}^\star}_{L^2([\underline y,\overline y])}^2 \Bigr)^{1/2}
  \le \sqrt{2}(\overline y-\underline y)^{3/2}L_{12}(x)\,\norm{\hat T_x-T_x^\star}_{\infty}.
\]
Combining this with \eqref{eq:transport-error-bound-explicit-x} gives
\[
  \mathrm{error}_{\nu,2}(x)
  \le \frac{2(\overline y-\underline y)^2L_{12}(x)}{\underline p(x)}\,\mathrm{error}_{P,2}(x)
  = H_f(x)\,\mathrm{error}_{P,2}(x).
\]
This proves part~\ref{part:monge-dual-rate}.

\item\label{step:monge-bias}
Lastly, we prove part~\ref{part:monge-bias}.
By Theorem~\ref{thm::inprodbound} and parts~\ref{part:monge-dual-optimality}--\ref{part:monge-dual-rate},
\[
  0 \le \theta_L-\tilde\theta_L
  \le \E\bigl[ \mathrm{error}_{P,2}(X)\, \mathrm{error}_{\nu,2}(X) \mid \mathcal D_1 \bigr]
  \le \E\bigl[ H_f(X)\, \mathrm{error}_{P,2}(X)^2 \mid \mathcal D_1 \bigr].
\]
Applying conditional Cauchy--Schwarz again,
\[
  \theta_L-\tilde\theta_L
  \le \Bigl( \E\bigl[ H_f(X)^2 \bigr] \Bigr)^{1/2}
     \Bigl( \E\bigl[ \mathrm{error}_{P,2}(X)^4 \mid \mathcal D_1 \bigr] \Bigr)^{1/2}.
\]
By definition,
\[
  \E\bigl[ H_f(X)^2 \bigr]
  = 4(\overline y-\underline y)^4 \E\left[ \left( \frac{L_{12}(X)}{\underline p(X)} \right)^2 \right] < \infty.
\]
Then if,
\[
  \E\bigl[ \mathrm{error}_{P,2}(X)^4 \bigr] = o(n^{-1}),
\]
we have
\[
  \Bigl( \E\bigl[ \mathrm{error}_{P,2}(X)^4 \mid \mathcal D_1 \bigr] \Bigr)^{1/2} = o_p(n^{-1/2}).
\]
Therefore
\[
  \theta_L-\tilde\theta_L = o_p(n^{-1/2}),
\]
which is equivalent to
\[
  \sqrt{n}\,\bigl(\tilde\theta_L-\theta_L\bigr)=o_p(1).
\]
This proves part~\ref{part:monge-bias}.

\end{enumerate}
\end{proof}
\endgroup

\subsection{A generic margin condition}\label{sec:general-margin}
\begingroup
\newcommand{\ErrP}{\mathrm{error}_{P,r}}
\newcommand{\Errnu}{\mathrm{error}_{\nu,s}}

In this subsection, we exploit a low-dimensional structure of the Kantorovich
duals for the seven examples in Section~\ref{subsec::examples} and develop a
generic margin condition. For any given \(x\), rather than the full conditional
feasible class \(\mathcal V_x\), we use the scalar-indexed dual families derived
in Appendix~\ref{sec: dual form}. We start by considering an abstract setting
that unifies Examples~\ref{ex::lee}--\ref{ex:CTE}. Let \(\mathcal U_x\subseteq\mathbb R\) be the
index set.

For each \(u\in\mathcal U_x\), \(\alpha_x^\star(u)\) records the
dual-family choice under the true conditional distributions, while
\(\widehat\alpha_x(u)\) records the corresponding choice after replacing those
distributions by first-stage estimates. For example, in the
Lee-bound case, \(u\) is the cutoff \(a\) indexing the dual family
\((\nu_{0,x}^{(a)},\nu_{1,x}^{(a)})\) in
\eqref{eq:lee-dual-family}. The population problem chooses \(a_x^\star\) as in
\eqref{eq:lee-dual-optimizer}, while the estimated cutoff is obtained directly
from \(\widehat q_0(x)\) and \(\widehat p_{1,1}(\cdot\mid x)\), as in
Section~\ref{subsec:structured-margins}:
\[
  \widehat a_x
  =\inf\left\{a\in\mathbb R:
    \int_{-\infty}^{a}\widehat p_{1,1}(y\mid x)\,dy
    \ge\widehat q_0(x)
  \right\}.
\]
In this case,
\[
  \alpha_x^\star(u)=\1\{u\le a_x^\star\},
  \qquad
  \widehat\alpha_x(u)=\1\{u\le \widehat a_x\},
\]
so \(u\in\mathcal U_x\) is the cutoff \(a\) indexing the dual family, and the
two rules record which side of the population and estimated cutoffs \(u\) falls
on.

We define the disagreement set
\[
  M_x=\{u\in\mathcal U_x:\widehat\alpha_x(u)\ne\alpha_x^\star(u)\}.
\]
We use \(\operatorname{Leb}\) to denote Lebesgue measure.

The proposition below provides generic conditions that yield a bound on the
dual error based on the primal error.

\begin{proposition}
\label{prop:structured-margin}
Let
\(\Psi_x,\widehat\Psi_x:\mathcal U_x\to\mathbb R\) be two
population and estimated score functions. Suppose that,
for almost every \(x\), there exist measurable functions
\(L_\nu(x),L_\Psi(x),m_b(x)<\infty\) and an exponent \(b>0\) such that the
following conditions hold.
\begin{enumerate}[label=\textnormal{(A\arabic*)},ref=\textnormal{(A\arabic*)}]
\item\label{cond:structured-dual-disagreement} The error in the dual variables is bounded by the size of the disagreement set
\[
  \operatorname{error}_{\nu,s}(x)\le L_\nu(x)\operatorname{Leb}(M_x).
\]
\item\label{cond:structured-score-disagreement} The disagreement set can be characterized by the score functions
\[
  M_x\subseteq
  \left\{u\in\mathcal U_x:0<|\Psi_x(u)|
  \le
  \sup_{v\in\mathcal U_x}|\widehat\Psi_x(v)-\Psi_x(v)|
  \right\}.
\]
\item\label{cond:structured-score-error} The error in the score is bounded by primal error
\[
  \sup_{u\in\mathcal U_x}|\widehat\Psi_x(u)-\Psi_x(u)|
  \le L_\Psi(x)\operatorname{error}_{P,r}(x).
\]
\item\label{cond:structured-margin-mass} For all \(s>0\)
\[
  \operatorname{Leb}\{u\in\mathcal U_x:0<|\Psi_x(u)|\le s\}\le m_b(x)s^b.
\]
\end{enumerate}
Then, with \(C_b(x)=L_\nu(x)m_b(x)L_\Psi(x)^b\),
\[
  \operatorname{error}_{\nu,s}(x)
  \le
  C_b(x)\operatorname{error}_{P,r}(x)^b.
\]
\end{proposition}

\begin{proof}
By \ref{cond:structured-score-disagreement} and \ref{cond:structured-margin-mass},
\[
\operatorname{Leb}(M_x)
\le
\operatorname{Leb}\!\left(
\left\{u: 0<|\Psi_x(u)|
\le \sup_{v\in\mathcal U_x}|\widehat\Psi_x(v)-\Psi_x(v)|
\right\}
\right)
\le
m_{b}(x)
\left(\sup_{v\in\mathcal U_x}|\widehat\Psi_x(v)-\Psi_x(v)|\right)^{b}.
\]
Using \ref{cond:structured-score-error} gives
\[
\operatorname{Leb}(M_x)
\le m_{b}(x)L_\Psi(x)^{b}\ErrP(x)^{b}.
\]
Multiplying by $L_\nu(x)$ and applying \ref{cond:structured-dual-disagreement} yields
\begin{equation}\label{eq:structured-dual-error-proof}
\Errnu(x)
\le L_\nu(x)\operatorname{Leb}(M_x)
\le L_\nu(x)m_{b}(x)L_\Psi(x)^{b}\ErrP(x)^{b}
= C_{b}(x)\ErrP(x)^{b}.
\end{equation}
\end{proof}

\endgroup

\subsection{Lee bounds under monotone selection}\label{sec:lee}
\begingroup
\newcommand{\Pbb}{\mathbb{P}}
\newcommand{\norm}[1]{\left\lVert #1 \right\rVert}
\newcommand{\abs}[1]{\left\lvert #1 \right\rvert}
\newcommand{\set}[1]{\left\{ #1 \right\}}
\newcommand{\Yset}{[\underline y,\overline y]}

We will assume that the potential outcomes have a bounded support
\(\mathcal Y=[\underline y,\overline y]\) throughout the section.
Recall the notation in Appendix~\ref{app:dual-lee}. Define
\[
q_0(x)=\Pbb\{S(0)=1\mid X=x\},\qquad
G_x(a)=\int_{\underline y}^{a}p_{1,1}(y\mid x)\,dy,
\qquad
q_1(x)=G_x(\overline y),
\]
and let \(\widehat q_0(x)\), \(\widehat p_{1,1}(\cdot\mid x)\), and
\[
\widehat G_x(a)=\int_{\underline y}^{a}
\widehat p_{1,1}(y\mid x)\,dy,
\qquad
\widehat q_1(x)=\widehat G_x(\overline y)
\]
be the corresponding first-stage estimates. Define
\begin{equation}\label{eq:lee-solution-set}
\mathcal A_x^\star=\{a\in\Yset:G_x(a)=q_0(x)\},\qquad
a_x^\star=\inf\mathcal A_x^\star,
\qquad
\widehat a_x=\inf\{a\in\Yset:\widehat G_x(a)\ge\widehat q_0(x)\}.
\end{equation}
The population and estimated duals are
\[
\nu^\star_{0,x}(r_0,s_0)=a_x^\star s_0-r_0,
\qquad
\nu^\star_{1,x}(r_1,s_1)=(r_1-a_x^\star s_1)_-,
\]
\[
\widehat\nu_{0,x}(r_0,s_0)=\widehat a_xs_0-r_0,
\qquad
\widehat\nu_{1,x}(r_1,s_1)=(r_1-\widehat a_xs_1)_-,
\]
with multipliers \(\lambda_x^\star=(a_x^\star)_+\) and
\(\widehat\lambda_x=(\widehat a_x)_+\). For \(a\in\Yset\), set
\[
\mu_0(x)=\E[Y(0)S(0)\mid X=x],\qquad
g_x(a)=a q_0(x)+\int_{\underline y}^{\overline y}
(y-a)_-p_{1,1}(y\mid x)\,dy.
\]
For \(a<b\),
\begin{equation}\label{eq:lee-objective-increment}
g_x(b)-g_x(a)=\int_a^b\{q_0(x)-G_x(c)\}\,dc.
\end{equation}
Finally, define
\[
\widetilde\beta_L
=\E[\widehat\nu_{0,X}\{Z(0)\}+\widehat\nu_{1,X}\{Z(1)\}\mid \mathcal D_1]
=\E[g_X(\widehat a_X)-\mu_0(X)\mid \mathcal D_1],
\]
\begin{equation}\label{eq:lee-error-P2}
\error_{P,2}(x)^2
=|\widehat q_0(x)-q_0(x)|^2
+\int_{\underline y}^{\overline y}
|\widehat p_{1,1}(y\mid x)-p_{1,1}(y\mid x)|^2\,dy,
\end{equation}
and
\begin{equation}\label{eq:lee-error-nu2}
\error_{\nu,2}(x)^2
=|\widehat a_x-a_x^\star|^2
+\int_{\underline y}^{\overline y}
|\widehat\nu_{1,x}(y,1)-\nu^\star_{1,x}(y,1)|^2\,dy.
\end{equation}

\begin{assumption}\label{ass:lee-setup}
For almost every \(x\), the following hold.
\begin{enumerate}[label=(\arabic*)]
\item The true distribution satisfies monotone selection
\(S(1)\ge S(0)\) almost surely conditional on \(X=x\).
\item For each \(w\in\{0,1\}\),
\[
\Pbb\{S(w)=1,\ Y(w)\notin\Yset\mid X=x\}=0.
\]
Moreover, \((y,x)\mapsto p_{1,1}(y\mid x)\) is a jointly measurable nonnegative
subdensity with \(p_{1,1}(\cdot\mid x)\in L^2(\Yset)\) and
\[
0\le q_0(x)\le q_1(x)\le1.
\]
\item Conditional on \(\mathcal D_1\), \(\widehat q_0(x)\) and
\((y,x)\mapsto\widehat p_{1,1}(y\mid x)\) are jointly measurable,
\(\widehat p_{1,1}(\cdot\mid x)\) is a nonnegative element of
\(L^2(\Yset)\), and
\[
0\le\widehat q_0(x)\le\widehat q_1(x)\le1.
\]
\end{enumerate}
\end{assumption}

\begin{assumption}\label{ass:lee-holder}
For almost every \(x\), \(\operatorname{Leb}(\mathcal A_x^\star)=0\).
There exist a measurable function \(m_b:\mathcal X\to[0,\infty)\) and an
exponent \(b>0\) such that, for almost every \(x\) and every \(u>0\),
\[
\operatorname{Leb}\{a\in\Yset:0<|G_x(a)-q_0(x)|\le u\}
\le m_b(x)u^b.
\]
\end{assumption}

\begin{theorem}\label{prop:lee-holder}
Suppose Assumptions~\ref{ass:lee-setup} and~\ref{ass:lee-holder} hold. Then,
for almost every \(x\),
\begin{equation}\label{eq:lee-dual-rate}
\error_{\nu,2}(x)
\le\sqrt2\,m_b(x)
\left(1+\overline y-\underline y\right)^{(b+1)/2}
\error_{P,2}(x)^b.
\end{equation}

The first-stage bias satisfies
\[
0\le\beta_L-\widetilde\beta_L
\le\sqrt2\left(1+\overline y-\underline y\right)^{(b+1)/2}
\E[m_b(X)\error_{P,2}(X)^{b+1}\mid \mathcal D_1].
\]
Consequently, if
\[
\E[m_b(X)\error_{P,2}(X)^{b+1}]=o(n^{-1/2}),
\]
then
\[
\sqrt n(\widetilde\beta_L-\beta_L)=o_p(1).
\]
\end{theorem}

\begin{proof}
Appendix~\ref{app:dual-lee} gives
\(\beta_L(x)=g_x(a_x^\star)-\mu_0(x)\) and the feasibility of the
estimated dual.

We apply Proposition~\ref{prop:structured-margin} with
\[
\mathcal U_x=\mathcal Y,\qquad
\Psi_x(u)=G_x(u)-q_0(x),\qquad
\widehat\Psi_x(u)=\widehat G_x(u)-\widehat q_0(x),
\]
and
\[
\alpha_x^\star(u)=\1\{u\le a_x^\star\},\qquad
\widehat\alpha_x(u)=\1\{u\le\widehat a_x\}.
\]
Since \(G_x\) is nondecreasing, \(\mathcal A_x^\star=G_x^{-1}\{q_0(x)\}\)
is an interval. Assumption~\ref{ass:lee-holder} therefore gives
\[
\operatorname{Leb}(\mathcal A_x^\star)=0
\quad\Longrightarrow\quad
\mathcal A_x^\star=\{a_x^\star\}.
\]
At this singleton, replace \(\widehat\alpha_x\) by
\[
\widehat\alpha_x(u)
\leftarrow
\begin{cases}
1,&u=a_x^\star,\\
\1\{u\le\widehat a_x\},&u\ne a_x^\star.
\end{cases}
\]
Because \(\operatorname{Leb}\{a_x^\star\}=0\), this modification is
null in \(L^2\), and
\[
M_x=
\begin{cases}
(a_x^\star,\widehat a_x],&\widehat a_x>a_x^\star,\\
(\widehat a_x,a_x^\star),&\widehat a_x<a_x^\star,\\
\varnothing,&\widehat a_x=a_x^\star,
\end{cases}
\qquad
\operatorname{Leb}(M_x)=|\widehat a_x-a_x^\star|.
\]

Assumption~\ref{ass:lee-setup} gives
\[
0=\widehat G_x(\underline y)
\le\widehat q_0(x)
\le\widehat q_1(x)=\widehat G_x(\overline y).
\]
By the definition of \(\widehat a_x\),
\[
\widehat G_x(\widehat a_x-)
\le\widehat q_0(x)
\le\widehat G_x(\widehat a_x).
\]
Here \(\widehat G_x(\underline y-):=0\).
Continuity of \(\widehat G_x\) therefore implies
\[
\widehat G_x(\widehat a_x)=\widehat q_0(x),
\qquad
\widehat\Psi_x(\widehat a_x)
=\widehat G_x(\widehat a_x)-\widehat q_0(x)=0.
\]
Consequently,
\[
|\Psi_x(\widehat a_x)|
=|\Psi_x(\widehat a_x)-\widehat\Psi_x(\widehat a_x)|
\le\sup_{u\in\mathcal Y}
|\widehat\Psi_x(u)-\Psi_x(u)|.
\]
Since \(G_x^{-1}\{q_0(x)\}=\{a_x^\star\}\), monotonicity gives, for
\(u\in M_x\),
\[
\begin{cases}
a_x^\star<u\le\widehat a_x
\ \Longrightarrow\ 
0<G_x(u)-q_0(x)\le G_x(\widehat a_x)-q_0(x),
&\widehat a_x>a_x^\star,\\
\widehat a_x<u<a_x^\star
\ \Longrightarrow\ 
0<q_0(x)-G_x(u)\le q_0(x)-G_x(\widehat a_x),
&\widehat a_x<a_x^\star.
\end{cases}
\]
Thus
\begin{equation}\label{eq:lee-score-disagreement}
0<|\Psi_x(u)|
\le |\Psi_x(\widehat a_x)|
\le\sup_{v\in\mathcal Y}
|\widehat\Psi_x(v)-\Psi_x(v)|,
\end{equation}
which verifies condition~\ref{cond:structured-score-disagreement}.

Cauchy--Schwarz and equation~\eqref{eq:lee-error-P2} give
\begin{equation}\label{eq:lee-score-error}
\begin{aligned}
\sup_{u\in\mathcal Y}
|\widehat\Psi_x(u)-\Psi_x(u)|
&\le |\widehat q_0(x)-q_0(x)|
+\sup_{u\in\mathcal Y}\left|
\int_{\underline y}^{u}
\{\widehat p_{1,1}(y\mid x)-p_{1,1}(y\mid x)\}\,dy\right|\\
&\le |\widehat q_0(x)-q_0(x)|
+\sqrt{\overline y-\underline y}\,
\|\widehat p_{1,1}(\cdot\mid x)-p_{1,1}(\cdot\mid x)\|_2\\
&\le\sqrt{1+\overline y-\underline y}
\left\{|\widehat q_0(x)-q_0(x)|^2
+\|\widehat p_{1,1}(\cdot\mid x)-p_{1,1}(\cdot\mid x)\|_2^2\right\}^{1/2}\\
&=\sqrt{1+\overline y-\underline y}\,\error_{P,2}(x).
\end{aligned}
\end{equation}
Thus condition~\ref{cond:structured-score-error} holds with
\(L_\Psi(x)=\sqrt{1+\overline y-\underline y}\). For \(y\in\mathcal Y\),
\[
\begin{aligned}
|\widehat\nu_{0,x}(y,1)-\nu^\star_{0,x}(y,1)|
&=|\widehat a_x-a_x^\star|,\\
|\widehat\nu_{1,x}(y,1)-\nu^\star_{1,x}(y,1)|
&=|(y-\widehat a_x)_--(y-a_x^\star)_-|\le |\widehat a_x-a_x^\star|.
\end{aligned}
\]
Hence, by equation~\eqref{eq:lee-error-nu2},
\begin{equation}\label{eq:lee-dual-disagreement}
\begin{aligned}
\error_{\nu,2}(x)^2
&\le (1+\overline y-\underline y)|\widehat a_x-a_x^\star|^2,\\
\error_{\nu,2}(x)
&\le\sqrt{1+\overline y-\underline y}\,|\widehat a_x-a_x^\star|
=\sqrt{1+\overline y-\underline y}\,\operatorname{Leb}(M_x).
\end{aligned}
\end{equation}
Thus condition~\ref{cond:structured-dual-disagreement} holds with
\(L_\nu(x)=\sqrt{1+\overline y-\underline y}\). Finally, Assumption~\ref{ass:lee-holder} is
condition~\ref{cond:structured-margin-mass}. Proposition~\ref{prop:structured-margin}
now yields
\[
\error_{\nu,2}(x)
\le m_b(x)
(1+\overline y-\underline y)^{(b+1)/2}\error_{P,2}(x)^b.
\]
This implies the bound in \eqref{eq:lee-dual-rate}.

Following the same argument as in
Theorem~\ref{thm::inprodbound},
\[
\begin{aligned}
0\le\beta_L-\widetilde\beta_L
&\le \E\!\left[
|\widehat q_0(X)-q_0(X)|\,|\widehat a_X-a_X^\star|
+\|\widehat p_{1,1}(\cdot\mid X)-p_{1,1}(\cdot\mid X)\|_2
\|\widehat\nu_{1,X}(\cdot,1)-\nu^\star_{1,X}(\cdot,1)\|_2
\,\middle|\,\mathcal D_1\right]\\
&\le \E[\error_{P,2}(X)\error_{\nu,2}(X)\mid \mathcal D_1]\\
&\le \sqrt2(1+\overline y-\underline y)^{(b+1)/2}
\E[m_b(X)\error_{P,2}(X)^{b+1}\mid \mathcal D_1].
\end{aligned}
\]
Under the stated rate condition, the law of iterated expectations and Markov's
inequality imply that this bound is \(o_p(n^{-1/2})\).
\end{proof}

\begin{assumption}\label{ass:lee-simple-zero}
There exist measurable functions \(\rho,\kappa:\mathcal X\to(0,\infty)\)
such that, for almost every \(x\),
\[
I_x=[a_x^\star-\rho(x),a_x^\star+\rho(x)]
\subset(\underline y,\overline y)
\]
and
\[
p_{1,1}(a\mid x)\ge\kappa(x)
\quad\text{for almost every }a\in I_x.
\]
\end{assumption}

\begin{corollary}\label{cor:lee-simple-zero}
Suppose Assumptions~\ref{ass:lee-setup} and~\ref{ass:lee-simple-zero} hold.
Then, for almost every \(x\),
\[
\error_{\nu,2}(x)\le B_{\nu,2}(x)\error_{P,2}(x),
\]
where
\[
B_{\nu,2}(x)
=\frac{1+\overline y-\underline y}{\kappa(x)}
\vee
\frac{(1+\overline y-\underline y)(\overline y-\underline y)}
{\kappa(x)\rho(x)}.
\]
Moreover,
\[
0\le\beta_L-\widetilde\beta_L
\le\E[B_{\beta,2}(X)\error_{P,2}(X)^2\mid \mathcal D_1],
\]
where
\[
B_{\beta,2}(x)
=\frac{1+\overline y-\underline y}{\kappa(x)}
\vee
\frac{(1+\overline y-\underline y)
(\overline y-\underline y)}{\kappa(x)\rho(x)}.
\]
Consequently, if
\[
\E[B_{\beta,2}(X)\error_{P,2}(X)^2]=o(n^{-1/2}),
\]
then \(\sqrt n(\widetilde\beta_L-\beta_L)=o_p(1)\).
\end{corollary}

\begin{proof}
Fix \(x\) and write
\[
\delta_x
=\sup_{u\in\mathcal Y}|\widehat\Psi_x(u)-\Psi_x(u)|.
\]
Equations~\eqref{eq:lee-score-error} and~\eqref{eq:lee-dual-disagreement}
in the proof of Theorem~\ref{prop:lee-holder} give
\begin{equation}\label{eq:lee-cor-score-error}
\delta_x
\le\sqrt{1+\overline y-\underline y}\,\error_{P,2}(x),
\end{equation}
and
\begin{equation}\label{eq:lee-cor-dual-disagreement}
\error_{\nu,2}(x)
\le\sqrt{1+\overline y-\underline y}\,\operatorname{Leb}(M_x),
\qquad
\operatorname{Leb}(M_x)=|\widehat a_x-a_x^\star|.
\end{equation}
For \(u\in I_x\), Assumption~\ref{ass:lee-simple-zero} implies
\begin{equation}\label{eq:lee-cor-local-growth}
|\Psi_x(u)|
=\left|\int_{a_x^\star}^{u}p_{1,1}(v\mid x)\,dv\right|
\ge\kappa(x)|u-a_x^\star|.
\end{equation}

If \(\delta_x<\kappa(x)\rho(x)\), then
\(\widehat\Psi_x(a_x^\star-\rho(x))<0\) and
\(\widehat\Psi_x(a_x^\star+\rho(x))>0\), so
\(\widehat a_x\in I_x\). Since
\(\widehat\Psi_x(\widehat a_x)=0\),
\eqref{eq:lee-cor-local-growth} gives
\[
\operatorname{Leb}(M_x)
=|\widehat a_x-a_x^\star|
\le\frac{|\Psi_x(\widehat a_x)|}{\kappa(x)}
\le\frac{\delta_x}{\kappa(x)}.
\]
If \(\delta_x\ge\kappa(x)\rho(x)\), then
\[
\operatorname{Leb}(M_x)
\le\overline y-\underline y
\le\frac{\overline y-\underline y}{\kappa(x)\rho(x)}\,\delta_x.
\]
Thus
\begin{equation}\label{eq:lee-cor-M-bound}
\operatorname{Leb}(M_x)
\le
\left\{
\frac{1}{\kappa(x)}
\vee
\frac{\overline y-\underline y}{\kappa(x)\rho(x)}
\right\}\delta_x.
\end{equation}
Combining \eqref{eq:lee-cor-score-error}--\eqref{eq:lee-cor-M-bound}
proves \(\error_{\nu,2}(x)\le B_{\nu,2}(x)\error_{P,2}(x)\).

Moreover, equations~\eqref{eq:lee-objective-increment}
and~\eqref{eq:lee-score-disagreement} give
\[
0\le g_x(a_x^\star)-g_x(\widehat a_x)
=\int_{M_x}|\Psi_x(u)|\,du
\le\delta_x\operatorname{Leb}(M_x).
\]
The two cases above therefore imply
\[
0\le g_x(a_x^\star)-g_x(\widehat a_x)
\le
\left\{
\frac{1}{\kappa(x)}
\vee
\frac{\overline y-\underline y}{\kappa(x)\rho(x)}
\right\}\delta_x^2
\le B_{\beta,2}(x)\error_{P,2}(x)^2.
\]
Averaging over \(X\) proves the bias bound. Under the stated rate condition,
the law of iterated expectations and Markov's inequality imply that this bound
is \(o_p(n^{-1/2})\), proving the final claim.
\end{proof}

\medskip
\noindent\textit{An alternative based on direct conditional-quantile estimation.}
The estimator analyzed in Theorem~\ref{prop:lee-holder} uses estimates of
\(q_0(x)\) and the treated-selected conditional subdensity
\(p_{1,1}(\cdot\mid x)\), whereas \citet{semenova2021generalized} uses estimates
of \(q_0(x)\), \(q_1(x)\), and the selected-treated conditional quantile
\(Q_x(\cdot)\). To make the comparison direct, we recast our dual-bound
estimator using the same nuisance parametrization as
\citet{semenova2021generalized}.
For \(q_1(x)>0\), define the selected-treated conditional CDF and its
generalized inverse by
\[
F_x(a)=\frac{G_x(a)}{q_1(x)},
\qquad
Q_x(u)=\inf\{a\in\Yset:F_x(a)\ge u\},
\qquad u\in[0,1],
\]
and let \(\eta(x)=q_0(x)/q_1(x)\). If \(q_1(x)=0\), then \(q_0(x)=0\);
in this case, set \(\eta(x)=0\) and \(Q_x(u)=\underline y\) for
\(u\in[0,1]\). With this convention, joint measurability of \(G_x\) implies
that \((x,u)\mapsto Q_x(u)\) is jointly measurable. Continuity of \(G_x\)
and \eqref{eq:lee-solution-set} imply
\begin{equation}\label{eq:lee-quantile-cutoff}
a_x^\star=Q_x\{\eta(x)\}.
\end{equation}

Conditional on \(\mathcal D_1\), let \(\widehat q_0(x)\) and
\(\widehat q_1(x)\) be estimates of \(q_0\) and \(q_1\), and let
\((x,u)\mapsto\widehat Q_x(u)\in\Yset\) be a jointly measurable direct
estimate of \(Q_x(u)\). Define
\[
\widehat\eta(x)=
\begin{cases}
\widehat q_0(x)/\widehat q_1(x),
&\widehat q_1(x)>0,\\
0,&\widehat q_1(x)=0,
\end{cases}
\qquad
\widehat a_x=
\begin{cases}
\widehat Q_x\{\widehat\eta(x)\},
&\widehat q_1(x)>0,\\
\underline y,&\widehat q_1(x)=0.
\end{cases}
\]
The corresponding duals and multiplier are
\begin{equation}\label{eq:lee-quantile-duals}
\widehat\nu_{0,x}(r_0,s_0)
 =\widehat a_xs_0-r_0,\qquad
\widehat\nu_{1,x}(r_1,s_1)
 =(r_1-\widehat a_xs_1)_-,\qquad
\widehat\lambda_x=(\widehat a_x)_+.
\end{equation}
Set
\[
e_a(x)=|\widehat a_x-a_x^\star|,
\]
and
\[
\widetilde\beta_L
=\E\!\left[
\widehat\nu_{0,X}\{Z(0)\}
+\widehat\nu_{1,X}\{Z(1)\}
\mid\mathcal D_1\right].
\]

With this notation, we can state the tightness theory for the new estimator.

\begin{theorem}\label{thm:lee-quantile-duals}
Suppose Assumption~\ref{ass:lee-setup}(1)--(2) holds and, conditional on
\(\mathcal D_1\),
\[
0\le\widehat q_0(x)\le\widehat q_1(x)\le1,
\qquad
\widehat Q_x(u)\in\Yset,
\qquad (x,u)\in\mathcal X\times[0,1].
\]
Suppose also that, for some \(\gamma>0\) and measurable
\(L_\gamma:\mathcal X\to[0,\infty)\),
\begin{equation}\label{eq:lee-quantile-holder}
|G_x(u)-G_x(a_x^\star)|
\le L_\gamma(x)|u-a_x^\star|^\gamma
\quad\text{for every }u\in\Yset
\end{equation}
for almost every \(x\). Then the duals in
\eqref{eq:lee-quantile-duals} are feasible. Moreover,
\begin{equation}\label{eq:lee-quantile-bias}
0\le\beta_L-\widetilde\beta_L
\le
\frac{1}{\gamma+1}
\E\!\left[L_\gamma(X)e_a(X)^{\gamma+1}\mid\mathcal D_1\right].
\end{equation}
Consequently, if
\[
\E\!\left[L_\gamma(X)e_a(X)^{\gamma+1}\right]=o(n^{-1/2}),
\]
then \(\sqrt n(\widetilde\beta_L-\beta_L)=o_p(1)\).
\end{theorem}

\begin{proof}
Equation~\eqref{eq:lee-dual-family} shows that the triple
\((\widehat\nu_{0,x},\widehat\nu_{1,x},
\widehat\lambda_x)\) defined in \eqref{eq:lee-quantile-duals} is
dual feasible whenever \(\widehat a_x\) is finite. Hence it is feasible, regardless of the accuracy of
\(\widehat Q_x\), and weak duality gives
\(\widetilde\beta_L\le\beta_L\).

Setting \(s=1\) in \eqref{eq:lee-dual-family} gives
\[
\nu_{0,x}^{(a)}(y,1)=a-y,
\qquad
\nu_{1,x}^{(a)}(y,1)=(y-a)_-.
\]
Since \(|u_--v_-|\le|u-v|\),
\begin{align*}
|\widehat\nu_{0,x}(y,1)-\nu_{0,x}^\star(y,1)|
&=|(\widehat a_x-y)-(a_x^\star-y)|=e_a(x),\\
|\widehat\nu_{1,x}(y,1)-\nu_{1,x}^\star(y,1)|
&=|(y-\widehat a_x)_--(y-a_x^\star)_-|\\
&\le|(y-\widehat a_x)-(y-a_x^\star)|=e_a(x).
\end{align*}

Since \(G_x(a_x^\star)=q_0(x)\), equation
\eqref{eq:lee-objective-increment} gives
\begin{align*}
0\le g_x(a_x^\star)-g_x(\widehat a_x)
&=\int_{\min\{a_x^\star,\widehat a_x\}}^
{\max\{a_x^\star,\widehat a_x\}}
|G_x(u)-G_x(a_x^\star)|\,du\\
&\le L_\gamma(x)\int_0^{e_a(x)}t^\gamma\,dt
=\frac{L_\gamma(x)}{\gamma+1}e_a(x)^{\gamma+1}.
\end{align*}
Averaging over \(X\) proves \eqref{eq:lee-quantile-bias}. By the law of
iterated expectations and Markov's inequality, the \(o_p(n^{-1/2})\)
conclusion follows.
\end{proof}


\begin{remark}[Comparison with Semenova (2025b)]
\label{rmk:lee-quantile-semenova-comparison}
For an apple-to-apple comparison, restrict both analyses to randomized
experiments with known propensity scores and standard one-sided monotonicity:
\begin{equation}\label{eq:lee-semenova-common-submodel}
(Z(0),Z(1))\perp W\mid X,\qquad
0<\underline\pi\le\pi(X)\le\overline\pi<1,\qquad
S(1)\ge S(0)\quad\text{a.s.},\qquad
0<q_0(x)<q_1(x).
\end{equation}
Semenova (2025b) also handles a weaker monotonicity assumption
\[
S(1)\ge S(0)\ \text{on }\mathcal X_+,\qquad
S(1)=S(0)\ \text{on }\mathcal X_0,\qquad
S(1)\le S(0)\ \text{on }\mathcal X_-,
\]
where the unknown partition
\((\mathcal X_+,\mathcal X_0,\mathcal X_-)\) is determined by the sign of
\(q_1(x)-q_0(x)\) and estimated from the data. This general case goes
beyond our framework, which only allows for conditional moment restrictions.

Both our result and Semenova (2025b) require bounded potential outcomes.
Both procedures then use
\[
\eta(x)=\frac{q_0(x)}{q_1(x)},
\qquad
a_x^\star=Q_x\{\eta(x)\}.
\]

\textbf{Semenova (2025b)'s assumptions. }Let \(f_{w,x}\) denote the density of
\(Y(w)\mid S(w)=1,X=x\), so
\[
f_{w,x}(y)=\frac{p_{w,1}(y\mid x)}{q_w(x)},
\qquad w\in\{0,1\}.
\]
Under the common assumptions in \eqref{eq:lee-semenova-common-submodel},
Semenova's Assumptions 3,
5, and 6 impose, uniformly in \(x\) and \(w\),
\begin{equation}\label{eq:semenova-density-regularity}
\epsilon\le q_w(x)\le1-\epsilon,\qquad
0<\underline f\le f_{w,x}(y)\le\overline f<\infty,\qquad
\|\partial_y f_{w,x}\|_\infty\le C.
\end{equation}
Only the selected-treated conditional quantile enters the bound under
\eqref{eq:lee-semenova-common-submodel}. Define
\[
Q_{1,x}(u)
=\inf\left\{y\in\Yset:
\int_{\underline y}^{y}f_{1,x}(v)\,dv\ge u\right\}.
\]
Let \(\widehat Q_x(u)\) denote an estimate of \(Q_{1,x}(u)\).
Semenova (2025b) defines
\[
\rho_n=\frac{1}{n^{1/4}\log n},
\qquad
\mathcal U_n=
\left[\frac{2\rho_n}{\kappa},1-\frac{2\rho_n}{\kappa}\right].
\]
Semenova (2025b) defines an event \(\mathcal E_n\) with
\(\Pr(\mathcal E_n)\to1\); see Definitions 6.4--6.5 on p.~22 therein for
the detailed construction of \(\mathcal E_n\). On \(\mathcal E_n\), for all
sufficiently large \(n\),
\begin{equation}\label{eq:semenova-estimated-q1-overlap}
\inf_{x\in\mathcal X}\widehat q_1(x)\ge\frac{\epsilon}{2},
\end{equation}
and
\begin{align}
\max_{w\in\{0,1\}}
\|\widehat q_w-q_w\|_{P_X,p}
&\le s_{n,p},
&&p\in\{1,2,\infty\},\notag\\
\sup_{u\in\mathcal U_n}
\|\widehat Q_X(u)-Q_{1,X}(u)\|_{P_X,p}
&\le q_{n,p},
&&p\in\{1,2\}.
\label{eq:semenova-fixed-index-quantile-rate}
\end{align}
Their Assumption 6 requires
\begin{equation}\label{eq:semenova-l1-sup-rates}
s_{n,1}+q_{n,1}=o(1),\qquad
s_{n,\infty}=o(\rho_n).
\end{equation}
It also requires
\begin{equation}\label{eq:semenova-l2-rates}
s_{n,2}+q_{n,2}=o(n^{-1/4}).
\end{equation}
Define
\(\tau(x)=q_1(x)-q_0(x)\) and
\(\widehat\tau(x)=\widehat q_1(x)-\widehat q_0(x)\).
Their Assumption 4 gives
\begin{equation}\label{eq:semenova-margin-rate}
\Pr\{0<|\tau(X)|\le 2\rho_n\}\le 2C\rho_n.
\end{equation}
\textbf{Our assumptions. }In addition to the common assumptions
\eqref{eq:lee-semenova-common-submodel},
Theorem~\ref{thm:lee-quantile-duals} requires
\begin{equation}\label{eq:lee-comparison-regularity}
p_{1,1}(\cdot\mid x)\in L^2(\Yset),\qquad
|G_x(u)-G_x(a_x^\star)|
\le L_\gamma(x)|u-a_x^\star|^\gamma
\end{equation}
and
\begin{equation}\label{eq:lee-comparison-rate}
\E[L_\gamma(X)e_a(X)^{\gamma+1}]=o(n^{-1/2}).
\end{equation}
In the following, we make a detailed comparison between our assumptions and
Semenova (2025b)'s assumptions.
\begin{itemize}
\item Our Theorem~\ref{thm:lee-quantile-duals} does not require any of
\eqref{eq:semenova-density-regularity}--\eqref{eq:semenova-margin-rate}.

\item Under Semenova's Assumptions 3--6, our condition
\eqref{eq:lee-comparison-regularity} holds. Indeed,
\eqref{eq:semenova-density-regularity} and
\(p_{1,1}(y\mid x)=q_1(x)f_{1,x}(y)\) imply
\[
p_{1,1}(\cdot\mid x)\in L^2(\Yset),
\qquad
|G_x(u)-G_x(a_x^\star)|
\le q_1(x)\overline f\,|u-a_x^\star|.
\]
Specifically, it holds with \(\gamma=1\) and
\(L_1(x)=q_1(x)\overline f\le\overline f\).
The converse is not true in general.

\item Semenova (2025b)'s assumptions imply our condition
\eqref{eq:lee-comparison-rate}. For ease of comparison, suppose additionally
that
\[
\Pr(\mathcal E_n)=1,\qquad
\widehat\eta(x)\in\mathcal U_n
\qquad\text{for every }x.
\]
To verify this, let
\[
\Delta_w(x)=\widehat q_w(x)-q_w(x).
\]
Hence, on the event \(\mathcal E_n\),
\begin{equation}\label{eq:semenova-eta-error}
|\widehat\eta(x)-\eta(x)|
=\frac{|\Delta_0(x)-\eta(x)\Delta_1(x)|}{\widehat q_1(x)}
\le\frac{2}{\epsilon}
\{|\Delta_0(x)|+|\Delta_1(x)|\}
\le\frac{4s_{n,\infty}}{\epsilon}.
\end{equation}
Let
\(H_{1,x}(y)=\int_{\underline y}^{y}f_{1,x}(z)\,dz\). The lower density
bound in \eqref{eq:semenova-density-regularity} makes \(H_{1,x}\) strictly
increasing, so \(H_{1,x}\{Q_{1,x}(u)\}=u\). Hence, for \(u\ge v\),
\[
u-v
=\int_{Q_{1,x}(v)}^{Q_{1,x}(u)}f_{1,x}(z)\,dz
\ge \underline f\{Q_{1,x}(u)-Q_{1,x}(v)\}.
\]
Interchanging \(u\) and \(v\) gives
\[
|Q_{1,x}(u)-Q_{1,x}(v)|
\le\frac{|u-v|}{\underline f},
\qquad u,v\in[0,1].
\]
In Step 2.d of the proof of Lemma A.11,
Semenova (2025b) proves the following bound:
\[
\left\|
\widehat Q_X\{\widehat\eta(X)\}
-Q_{1,X}\{\widehat\eta(X)\}
\right\|_{P_X,2}\le q_{n,2},
\]
Then, we have
\[
\widehat a_x=\widehat Q_x\{\widehat\eta(x)\}.
\]
Then
\[
\begin{aligned}
e_a(x)
&=|\widehat a_x-a_x^\star|\\
&=|\widehat Q_x\{\widehat\eta(x)\}
  -Q_{1,x}\{\eta(x)\}|\\
&\le
|\widehat Q_x\{\widehat\eta(x)\}
  -Q_{1,x}\{\widehat\eta(x)\}|
+|Q_{1,x}\{\widehat\eta(x)\}-Q_{1,x}\{\eta(x)\}|\\
&\le
|\widehat Q_x\{\widehat\eta(x)\}
  -Q_{1,x}\{\widehat\eta(x)\}|
+\frac{|\widehat\eta(x)-\eta(x)|}{\underline f}\\
&\le
|\widehat Q_x\{\widehat\eta(x)\}
  -Q_{1,x}\{\widehat\eta(x)\}|
+\frac{4s_{n,\infty}}{\epsilon\underline f}.
\end{aligned}
\]
Combining \eqref{eq:semenova-l2-rates} and
\eqref{eq:semenova-eta-error}, on \(\mathcal E_n\) the first-stage rates give
\[
\E[L_1(X)e_a(X)^2]
\lesssim s_{n,\infty}^2+q_{n,2}^2
=o(n^{-1/2}).
\]
The converse is not true in general.

\end{itemize}

\end{remark}

\endgroup

\subsection{\texorpdfstring{Distribution treatment effect}{Distribution treatment effect}}\label{sec:makarov}
\begingroup
\newcommand{\ind}[1]{\mathbf{1}\!\left\{#1\right\}}
\newcommand{\thetL}{\theta_L}
\newcommand{\tildethetL}{\widetilde\theta_L}
\newcommand{\Gx}{\mathcal{G}_x}
\newcommand{\ErrP}{\mathrm{error}_{P,r}}
\newcommand{\ErrNu}{\mathrm{error}_{\nu,s}}

Fix \(r\in(1,\infty]\), and let \(s\in[1,\infty)\) satisfy
\(1/r+1/s=1\). We use the general errors \(\ErrP(x)\) and \(\ErrNu(x)\)
defined in Section~\ref{subsec::tightness}, with the convention
\(b/\infty=0\).

Without loss of generality, consider Example~\ref{ex::makarov} at \(t=0\):
\[
\theta(P)=\E_P\!\left[\ind{Y(1)\le Y(0)}\right],
\qquad
f(y_0,y_1,x)=\ind{y_1\le y_0}.
\]
When \(t\neq 0\), we can redefine \(Y(1)\) as \(Y(1)-t\), and the estimand
reduces to the case with \(t=0\). For
\(k\in\{0,1\}\), let \(p_k^\star(\cdot\mid x)\) and
\(\widehat p_k(\cdot\mid x)\) be the true and estimated conditional densities
on \(\mathbb R\), and define
\begin{equation}
\label{eq:makarov-cdfs}
\begin{aligned}
F_{k,x}(a)=\int_{-\infty}^{a}p_k^\star(y\mid x)\,dy,
\qquad
\widehat F_{k,x}(a)=\int_{-\infty}^{a}\widehat p_k(y\mid x)\,dy.
\end{aligned}
\end{equation}
The true and estimated CDF gaps are
\begin{equation}
\label{eq:makarov-cdf-gaps}
\begin{aligned}
\Delta_x(a)=F_{1,x}(a)-F_{0,x}(a),
\qquad
\widehat\Delta_x(a)=\widehat F_{1,x}(a)-\widehat F_{0,x}(a),
\end{aligned}
\end{equation}
and the selected thresholds are
\[
a_x^\star=\inf\argmax_{a\in\mathbb R}\Delta_x(a),
\qquad
\widehat a_x=\inf\argmax_{a\in\mathbb R}\widehat\Delta_x(a).
\]
Define
\begin{equation}\label{eq:makarov-selectors}
\widehat\alpha_x(u)=\ind{u\le\widehat a_x},
\qquad
\alpha_x^\star(u)=\ind{u\le a_x^\star},
\qquad u\in\mathbb R.
\end{equation}
Set
\begin{equation}\label{eq:makarov-selector-scores}
\mathcal U_x=\mathbb R,
\qquad
\Psi_x(u)=\Delta_x'(u),
\qquad
\widehat\Psi_x(u)=\widehat\Delta_x'(u).
\end{equation}
For \(a\in\mathbb R\), use the duals in \eqref{eq:makarov-dual-family} from
Appendix~\ref{app:dual-makarov}, with \(t=0\):
\[
\nu_{0,x}^{(a)}(y_0)=-\ind{y_0<a},
\qquad
\nu_{1,x}^{(a)}(y_1)=\ind{y_1\le a}.
\]
Define
\[
\nu_{0,x}^\star(y_0)=-\ind{y_0<a_x^\star},
\qquad
\nu_{1,x}^\star(y_1)=\ind{y_1\le a_x^\star},
\]
and
\[
\hat\nu_{0,x}(y_0)=-\ind{y_0<\widehat a_x},
\qquad
\hat\nu_{1,x}(y_1)=\ind{y_1\le\widehat a_x}.
\]
For any duals \((\nu_{0,x},\nu_{1,x})\), let
\[
\Gx(\nu_{0,x},\nu_{1,x})
=
\int_{\mathbb R}\nu_{0,x}(y)p_0^\star(y\mid x)\,dy
+
\int_{\mathbb R}\nu_{1,x}(y)p_1^\star(y\mid x)\,dy.
\]
Because the conditional distributions have densities,
\(F_{0,x}(a-)=F_{0,x}(a)\). Equation~(\ref{eq:makarov-sharp-lower-bound}) in Appendix~\ref{app:dual-makarov} therefore gives
\[
\thetL(x)
=
\max\left\{\sup_{a\in\mathbb R}\Delta_x(a),0\right\}
=
\sup_{a\in\mathbb R}\Delta_x(a)
=
\Gx(\nu_{0,x}^\star,\nu_{1,x}^\star),
\]
where the second equality follows from
\(\lim_{a\to-\infty}\Delta_x(a)=0\).
The same formula with the estimated conditional distributions shows that
\((\hat\nu_{0,x},\hat\nu_{1,x})\) is optimal for the estimated conditional problem.

Set
\[
\thetL=\E[\thetL(X)],
\qquad
\tildethetL
=
\E\!\left[
\hat\nu_{0,X}(Y(0))+\hat\nu_{1,X}(Y(1))
\mid \mathcal D_1
\right],
\]
where \(X\) is a fresh covariate draw independent of \(\mathcal D_1\).

Define the interval on which the population and estimated selectors disagree:
\begin{equation}\label{eq:makarov-selector-Mx}
M_x
=
\begin{cases}
(a_x^\star,\widehat a_x],&a_x^\star<\widehat a_x,\\
(\widehat a_x,a_x^\star),&\widehat a_x<a_x^\star,\\
\varnothing,&\widehat a_x=a_x^\star.
\end{cases}
\end{equation}
\begin{assumption}\label{asm:makarov-selector-margin}
There exist an exponent \(b>0\) and a measurable function
\(m_b:\mathcal X\to(0,\infty)\) such that, for almost every \(x\), the
following hold.
\begin{enumerate}[label=\textnormal{(\arabic*)},ref=\textnormal{(\arabic*)},leftmargin=2.8em]
\item\label{cond:makarov-unbounded-density}
Conditional on \(\mathcal D_1\), each
first-stage estimator returns a jointly measurable, nonnegative density
\(\widehat p_k(\cdot\mid x)\) on \(\mathbb R\) that integrates to one.
\item\label{cond:makarov-single-peaked}
The argmax sets defining \(a_x^\star\) and \(\widehat a_x\) are nonempty,
and the selected ones are finite. Moreover, for
Lebesgue-almost every \(u\in\mathbb R\),
\[
\begin{array}{lll}
\Delta_x'(u)>0&\text{if }u<a_x^\star,
&\Delta_x'(u)<0\text{ if }u>a_x^\star,\\
\widehat\Delta_x'(u)>0&\text{if }u<\widehat a_x,
&\widehat\Delta_x'(u)<0\text{ if }u>\widehat a_x.
\end{array}
\]
\item\label{cond:makarov-derivative-margin} For every \(t>0\),
\begin{equation}\label{eq:makarov-derivative-margin}
\operatorname{Leb}\bigl\{u\in M_x:0<|\Delta_x'(u)|\le t\bigr\}
\le m_b(x)t^b.
\end{equation}
\end{enumerate}
\end{assumption}

\begin{remark}
The margin condition in
Condition~\ref{cond:makarov-derivative-margin} of Assumption~\ref{asm:makarov-selector-margin} holds with
\(m_b(x)=2\kappa(x)^{-b}\) whenever
\[
|\Delta_x'(u)|\ge\kappa(x)|u-a_x^\star|^{1/b}
\qquad\text{for every }u\in M_x.
\]
Indeed,
\[
\{u\in M_x:0<|\Delta_x'(u)|\le t\}
\subseteq
\{u\in M_x:0<|u-a_x^\star|\le (t/\kappa(x))^b\},
\]
and the set on the right has Lebesgue measure at most
\(2\kappa(x)^{-b}t^b\).
\end{remark}

\begin{theorem}\label{thm:makarov-selector-margin}
If Assumption~\ref{asm:makarov-selector-margin} holds, then, for almost every \(x\),
\begin{equation}\label{eq:makarov-selector-dual-rate}
\ErrNu(x)
\le
2^{\frac1s+\frac{b+1}{s(b/r+1)}}
m_b(x)^{\frac{1}{s(b/r+1)}}
\ErrP(x)^{\frac{b}{s(b/r+1)}}.
\end{equation}
Moreover,
\begin{equation}\label{eq:makarov-selector-bias}
0\le\thetL-\tildethetL
\le
\E\!\left[
2^{\frac1s+\frac{b+1}{s(b/r+1)}}
m_b(X)^{\frac{1}{s(b/r+1)}}
\ErrP(X)^{\frac{b+1}{b/r+1}}
\mathrel{\big|}\mathcal D_1
\right].
\end{equation}
Consequently, if
\[
\E\!\left[
m_b(X)^{\frac{1}{s(b/r+1)}}
\ErrP(X)^{\frac{b+1}{b/r+1}}
\right]=o(n^{-1/2}),
\]
then \(\sqrt n(\tildethetL-\thetL)=o_p(1)\).
\end{theorem}

\begin{proof}
Fix \(x\).
For \(k\in\{0,1\}\),
\begin{equation}\label{eq:makarov-selector-dual-difference}
|\hat\nu_{k,x}(y)-\nu_{k,x}^\star(y)|
=
\ind{a_x^\star\wedge\widehat a_x<y<a_x^\star\vee\widehat a_x}
\quad\text{for almost every }y\in\mathbb R.
\end{equation}
Hence
\begin{equation}\label{eq:makarov-selector-dual-identity}
\begin{aligned}
\ErrNu(x)^s
&=\sum_{k=0}^1\int_{\mathbb R}
|\hat\nu_{k,x}(y)-\nu_{k,x}^\star(y)|^s\,dy\\
&=\sum_{k=0}^1\int_{\mathbb R}
\ind{a_x^\star\wedge\widehat a_x<y<a_x^\star\vee\widehat a_x}\,dy\\
&=2|\widehat a_x-a_x^\star|
=2\operatorname{Leb}(M_x).
\end{aligned}
\end{equation}

Condition~\ref{cond:makarov-single-peaked} of
Assumption~\ref{asm:makarov-selector-margin} and
\eqref{eq:makarov-selector-Mx} imply that, for almost every \(u\in M_x\),
\[
\begin{cases}
a_x^\star<u<\widehat a_x
\ \Longrightarrow\ 
\Psi_x(u)<0<\widehat\Psi_x(u),
&a_x^\star<\widehat a_x,\\
\widehat a_x<u<a_x^\star
\ \Longrightarrow\ 
\widehat\Psi_x(u)<0<\Psi_x(u),
&\widehat a_x<a_x^\star.
\end{cases}
\]
Consequently,
\begin{equation}\label{eq:makarov-selector-sign-disagreement}
\Psi_x(u)\widehat\Psi_x(u)\le0,
\qquad
0<|\Psi_x(u)|
\le |\Psi_x(u)|+|\widehat\Psi_x(u)|
=|\widehat\Psi_x(u)-\Psi_x(u)|.
\end{equation}
Set \(e_x(u)=\widehat\Psi_x(u)-\Psi_x(u)\). For every \(t>0\),
\eqref{eq:makarov-selector-sign-disagreement} gives
\begin{equation}\label{eq:makarov-selector-split}
M_x
\subseteq
\{u\in M_x:0<|\Psi_x(u)|\le t\}
\cup
\{u\in M_x:|e_x(u)|>t\}.
\end{equation}

Suppose first that \(r<\infty\). Since
\(e_x=(\widehat p_1-p_1^\star)-(\widehat p_0-p_0^\star)\),
\begin{equation}\label{eq:makarov-selector-Lr-score-error}
\int_{\mathbb R}|e_x(u)|^r\,du
\le 2^{r-1}\error_{P,r}(x)^r.
\end{equation}
Equations~\eqref{eq:makarov-selector-split} and
\eqref{eq:makarov-selector-Lr-score-error}, together with
Condition~\ref{cond:makarov-derivative-margin}, yield
\begin{equation}\label{eq:makarov-selector-Lr-Mx-bound}
\begin{aligned}
\operatorname{Leb}(M_x)
&\le
\operatorname{Leb}\{u\in M_x:0<|\Psi_x(u)|\le t\}
+\operatorname{Leb}\{u\in M_x:|e_x(u)|>t\}\\
&\le m_b(x)t^b
+\frac{1}{t^r}\int_{\mathbb R}|e_x(u)|^r\,du\\
&\le m_b(x)t^b
+\frac{2^{r-1}\error_{P,r}(x)^r}{t^r}.
\end{aligned}
\end{equation}
If \(\error_{P,r}(x)>0\), take
\[
t=\left\{\frac{2^{r-1}\error_{P,r}(x)^r}{m_b(x)}\right\}^{1/(b+r)}.
\]
Then
\begin{equation}\label{eq:makarov-selector-Lr-Mx-rate}
\operatorname{Leb}(M_x)
\le
2^{\frac{r(b+1)}{b+r}}
m_b(x)^{\frac{r}{b+r}}
\error_{P,r}(x)^{\frac{br}{b+r}}.
\end{equation}
If \(\error_{P,r}(x)=0\), the same bound follows by letting \(t\downarrow0\).
Combining \eqref{eq:makarov-selector-dual-identity} and
\eqref{eq:makarov-selector-Lr-Mx-rate}, and using
\[
\frac{r-1}{r}=\frac1s,
\qquad
\frac{r-1}{b+r}=\frac{1}{s(b/r+1)},
\qquad
\frac{b(r-1)}{b+r}=\frac{b}{s(b/r+1)},
\]
proves \eqref{eq:makarov-selector-dual-rate} for \(r<\infty\).

If \(r=\infty\), then \(s=1\) and
\[
\|e_x\|_\infty\le2\error_{P,\infty}(x).
\]
Equations~\eqref{eq:makarov-selector-sign-disagreement} and
\eqref{eq:makarov-derivative-margin} therefore give
\[
\operatorname{Leb}(M_x)
\le 2^b m_b(x)\error_{P,\infty}(x)^b.
\]
Equation~\eqref{eq:makarov-selector-dual-identity} gives the sharper bound
\[
\error_{\nu,1}(x)
\le2^{b+1}m_b(x)\error_{P,\infty}(x)^b
\le2^{b+2}m_b(x)\error_{P,\infty}(x)^b.
\]
Since the right-hand side of \eqref{eq:makarov-selector-dual-rate} equals
\(2^{b+2}m_b(x)\error_{P,\infty}(x)^b\) when \(r=\infty\), this proves
\eqref{eq:makarov-selector-dual-rate} in that case as well.

Theorem~\ref{thm::inprodbound} gives
\[
0\le\thetL-\tildethetL
\le\E[\ErrP(X)\ErrNu(X)\mid \mathcal D_1].
\]
Substituting \eqref{eq:makarov-selector-dual-rate} proves
\eqref{eq:makarov-selector-bias}. Under the stated rate condition, the law
of iterated expectations and Markov's inequality imply that the right-hand side
is \(o_p(n^{-1/2})\), proving the final claim.
\end{proof}

\endgroup

\subsection{Bound on positive treatment effects}\label{sec:pte}
\begingroup
\newcommand{\ind}{\mathbf{1}}
\newcommand{\ErrP}{\mathrm{error}_{P,2}}
\newcommand{\Errnu}{\mathrm{error}_{\nu,2}}
\newcommand{\thetL}{\theta_L}
\newcommand{\thetLt}{\widetilde\theta_L}
\newcommand{\epsF}{\varepsilon_F}
\newcommand{\lam}{\operatorname{Leb}}

For \(w\in\{0,1\}\), let \(p_w^\star(\cdot\mid x)\) and
\(\widehat p_w(\cdot\mid x)\) denote the true and estimated conditional densities
of \(Y(w)\mid X=x\) on \([\underline y,\overline y]\). Define
\begin{equation}\label{eq:pte-cdfs}
F_{w,x}(t)=\int_{\underline y}^t p_w^\star(y\mid x)\,dy,
\qquad
\widehat F_{w,x}(t)=\int_{\underline y}^t \widehat p_w(y\mid x)\,dy,
\qquad w\in\{0,1\},
\end{equation}
and
\begin{equation}\label{eq:pte-cdf-gaps}
\Delta_x(t)=F_{0,x}(t)-F_{1,x}(t),
\qquad
\widehat\Delta_x(t)=\widehat F_{0,x}(t)-\widehat F_{1,x}(t).
\end{equation}

Recall Appendix~\ref{app:dual-positive-ite}. It gives
\begin{equation}\label{eq:pte-sharp-bound}
\thetL(x)=\int_{\underline y}^{\overline y}\Delta_x(t)_+\,dt
\end{equation}

Define
\begin{equation}\label{eq:pte-selectors}
\widehat\alpha_x(t)=\ind\{\widehat\Delta_x(t)>0\},
\qquad
\alpha_x^\star(t)=\ind\{\Delta_x(t)>0\},
\qquad t\in[\underline y,\overline y].
\end{equation}

We will apply
Proposition~\ref{prop:structured-margin} with
\[
\mathcal U_x=[\underline y,\overline y],
\qquad
\Psi_x(t)=\Delta_x(t),
\qquad
\widehat\Psi_x(t)=\widehat\Delta_x(t),
\]
where \(\operatorname{Leb}\) denotes Lebesgue measure and the selectors are those in
\eqref{eq:pte-selectors}.
The corresponding population and estimated duals are
\begin{equation}\label{eq:pte-duals}
\begin{array}{ll}
\displaystyle
\nu_{0,x}^\star(y)=\int_y^{\overline y}\alpha_x^\star(t)\,dt,
&
\displaystyle
\nu_{1,x}^\star(y)=-\int_y^{\overline y}\alpha_x^\star(t)\,dt,
\\[1.2em]
\displaystyle
\widehat\nu_{0,x}(y)=\int_y^{\overline y}\widehat\alpha_x(t)\,dt,
&
\displaystyle
\widehat\nu_{1,x}(y)=-\int_y^{\overline y}\widehat\alpha_x(t)\,dt.
\end{array}
\end{equation}
By Appendix~\ref{app:dual-positive-ite}, \((\nu_{0,x}^\star,\nu_{1,x}^\star)\)
is optimal for the true conditional problem and
\((\widehat\nu_{0,x},\widehat\nu_{1,x})\) is optimal for the fitted conditional
problem.

Set
\begin{equation}\label{eq:pte-errors}
\begin{aligned}
\ErrP(x)^2
&=\sum_{w=0}^1\int_{\underline y}^{\overline y}
\bigl(\widehat p_w(y\mid x)-p_w^\star(y\mid x)\bigr)^2\,dy,
\\
\Errnu(x)^2
&=\sum_{w=0}^1\int_{\underline y}^{\overline y}
\bigl(\widehat\nu_{w,x}(y)-\nu_{w,x}^\star(y)\bigr)^2\,dy.
\end{aligned}
\end{equation}
Finally, let \(\thetL=\E[\thetL(X)]\), and define the effective estimand
\begin{equation}\label{eq:pte-effective-estimand}
\thetLt=\E\bigl[\widehat\nu_{0,X}(Y(0))+\widehat\nu_{1,X}(Y(1))\mid \mathcal D_1\bigr],
\end{equation}
where \(X\) is a fresh draw from the marginal covariate distribution, independent
of \(\mathcal D_1\).

\begin{assumption}\label{asm:PTE}
For almost every \(x\),
\[
\lam\{t\in[\underline y,\overline y]:\Delta_x(t)=0\}=0.
\]

There exist a measurable function
\(m_b:\mathcal X\to[0,\infty)\) and an exponent \(b>0\) such that,
for almost every \(x\) and every \(s>0\),
\begin{equation}\label{eq:pte-margin}
\lam\{t\in[\underline y,\overline y]:0<|\Delta_x(t)|\le s\}
\le m_b(x)s^b,
\end{equation}
where \(\lam\) denotes Lebesgue measure on \([\underline y,\overline y]\).
\end{assumption}

\begin{theorem}\label{thm:pte-stability}
If Assumptions~\ref{asm:common-density-setup} and~\ref{asm:PTE} hold, then, for almost every \(x\),
\begin{equation}\label{eq:pte-dual-rate}
\Errnu(x)\le
\bigl(2(\overline y-\underline y)\bigr)^{(b+1)/2}
m_b(x)\ErrP(x)^b.
\end{equation}
Moreover,
\begin{equation}\label{eq:pte-bias-bound}
0\le
\thetL-\thetLt
\le
\bigl(2(\overline y-\underline y)\bigr)^{(b+1)/2}
\E\bigl[m_b(X)\ErrP(X)^{b+1}\mid \mathcal D_1\bigr].
\end{equation}
Consequently, if
\[
\E\bigl[m_b(X)\ErrP(X)^{b+1}\bigr]=o(n^{-1/2}),
\]
then
\[
\sqrt n\,(\thetLt-\thetL)=o_p(1).
\]
\end{theorem}

\begin{proof}
Fix \(x\), and let
\[
M_x=\{t\in[\underline y,\overline y]:
\widehat\alpha_x(t)\neq\alpha_x^\star(t),\ \Delta_x(t)\neq0\}.
\]

Appendix~\ref{app:dual-positive-ite} gives the dual formula.
By \eqref{eq:pte-selectors}, \(t\in M_x\) implies
\(\Delta_x(t)\neq0\). Since the signs of \(\Delta_x(t)\) and
\(\widehat\Delta_x(t)\) disagree on \(M_x\), if \(\Delta_x(t)>0\), then
\(\widehat\Delta_x(t)\le0\), and hence
\[
|\widehat\Delta_x(t)-\Delta_x(t)|
=\Delta_x(t)-\widehat\Delta_x(t)
\ge \Delta_x(t)
=|\Delta_x(t)|.
\]
If \(\Delta_x(t)<0\), then \(\widehat\Delta_x(t)>0\), and hence
\[
|\widehat\Delta_x(t)-\Delta_x(t)|
=\widehat\Delta_x(t)-\Delta_x(t)
\ge -\Delta_x(t)
=|\Delta_x(t)|.
\]
Therefore,
\begin{equation}\label{eq:pte-disagreement-tube-step}
t\in M_x
\quad\Longrightarrow\quad
0<|\Delta_x(t)|
\le |\widehat\Delta_x(t)-\Delta_x(t)|.
\end{equation}
Define
\begin{equation}\label{eq:epsF-def}
\epsF(x)=\|F_{0,x}-\widehat F_{0,x}\|_\infty
+\|F_{1,x}-\widehat F_{1,x}\|_\infty .
\end{equation}
For each \(w\in\{0,1\}\), Cauchy--Schwarz gives
\[
\|F_{w,x}-\widehat F_{w,x}\|_\infty
\le
\sqrt{\overline y-\underline y}
\left\{\int_{\underline y}^{\overline y}
\bigl(p_w^\star(y\mid x)-\widehat p_w(y\mid x)\bigr)^2\,dy
\right\}^{1/2}.
\]
Summing over \(w\) and applying Cauchy--Schwarz once more yields
\begin{equation}\label{eq:epsF-bound}
\epsF(x)\le \sqrt{2(\overline y-\underline y)}\,\ErrP(x).
\end{equation}
By \eqref{eq:pte-cdf-gaps},
\begin{equation}\label{eq:Delta-sup-bound}
\|\Delta_x-\widehat\Delta_x\|_\infty\le\epsF(x).
\end{equation}
Thus condition~\ref{cond:structured-score-error} of Proposition~\ref{prop:structured-margin}
holds for the score \(\Psi_x^{\mathrm{PTE}}(t)=\Delta_x(t)\), with
\(L_\Psi(x)=\sqrt{2(\overline y-\underline y)}\).

Combining \eqref{eq:pte-disagreement-tube-step} and \eqref{eq:Delta-sup-bound} gives
\[
M_x\subseteq\{t\in[\underline y,\overline y]:0<|\Delta_x(t)|\le\epsF(x)\}.
\]
Combining this inclusion with the margin condition \eqref{eq:pte-margin} gives
\begin{equation}\label{eq:mu-bound}
\lam(M_x)\le m_b(x)\epsF(x)^b,
\end{equation}
and therefore, by \eqref{eq:epsF-bound},
\begin{equation}\label{eq:mu-bound-ErrP}
\lam(M_x)\le
m_b(x)
\bigl(\sqrt{2(\overline y-\underline y)}\,\ErrP(x)\bigr)^b.
\end{equation}

By \eqref{eq:pte-duals}, for every \(y\in[\underline y,\overline y]\),
\[
\widehat\nu_{0,x}(y)-\nu_{0,x}^\star(y)
=
\int_y^{\overline y}
\bigl(\widehat\alpha_x(t)-\alpha_x^\star(t)\bigr)\,dt,
\qquad
\widehat\nu_{1,x}(y)-\nu_{1,x}^\star(y)
=
-\int_y^{\overline y}
\bigl(\widehat\alpha_x(t)-\alpha_x^\star(t)\bigr)\,dt.
\]
By Assumption~\ref{asm:PTE} and the definition of \(M_x\),
\[
\bigl|\widehat\alpha_x(t)-\alpha_x^\star(t)\bigr|
=\ind\{t\in M_x\}
\qquad\text{for almost every }t.
\]
Therefore,
\[
\begin{aligned}
|\widehat\nu_{0,x}(y)-\nu_{0,x}^\star(y)|
&\le
\int_y^{\overline y}\ind\{t\in M_x\}\,dt
\le\lam(M_x),
\\
|\widehat\nu_{1,x}(y)-\nu_{1,x}^\star(y)|
&\le
\int_y^{\overline y}\ind\{t\in M_x\}\,dt
\le\lam(M_x).
\end{aligned}
\]
Hence, by \eqref{eq:pte-errors},
\[
\begin{aligned}
\Errnu(x)^2
&=
\sum_{w=0}^1\int_{\underline y}^{\overline y}
\bigl(\widehat\nu_{w,x}(y)-\nu_{w,x}^\star(y)\bigr)^2\,dy
\\
&\le
\sum_{w=0}^1\int_{\underline y}^{\overline y}
\lam(M_x)^2\,dy
\\
&=2(\overline y-\underline y)\,\lam(M_x)^2.
\end{aligned}
\]
This verifies condition~\ref{cond:structured-dual-disagreement} of
Proposition~\ref{prop:structured-margin} with
\(L_\nu(x)=\sqrt{2(\overline y-\underline y)}\). Applying
Proposition~\ref{prop:structured-margin}, using \eqref{eq:mu-bound-ErrP}, gives
\eqref{eq:pte-dual-rate}.

The dual optimality recalled from Appendix~\ref{app:dual-positive-ite} and the
product-bias inequality in Theorem~\ref{thm::inprodbound} give
\eqref{eq:pte-bias-bound}, because
\[
L_\nu(x)m_b(x)L_\Psi(x)^b
=\bigl(2(\overline y-\underline y)\bigr)^{(b+1)/2}m_b(x).
\]
The final \(o_p(n^{-1/2})\) statement follows from
\eqref{eq:pte-bias-bound}, the law of iterated expectations, and Markov's
inequality.
\end{proof}
\endgroup

\subsection{Baseline-disadvantaged subgroup treatment effect}
\label{app:subgroup-proof}
\begingroup
\newcommand{\ind}{\mathbf{1}}
\newcommand{\ErrP}{\mathrm{error}_{P,2}}
\newcommand{\Errnu}{\mathrm{error}_{\nu,2}}
\newcommand{\betat}{\widetilde\beta_L}
\newcommand{\lam}{\operatorname{Leb}}

For \(w\in\{0,1\}\), let \(p_w^\star(\cdot\mid x)\) and
\(\widehat p_w(\cdot\mid x)\) denote the true and estimated conditional densities
of \(Y(w)\mid X=x\) on \([\underline y,\overline y]\). Define
\begin{equation}\label{eq:subgroup-cdfs}
F_{w,x}(t)=\int_{\underline y}^t p_w^\star(y\mid x)\,dy,
\qquad
\widehat F_{w,x}(t)=\int_{\underline y}^t\widehat p_w(y\mid x)\,dy,
\qquad w\in\{0,1\},
\end{equation}

Recall Appendix~\ref{app:dual-cte}. For
\[
\beta(P)=\E_P\!\left[Y(1)\ind\{Y(0)\le c\}\right],
\]
the conditional sharp lower bound is
\begin{equation}\label{eq:subgroup-conditional-bound}
\beta_L(x)=\int_0^{F_{0,x}(c)}F_{1,x}^{-1}(u)\,du.
\end{equation}
Define
\begin{equation}\label{eq:subgroup-cutoffs}
\mathcal T_x^\star
=\{\tau\in[\underline y,\overline y]:F_{1,x}(\tau)=F_{0,x}(c)\},
\qquad
\widehat a_x
=\inf\{\tau\in[\underline y,\overline y]:
\widehat F_{1,x}(\tau)\ge\widehat F_{0,x}(c)\}.
\end{equation}
Set
\begin{equation}\label{eq:subgroup-cutoff-choice}
a_x^\star=\inf\mathcal T_x^\star.
\end{equation}
Since \(F_{1,x}\) is continuous, \(\mathcal T_x^\star\) is closed, so
\(a_x^\star\in\mathcal T_x^\star\).

Define
\begin{equation}\label{eq:subgroup-selectors}
\widehat\alpha_x(u)=\ind\{u\le\widehat a_x\},
\qquad
\alpha_x^\star(u)=\ind\{u\le a_x^\star\},
\qquad u\in[\underline y,\overline y].
\end{equation}
We will apply Proposition~\ref{prop:structured-margin} with
\[
\mathcal U_x=[\underline y,\overline y],
\qquad
\Psi_x(u)=F_{1,x}(u)-F_{0,x}(c),
\qquad
\widehat\Psi_x(u)=\widehat F_{1,x}(u)-\widehat F_{0,x}(c),
\]
where \(\operatorname{Leb}\) denotes Lebesgue measure and the selectors are those in
\eqref{eq:subgroup-selectors}.

The population and estimated duals from Appendix~\ref{app:dual-cte} are
\begin{equation}\label{eq:subgroup-dual-family}
\begin{array}{ll}
\nu_{0,x}^\star(y_0)=-a_x^\star\ind\{y_0>c\},
&
\nu_{1,x}^\star(y_1)=\min\{y_1,a_x^\star\},
\\[0.5em]
\widehat\nu_{0,x}(y_0)=-\widehat a_x\ind\{y_0>c\},
&
\widehat\nu_{1,x}(y_1)=\min\{y_1,\widehat a_x\}.
\end{array}
\end{equation}
Appendix~\ref{app:dual-cte} shows that
\((\nu_{0,x}^\star,\nu_{1,x}^\star)\) is optimal for the true conditional
problem, and the same formula with hats shows that
\((\widehat\nu_{0,x},\widehat\nu_{1,x})\) is optimal for the fitted conditional
problem.

Set
\begin{equation}\label{eq:subgroup-errors}
\begin{aligned}
\ErrP(x)^2
&=\sum_{w=0}^1\int_{\underline y}^{\overline y}
\bigl(\widehat p_w(y\mid x)-p_w^\star(y\mid x)\bigr)^2\,dy,
\\
\Errnu(x)^2
&=\sum_{w=0}^1\int_{\underline y}^{\overline y}
\bigl(\widehat\nu_{w,x}(y)-\nu_{w,x}^\star(y)\bigr)^2\,dy.
\end{aligned}
\end{equation}
Finally, let \(\beta_L=\E[\beta_L(X)]\), and define
\begin{equation}\label{eq:subgroup-effective-estimand}
\betat
=\E\!\left[
\widehat\nu_{0,X}(Y(0))+\widehat\nu_{1,X}(Y(1))
\mid \mathcal D_1
\right],
\end{equation}
where \(X\) is a fresh draw from the marginal covariate distribution, independent
of \(\mathcal D_1\).

\begin{assumption}\label{asm:subgroup}
The threshold satisfies \(c\in[\underline y,\overline y]\), and, for
almost every \(x\),
\[
\lam(\mathcal T_x^\star)=0.
\]
There exist a measurable function
\(m_b^{\mathrm{sub}}:\mathcal X\to[0,\infty)\) and an exponent \(b>0\) such that,
for almost every \(x\) and every \(s>0\),
\begin{equation}\label{eq:subgroup-margin}
\lam\{\tau\in[\underline y,\overline y]:0<|F_{1,x}(\tau)-F_{0,x}(c)|\le s\}
\le m_b^{\mathrm{sub}}(x)s^b,
\end{equation}
where \(\lam\) denotes Lebesgue measure on \([\underline y,\overline y]\).
\end{assumption}

\begin{theorem}\label{thm:subgroup-stability}
If Assumptions~\ref{asm:common-density-setup} and~\ref{asm:subgroup} hold, then, for almost every \(x\),
\begin{equation}\label{eq:subgroup-dual-error}
\Errnu(x)
\le
\bigl(2(\overline y-\underline y)\bigr)^{(b+1)/2}
m_b^{\mathrm{sub}}(x)\ErrP(x)^b.
\end{equation}
Moreover,
\begin{equation}\label{eq:subgroup-pop-bias}
0\le\beta_L-\betat
\le
\bigl(2(\overline y-\underline y)\bigr)^{(b+1)/2}
\E\!\left[m_b^{\mathrm{sub}}(X)\ErrP(X)^{b+1}\mid \mathcal D_1\right].
\end{equation}
Consequently, if
\[
\E\!\left[m_b^{\mathrm{sub}}(X)^2\ErrP(X)^{2b+2}\right]=o(n^{-1}),
\]
then
\[
\sqrt n\,(\betat-\beta_L)=o_p(1).
\]
\end{theorem}

\begin{proof}
Fix \(x\), and let
\begin{equation}\label{eq:subgroup-Mx-definition}
M_x=
\{u\in[\underline y,\overline y]:
\widehat\alpha_x(u)\neq\alpha_x^\star(u)\}
\setminus\mathcal T_x^\star.
\end{equation}
Equations~\eqref{eq:subgroup-selectors}
and~\eqref{eq:subgroup-Mx-definition} give
\[
M_x=
\begin{cases}
(a_x^\star,\widehat a_x]\setminus\mathcal T_x^\star,
& a_x^\star<\widehat a_x,\\
(\widehat a_x,a_x^\star]\setminus\mathcal T_x^\star,
& \widehat a_x<a_x^\star,\\
\varnothing, & \widehat a_x=a_x^\star.
\end{cases}
\]
By Assumption~\ref{asm:subgroup}, \(\lam(\mathcal T_x^\star)=0\). Therefore,
\begin{equation}\label{eq:subgroup-cutoff-length}
\lam(M_x)=|\widehat a_x-a_x^\star|.
\end{equation}
Define
\begin{equation}\label{eq:subgroup-score-sup-error}
\varepsilon_F(x)
=\sup_{u\in[\underline y,\overline y]}
|\widehat\Psi_x(u)-\Psi_x(u)|.
\end{equation}
For each \(w\in\{0,1\}\), Cauchy--Schwarz gives
\[
\|F_{w,x}-\widehat F_{w,x}\|_\infty
\le
\sqrt{\overline y-\underline y}
\left\{\int_{\underline y}^{\overline y}
\bigl(p_w^\star(y\mid x)-\widehat p_w(y\mid x)\bigr)^2\,dy
\right\}^{1/2}.
\]
By the definitions of \(\Psi_x\) and \(\widehat\Psi_x\),
\[
\varepsilon_F(x)
\le
\|F_{1,x}-\widehat F_{1,x}\|_\infty
+|F_{0,x}(c)-\widehat F_{0,x}(c)|.
\]
Since \(|F_{0,x}(c)-\widehat F_{0,x}(c)|
\le \|F_{0,x}-\widehat F_{0,x}\|_\infty\), summing the bounds for \(w=0,1\) and
applying Cauchy--Schwarz once more gives
\begin{equation}\label{eq:subgroup-score-perturbation}
\varepsilon_F(x)
\le\sqrt{2(\overline y-\underline y)}\,\ErrP(x).
\end{equation}
This verifies condition~\ref{cond:structured-score-error} of
Proposition~\ref{prop:structured-margin} with
\[
L_\Psi(x)=\sqrt{2(\overline y-\underline y)}.
\]

Suppose first that \(\widehat a_x\ge a_x^\star\). By the definition of
\(M_x\), every \(u\in M_x\) lies outside \(\mathcal T_x^\star\). By
definition~\eqref{eq:subgroup-Mx-definition} of \(M_x\),
\[
a_x^\star<u\le\widehat a_x.
\]
By \eqref{eq:subgroup-cutoffs} and \eqref{eq:subgroup-cutoff-choice},
\(F_{1,x}(a_x^\star)=F_{0,x}(c)\). Since \(F_{1,x}\) is nondecreasing,
\(u>a_x^\star\) gives
\(F_{1,x}(u)\ge F_{0,x}(c)\); the inequality is strict because
\(u\notin\mathcal T_x^\star\). On the other hand, the definition of
\(\widehat a_x\) in \eqref{eq:subgroup-cutoffs} and the continuity of
\(\widehat F_{1,x}\) give
\(\widehat F_{1,x}(u)\le\widehat F_{0,x}(c)\) for
\(u\le\widehat a_x\). Hence
\[
F_{1,x}(u)-F_{0,x}(c)>0,
\qquad
\widehat F_{1,x}(u)-\widehat F_{0,x}(c)\le0.
\]
Therefore,
\begin{equation}\label{eq:subgroup-pointwise-score-bound}
0<|F_{1,x}(u)-F_{0,x}(c)|
\le
|F_{1,x}(u)-F_{0,x}(c)-\widehat F_{1,x}(u)+\widehat F_{0,x}(c)|
\le\varepsilon_F(x).
\end{equation}
If \(\widehat a_x<a_x^\star\), the same argument gives
\(\widehat a_x<u<a_x^\star\). Monotonicity of \(F_{1,x}\), together
with \(u\notin\mathcal T_x^\star\), gives
\(F_{1,x}(u)-F_{0,x}(c)<0\), while the definition of
\(\widehat a_x\) gives
\(\widehat F_{1,x}(u)-\widehat F_{0,x}(c)\ge0\). Thus,
\eqref{eq:subgroup-pointwise-score-bound} continues to hold. Hence
\[
M_x\subseteq
\{u\in[\underline y,\overline y]:
0<|\Psi_x(u)|\le\varepsilon_F(x)\}.
\]
Together with \eqref{eq:subgroup-score-sup-error}, this verifies
condition~\ref{cond:structured-score-disagreement} of Proposition~\ref{prop:structured-margin}.
The margin condition in Assumption~\ref{asm:subgroup} is condition~\ref{cond:structured-margin-mass}
with \(m_b(x)=m_b^{\mathrm{sub}}(x)\). Combining
conditions~\ref{cond:structured-score-disagreement} and \ref{cond:structured-margin-mass} with
\eqref{eq:subgroup-cutoff-length} gives
\begin{equation}\label{eq:subgroup-cutoff-stability}
|\widehat a_x-a_x^\star|
\le
m_b^{\mathrm{sub}}(x)\varepsilon_F(x)^b
\le
m_b^{\mathrm{sub}}(x)
\bigl(\sqrt{2(\overline y-\underline y)}\,\ErrP(x)\bigr)^b,
\end{equation}
where the second inequality follows from
\eqref{eq:subgroup-score-perturbation}.

By \eqref{eq:subgroup-dual-family}, for every
\(y\in[\underline y,\overline y]\),
\[
|\widehat\nu_{0,x}(y)-\nu_{0,x}^\star(y)|
=\ind\{y>c\}|\widehat a_x-a_x^\star|
\le|\widehat a_x-a_x^\star|,
\]
and, since \(\tau\mapsto\min\{y,\tau\}\) is \(1\)-Lipschitz,
\[
|\widehat\nu_{1,x}(y)-\nu_{1,x}^\star(y)|
\le|\widehat a_x-a_x^\star|.
\]
Consequently,
\[
\Errnu(x)^2
\le
2(\overline y-\underline y)|\widehat a_x-a_x^\star|^2.
\]
By \eqref{eq:subgroup-cutoff-length},
\[
\Errnu(x)
\le
\sqrt{2(\overline y-\underline y)}\,\lam(M_x).
\]
This verifies condition~\ref{cond:structured-dual-disagreement} of
Proposition~\ref{prop:structured-margin} with
\[
L_\nu(x)=\sqrt{2(\overline y-\underline y)}.
\]
Thus conditions~\ref{cond:structured-dual-disagreement}--\ref{cond:structured-margin-mass} of
Proposition~\ref{prop:structured-margin} hold with
\[
L_\nu(x)=L_\Psi(x)=\sqrt{2(\overline y-\underline y)},
\qquad
m_b(x)=m_b^{\mathrm{sub}}(x).
\]
Applying Proposition~\ref{prop:structured-margin} gives
\[
\Errnu(x)
\le
L_\nu(x)m_b^{\mathrm{sub}}(x)L_\Psi(x)^b\ErrP(x)^b
=
\bigl(2(\overline y-\underline y)\bigr)^{(b+1)/2}
m_b^{\mathrm{sub}}(x)\ErrP(x)^b,
\]
which proves \eqref{eq:subgroup-dual-error}.

The dual optimality from Appendix~\ref{app:dual-cte} and
Theorem~\ref{thm::inprodbound}, combined with
\eqref{eq:subgroup-dual-error}, give
\[
0\le\beta_L-\betat
\le
\bigl(2(\overline y-\underline y)\bigr)^{(b+1)/2}
\E\!\left[m_b^{\mathrm{sub}}(X)\ErrP(X)^{b+1}\mid \mathcal D_1\right],
\]
which proves \eqref{eq:subgroup-pop-bias}. The final \(o_p(n^{-1/2})\)
statement follows from \eqref{eq:subgroup-pop-bias}, Cauchy--Schwarz, and
Markov's inequality.
\end{proof}

\endgroup
\subsection{Quantile of ITE}
\label{sec:qte-proof}
\begingroup
\newcommand{\ind}{\mathbf{1}}
\newcommand{\ErrP}{\mathrm{error}_{P,r}}
\newcommand{\ErrNuC}{\mathrm{error}_{\nu,s}^{c}}
\newcommand{\lam}{\operatorname{Leb}}

Fix \(\alpha\in(0,1)\) and \(r\in(1,\infty]\), and let
\(s\in[1,\infty)\) satisfy \(1/r+1/s=1\). We use the general primal error
\(\ErrP(x)\) defined in Section~\ref{subsec::tightness}, with the convention
\(b/\infty=0\). As in Remark~\ref{rmk:quasilinear}, the lower bound
for the \(\alpha\)-quantile of \(Y(1)-Y(0)\) can be obtained by inverting the
linearized problems
\begin{equation}\label{eq:qte-linearized-cost}
f_c(y_0,y_1,x)=\alpha-\ind\{y_1-y_0\le c\},
\qquad c\in\mathbb R.
\end{equation}
Extend the conditional CDFs \(F_{w,x}\) and \(\widehat F_{w,x}\) to all of
\(\mathbb R\) by setting them equal to \(0\) below \(\underline y\) and \(1\)
above \(\overline y\), and extend the corresponding densities by zero outside
\([\underline y,\overline y]\). For each \(c\in\mathbb R\), define
\begin{equation}\label{eq:qte-cdf-gaps}
\Delta_{x,c}(a)=F_{1,x}(a+c)-F_{0,x}(a),
\qquad
\widehat\Delta_{x,c}(a)=\widehat F_{1,x}(a+c)-\widehat F_{0,x}(a),
\qquad a\in[\underline y,\overline y].
\end{equation}
Define
\begin{equation}\label{eq:qte-cutoffs}
a_{x,c}^\star
=\inf\argmin_{a\in[\underline y,\overline y]}\Delta_{x,c}(a),
\qquad
\widehat a_{x,c}
=\inf\argmin_{a\in[\underline y,\overline y]}\widehat\Delta_{x,c}(a).
\end{equation}
Appendix~\ref{app:dual-qte} gives the lower-bound dual family
\begin{equation}\label{eq:qte-duals}
\begin{array}{ll}
\nu_{0,x}^{\star,c}(y_0)=\alpha-\ind\{y_0\ge a_{x,c}^\star\},
&
\nu_{1,x}^{\star,c}(y_1)=-\ind\{y_1\le a_{x,c}^\star+c\},
\\[0.5em]
\widehat\nu_{0,x}^{c}(y_0)=\alpha-\ind\{y_0\ge\widehat a_{x,c}\},
&
\widehat\nu_{1,x}^{c}(y_1)=-\ind\{y_1\le\widehat a_{x,c}+c\}.
\end{array}
\end{equation}
Thus
\[
\theta_{L,c}(x)
=
\alpha-1-\inf_{a\in[\underline y,\overline y]}\Delta_{x,c}(a)
\]
is the conditional sharp lower bound for the linearized problem. Set
\[
\theta_{L,c}=\E[\theta_{L,c}(X)]
\]
and
\begin{equation}\label{eq:qte-effective-bound}
\widetilde\theta_{L,c}
=
\E\!\left[
\widehat\nu_{0,X}^{c}(Y(0))+\widehat\nu_{1,X}^{c}(Y(1))
\mid \mathcal D_1
\right].
\end{equation}
For each \(c\in\mathbb R\), define the corresponding dual error by
\[
\ErrNuC(x)
=
\left\{
\sum_{w=0}^1\int_{\underline y}^{\overline y}
|\widehat\nu_{w,x}^{c}(y)-\nu_{w,x}^{\star,c}(y)|^s\,dy
\right\}^{1/s}.
\]

\begin{assumption}\label{asm:qte-selector-margin}
There exist an exponent \(b>0\) and measurable functions
\(\{m_{b,c}:\mathcal X\to(0,\infty):c\in\mathbb R\}\), and a measurable
envelope \(m_b:\mathcal X\to(0,\infty)\) satisfying
\[
m_{b,c}(x)\le m_b(x)\qquad\text{for all }c\in\mathbb R,
\]
such that, for almost every \(x\) and every
\(c\in\mathbb R\), the following hold.
\begin{enumerate}[label=\textnormal{(\arabic*)},leftmargin=2.8em]
\item \(a_{x,c}^\star,\widehat a_{x,c}\in[\underline y,\overline y]\).
Moreover, for Lebesgue-almost every
\(a\in[\underline y,\overline y]\),
\[
\begin{array}{lll}
\partial_a\Delta_{x,c}(a)<0&\text{if }a<a_{x,c}^\star,
&\partial_a\Delta_{x,c}(a)>0\text{ if }a>a_{x,c}^\star,\\
\partial_a\widehat\Delta_{x,c}(a)<0&\text{if }a<\widehat a_{x,c},
&\partial_a\widehat\Delta_{x,c}(a)>0\text{ if }a>\widehat a_{x,c}.
\end{array}
\]
\item For every \(s>0\),
\begin{equation}\label{eq:qte-margin}
\lam\{a\in[\underline y,\overline y]:
0<|\partial_a\Delta_{x,c}(a)|\le s\}
\le m_{b,c}(x)s^b.
\end{equation}
\end{enumerate}
\end{assumption}

\begin{theorem}\label{thm:qte-inversion}
Suppose Assumptions~\ref{asm:common-density-setup}
and~\ref{asm:qte-selector-margin} hold. Let
\begin{equation}\label{eq:qte-general-r-constant}
A_{b,r,s}
=
2^{\frac1s+\frac{b+1}{s(b/r+1)}}.
\end{equation}
Then, for almost every \(x\) and every \(c\in\mathbb R\),
\begin{equation}\label{eq:qte-general-r-dual-rate}
\ErrNuC(x)
\le
A_{b,r,s}
m_{b,c}(x)^{\frac{1}{s(b/r+1)}}
\ErrP(x)^{\frac{b}{s(b/r+1)}}.
\end{equation}
Moreover, for every \(c\in\mathbb R\),
\begin{equation}\label{eq:qte-fixed-c-bias}
0\le\theta_{L,c}-\widetilde\theta_{L,c}
\le
\E\!\left[
A_{b,r,s}
m_{b,c}(X)^{\frac{1}{s(b/r+1)}}
\ErrP(X)^{\frac{b+1}{b/r+1}}
\mathrel{\big|}\mathcal D_1
\right].
\end{equation}

Define the population and plug-in inverted bounds over \(\mathbb R\) by
\[
q_L=\inf\{c\in\mathbb R:\theta_{L,c}\le0\},
\qquad
\widetilde q_L=\inf\{c\in\mathbb R:\widetilde\theta_{L,c}\le0\}.
\]
Assume the infimum defining \(q_L\) is attained and that, for some
\(\ell_q>0\) and \(\eta>0\),
\begin{equation}\label{eq:qte-inverse-margin}
\theta_{L,c}\ge \ell_q(q_L-c)
\qquad
\text{for all }c\in[q_L-\eta,q_L].
\end{equation}
Set
\begin{equation}\label{eq:qte-uniform-bias}
R_{n,r}
=
\E\!\left[
A_{b,r,s}
m_b(X)^{\frac{1}{s(b/r+1)}}
\ErrP(X)^{\frac{b+1}{b/r+1}}
\mathrel{\big|}\mathcal D_1
\right].
\end{equation}
If \(R_{n,r}<\ell_q\eta\), then
\[
0\le q_L-\widetilde q_L
\le
\frac{R_{n,r}}{\ell_q}.
\]
Consequently, if
\begin{equation}\label{eq:qte-inversion-rate}
\E\!\left[
m_b(X)^{\frac{1}{s(b/r+1)}}
\ErrP(X)^{\frac{b+1}{b/r+1}}
\right]
=o(n^{-1/2}),
\end{equation}
then \(\sqrt{n}(q_L-\widetilde q_L)=o_P(1)\).
\end{theorem}

\begin{proof}
Fix \(c\in\mathbb R\). Apply the argument used to prove
Theorem~\ref{thm:makarov-selector-margin} to the single-peaked criterion
\(-\Delta_{x,c}\). Write
\[
\delta_{w,x}(y)
=\widehat p_w(y\mid x)-p_w^\star(y\mid x),
\qquad w\in\{0,1\},
\]
and extend \(\delta_{w,x}\) by zero outside
\([\underline y,\overline y]\). The population and estimated scores then
differ by
\[
e_{x,c}(a)=-\delta_{1,x}(a+c)+\delta_{0,x}(a).
\]
For \(r<\infty\), the inequality
\(|u+v|^r\le2^{r-1}(|u|^r+|v|^r)\) and the change of variables
\(y=a+c\) give
\begin{equation}\label{eq:qte-general-r-score-error}
\begin{aligned}
\int_{\underline y}^{\overline y}|e_{x,c}(a)|^r\,da
&\le
2^{r-1}\left\{
\int_{\underline y}^{\overline y}|\delta_{1,x}(a+c)|^r\,da
+\int_{\underline y}^{\overline y}|\delta_{0,x}(a)|^r\,da
\right\}\\
&=
2^{r-1}\left\{
\int_{\underline y+c}^{\overline y+c}|\delta_{1,x}(y)|^r\,dy
+\int_{\underline y}^{\overline y}|\delta_{0,x}(y)|^r\,dy
\right\}\\
&\le
2^{r-1}\sum_{w=0}^1
\int_{\underline y}^{\overline y}|\delta_{w,x}(y)|^r\,dy
=2^{r-1}\ErrP(x)^r.
\end{aligned}
\end{equation}
The last inequality follows because \(\delta_{1,x}(y)=0\) outside
\([\underline y,\overline y]\), so its integral over the shifted interval
\([\underline y+c,\overline y+c]\) cannot exceed its integral over the full
outcome support. The last equality is the definition of \(\ErrP(x)\).
For \(r=\infty\), the corresponding bound is
\begin{equation}\label{eq:qte-infinity-score-error}
\|e_{x,c}\|_\infty\le2\ErrP(x).
\end{equation}
Each dual coordinate in \eqref{eq:qte-duals} differs on the intersection of
the outcome support with an interval of length
\(|\widehat a_{x,c}-a_{x,c}^\star|\). Hence
\begin{equation}\label{eq:qte-dual-error-bound}
\ErrNuC(x)^s
\le2|\widehat a_{x,c}-a_{x,c}^\star|.
\end{equation}
Combining \eqref{eq:qte-margin}--\eqref{eq:qte-dual-error-bound} with the
threshold-splitting argument in the proof of
Theorem~\ref{thm:makarov-selector-margin} proves
\eqref{eq:qte-general-r-dual-rate}. Theorem~\ref{thm::inprodbound} then yields
\eqref{eq:qte-fixed-c-bias}.

Because \(m_{b,c}\le m_b\), \eqref{eq:qte-fixed-c-bias} implies
\begin{equation}\label{eq:qte-uniform-bias-bound}
0\le
\sup_{c\in\mathbb R}
\{\theta_{L,c}-\widetilde\theta_{L,c}\}
\le R_{n,r}.
\end{equation}

By \eqref{eq:qte-fixed-c-bias}, \(\widetilde\theta_{L,c}\le\theta_{L,c}\) for
all \(c\in\mathbb R\), so \(\widetilde q_L\le q_L\). If
\(c\in[q_L-\eta,q_L]\) and
\[
c<q_L-\frac{R_{n,r}}{\ell_q},
\]
then \eqref{eq:qte-inverse-margin} gives
\[
\theta_{L,c}>\ell_q\frac{R_{n,r}}{\ell_q}
=R_{n,r}.
\]
If \(c\le q_L-\eta\), monotonicity and
\eqref{eq:qte-inverse-margin} at \(q_L-\eta\) give
\[
\theta_{L,c}
\ge
\theta_{L,q_L-\eta}
\ge
\ell_q\eta
>
R_{n,r}.
\]
Equation~\eqref{eq:qte-uniform-bias-bound} therefore implies
\[
\widetilde\theta_{L,c}
\ge
\theta_{L,c}-R_{n,r}
>0,
\]
so such a \(c\) cannot belong to
\[
\{c'\in\mathbb R:\widetilde\theta_{L,c'}\le0\}.
\]
Thus \(\widetilde q_L\ge q_L-R_{n,r}/\ell_q\), which proves the
inversion bound. Under the rate condition in \eqref{eq:qte-inversion-rate}, the law of iterated
expectations and Markov's inequality give
\(R_{n,r}=o_p(n^{-1/2})\), which also implies
\(R_{n,r}<\ell_q\eta\) with probability approaching one and proves
the final claim.
\end{proof}

\endgroup
\section{Theory of discrete potential outcomes}
\label{sec:discrete-margin-comparison}
\begingroup
In this section, we prove the results in Section~\ref{sec: refined theory for discrete outcomes}. We show that our bias bound nests the bound
in \citet{semenova2023aggregated} when the conditional potential-outcome
distributions are estimated consistently and uniformly over the covariate
space. When this assumption fails, our bound can continue to yield
statistically negligible bias provided that the primal error is small on
average over the covariate space. We provide examples with simple
data-generating processes and first-stage estimators for which the uniform
rate required by \citet{semenova2023aggregated} fails, whereas our integrated
bound is \(o_p(n^{-1/2})\).
Throughout this appendix, the conditional coupling is unrestricted apart
from its two prescribed marginals; additional conditional restrictions would
change the dual feasible set.
\subsection{Proof of Theorem~\ref{prop:discrete-lower-level}}
\label{sec:discrete-generic-margin}

We begin by proving an upper bound on \(D_2(x)\).

\begin{lemma}[Bound on the dual diameter]
\label{lem:discrete-dual-diameter}
Under the cost-oscillation condition
\eqref{eq:discrete-cost-oscillation-bound}, the diameter defined in
\eqref{eq:discrete-l2-vertex-diameter} satisfies
\begin{equation}\label{eq:discrete-dual-diameter-bound}
D_2(x)
\le
2B\sqrt{2m-1}.
\end{equation}
\end{lemma}

\begin{proof}
Fix \(x\) and a vertex
\((\nu_{0,x}^t,\nu_{1,x}^t)\) of \(\mathcal V_{f,x}^0\).
Let \(G_t\) have nodes \(\{(0,j),(1,i):i,j\in[m]\}\) and edges
\[
((0,j),(1,i))\in E_t
\quad\Longleftrightarrow\quad
\nu_{0,x,j}^t+\nu_{1,x,i}^t=f(y_j,y_i,x).
\]
If \(G_t\) were disconnected, choose a component \(C\) not containing
\((0,1)\), and set
\[
d_{0,j}=\1\{(0,j)\in C\},
\qquad
d_{1,i}=-\1\{(1,i)\in C\}.
\]
Then \(d_{0,1}=0\) and
\(d_{0,j}+d_{1,i}=0\) on every edge in \(E_t\). Choose
\[
0<\varepsilon<
\min_{\substack{f(y_j,y_i,x)-\nu_{0,x,j}^t-\nu_{1,x,i}^t>0:\
d_{0,j}+d_{1,i}\ne0}}
\frac{f(y_j,y_i,x)-\nu_{0,x,j}^t-\nu_{1,x,i}^t}
{|d_{0,j}+d_{1,i}|},
\]
where the minimum of the empty set is \(\infty\). Define
\[
\nu_{0,x}^{t,\pm}=\nu_{0,x}^t\pm\varepsilon d_0,
\qquad
\nu_{1,x}^{t,\pm}=\nu_{1,x}^t\pm\varepsilon d_1.
\]
For every \(i,j\),
\[
\nu_{0,x,j}^{t,\pm}+\nu_{1,x,i}^{t,\pm}
=\nu_{0,x,j}^t+\nu_{1,x,i}^t
\pm\varepsilon(d_{0,j}+d_{1,i})
\le f(y_j,y_i,x),
\qquad
\nu_{0,x,1}^{t,\pm}=0.
\]
Indeed, if \(((0,j),(1,i))\in E_t\), the original inequality binds and
\(d_{0,j}+d_{1,i}=0\), so the perturbed inequality also binds. If
\(((0,j),(1,i))\notin E_t\), then
\[
f(y_j,y_i,x)-\nu_{0,x,j}^t-\nu_{1,x,i}^t>0.
\]
When \(d_{0,j}+d_{1,i}=0\), the left-hand side is unchanged; otherwise,
the choice of \(\varepsilon\) gives
\[
\nu_{0,x,j}^t+\nu_{1,x,i}^t
\pm\varepsilon(d_{0,j}+d_{1,i})
\le
\nu_{0,x,j}^t+\nu_{1,x,i}^t
+\varepsilon|d_{0,j}+d_{1,i}|
<f(y_j,y_i,x).
\]
Thus both points are feasible, and
\[
(\nu_{0,x}^{t,+},\nu_{1,x}^{t,+})
\ne
(\nu_{0,x}^{t,-},\nu_{1,x}^{t,-}),
\qquad
(\nu_{0,x}^t,\nu_{1,x}^t)
=\frac{
(\nu_{0,x}^{t,+},\nu_{1,x}^{t,+})
+(\nu_{0,x}^{t,-},\nu_{1,x}^{t,-})
}{2},
\]
contradicting extremality. Hence \(G_t\) is connected.

Because \(G_t\) is connected, for every \(j\) there is an \(i_j\) such that
\[
((0,j),(1,i_j))\in E_t.
\]
Feasibility and this binding edge give
\[
\nu_{0,x,j}^t
\le \min_i\{f(y_j,y_i,x)-\nu_{1,x,i}^t\}
\le f(y_j,y_{i_j},x)-\nu_{1,x,i_j}^t
=\nu_{0,x,j}^t.
\]
Hence
\[
\nu_{0,x,j}^t
=\min_i\{f(y_j,y_i,x)-\nu_{1,x,i}^t\}.
\]

For any \(A,B\in\mathbb R^m\), let
\[
\underline d=\min_i(A_i-B_i),
\qquad
\overline d=\max_i(A_i-B_i).
\]
Since
\[
B_i+\underline d\le A_i\le B_i+\overline d
\qquad\text{for every }i,
\]
taking minima yields
\[
\underline d
\le \min_i A_i-\min_i B_i
\le \overline d.
\]
Apply this inequality with
\[
A_i=f(y_j,y_i,x)-\nu_{1,x,i}^t,
\qquad
B_i=f(y_1,y_i,x)-\nu_{1,x,i}^t.
\]
Using
\(\min_iA_i=\nu_{0,x,j}^t\) and
\(\min_iB_i=\nu_{0,x,1}^t=0\), we obtain
\[
\min_i\{f(y_j,y_i,x)-f(y_1,y_i,x)\}
\le \nu_{0,x,j}^t
\le \max_i\{f(y_j,y_i,x)-f(y_1,y_i,x)\}.
\]
Therefore, \(|\nu_{0,x,j}^t|\le B\). Moreover, feasibility at \(j=1\)
implies
\[
\nu_{1,x,i}^t
\le f(y_1,y_i,x)
\le \max_{k,\ell}f(y_\ell,y_k,x).
\]
Because every
\(\nu_{1,x,i}^t\) is incident to a binding inequality, for some \(j\),
\[
\nu_{1,x,i}^t=f(y_j,y_i,x)-\nu_{0,x,j}^t
\ge \min_{k,\ell}f(y_\ell,y_k,x)-B.
\]
Thus, across any two vertices, each nonnormalized \(\nu_0\) coordinate
and each \(\nu_1\) coordinate differ by at most \(2B\). Since
\(\nu_{0,x,1}^t=0\) at every vertex,
\[
\left\{
\|\nu_{0,x}^t-\nu_{0,x}^{t'}\|_2^2
+\|\nu_{1,x}^t-\nu_{1,x}^{t'}\|_2^2
\right\}^{1/2}
\le 2B\sqrt{(m-1)+m}
=2B\sqrt{2m-1}.
\]
Taking the maximum over \(t,t'\) proves
\eqref{eq:discrete-dual-diameter-bound}.
\end{proof}

\begin{proof}[Proof of Theorem~\ref{prop:discrete-lower-level}]
By \eqref{eq:discrete-primal-l2-error},
$\|\widehat\mu(x)-\mu(x)\|_2=\error_{P,2}(x)$.
Note that, for each $t\in\mathcal T$, $\phi_x(t;\mu(x))$ is linear in $\mu(x)$.
Hence, for every $t,t'\in\mathcal T$ with $t\ne t'$, the
Cauchy--Schwarz inequality gives
\[
\begin{aligned}
&
\left|
\left\{\phi_x(t';\widehat\mu(x))-\phi_x(t;\widehat\mu(x))\right\}
-
\left\{\phi_x(t';\mu(x))-\phi_x(t;\mu(x))\right\}
\right|\\
&\quad\le
\left\{
\|\nu_{0,x}^t-\nu_{0,x}^{t'}\|_2^2
+\|\nu_{1,x}^t-\nu_{1,x}^{t'}\|_2^2
\right\}^{1/2}
\|\widehat\mu(x)-\mu(x)\|_2
\le D_2(x)\error_{P,2}(x).
\end{aligned}
\]
If \(\widehat t_x\in\mathcal T_x^\star\), then
\(\error_{\nu,2}(x)=0\), so \eqref{eq:causal-vertex-direct-bound} is
immediate. Otherwise, since a nonoptimal vertex of an LP with a finite
optimum has a strictly improving adjacent vertex, there is a vertex \(t_x^+\)
adjacent to \(\widehat t_x\) satisfying
\[
\phi_x(\widehat t_x;\mu(x))
-\phi_x(t_x^+;\mu(x))>0.
\]
On the other hand,
\[
\phi_x(\widehat t_x;\widehat\mu(x))
-\phi_x(t_x^+;\widehat\mu(x))\le0,
\]
where the inequality follows from the optimality of
$\widehat t_x$ for the estimated objective. Therefore,
\[
\begin{aligned}
0<
\left|\phi_x(\widehat t_x;\mu(x))
-\phi_x(t_x^+;\mu(x))\right|
&\le
\left|
\left\{\phi_x(\widehat t_x;\widehat\mu(x))
-\phi_x(t_x^+;\widehat\mu(x))\right\}
-
\left\{\phi_x(\widehat t_x;\mu(x))
-\phi_x(t_x^+;\mu(x))\right\}
\right|
\\
&\le D_2(x)\error_{P,2}(x).
\end{aligned}
\]
By the definition of $\omega_x$ in
\eqref{eq:causal-vertex-margin-modulus}, the adjacent pair
$(t_x^+,\widehat t_x)$ witnesses the event defining
\(\omega_x(D_2(x)\error_{P,2}(x))\). Hence the indicator equals one, and
\[
\error_{\nu,2}(x)
\le D_2(x)
\le
D_2(x)\omega_x\!\left(D_2(x)\error_{P,2}(x)\right).
\]
This proves \eqref{eq:causal-vertex-direct-bound}.

It remains to prove \eqref{eq:causal-vertex-envelope-bias}. By
\eqref{eq:causal-vertex-direct-bound},
Lemma~\ref{lem:discrete-dual-diameter}, and the monotonicity of
\(\omega_x\),
\[
\begin{aligned}
\error_{P,2}(x)\error_{\nu,2}(x)
&\le
\error_{P,2}(x)D_2(x)
\omega_x\!\left(D_2(x)\error_{P,2}(x)\right)\\
&\le
\overline{\error}_{P,2}(x)
\omega_x\!\left(\overline{\error}_{P,2}(x)\right).
\end{aligned}
\]
Choose a population-optimal vertex
\(\nu_x^\star=(\nu_{0,x}^\star,\nu_{1,x}^\star)\) attaining the infimum in
the definition of \(\error_{\nu,2}(x)\), and write
\(\widehat\nu_x=(\widehat\nu_{0,x},\widehat\nu_{1,x})\). Empirical optimality gives
\(\widehat\mu(x)^\top(\nu_x^\star-\widehat\nu_x)\le0\). Therefore,
\begin{equation}\label{eq:discrete-pointwise-l2-bias}
\begin{aligned}
0\le\theta_L(x)-\widetilde\theta_L(x)
&=\mu(x)^\top(\nu_x^\star-\widehat\nu_x)\\
&=\{\mu(x)-\widehat\mu(x)\}^\top
  (\nu_x^\star-\widehat\nu_x)
  +\widehat\mu(x)^\top(\nu_x^\star-\widehat\nu_x)\\
&\le\error_{P,2}(x)\error_{\nu,2}(x).
\end{aligned}
\end{equation}
Taking the conditional expectation over \(X\) and using the preceding
envelope inequality proves \eqref{eq:causal-vertex-envelope-bias}.
\end{proof}

\subsection{Proof of Corollary~\ref{thm:discrete-bias-under-semenova}}
\label{sec:semenova-margin-comparison}

The margin condition in \citet{semenova2023aggregated} considers
\[
\psi_0=E\left[\min_{t\in T}\phi\big(t,\nu_0(X)\big)\right],
\]
where $T$ is finite and $\nu_0(X)$ is a nuisance vector. Assumption~4.4 in
\citet{semenova2023aggregated} is the case $b=1$ of the condition that, for
some $\bar B,\delta\in(0,\infty)$ and $b\in(0,1]$,
\begin{equation}\label{eq:semenova-margin}
\sup_{t,t'\in T:\,t\ne t'}
\Pr\left(
0\le \phi\big(t,\nu_0(X)\big)-\phi\big(t',\nu_0(X)\big)\le s
\right)
\le \bar B s^b,
\qquad s\in(0,\delta).
\end{equation}
\citet{semenova2023aggregated} formally imposes \(b=1\) and lets
\(\mathcal T_n^\nu\) be a shrinking neighborhood of \(\nu_0\) satisfying
\begin{equation}\label{eq:semenova-uniform-nuisance-rate}
\sup_{\nu\in\mathcal T_n^\nu}\sup_{x\in\mathcal X}
\|\nu(x)-\nu_0(x)\|_2\le\nu_n^\infty,
\qquad
\nu_n^\infty=o(n^{-1/4}).
\end{equation}
Here the superscript \(\infty\) denotes uniformity over \(x\); the
finite-dimensional nuisance norm is Euclidean.
For the cross-fitted first-stage estimates, Assumption~4.2 in
\citet{semenova2023aggregated} requires
\[
\Pr\!\left\{
\widehat\nu_k\in\mathcal T_n^\nu
\ \text{for every }k\in[K]
\right\}\to1.
\]
For a first-stage estimate \(\widehat\nu\), let
\(\widehat t(x)\in\argmin_{t\in T}\phi(t,\widehat\nu(x))\), and define its
conditional selector regret by
\[
\mathcal B(\widehat\nu)
=E\!\left[
\phi\{\widehat t(X),\nu_0(X)\}
-\min_{t\in T}\phi\{t,\nu_0(X)\}
\,\middle|\,\widehat\nu
\right].
\]
Equations~(A.10)--(A.11) in \citet{semenova2023aggregated} imply that, on
\(\{\widehat\nu\in\mathcal T_n^\nu\}\),
\begin{equation}\label{eq:semenova-selection-bias-bound}
0\le\mathcal B(\widehat\nu)\preceq(\nu_n^\infty)^2.
\end{equation}
Thus, \(\mathcal T_n^\nu\) is contained in the uniform-error ball
\[
\left\{
\nu:
\sup_{x\in\mathcal X}
\|\nu(x)-\nu_0(x)\|_2
\le \nu_n^\infty
\right\}.
\]

For the class of partially identified causal estimands considered in
this paper, the nuisance vector in \citet{semenova2023aggregated} satisfies
\(\nu_0(x)=\mu(x)\), where \(\mu(x)\) is defined in
\eqref{eq:discrete-conditional-marginal-vector}.

\begin{proof}[Proof of Corollary~\ref{thm:discrete-bias-under-semenova}]
By \eqref{eq:discrete-vertex-objective},
\[
\nabla_\eta\phi_x(t;\eta)
=-\big(\nu_{0,x}^t,\nu_{1,x}^t\big).
\]
Condition~\ref{cond:semenova-gradient} gives
\begin{equation}\label{eq:semenova-dual-vertex-norm-bound}
\sup_{x\in\mathcal X,\,t\in\mathcal T}
\big\|\big(\nu_{0,x}^t,\nu_{1,x}^t\big)\big\|_2
\le B_\phi.
\end{equation}
The triangle inequality therefore yields the sharper Euclidean diameter
bound
\begin{equation}\label{eq:semenova-dual-d2-bound}
D_2(x)\le 2B_\phi
\qquad\text{uniformly in }x.
\end{equation}

For \(t,t'\in\mathcal T\), write
\[
\Delta_{t,t'}(x)
=\phi_x(t';\mu(x))-\phi_x(t;\mu(x)).
\]
For every \(z\in(0,\delta)\), the definition of \(\omega_x\), the union
bound, and the displayed specialization of
\eqref{eq:semenova-margin} give
\[
\begin{aligned}
\E\{\omega_X(z)\}
&\le
\sum_{\substack{t,t'\in\mathcal T\\t\ne t'}}
\Pr\{0<|\Delta_{t,t'}(X)|\le z\}\\
&\le
\sum_{\substack{t,t'\in\mathcal T\\t\ne t'}}
\left[
\Pr\{0\le\Delta_{t,t'}(X)\le z\}
+\Pr\{0\le-\Delta_{t,t'}(X)\le z\}
\right]\\
&\le
2|\mathcal T|(|\mathcal T|-1)\bar B z.
\end{aligned}
\]
Since \(2B_\phi\nu_n^\infty<\delta\) for all sufficiently large \(n\),
on the event
\(\sup_x\|\widehat\mu(x)-\mu(x)\|_2\le\nu_n^\infty\),
the pointwise bound \eqref{eq:discrete-pointwise-l2-bias},
\eqref{eq:causal-vertex-direct-bound}, and
\eqref{eq:semenova-dual-d2-bound} give
\[
\begin{aligned}
0\le\theta_L-\widetilde\theta_L
&\le
\E\!\left[
\error_{P,2}(X)D_2(X)
\omega_X\!\left(D_2(X)\error_{P,2}(X)\right)
\mid\mcD_1
\right]
\\
&\le
2B_\phi\nu_n^\infty
\E\!\left\{
\omega_X\!\left(2B_\phi\nu_n^\infty\right)
\right\}
\\
&\le
2|\mathcal T|(|\mathcal T|-1)\bar B
\left(2B_\phi\nu_n^\infty\right)^2
\end{aligned}
\]
Because a polyhedron in dimension \(2m-1\) defined by \(m^2\) inequalities
has at most \(\binom{m^2}{2m-1}\) vertices, the last display is bounded by
\(C(\nu_n^\infty)^2\) for a constant \(C\) with the stated dependence.
Thus,
\eqref{eq:prop-d2-bias-under-semenova} is proved. The final conclusion in
\eqref{eq:prop-d2-root-n-bias-under-semenova} follows from
condition~\ref{cond:semenova-uniform} and
\((\nu_n^\infty)^2=o(n^{-1/2})\).
\end{proof}

\begin{example}[Logistic regression with an unbounded covariate]
\label{ex:logistic-unbounded-covariate}
Consider an estimand with \(f\) satisfying
\eqref{eq:discrete-cost-oscillation-bound}. Suppose
\(Y(0),Y(1)\in\{0,1\}\), let \(X\sim N(0,1)\), and write
\(p_w(x)=P\{Y(w)=1\mid X=x\}\). Assume
\[
p_0(x)=\frac14,
\qquad
p_1(x)=\frac34
\qquad\text{for every }x.
\]
Thus, the true conditional marginal vector \(\mu\) is constant in \(x\).
Assume additionally that \(f\) is constant in \(x\). The
inequalities defining \(\mathcal V_{f,x}^0\) are then identical for every
\(x\). Write the resulting common polyhedron as \(\mathcal V_f^0\), and index
its vertices by the same finite set \(\mathcal T\), so that
\((\nu_{0,x}^t,\nu_{1,x}^t)=(\nu_0^t,\nu_1^t)\) for every \(x\) and
\(t\in\mathcal T\). Since \(\mu(x)=\mu\), writing
\(\mu=(\mu_0,\mu_1)\), the definition of \(\phi_x\) gives
\[
\phi_x(t;\mu(x))
=-\mu_0^\top\nu_0^t-\mu_1^\top\nu_1^t
=:\phi(t;\mu),
\qquad t\in\mathcal T.
\]
By \eqref{eq:causal-vertex-margin-modulus}, \(\omega_x(z)\) indicates whether
an adjacent ordered pair has a strictly positive gap no larger than \(z\).
To isolate the role of uniform first-stage convergence, assume that distinct
vertices have distinct population objective values. The finite collection of
pairwise gaps then has a strictly positive minimum. Define
\[
\gamma_f
=
\min_{\substack{t,t'\in\mathcal T\\t\ne t'}}
|\phi(t';\mu)-\phi(t;\mu)|
>0.
\]
Such a pair can exist only when \(z\ge\gamma_f\). Hence,
\[
\omega_x(z)\le \1\{z\ge\gamma_f\}.
\]

Under random assignment, fit separate, correctly specified logistic
regressions for the two conditional marginals,
\[
\widehat p_{w,n}(x)
=\Lambda\!\left(\widehat\alpha_{w,n}+\widehat\beta_{w,n}x\right),
\qquad
\Lambda(s)=\frac{1}{1+e^{-s}},
\]
where the true intercepts are
\(\alpha_0=-\log 3\) and \(\alpha_1=\log 3\), and both true slopes are
zero. Define
\[
a_n=\max_{w\in\{0,1\}}
|\widehat\alpha_{w,n}-\alpha_w|,
\qquad
b_n=\max_{w\in\{0,1\}}|\widehat\beta_{w,n}|.
\]
Define the fitted coefficients arbitrarily on the event that an unpenalized
maximum likelihood estimate is not finite. Under the usual fixed-dimensional
asymptotics, that event has probability approaching zero and
\(a_n,b_n=O_p(n^{-1/2})\).
Moreover, \(P(b_n>0)\to1\). Because the logistic link is
\(1/4\)-Lipschitz, writing
\(\delta_{w,n}(x)=\widehat p_{w,n}(x)-p_w(x)\), we have
\[
\begin{aligned}
\error_{P,2}(x)^2
&=2\delta_{0,n}(x)^2+2\delta_{1,n}(x)^2,\\
\error_{P,2}(x)
&\le\frac12\{a_n+b_n|x|\},\\
\left\{
\E\!\left[\error_{P,2}(X)^2\mid\mcD_1\right]
\right\}^{1/2}
&\le\frac12(a_n+b_n)=O_p(n^{-1/2}).
\end{aligned}
\]
In contrast, on the event \(b_n>0\), the unbounded support of \(X\) implies
\[
\operatorname*{ess\,sup}_{x}\error_{P,2}(x)
\ge\frac{3\sqrt2}{4}.
\]
Thus, the result in \citet{semenova2023aggregated} does not apply to this
example.

Now turning to our bound in
Theorem~\ref{prop:discrete-lower-level}, because \(m=2\),
set \(C_f=2\sqrt3B\), so that
\(\overline{\error}_{P,2}(x)=C_f\error_{P,2}(x)\). Therefore,
\[
\omega_x\!\left(\overline{\error}_{P,2}(x)\right)
\le
\1\!\left\{
C_f\error_{P,2}(x)\ge\gamma_f
\right\}.
\]
Because \(a_n=O_p(n^{-1/2})\),
\(P\{a_n\le\gamma_f/C_f\}\to1\). On this event, the Lipschitz bound above implies
\[
\omega_x\!\left(\overline{\error}_{P,2}(x)\right)>0
\quad\Longrightarrow\quad
|x|\ge\frac{\gamma_f}{C_fb_n},
\]
with the event on the left empty when \(b_n=0\). When \(b_n>0\),
using \(\error_{P,2}(x)\le3/2\), the Gaussian tail bound and
Theorem~\ref{prop:discrete-lower-level} yield
\[
\begin{aligned}
0\le\theta_L-\widetilde\theta_L
&\le
\E\!\left[
\overline{\error}_{P,2}(X)
\omega_X\!\left(\overline{\error}_{P,2}(X)\right)
\mid\mcD_1
\right]\\
&\le
\frac{3C_f}{2}\Pr\!\left\{|X|\ge\frac{\gamma_f}{C_fb_n}\right\}
\le
3C_f\exp\!\left(-\frac{\gamma_f^2}{2C_f^2b_n^2}\right)
=o_p\!\left(n^{-1/2}\right).
\end{aligned}
\]
\end{example}

\begin{example}[Regression trees with a bounded covariate]
\label{ex:honest-tree-integrated-loss}
Consider the binary-outcome setting of
Example~\ref{ex:logistic-unbounded-covariate}, but now let
\(X\sim\operatorname{Unif}[0,1]\) and let both conditional marginals be
constant:
\[
p_w(x)=p_w\in(0,1),
\qquad w\in\{0,1\}.
\]
For each arm, estimate \(p_w\) using binary-response regression CART+
with leaf proportions under the repeated-response sampling structure in
Definition~5.1 of \citet{cattaneo2024pointwise}; this is their
phenomenological repeated-response construction rather than an ordinary
one-response-per-unit randomized experiment. Let \(N_{w,n}\) be the
arm-specific training-sample size, and choose maximal depth
\(K_n\asymp\log\log n\), with the lower constant sufficiently large for
their Theorem~5.2. That theorem applies to this constant regression model
and implies that, for some \(c>0\),
\[
\liminf_{n\to\infty}
P\!\left\{
\sup_{x\in[0,1]}
|\widehat p_{w,n}(x)-p_w|>c
\right\}>0.
\]
Consequently,
\(\sup_x\error_{P,2}(x)\) does not converge to zero in
probability. In particular, no deterministic
\(\nu_n^\infty=o(n^{-1/4})\) can satisfy condition~\ref{cond:semenova-uniform} of
Corollary~\ref{thm:discrete-bias-under-semenova}. Thus,
\citet{semenova2023aggregated}'s result does not apply to this example.

The average error is substantially smaller. Equation~(12) of
\citet{cattaneo2024pointwise} gives, for each arm,
\[
E\!\left[
\int_0^1
\{\widehat p_{w,n}(x)-p_w\}^2\,dP_X(x)
\,\middle|\,N_{w,n}
\right]
\le
\frac{2^{K_n+1}p_w(1-p_w)}{N_{w,n}+1}.
\]
Assume assignment is independent of \(X\), with constant probabilities
\(\pi_w\in(0,1)\), so the covariate distribution in each arm is \(P_X\) and
\(N_{w,n}/n\to\pi_w\) in probability. Moreover, with
\(\delta_{w,n}(x)=\widehat p_{w,n}(x)-p_w\),
\[
\error_{P,2}(x)^2
=2\delta_{0,n}(x)^2+2\delta_{1,n}(x)^2.
\]
Equation~(12) and Markov's inequality therefore give
\[
E\!\left[
\error_{P,2}(X)^2\mid\mcD_1
\right]
=O_p\!\left(\frac{\operatorname{polylog}(n)}{n}\right).
\]

To translate this difference into the first-stage bias, suppose the cost is
constant in \(x\) and satisfies
\eqref{eq:discrete-cost-oscillation-bound}. As in
Example~\ref{ex:logistic-unbounded-covariate}, let
\(\gamma_f>0\) be the smallest pairwise population objective gap, again
assuming that distinct vertices have distinct population objective values so
that the comparison isolates the uniform first-stage requirement. With
\(C_f=2B\sqrt{2m-1}\),
\[
\omega_x\!\left(C_f\error_{P,2}(x)\right)
\le
\1\!\left\{
C_f\error_{P,2}(x)\ge\gamma_f
\right\}.
\]
Theorem~\ref{prop:discrete-lower-level} and
\(a\1\{a\ge\gamma_f\}\le a^2/\gamma_f\) then yield
\[
\begin{aligned}
0\le\theta_L-\widetilde\theta_L
&\le
E\!\left[
C_f\error_{P,2}(X)
\1\!\left\{
C_f\error_{P,2}(X)\ge\gamma_f
\right\}
\mid\mcD_1
\right]\\
&\le
\frac{C_f^2}{\gamma_f}
E\!\left[
\error_{P,2}(X)^2\mid\mcD_1
\right]
=O_p\!\left(\frac{\operatorname{polylog}(n)}{n}\right)
=o_p(n^{-1/2}).
\end{aligned}
\]
Hence the uniform bound required by \citet{semenova2023aggregated} can be
qualitatively conservative even when the \(L^2\) error is small
enough for our bias bound to be negligible at the root-\(n\) scale.
\end{example}

\endgroup

\section{Proofs of results in other sections}

\subsection{Proof of Theorem \ref{thm::kantorovich}}
\label{sec: comp proof_sec2}
In this section, we first prove strong duality for the constrained optimal transport formulation in the absence of covariates. We then show that the problem separates in $X$ and that the Kantorovich duals can be constructed by conditioning on $X$. Finally, we present primitive conditions that justify the measurability of conditional Kantorovich duals with respect to $X$ so that $\theta_L(X)$ is a random variable.

\subsubsection{Strong duality without covariates}

The standard Monge-Kantorovich optimal transport problem can be formulated as: 
\begin{equation*}
    \theta_L = \inf_{P} \E_P[f(Y(0), Y(1))] \suchthat P_{Y(1)} = P_{Y(1)}\opt \text{ and } P_{Y(0)} = P_{Y(0)}\opt.
\end{equation*}
We state a version of Kantorovich strong duality below for completeness. The proof can be found in \cite{villani2009optimal,zaev2015monge}.
\begin{definition}
Let $Z_0$ and $Z_1$ be Polish spaces, let $Z=Z_0\times Z_1$, and write
\[
P_i=P_{Y(i)}\opt,\qquad i\in\{0,1\}.
\]
Define
\[
C_L(P_i)=C(Z_i)\cap L^1(P_i),\qquad i\in\{0,1\}.
\]
Here $C(Z_i)$ denotes the space of real-valued continuous functions on $Z_i$, and $L^1(P_i)$ denotes the space of real-valued Borel-measurable functions $g$ on $Z_i$ such that $\int |g|\,dP_i<\infty$.
For $P\opt=(P_0,P_1)$, define
\[
C_L(P\opt)
=
\left\{
h\in C(Z_0\times Z_1):
\begin{array}{l}
\text{there exist nonnegative }f_0\in C_L(P_0)
\text{ and }f_1\in C_L(P_1)\\
\text{such that }
|h(y_0,y_1)|
\le f_0(y_0)+f_1(y_1)
\text{ for every }(y_0,y_1)\in Z_0\times Z_1
\end{array}
\right\}.
\]
\end{definition}
\begin{theorem}
\label{thm: original OT}
Let $Z_0,Z_1, Z=Z_0 \times Z_1$ be Polish spaces, $P_{Y(0)}\opt,P_{Y(1)}\opt$ be two probability measures on $Z_0$ and $Z_1$, $f \in C_L(P\opt)$. Define the feasible set as
    \begin{equation}
        \mcQ =\left\{P\in\mcQ_{0}: P_{Y(1)} = P_{Y(1)}\opt \text{ and } P_{Y(0)} = P_{Y(0)}\opt\right\} 
    \end{equation}
    where $\mcQ_0$ denotes the set of all probability measures on $Z$. Then strong duality holds, that is,
    \begin{equation}
    \label{eqn::unregunconsdual_ot_app}
        \inf _{P\in\mcQ} \E_P[f(Y(0), Y(1))]=\sup _{\nu_0+\nu_1\leq f} \E_{P_{Y(0)}\opt}[\nu_0(Y(0))]+ \E_{P_{Y(1)}\opt}[\nu_1(Y(1))]
    \end{equation}
where $\nu_0\in C_L(P_{Y(0)}\opt),\nu_1\in C_L(P_{Y(1)}\opt)$.
\end{theorem}
\begin{remark}
\label{rmk:semicontinuous}
    The assumption of continuity could be weakened to lower semi-continuity without constraints; see Section 5 of \cite{villani2009optimal} for details. Here we state the stronger version because it is needed for our next theorem on the constrained optimal transport problems.
\end{remark}
With extra constraints, we can similarly derive the following duality theorem: 
\begin{theorem}
\label{thm: ineq OT}
    Let $Z_0,Z_1, Z=Z_0 \times Z_1$ be Polish spaces, $P_{Y(0)}\opt,P_{Y(1)}\opt$ be two probability measures on $Z_0$ and $Z_1$, $f \in C_L(P\opt)$, and let $W$ be a convex cone contained in $C_L(P\opt)$. Define the feasible set
    \begin{equation}
        \mcQ_W =\left\{P\in\mcQ_{0}:P_{Y(1)} = P_{Y(1)}\opt \text{ and } P_{Y(0)} = P_{Y(0)}\opt, \E_P[w(Y(0),Y(1))]\leq0, \forall w \in W\right\} 
    \end{equation}
    where $\mcQ_0$ denotes the set of all probability measures on $Z$.  Assume that $\mcQ_W$ is not empty, then the minimum of $\inf _{P\in\mcQ_W} \E_P[f(Y(0), Y(1))]$ can be achieved and strong duality holds in the sense that
    \begin{equation}
    \label{eqn::unregdual_ot}
        \inf _{P\in\mcQ_W} \E_P[f(Y(0), Y(1))]=\sup_{w\in W}\sup _{\nu_0+\nu_1-w \leq f} \E_{P_{Y(0)}\opt}[\nu_0(Y(0))]+ \E_{P_{Y(1)}\opt}[\nu_1(Y(1))]
    \end{equation}
where $\nu_0\in C_L(P_{Y(0)}\opt),\nu_1\in C_L(P_{Y(1)}\opt)$.
\end{theorem}
To prove Theorem \ref{thm: ineq OT}, we need a general version of the minimax theorem.

\begin{theorem}[Theorem 2.4.1 in \cite{adams1999function}]
\label{thm: minimax}
    Let $K$ be a compact convex subset of a Hausdorff topological vector space, $Y$ be a convex subset of an arbitrary vector space, and $h$ be a real-valued function $(\leq+\infty)$ on $K \times Y$, which is lower semicontinuous in $x$ for each fixed $y$, convex in $x\in K$, and concave in $y\in Y$. Then
$$
\min_{x \in K} \sup _{y \in Y} h(x, y)=\sup_{y \in Y} \min _{x \in K} h(x, y)
$$
\end{theorem}
With Theorem \ref{thm: original OT} and \ref{thm: minimax}, we can prove Theorem \ref{thm: ineq OT}.
\begin{proof}[Proof of Theorem \ref{thm: ineq OT}]
    First, it's straightforward to prove the LHS is at least as large as the RHS:
    \begin{equation*}
\begin{aligned}
\inf_{P\in\mcQ_W}\E_P[f(Y(0),Y(1))]
&\ge
\inf_{P\in\mcQ_W}
\sup_{\substack{w\in W\\ \nu_0+\nu_1-w\le f}}\E_P\!\left[\nu_0(Y(0))+\nu_1(Y(1))-w(Y(0),Y(1))\right]\\&\ge\inf_{P\in\mcQ_W}\sup_{\substack{w\in W\\ \nu_0+\nu_1-w\le f}}\left\{\E_{P_{Y(0)}\opt}[\nu_0(Y(0))]+\E_{P_{Y(1)}\opt}[\nu_1(Y(1))]\right\}\\&=\sup_{w\in W}\sup_{\nu_0+\nu_1-w\le f}\left\{\E_{P_{Y(0)}\opt}[\nu_0(Y(0))]+\E_{P_{Y(1)}\opt}[\nu_1(Y(1))]\right\}.\end{aligned}
\end{equation*}
To prove the other direction, we note that 
\begin{equation}
\label{eqn:duality second}
\begin{aligned}
    &\sup_{w\in W}\sup_{\nu_0+\nu_1-w\le f} \E_{P_{Y(0)}\opt}[\nu_0(Y(0))]+ \E_{P_{Y(1)}\opt}[\nu_1(Y(1))]\\
    =&\sup _{w \in W} \sup _{\nu_0+\nu_1\leq f+w} \E_{P_{Y(0)}\opt}[\nu_0(Y(0))]+ \E_{P_{Y(1)}\opt}[\nu_1(Y(1))]\\
    =&\sup _{w \in W} \inf _{P \in \mcQ} \E_{P}[f(Y(0), Y(1))+w(Y(0), Y(1))]
\end{aligned}
\end{equation}
where the last equality is obtained by applying Theorem \ref{thm: original OT} with $f$ replaced by $f + w$. Now we can apply the minimax theorem to interchange the supremum and infimum. Specifically, let $K=\mcQ,Y=W,h(P,w) =\E_{P}[f(Y(0), Y(1))+w(Y(0), Y(1))] $ in Theorem \ref{thm: minimax}.
It is a well-known consequence of the Prokhorov theorem that the set $\mcQ$ is compact
under the topology of weak convergence, and $\mcQ_W$ is a closed set of $\mcQ$, thus is also compact. The functional $h$ is linear in both arguments and thus it is convex in the first argument and concave in the second argument. By Corollary 1.5 of \cite{zaev2015monge},  for each fixed $w\in W$, the map $P\mapsto h(P,w)$ is continuous. Compactness and continuity together imply the existence of the solution. Moreover, the assumptions of Theorem \ref{thm: minimax} are satisfied and therefore
\begin{equation*}
    \sup _{w \in W} \inf _{P \in \mcQ} \E_P[f(Y(0), Y(1))+w(Y(0), Y(1))] = \inf _{P \in \mcQ} \sup _{w \in W} \E_P[f(Y(0), Y(1))+w(Y(0), Y(1))]
\end{equation*}
For $P\notin\mcQ_W$, there exists $w_1\in W$ such that $\E_P[ w_1(Y(0), Y(1))]>0$ by definition. Since $W$ is a convex cone, we know $\alpha w_1\in W$ for any $\alpha\geq 0$. Letting $\alpha\rightarrow \infty$ we see that $\E_P[f(Y(0), Y(1))+\alpha w_1(Y(0), Y(1))]\rightarrow \infty$ thus $\sup_{w\in W}\E_P[f(Y(0), Y(1))+w(Y(0), Y(1))]=\infty$. This implies 
\[\inf _{P \in \mcQ} \sup _{w \in W} \E_P[f(Y(0), Y(1))+w(Y(0), Y(1))] = \inf _{P \in \mcQ_W} \sup _{w \in W} \E_P[f(Y(0), Y(1))+w(Y(0), Y(1))].\]
Putting two pieces together, we obtain that
\begin{align*}
    &\sup _{w \in W} \inf _{P \in \mcQ}\E_P[f(Y(0), Y(1))+w(Y(0), Y(1))]\\
    & = \inf_{P \in \mcQ_W}\sup_{w\in W}\E_P[f(Y(0), Y(1))+w(Y(0), Y(1))]\\
    & = \inf_{P \in \mcQ_W}\E_P[f(Y(0), Y(1))].
\end{align*}
where the last equality holds by the simple fact that $0\in W$ and $\E_P[w(Y(0), Y(1))] \le 0$ for any $w\in W$ and $P \in Q_W$. Combining this identity with (\ref{eqn:duality second}), the proof of the other direction is completed.
\end{proof}

\subsubsection{Separability in $X$}
Recall that the problem with covariates is  
\begin{equation}
\label{eqn:dual original appendix}
    \theta_L = \inf_{P \in \mcP} \E_P[f(Y(0), Y(1), X)] \suchthat P_{Y(1), X} = P_{Y(1), X}\opt \text{ and } P_{Y(0), X} = P_{Y(0), X}\opt,
\end{equation}
where
\begin{equation*}
    \mcP = \bigg \{\text{joint distributions $P$ over } \mcY^2 \times \mcX \suchthat \E_P[w(Y(0), Y(1)) \mid X = x] \le 0 \,\,\, \forall w \in \mcW_x, x \in \mcX \bigg \}, 
\end{equation*}
and $\mcW_x = \{w_{x,1},\cdots,w_{x,L}\}$ is a finite collection of functions. In particular, by the linearity of expectation, we know that $\E_P[w(Y(0), Y(1)) \mid X = x] \le 0 \,\,\, \forall w \in \mcW_x,$ is equivalent to $\E_P[w(Y(0), Y(1)) \mid X = x] \le 0 \,\,\, \forall w \in W_x,$ where $W_x$ is the convex cone spanned by $\{w_{x,1},\cdots,w_{x,L}\}$. Now we are ready to prove a strong duality result for equation (\ref{eqn:dual original appendix}).

\begin{theorem}
\label{thm:conditioning_app}
    For $\theta_L$ as defined in \eqref{eqn:dual original appendix}, and for fixed x, we define $\theta_L(x)$ as
    \begin{equation}
    \label{eqn: thetaLx}
\begin{aligned}
    \theta_L(x) = &\inf_{P_x} \E_{P_x}[f(Y(0), Y(1), x)]  \\
    &\suchthat (P_x)_{Y(0)} = P_{Y(0)| X=x}\opt , (P_x)_{Y(1)} = P_{Y(1)| X=x}\opt , \\
    &\,\,\,\,\,\,\,\,\,\,\,\, \E_{P_x}[w(Y(0), Y(1))]\leq 0, \forall w \in W_x,
\end{aligned}
\end{equation}
Assume that for each $x$, $f(Y(0),Y(1),x)\in C_L(P\opt)$ and $W_x\subset C_L(P\opt)$. Assume further that $x\mapsto\theta_L(x)$ is measurable and $P_X\opt$-integrable, and that there exists a measurable probability kernel
\[
x\longmapsto P_x^{\mathrm{opt}}\in\mathcal P(\mcY^2)
\]
such that $P_x^{\mathrm{opt}}$ solves \eqref{eqn: thetaLx} for every $x$. Then
\[
\theta_L=\E_{P_X\opt}[\theta_L(X)].
\]

Moreover, let
\begin{equation}
    {(\nu_{0,x}\opt,\nu_{1,x}\opt) \in \argmax_{(\nu_{0,x},\nu_{1,x}) \in \mcV_x}} \E_{P\opt_{Y(0) \mid X = x}}[\nu_{0,x}(Y(0))] + \E_{P\opt_{Y(1) \mid X = x}}[\nu_{1,x}(Y(1))],
\end{equation}
then $\theta_L = \E_{P\opt}[\nu_{0, X}\opt(Y(0)) + \nu_{1, X}\opt(Y(1))]$, if $\nu_{0, X}\opt(Y(0)), \nu_{1, X}\opt(Y(1))$ are measurable with respect to $(X, Y(0), Y(1))$ and integrable under $P\opt$.
\end{theorem}
\begin{proof}
    We denote by $\mcQ_x$ the set of couplings
$P_x\in\mathcal P(\mcY^2)$ that are feasible for
\eqref{eqn: thetaLx}. Note that 
    \begin{align*}\theta_L
    &=\inf_{P\in\mcP}\E_P[f(Y(0), Y(1), X)] &\suchthat P_{Y(k), X} = P_{Y(k), X}\opt \text{ for } k \in \{0,1\} \\
    &=\inf_{P\in\mcP}\E_{P_{X}}\E_{P_{Y(0),Y(1)|X}}[f(Y(0), Y(1), X)] &\suchthat P_{Y(k), X} = P_{Y(k), X}\opt \text{ for } k \in \{0,1\}. \end{align*}
    By the constraint $P_{Y(1), X} = P_{Y(1), X}\opt \text{ and } P_{Y(0), X} = P_{Y(0), X}\opt$, we know that for each $P\in\mcP$, we will have $P_{X} = P_X\opt$ and $P_{Y(0), Y(1)|X=x}\in\mcQ_x$. By the measurable-kernel assumption, $x\mapsto P_x^{\mathrm{opt}}$ is a measurable probability kernel, $P_x^{\mathrm{opt}}\in\mcQ_x$, and $P_x^{\mathrm{opt}}$ solves \eqref{eqn: thetaLx} for every $x$. As a result, 
    \[\theta_L(x)
=
\inf_{P_x\in\mcQ_x}
\E_{P_x}[f(Y(0),Y(1),x)].\]
    Then $\theta_L(X)$ is measurable and hence
    \begin{equation}
    \begin{aligned}
        &\theta_L\geq \E_{P_{X}\opt}[\theta_L(X)],
    \end{aligned}
    \end{equation}
     To prove equality, define \[P^{\mathrm{opt}}(dx,dy_0,dy_1) =P_X\opt(dx)\,P_x^{\mathrm{opt}}(dy_0,dy_1).\] Because $x\mapsto P_x^{\mathrm{opt}}$ is a measurable probability kernel and $P_x^{\mathrm{opt}}\in\mcQ_x$ for every $x$, $P^{\mathrm{opt}}$ is a valid feasible joint distribution. Thus, \[\theta_L \le \E_{P^{\text{opt}}_{X, Y(0),Y(1)}}[f(Y(0), Y(1), X)] = \E_{P_{X}\opt}[\theta_L(X)].\] Therefore, $\theta_L = \E_{P_X\opt}[\theta_L(X)]$.

    For the dual representation, fix $x$. The conditional optimization problem in \eqref{eqn: thetaLx} is exactly the constrained optimal transport problem in Theorem \ref{thm: ineq OT}: its cost function is $f(\cdot,\cdot,x)$, its two marginals are $P_{Y(0)\mid X=x}\opt$ and $P_{Y(1)\mid X=x}\opt$, and its additional restrictions are generated by the cone $W_x$. Strong duality therefore equates its primal value $\theta_L(x)$ with the supremum of the corresponding conditional dual objective over $\mcV_x$. By the additional dual-attainment assumption, the selected pair $(\nu_{0,x}\opt,\nu_{1,x}\opt)$ attains that supremum. Hence
    \[
    \theta_L(x) = \E_{P_{Y(0)| X=x}\opt}[\nu_{0,x}\opt(Y(0))]+\E_{P_{Y(1)| X=x}\opt}[\nu_{1,x}\opt(Y(1))].
    \]
    Since $\nu_{0, X}\opt(Y(0)), \nu_{1, X}\opt(Y(1))$ are measurable with respect to $(X, Y(0), Y(1))$ and integrable under $P\opt$, apply the law of iterated expectation 
    $$\theta_L = \E_{P_{X}\opt}[\theta_L(X)] = \E_{P_{X}\opt}[\E_{P_{Y(0)| X}}\opt[\nu_{0,x}\opt(Y(0))]+\E_{P_{Y(1)| X}}\opt[\nu_{1,x}\opt(Y(1))]] =\E_{P\opt}[\nu_{0,X}\opt(Y(0))+\nu_{1,X}\opt(Y(1))],$$
    this finishes the proof.
    
\end{proof}

\subsubsection{Proof of Theorem \ref{thm::kantorovich}}
    Weak duality is straightforward from the fact that $(\nu_0,\nu_1)\in\mcV$ implies
    $$
    \nu_{0,x}(y_0) + \nu_{1,x}(y_1) 
    \le f(y_0, y_1, x) + \sum_{\ell=1}^{L} \lambda_{x,\ell} \cdot w_{x, \ell}(y_0, y_1) 
    $$
    and as a result, {for any $P\in\mcP$ satisfying the
    marginal constraints in \eqref{eq::init_thetaL_def},}
    \begin{equation}
        \begin{aligned}
            &g(\nu) = \E_{P\opt}[\nu_{0,X}(Y(0)) + \nu_{1,X}(Y(1)) ]= \E_{P}[\nu_{0,X}(Y(0)) + \nu_{1,X}(Y(1)) ]\\
            \leq& \E_{P}\left[f(Y(0),Y(1),X)+ \sum_{\ell=1}^{L} \lambda_{X,\ell} \cdot w_{X, \ell}(Y(0),Y(1))\right]\\
            =&\E_{P}[f(Y(0),Y(1),X)]+ \E_{P_X}\left[\sum_{\ell=1}^{L} \lambda_{X,\ell} \cdot \E_P[w_{X, \ell}(Y(0),Y(1))|X]\right]\\
            \leq& \E_{P}[f(Y(0),Y(1),X)].
        \end{aligned}
    \end{equation}
    {Since this holds for every such $P$, taking the infimum
    over the feasible set in \eqref{eq::init_thetaL_def} gives
    $g(\nu)\leq\theta_L$.}
    Strong duality follows from Theorem \ref{thm:conditioning_app}. Later, in Section \ref{sec: general measurability}, we will show that under Assumption \ref{asm:measureregularity}, Theorems \ref{thm:value-meas} and \ref{thm:primal-meas} provide, respectively, measurability of $x\mapsto\theta_L(x)$ and a measurable primal optimizer kernel, while Theorems \ref{thm:dual-attainment} and \ref{thm:dual-meas} provide dual attainment and a measurable selection of dual optimizers. The integrability conditions remain as imposed in Theorem \ref{thm:conditioning_app}.

\begin{remark}
\label{rmk:discussion example}
    Here we remark on how our examples could satisfy the regularity conditions on $f$ and $W_x$. 
    \begin{itemize}
        \item For Example \ref{ex::fh}, we can redefine $Y(0)$ and $Y(1)$ as $I(Y(0)\leq y_0)$ and $I(Y(1)\leq y_1)$. Then the problem becomes discrete. Clearly, the objective function $f(Y(0), Y(1)) = I(Y(0) = Y(1) = 1)$ is bounded and continuous under the discrete topology.
        \item For Example \ref{ex::varite}, the objective function $f(Y(0), Y(1)) = (Y(1) - Y(0))^2$ is clearly continuous under the standard Euclidean topology in $\R^2$. It is bounded by $2(Y(0)^2+Y(1)^2)$ which satisfies the integrability assumption if $Y(0), Y(1)$ have finite second moments.
        {\item For Example \ref{ex::lee}, use the augmented potential outcome
        $Z(w)=\bigl(Y(w)S(w),S(w)\bigr)$ on
        $\mcZ=\{(0,0)\}\cup(\R\times\{1\})$, equipped with the subspace topology
        inherited from $\R\times\{0,1\}$. Writing $z_w=(r_w,s_w)$, use the
        continuous cost $f^{\mathrm{Lee}}(z_0,z_1)=s_0(r_1-r_0)$ and the
        conditional moment function
        $w_x^{\mathrm{Lee}}(z_0,z_1)=\I\{s_0>s_1\}$. The latter is continuous
        because the selection coordinate has the discrete topology, and its
        nonnegativity makes the conditional restriction
        $\E[w_x^{\mathrm{Lee}}\{Z(0),Z(1)\}\mid X=x]\le0$ equivalent to monotone
        selection. Moreover,
        $|f^{\mathrm{Lee}}(z_0,z_1)|\le |r_0|+|r_1|$, so the integrability
        condition follows from finite first moments of the selected outcomes.
        Appendix \ref{app:dual-lee} exhibits the finite multiplier
        $\lambda_x^{(a)}=(a)_+$ and establishes strong duality within the
        generic moment-constrained formulation.}
        {\item For Example \ref{ex::makarov}, \(\mathcal Y=\mathbb R\)
        is Polish and the cost
        \(f_t(y_0,y_1)=\I\{y_1-y_0\le t\}\) is bounded and Borel measurable.
        Thus the primal objective is integrable without a moment condition, and
        the conditional marginals admit regular conditional laws. The duals in
        \eqref{eq:makarov-dual-family} are bounded Borel functions and hence are
        measurable and integrable. Finally,
        \eqref{eq:makarov-sharp-lower-bound} establishes the required conditional
        primal--dual equality directly.}
\item Examples \ref{ex::positive_ite}--\ref{ex:QTE} can be reasoned similarly as above. 
    \end{itemize}

\end{remark}

\subsubsection{Primitive assumptions for measurability with discrete potential outcomes}
In Theorem \ref{thm:conditioning_app}, we assume that the family $x\mapsto P_x^{\mathrm{opt}}$ is a measurable probability kernel that selects a pointwise primal optimizer and that the dual solutions $\nu_{0,X}\opt(Y(0)),\nu_{1,X}\opt(Y(1))$ are measurable with respect to $(X, Y(0), Y(1))$ and integrable on the product spaces. The integrability assumption ensures that the bound is finite, so we skip its discussion. Instead, in this section, we provide primitive conditions for measurability.

We remind readers that $x\mapsto P_x^{\mathrm{opt}}$ is a probability kernel if and only if
\begin{enumerate}
    \item For any fixed $x$, $P^{\text{opt}}_{x}(\cdot)$ is a probability distribution.
    \item For any fixed $A\in\mcF$, $P^{\text{opt}}_{x}(A)$ is a measurable function with respect to $x$, where $\mcF$ is the $\sigma$-algebra on the product space $\mcY^2$.
\end{enumerate}

We prove the following result that the measurability assumptions are satisfied when $\mathcal{X}$ is Euclidean and $\mathcal{Y}$ is discrete.

\begin{proposition}
\label{prop: measurability}
    Assume \(\mcX=\R^d\) and
\(\mcY=\{y_1,\ldots,y_K\}\), both with their Borel sigma-algebras.
For every \(j,k\in[K]\), \(\ell\in[L]\), and \(r\in\{0,1\}\), assume
that
\[
x\mapsto f(y_j,y_k,x),\quad
x\mapsto w_{x,\ell}(y_j,y_k),\quad
x\mapsto P^\star(Y(r)=y_j\mid X=x)
\]
are Borel. Assume also that the conditional primal problem is feasible
and has finite value for every x. Then one can choose primal and dual
optimizers whose coordinates are Borel functions of x. Consequently,
the measurability requirements of Theorem \ref{thm:conditioning_app} hold.
\end{proposition}
To prove Proposition \ref{prop: measurability}, we will need the following results from the theory of linear programming \citep{matouvsek2007understanding}:
\begin{definition}
    A basic feasible solution of the linear program
$$
\text { max } c^T x \quad \text { s.t. } A x=b \text { and } x \geq \mathbf{0}
$$
is a feasible solution $x \in \mathbb{R}^n$ for which there exists an $m$-element set $B \subseteq\{1,2, \ldots, n\}$ such that
\begin{itemize}
    \item the (square) matrix $A_B$ is nonsingular, i.e., the columns indexed by $B$ are linearly independent,
    \item $x_j=0$ for all $j \notin B$.
\end{itemize}
\end{definition}
\begin{lemma}[Theorem 4.2.3 of \cite{matouvsek2007understanding}]
\label{lem: BFS}
    Consider the following linear program 
    $$
\text { max } c^T x \quad\text { s.t. } A x=b \text { and } x \geq \mathbf{0}.
$$
\begin{enumerate}
    \item If there is at least one feasible solution and the objective function is bounded from above on the set of all feasible solutions, then there exists an optimal solution.
    \item If an optimal solution exists, then there is a basic feasible solution that is optimal.
\end{enumerate}
\end{lemma}
\begin{proof}[Proof of Proposition \ref{prop: measurability}]
Vectorize a conditional coupling as $p\in\R^{K^2}$. Delete one fixed
redundant marginal equation, for example the final equation for the
$Y(1)$ marginal. The remaining $2K-1$ marginal equations have a
fixed full-row-rank coefficient matrix $A_0$. The conditional primal
can therefore be written as
$$
\min_{p\ge0} c(x)^Tp
\quad\text{s.t.}\quad
A_0p=b_0(x),\qquad C(x)p\le0,
$$
where every entry of $c(x)$, $b_0(x)$, and $C(x)$ is Borel.

Introduce slack variables $s\in\R_+^L$, set $q=(p^T,s^T)^T$, and
write
$$
\min_{q\in\R^N}\ \widetilde c(x)^Tq
\quad\text{s.t.}\quad
\widetilde A(x)q=\widetilde b(x),\qquad q\ge0,
$$
where
$$
N=K^2+L,\qquad r=2K-1+L,
$$
and
$$
\widetilde A(x)
=
\begin{bmatrix}
A_0&0\\ C(x)&I_L
\end{bmatrix},\qquad
\widetilde b(x)=
\begin{bmatrix}b_0(x)\\0\end{bmatrix},\qquad
\widetilde c(x)=
\begin{bmatrix}c(x)\\0\end{bmatrix}.
$$
The matrix $\widetilde A(x)\in\R^{r\times N}$ has full row rank for
every $x$: the marginal block has rank $2K-1$, and the final $L$
rows are independent because of the identity block. Thus the basis size
$r$ is fixed and does not depend on $x$.

Enumerate the finitely many $r$-element subsets
$B_1,\ldots,B_M$ of $[N]$. For each $j$, on the Borel set
$$
G_j=\left\{x:
\det\widetilde A_{B_j}(x)\ne0,\quad
\widetilde A_{B_j}(x)^{-1}\widetilde b(x)\ge0
\right\},
$$
define the full vector
$$
q^{(j)}_{B_j}(x)
=\widetilde A_{B_j}(x)^{-1}\widetilde b(x),
\qquad
q^{(j)}_{B_j^c}(x)=0.
$$
Set
$$
\phi_j(x)=
\begin{cases}
\widetilde c(x)^Tq^{(j)}(x),&x\in G_j,\\
+\infty,&x\notin G_j.
\end{cases}
$$
Matrix inversion on the nonsingular set is Borel, so each $G_j$,
$q^{(j)}$, and $\phi_j$ is Borel. Let $j^\star(x)$ be the
smallest index minimizing $\phi_j(x)$. Because the collection is finite,
$j^\star(x)$ is Borel. Existence of an optimal basic feasible solution
shows that $q^\star(x)=q^{(j^\star(x))}(x)$ is optimal. Its first
$K^2$ coordinates give the required Borel primal coupling.

Convert the finite-dimensional dual to standard form by splitting each
free variable into positive and negative parts and adding slack variables.
Its dimensions are again fixed, and all coefficient maps are Borel.
Applying the same finite, tie-broken basis enumeration gives a Borel dual
optimizer. This proves the proposition.
\end{proof}

\subsection{Measurability results for general potential outcomes}
\label{sec: general measurability}
In this section, we extend the measurability result to general spaces. For clarity, we restate the problem in an equivalent form below. 

Let $\mcX, \mcY$ be Polish spaces with their Borel $\sigma$-algebra. Let $\mcP(\mcY^2)$ be the space of probability distributions on $\mcY^2$ endowed with weak topology.
For each $x\in X$ we are given the conditional marginal distribution
\[
P^\star_{Y(0)\mid X=x}\in\mathcal P(\mcY),\qquad P^\star_{Y(1)\mid X=x}\in\mathcal P(\mcY),
\]
a cost $f(Y(0),Y(1),X):\mcY\times \mcY\times \mcX\to\R$, and sets of inequality constraints $\{w_{x,\ell}\}_{\ell=1}^L$ and equality constraints $\{h_{x,k}\}_{k=1}^K$. For each $x\in X$ consider the primal optimal transport problem
\begin{align}P^{\textrm{opt}}_{Y(0),Y(1)\mid X=x}\in\argmin_{P_x\in\mcP(\mcY^2)}\;&\E_{P_x}[f(Y(0),Y(1),x)]\label{eq:cond-OT}\\ \text{s.t. }\quad&(P_x)_{Y(0)}=P^\star_{Y(0)\mid X=x},\qquad(P_x)_{Y(1)}=P^\star_{Y(1)\mid X=x},\nonumber\\ &\E_{P_x}[w_{x,\ell}(Y(0),Y(1))]\le0, \qquad \ell=1,\ldots,L,\nonumber\\
&\E_{P_x}[h_{x,k}(Y(0),Y(1))]=0,
\qquad k=1,\ldots,K.\nonumber
\end{align}
Similar to before, we denote the feasible set of \eqref{eq:cond-OT} as $\mcQ_x$, the set of distributions that satisfy the marginal constraints as
\begin{equation}\label{eq:tilde-Qx-definition}
\widetilde{\mcQ}_x=\left\{P_x\in\mcP(\mcY^2):(P_x)_{Y(0)}=P^\star_{Y(0)\mid X=x},\quad(P_x)_{Y(1)}=P^\star_{Y(1)\mid X=x}
\right\}.
\end{equation}
and the optimal value as $\theta_L(x)$.
Introduce inequality multipliers $\lambda=(\lambda_\ell)_{\ell=1}^L\in\R_+^L$ and equality multipliers
$\gamma=(\gamma_k)_{k=1}^K\in\R^K$, and define the augmented cost
\begin{equation}\label{eq:aug-cost}
c_{x,\lambda, \gamma}(y_0,y_1)
=f(y_0,y_1,x)+\sum_{\ell=1}^L \lambda_\ell\,w_{x,\ell}(y_0,y_1)
+\sum_{k=1}^K \gamma_k\,h_{x,k}(y_0,y_1).
\end{equation}
The dual formulation of \eqref{eq:cond-OT} can be written as
\begin{align}
(\nu_{0,x}^\star,\nu_{1,x}^\star,\lambda_x^\star,\gamma_x^\star)
\in\argmax_{\substack{\nu_{0,x},\nu_{1,x},\
\lambda\in\R_+^L,\ \gamma\in\R^K}}
&\ \E_{P^\star_{Y(0)\mid X=x}}[\nu_{0,x}(Y(0))]
+\E_{P^\star_{Y(1)\mid X=x}}[\nu_{1,x}(Y(1))]
\label{eq:cond-OT-dual}\\
\text{s.t. }\quad
&\nu_{0,x}(y_0)+\nu_{1,x}(y_1)
\le c_{x,\lambda,\gamma}(y_0,y_1),
\quad \forall(y_0,y_1)\in\mcY^2.\nonumber
\end{align}
We want to prove the following claims:
\begin{itemize}
    \item There exists a selection of optimizers $P^{\textrm{opt}}_{Y(0), Y(1)|X=x}$ for the primal problem \ref{eq:cond-OT} such that for any $A\in \mathcal{F}$, $x\mapsto P^{\textrm{opt}}_{Y(0), Y(1)|X=x}(A)$ is a measurable function, where $\mathcal{F}$ is the Borel $\sigma$ algebra on $\mcY^2$.
    \item There exists a selection of optimizers {$\nu_{0,x}^\star,\nu_{1,x}^\star$} for the dual problem \ref{eq:cond-OT-dual} such that $(x, Y(0), Y(1)) \mapsto {(\nu_{0,x}^\star(Y(0)),\nu_{1,x}^\star(Y(1)))}$ is measurable with respect to the Borel $\sigma$ algebra on the product space $\mcX\times \mcY\times \mcY$
\end{itemize}

\begin{assumption}\label{asm:measureregularity}
Assume:
\begin{enumerate}
\item (\emph{Compactness}) $\mcX$ and $\mcY$ are Polish spaces, and $\mcY$ is compact.
\item(\emph{Marginal measurability}) The maps
$x\mapsto P^\star_{Y(0)\mid X=x}$ and $x\mapsto P^\star_{Y(1)\mid X=x}$
are Borel measurable into $\mcP(\mcY)$ endowed with the weak topology.
\item (\emph{Cost regularity}) The map $(x,y_0,y_1)\mapsto f(y_0,y_1,x)$ is Borel measurable and uniformly bounded on $\mcY^2 \times \mcX$. 
For each fixed $x$, the function $(y_0,y_1)\mapsto f(y_0,y_1,x)\in C_L(P^\star)$. 
\item (\emph{Constraint regularity}) For each $\ell=1,\dots,L$ and $k=1, \cdots, K$,
$(x,y_0,y_1)\mapsto w_{x,\ell}(y_0,y_1)$ and $(x,y_0,y_1)\mapsto h_{x,k}(y_0,y_1)$ are Borel measurable, and for each fixed $x$,
$(y_0,y_1)\mapsto w_{x,\ell}(y_0,y_1)\in C_L(P^\star), (y_0,y_1)\mapsto h_{x,k}(y_0,y_1)\in C_L(P^\star)$.
\item (\emph{Feasibility}) For each $x$, the feasible set $\mcQ_x$ in \eqref{eq:cond-OT} is nonempty.
\item (\emph{Slater condition}) For each $x$, there exists a distribution $\bar P_x$ on $\mcY^2$ for the pair $(Y(0),Y(1))$ such that $\bar P_{x}|_{Y(0)}=P^\star_{Y(0)|X=x}, \bar P_{x}|_{Y(1)}=P^\star_{Y(1)|X=x}$, and
\begin{align*}
    &\E_{\bar P_x}[w_{x,\ell}(Y(0),Y(1))]<0,\forall \ell=1,\cdots, L, \\
    &\E_{\bar P_x}[h_{x,k}(Y(0),Y(1))]=0,\forall k=1,\cdots, K
\end{align*}
Further, assume that $0\in\operatorname{ri}(H_x)$ where
\[
H_x
=
\left\{
\left(
\E_P[h_{x,1}],\ldots,\E_P[h_{x,K}]
\right):
P\in\widetilde{\mathcal Q}_x
\right\},
\]
\end{enumerate}
\end{assumption}
Below, we will prove the two claims about primal measurability (Theorem \ref{thm:primal-meas}) and dual measurability (Theorem \ref{thm:dual-meas}) separately. 
\subsubsection{Primal Measurability}
We start with some technical lemmas in measure theory.
\begin{lemma}[Theorem 18.18 of \cite{kechris2012classical}]\label{lem:selection}
Let $A_1,A_2$ be Polish spaces and $B\subset A_1\times A_2$ a Borel set such that for all $x\in A_1$,
the sections $B_x=\{y\in A_2:(x,y)\in B\}$ are $\sigma$-compact. Then the projection
$\pi_{A_1}(B)=\{x\in A_1:\exists y\in A_2,\ (x,y)\in B\}$ is Borel and there exists a Borel map
$f:\pi_{A_1}(B)\to A_2$ with $(x,f(x))\in B$ for all $x\in\pi_{A_1}(B)$.
\end{lemma}

\begin{lemma}[Theorem 2.3 of \cite{gaudard1989sigma}]\label{lem:eval-weak}
Let $(Z,\mathcal{Z})$ be a measurable space with $Z$ separable. Let $\mcP(Z)$ be the set of probability
measures on $(Z,\mathcal{Z})$. Let $\Sigma$ be the $\sigma$-algebra on $\mcP(Z)$ generated by evaluation maps
$P\mapsto P(D)$ for $D\in\mathcal{Z}$, and let $\mathcal{B}_{\mathrm{w}}$ be the Borel $\sigma$-algebra of the weak topology
on $\mcP(Z)$. Then $\Sigma=\mathcal{B}_{\mathrm{w}}$.
\end{lemma}

\begin{lemma}[Theorem 17.25 of \cite{kechris2012classical}]\label{lem:int-borel}
Let $\mcX$ and $Z$ be Polish spaces, and endow $\mcP(Z)$ with the weak topology. Let $g:\mcX\times Z\to\R$ be bounded and Borel measurable, where $\mcX\times Z$ has the product topology.
Define $I_g:\mcX\times \mcP(Z)\to\R$ by
\[
I_g(x,P)=\int_Z g(x,z)\,P(dz).
\]
Here $\mcX\times \mcP(Z)$ is endowed with the product topology generated by the given topology on $\mcX$ and the weak topology on $\mcP(Z)$. Then $I_g$ is Borel measurable with respect to the corresponding Borel $\sigma$-algebra.
\end{lemma}

\begin{proposition}\label{prop:feasible}
Under Assumption \ref{asm:measureregularity}
\begin{enumerate}
\item $\mcQ_x$ is compact in $\mcP(\mcY^2)$ for each $x$.
\item\label{item:feasible-graph-borel} The graph {$\mathrm{Gr}(\mcQ)=\{(x,P):P\in\mcQ_x\}$}$\subseteq \mcX\times\mcP(\mcY^2)$ is Borel.
\end{enumerate}
\end{proposition}

\begin{proof}
For compactness, since $\mcY$ is compact, $\mcP(\mcY^2)$ is also compact under the weak topology.
For fixed $x$, recall that the set of couplings with the given marginals is
\[
\tilde{\mcQ}_x
=\{P\in\mcP(\mcY^2):P|_{Y(0)}=P^\star_{Y(0)\mid X=x},\ P|_{Y(1)}=P^\star_{Y(1)\mid X=x}\}.
\]
This set is closed in $\mcP(\mcY^2)$, hence compact.

By Assumption \ref{asm:measureregularity}, for fixed $x$ each $w_{x,\ell}$ and $h_{x,k}$ is continuous on compact $\mcY^2$, hence bounded.
Therefore the maps $P\mapsto \int w_{x,\ell}\,dP$ and $P\mapsto \int h_{x,k}\,dP$ are continuous on $\mcP(\mcY^2)$.
Thus the constraint sets $\{P:\int w_{x,\ell}\,dP\le 0\}$ and $\{P:\int h_{x,k}\,dP=0\}$ are closed.
Intersecting these closed sets with the compact $\tilde{\mcQ}_x$ gives that {$\mcQ_x$} is compact and nonempty.

To prove claim~\ref{item:feasible-graph-borel}, let $\{\varphi_n\}_{n\ge 1}\subset C(\mcY)$ be a countable set that separates probability measures on $\mcY$
(e.g.\ a dense subset of $C(\mcY)$ in the sup norm).
Then the marginal constraints are equivalent to the countable family of moment equalities
\[
\int \varphi_n(y_0)\,dP=\int \varphi_n(y_0)\,dP^\star_{Y(0)\mid X=x},\qquad
\int \varphi_n(y_1)\,dP=\int \varphi_n\,dP^\star_{Y(1)\mid X=x},\qquad n\ge 1.
\]
For each $r\in\{0,1\}$, the map $(x,P)\mapsto \int \varphi_n(y_r)\,dP$ is continuous in $P$ and Borel in $(x,P)$, while
$x\mapsto \int \varphi_n\,dP^\star_{Y(r)\mid X=x}$ is Borel by Assumption \ref{asm:measureregularity} and continuity of $\rho\mapsto\int\varphi_n\,d\rho$.
Hence the set of $(x,P)$ satisfying both marginal constraints is Borel.

{For \(g\in\{w_{\cdot,\ell},h_{\cdot,k}\}\), define the
bounded Borel truncation \(g_m=(-m)\vee(g\wedge m)\). Lemma
\ref{lem:int-borel} shows that
\((x,P)\mapsto\int g_m(x,z)\,P(dz)\) is Borel. For every \((x,P)\),
sectionwise boundedness gives
\(\int g_m(x,z)\,P(dz)\to\int g(x,z)\,P(dz)\); hence the latter map is
Borel as a pointwise limit.} Therefore each map
$(x,P)\mapsto\int w_{x,\ell}\,dP$ and
$(x,P)\mapsto\int h_{x,k}\,dP$ is Borel.
Therefore the inequality/equality constraint sets are Borel.
Intersecting finitely many such Borel sets with the marginal-constraint Borel set yields that {$\mathrm{Gr}(\mcQ)$} is Borel.
\end{proof}
\begin{theorem}\label{thm:value-meas}
Under Assumption \ref{asm:measureregularity}, the map $x\mapsto \theta_L(x)$ is Borel measurable.
\end{theorem}
\begin{proof}
Define the objective map
\[
F:\mcX\times \mcP(\mcY^2)\to\R,\qquad F(x,P)=\E_P[f(Y(0),Y(1),x)]
\]
By Assumption \ref{asm:measureregularity}, $f$ is uniformly bounded on $\mcY^2\times \mcX$, and by Lemma \ref{lem:int-borel} the map $F$ is Borel.

Fix $t\in\R$ and define the Borel set
\[
B_t=\{(x,P): {P\in\mcQ_x},\ \ F(x,P)<t\}\subseteq \mcX\times \mcP(\mcY^2).
\]
By Proposition \ref{prop:feasible}, {$\mathrm{Gr}(\mcQ)$} is Borel, and since $F$ is Borel, $B_t$ is Borel.

For each $x$, the section $(B_t)_x=\{P\in{\mcQ_x}: F(x,P)<t\}$ is $\sigma$-compact: indeed,
\[
(B_t)_x=\bigcup_{m=1}^\infty \bigl\{P\in{\mcQ_x}: F(x,P)\le t-\tfrac1m\bigr\},
\]
and each set in the union is compact as a closed subset of the compact {$\mcQ_x$} under the weak topology.

By Lemma \ref{lem:selection}, the projection $\pi_{\mcX}(B_t)=\{x:\exists {P\in\mcQ_x} \text{ with } F(x,P)<t\}$ is Borel.
But $\pi_{\mcX}(B_t)=\{x:\theta_L(x)<t\}$ by definition of $\theta_L$.
Therefore all strict sublevel sets $\{x:\theta_L(x)<t\}$ are Borel, hence $\theta_L$ is Borel measurable.
\end{proof}

Now we are ready to prove the primal measurability.

\begin{theorem}\label{thm:primal-meas}
Under Assumption \ref{asm:measureregularity}, there exist a selection of optimizer $P^{\textrm{opt}}_{Y(0), Y(1)|X=x}$ for the primal problem \eqref{eq:cond-OT} such that the mapping $x\mapsto P^{\mathrm{opt}}_{Y(0),Y(1)\mid X=x}$ is Borel measurable. Moreover, for any $A\in \mathcal{F}$, $x\mapsto P^{\textrm{opt}}_{Y(0), Y(1)|X=x}(A)$ is a measurable function.
\end{theorem}

\begin{proof}
Consider the set of optimal pairs
\[
M=\{(x,P): P\in\mcQ_x,\ \ \E_P[f(Y(0),Y(1),x)]=\theta_L(x)\}\subseteq \mcX\times\mcP(\mcY^2),
\]
which is Borel because {$\mathrm{Gr}(\mcQ)$} is Borel (Proposition \ref{prop:feasible}), $F$ is Borel (Lemma \ref{lem:int-borel}),
and $\theta_L$ is Borel (Theorem \ref{thm:value-meas}).

For each $x$, the section $M_x=\{P\in\mcQ_x:\E_P[f(Y(0),Y(1),x)]=\theta_L(x)\}$ is a closed subset of the compact {$\mcQ_x$}, hence compact.
Applying Lemma \ref{lem:selection}, we obtain a Borel selector $x\mapsto P^{\mathrm{opt}}_{Y(0),Y(1)\mid X=x}$ with $(x,P^{\mathrm{opt}}_{Y(0),Y(1)\mid X=x})\in M$.

Finally, since $\mcY^2$ is separable, Lemma \ref{lem:eval-weak} implies that evaluation maps $P\mapsto P(A)$ are Borel on $\mcP(\mcY^2)$
for each Borel $A\subseteq\mcY^2$. Therefore $x\mapsto P^{\mathrm{opt}}_{Y(0),Y(1)\mid X=x}(A)$ is Borel as a composition of Borel mappings.
\end{proof}
\subsubsection{Dual attainability}
Here, we first show the existence of the optimal dual variables. 
\begin{lemma}[Theorem 23.4 of \cite{rockafellar1997convex}]\label{lem:subgrad}
    If $u:\R^d\to(-\infty,+\infty]$ is proper and convex, and if $z_0$ is in the relative interior of $\dom(u)$, then the subgradient $\partial u(z_0)$ is nonempty.
\end{lemma}
\begin{lemma}[Theorem 5.10 of \cite{villani2009optimal}]\label{lem:dual attain unconstrain}
    If $f(\cdot,\cdot, x)\in C_L(P^\star)$, then the unconstrained Kantorovich problem
\begin{align}
{\sup_{\nu_{0,x},\nu_{1,x}}}\;&\ \E_{P^\star_{Y(0)\mid X=x}}[{\nu_{0,x}}(Y(0))] + \E_{P^\star_{Y(1)\mid X=x}}[{\nu_{1,x}}(Y(1))]
\\
\text{s.t. }&\nu_{0,x}(y_0) + \nu_{1,x}(y_1) 
    \le f(y_0, y_1, x) , \quad \forall y_0, y_1 \in \mcY.\nonumber
\end{align}
has a dual maximizer in $C_L(P^\star_{Y(0)})\times C_L(P^\star_{Y(1)})$.
\end{lemma}
\begin{theorem}
\label{thm:dual-attainment}
Under Assumption \ref{asm:measureregularity}, there exist
\[
\lambda_x=(\lambda_{x,1},\dots,\lambda_{x,L})\in\R_+^L,
\quad
\gamma_x=(\gamma_{x,1},\dots,\gamma_{x,K})\in\R^K,\quad 
\nu_{0,x},\nu_{1,x}\in C(\mcY)
\]
such that
\begin{equation*}
\label{eq:dual-feasibility}
\nu_{0,x}(y_0)+\nu_{1,x}(y_1)
\le
f_x(y_0,y_1)
+\sum_{\ell=1}^L \lambda_{x,\ell}w_{x,\ell}(y_0,y_1)
+\sum_{k=1}^K \gamma_{x,k}h_{x,k}(y_0,y_1)
\end{equation*}
for all $(y_0,y_1)\in\mcY^2$, and
\begin{equation*}
    \E_{P^\star_{Y(0)\mid X=x}}[{\nu_{0,x}}(Y(0))] + \E_{P^\star_{Y(1)\mid X=x}}[{\nu_{1,x}}(Y(1))] = \theta_L(x)
\end{equation*}
\end{theorem}
\begin{proof}
For $P\in {\tilde{\mcQ}_x}$, with {$\tilde{\mcQ}_x$ defined in \eqref{eq:tilde-Qx-definition}}, define the objective value $I_x(P) = \E_P[f(Y(0),Y(1),x)]$, the constraints moment vector
$$
m_x(P)
=
\Bigl(
\E_P[w_{x,1}],\dots,\E_P[w_{x,L}],
\E_P[h_{x,1}],\dots,\E_P[h_{x,K}]
\Bigr)\in\R^{L+K},
$$
and {define the attainable moment set $M_x=\{m_x(P):P\in\tilde{\mcQ}_x\}$.} Define the feasible cone $\mathcal C=(-\infty,0]^L\times\{0\}^K\subset\R^{L+K},$ then the constraint is equivalent to $m_x(P)\in \mathcal{C}$. Define
\begin{equation}
\label{eq:perturbation-value}
 u_x(z)
=
\inf\Bigl\{I_x(P): P\in\tilde{\mcQ}_x,\ m_x(P)+z\in\mathcal C\Bigr\},
\qquad z\in\R^{L+K}.
\end{equation}
By definition,
$u_x(0)=\theta_L(x).
$
Its effective domain is
\[
\dom(u_x)
=
\Bigl\{z\in\R^{L+K}:\exists P\in\tilde{\mcQ}_x\text{ such that }m_x(P)+z\in\mathcal C\Bigr\}
=
\mathcal C-M_x.
\]
Hence the strict feasible assumption in Assumption \ref{asm:measureregularity} implies that $0$ is in the relative interior of $\dom(u_x)$. 
\begin{enumerate}[label=\textbf{Step \arabic*:}, leftmargin=*, align=left]
\item \textbf{Existence of subgradient.}
Below, we will show that $u_x$ is a proper convex function. 

The properness comes from the fact that $u_x(0) = \theta_L(x)<\infty$. 
For convexity, take $z^1,z^2\in\R^{L+K}$, $t\in[0,1]$, and $\varepsilon>0$. Choose {$P_1,P_2\in\tilde{\mcQ}_x$} such that
\[
m_x(P_i)+z^i\in\mathcal C,
\qquad
I_x(P_i)\le u_x(z^i)+\varepsilon,
\qquad i=1,2.
\]
Set $P_t=tP_1+(1-t)P_2$. Since {$\tilde{\mcQ}_x$} is convex, {$P_t\in\tilde{\mcQ}_x$}. Since $m_x$ is affine,
\[
m_x(P_t)+tz^1+(1-t)z^2
=
 t\bigl(m_x(P_1)+z^1\bigr)+(1-t)\bigl(m_x(P_2)+z^2\bigr)\in\mathcal C,
\]
because $\mathcal C$ is convex. Therefore $P_t$ is feasible for the perturbation $tz^1+(1-t)z^2$. Since $I_x$ is affine in $P$,
\[
u_x\bigl(tz^1+(1-t)z^2\bigr)
\le I_x(P_t)
= tI_x(P_1)+(1-t)I_x(P_2)
\le t u_x(z^1)+(1-t) u_x(z^2)+\varepsilon.
\]
Letting $\varepsilon\downarrow 0$ proves convexity. 

Using Lemma \ref{lem:subgrad}, the subgradient $\partial u_x(0)$ is nonempty. Choose
$
\eta_x\in\partial u_x(0).
$
Then
\begin{equation}
\label{eq:subgradient-ineq}
u_x(z)\ge u_x(0)+\langle \eta_x,z\rangle
= \theta_L(x)+\langle \eta_x,z\rangle
\qquad \forall z\in\R^{L+K}.
\end{equation}
\item \textbf{Subgradient belongs to the dual cone.}
We claim that
\[
\eta_x\in\mathcal C^{\circ}=\Bigl\{\eta\in\R^{L+K}:\langle \eta,c\rangle\le 0\ \forall c\in\mathcal C\Bigr\}.
\]
Take an arbitrary $c\in\mathcal C$. If $P$ is feasible for $u_x(0)$, then $m_x(P)\in\mathcal C$. Since $\mathcal C$ is a convex cone, $m_x(P)+c\in\mathcal C.$ Hence every feasible point for $u_x(0)$ is also feasible for $u_x(c)$, and therefore
$u_x(c)\le u_x(0)=\theta_L(x).$ On the other hand, applying \eqref{eq:subgradient-ineq} with $z=c$ gives
\[
u_x(c)\ge \theta_L(x)+\langle \eta_x,c\rangle.
\]
Combining the two inequalities yields
$
\theta_L(x)+\langle \eta_x,c\rangle\le u_x(c)\le \theta_L(x),
$
so
$
\langle \eta_x,c\rangle\le 0, \forall c\in\mathcal C.
$
Thus $\eta_x\in\mathcal C^{\circ}$.

A direct computation of the polar cone gives
$
\mathcal C^{\circ}=\R_+^L\times\R^K.
$
Accordingly, we write
$
\eta_x=(\lambda_x,\gamma_x),
$
where
\[
\lambda_x=(\lambda_{x,1},\dots,\lambda_{x,L})\in\R_+^L,
\qquad
\gamma_x=(\gamma_{x,1},\dots,\gamma_{x,K})\in\R^K.
\]
\item \textbf{$\eta_x$ attains the Lagrangian relaxation.}
Take arbitrary $P\in\tilde{\mcQ}_x$. Set
$
z=-m_x(P).
$
Then
$
m_x(P)+z=0\in\mathcal C,
$
so $P$ is feasible for the perturbed problem $u_x(-m_x(P))$. Hence
$
u_x\bigl(-m_x(P)\bigr)\le I_x(P).
$
Applying \eqref{eq:subgradient-ineq} with $z=-m_x(P)$ gives
\[
u_x\bigl(-m_x(P)\bigr)
\ge
\theta_L(x)-\langle \eta_x,m_x(P)\rangle.
\]
Therefore
\[
I_x(P)+\langle \eta_x,m_x(P)\rangle\ge \theta_L(x)
\qquad\forall P\in\tilde{\mcQ}_x.
\]
Taking the infimum over all $P\in\tilde{\mcQ}_x$ yields
\begin{equation}
\label{eq:lagrangian-lower-bound}
\inf_{P\in\tilde{\mcQ}_x}\Bigl(I_x(P)+\langle \eta_x,m_x(P)\rangle\Bigr)\ge \theta_L(x).
\end{equation}

We now prove the reverse inequality. If $P\in\tilde{\mcQ}_x$ is primal feasible, then $m_x(P)\in\mathcal C$. Since $\eta_x\in\mathcal C^{\circ}$,
$
\langle \eta_x,m_x(P)\rangle\le 0.
$
Hence
\[
I_x(P)+\langle \eta_x,m_x(P)\rangle\le I_x(P).
\]
Taking the infimum over all primal-feasible $P$ gives
\[
\inf_{P\in\tilde{\mcQ}_x}\Bigl(I_x(P)+\langle \eta_x,m_x(P)\rangle\Bigr)
\le
\inf\Bigl\{I_x(P):P\in\tilde{\mcQ}_x,\ m_x(P)\in\mathcal C\Bigr\}
=\theta_L(x).
\]
Combining this with \eqref{eq:lagrangian-lower-bound}, we obtain
\[
\inf_{P\in\tilde{\mcQ}_x}\Bigl(I_x(P)+\langle \eta_x,m_x(P)\rangle\Bigr)=\theta_L(x).
\]

Now define the augmented cost
\begin{equation}
\label{eq:augmented-cost-proof}
c_{x,\eta_x}(y_0,y_1)
=
f_x(y_0,y_1)
+\sum_{\ell=1}^L \lambda_{x,\ell}w_{x,\ell}(y_0,y_1)
+\sum_{k=1}^K \gamma_{x,k}h_{x,k}(y_0,y_1).
\end{equation}
By the definitions of $I_x$ and $m_x$,
$
I_x(P)+\langle \eta_x,m_x(P)\rangle = \E_P[ c_{x,\eta_x}].
$
Hence
\begin{equation}
\label{eq:unconstrained-augmented-value}
\inf_{P\in\tilde{\mcQ}_x}\E_P[ c_{x,\eta_x}]=\theta_L(x).
\end{equation}
\item \textbf{Dual attainability.}
Because $f_x,w_{x,\ell},h_{x,k}$ are continuous on the compact space $\mcY^2$, the augmented cost $c_{x,\eta_x}$ defined in \eqref{eq:augmented-cost-proof} belongs to $C(\mcY^2)$. Therefore apply the dual attainment of unconstrained optimal transport problem (Lemma \ref{lem:dual attain unconstrain}), we know that there exist continuous functions
$
\nu_{0,x},\nu_{1,x}\in C(\mcY)
$
such that
\[
\nu_{0,x}(y_0)+\nu_{1,x}(y_1)\le c_{x,\eta_x}(y_0,y_1)
\qquad\forall (y_0,y_1)\in\mcY^2,
\]
and
\[
\E_{P^\star_{Y(0)\mid X=x}}[{\nu_{0,x}}(Y(0))] + \E_{P^\star_{Y(1)\mid X=x}}[{\nu_{1,x}}(Y(1))] = \inf_{P\in \tilde\mcQ_x}\E_P[c_{x,\eta_x}] = \theta_L(x),
\]
which proves the theorem.
\end{enumerate}
\end{proof}

\subsubsection{Dual Measurability}

Fix a continuous cost $c:\mcY^2\to\R$ and $\psi\in C(\mcY)$. Define the $c$-transform
\[
\psi^c(y_0)=\inf_{y_1\in\mcY}\{c(y_0,y_1)-\psi(y_1)\}\in C(\mcY),
\]
and for $\varphi\in C(\mcY)$ define $\varphi^c(y_1)=\inf_{y_0\in\mcY}\{c(y_0,y_1)-\varphi(y_0)\}$.

\begin{lemma}\label{lem:ctrans-bounds}
Let $c\in C(\mcY^2)$ and $\psi\in C(\mcY)$.
Then $\psi^c\in C(\mcY)$ and for all $y_0,y_0'\in\mcY$,
\begin{equation}\label{eq:ctrans-modulus}
\big|\psi^c(y_0)-\psi^c(y_0')\big|
\le \sup_{y_1\in\mcY}\big|c(y_0,y_1)-c(y_0',y_1)\big|.
\end{equation}
Similarly, for all $y_1,y_1'\in\mcY$,
\[
\big|\varphi^c(y_1)-\varphi^c(y_1')\big|
\le {\sup_{y_0\in\mcY}\big|c(y_0,y_1)-c(y_0,y_1')\big|}.
\]
\end{lemma}
\begin{proof}
We prove the bound for $\psi^c$; the proof for $\varphi^c$ is identical by symmetry. 
Fix $y_0, y_0' \in \mcY$. By definition of the infimum, for any $y_1 \in \mcY$, we have the trivial inequality:
\[
\psi^c(y_0') \le c(y_0', y_1) - \psi(y_1).
\]
We can add and subtract $c(y_0, y_1)$ on the right-hand side to obtain:
\[
\psi^c(y_0') \le \big(c(y_0, y_1) - \psi(y_1)\big) + \big(c(y_0', y_1) - c(y_0, y_1)\big).
\]
Taking the supremum over $y_1$ for the second term, we get:
\[
\psi^c(y_0') \le \big(c(y_0, y_1) - \psi(y_1)\big) + \sup_{y_1 \in \mcY} \big|c(y_0', y_1) - c(y_0, y_1)\big|.
\]
Since this holds for all $y_1 \in \mcY$, we can take the infimum over $y_1$ on the right-hand side, which yields:
\[
\psi^c(y_0') \le \psi^c(y_0) + \sup_{y_1 \in \mcY} \big|c(y_0', y_1) - c(y_0, y_1)\big|.
\]
Rearranging gives $\psi^c(y_0') - \psi^c(y_0) \le \sup_{y_1 \in \mcY} \big|c(y_0, y_1) - c(y_0', y_1)\big|$. Reversing the roles of $y_0$ and $y_0'$ yields the symmetric inequality, which gives the desired absolute value bound.

Finally, because $\mcY$ is compact and $c \in C(\mcY^2)$, $c$ is uniformly continuous. Equation \eqref{eq:ctrans-modulus} implies that $\psi^c$ shares the modulus of continuity of $c$, and hence $\psi^c \in C(\mcY)$.
\end{proof}
\begin{lemma}\label{lem:ctrans-lip}
    The \(c\)-transform is Lipschitz in the supremum norm:
for any \(\psi,\tilde\psi\in C(\mcY)\),
\[
\|\psi^c-\tilde\psi^c\|_\infty\le \|\psi-\tilde\psi\|_\infty.
\]
\end{lemma}
\begin{proof}
Let $\psi, \tilde\psi \in C(\mcY)$. For any $y_1 \in \mcY$, the definition of the supremum norm implies:
\[
\psi(y_1) \le \tilde\psi(y_1) + \|\psi - \tilde\psi\|_\infty.
\]
Substituting this into the definition of the $c$-transform evaluated at an arbitrary $y_0 \in \mcY$, we have:
\begin{align*}
\psi^c(y_0) &= \inf_{y_1 \in \mcY} \big\{ c(y_0, y_1) - \psi(y_1) \big\} \\
&\ge \inf_{y_1 \in \mcY} \big\{ c(y_0, y_1) - \tilde\psi(y_1) - \|\psi - \tilde\psi\|_\infty \big\} \\
&= \tilde\psi^c(y_0) - \|\psi - \tilde\psi\|_\infty.
\end{align*}
This implies $\tilde\psi^c(y_0) - \psi^c(y_0) \le \|\psi - \tilde\psi\|_\infty$. 
Reversing the roles of $\psi$ and $\tilde\psi$ gives $\psi^c(y_0) - \tilde\psi^c(y_0) \le \|\psi - \tilde\psi\|_\infty$. 
Therefore, for all $y_0 \in \mcY$, we have 
\[
\big|\psi^c(y_0) - \tilde\psi^c(y_0)\big| \le \|\psi - \tilde\psi\|_\infty.
\]
Taking the supremum over $y_0 \in \mcY$ yields the result.
\end{proof}
\begin{lemma}
\label{lem:normalization-dual}
Let
$c(y_0,y_1) \in C(Y^2)$ and $(\tilde \nu_0,\tilde \nu_1) \in C(Y) \times C(Y)$ satisfy
\[
\tilde \nu_0(y_0) + \tilde \nu_1(y_1) \le c(y_0,y_1),
\qquad \forall (y_0,y_1) \in Y^2.
\]
Fix $y^\star \in Y$, and define
\(
\hat \nu_0 = \tilde \nu_1^{c},
\hat \nu_1 = \hat \nu_0^{c},
\)
and
\(
a = \hat \nu_1(y^\star),
\nu_0 = \hat \nu_0 + a,
\nu_1 = \hat \nu_1 - a.
\)
Then $(\nu_0,\nu_1) \in C(Y)\times C(Y)$ satisfies
\[
\nu_1(y^\star)=0,
\qquad
\nu_0 = \nu_1^{c},
\qquad
\nu_1 = \nu_0^{c}, 
\qquad
\nu_0+\nu_1\geq \tilde\nu_0 + \tilde\nu_1
\]
\end{lemma}

\begin{proof}
Since $\tilde \nu_0 + \tilde \nu_1 \le c$, for every $y_0 \in Y$,
\[
\tilde \nu_0(y_0)
\le
\inf_{y_1 \in Y} \bigl\{ c(y_0,y_1) - \tilde \nu_1(y_1) \bigr\}
=
\tilde \nu_1^{c}(y_0)
=
\hat \nu_0(y_0).
\]
Likewise, since $\hat \nu_0 + \tilde \nu_1 \le c$ by definition of the $c$-transform,
\[
\tilde \nu_1 \le \hat \nu_0^{c} = \hat \nu_1.
\]
Hence $\nu_0 + \nu_1 = \hat \nu_0 + \hat \nu_1 \geq \tilde \nu_0 + \tilde \nu_1.$ Because $Y$ is compact and $c$ is continuous, both $\hat \nu_0$ and $\hat \nu_1$
belong to $C(Y)$. By construction, $\hat \nu_1 = \hat \nu_0^{c}$.
To show that $\hat \nu_0 = \hat \nu_1^{c}$, note that
$\tilde \nu_1 \le \hat \nu_1$, and by definition of $c$-transform, if $f\leq g$, then $f^c=\inf(c-f)\geq\inf(c-g)=g^c$, so
$
\hat \nu_1^{c}
\le
\tilde \nu_1^{c}
=
\hat \nu_0.
$
On the other hand, for any function $g$, by the definition of c-transform, we have 
$
g(y_0)\leq c(y_0,y_1) - g^c(y_1),
$
taking the infimum over $y_1$ on the right-hand side, 
one has $g \le g^{cc}$; applying this to
$g=\hat \nu_0$ gives
$
\hat \nu_0
\le
\hat \nu_0^{cc}
=
\hat \nu_1^{c}.
$
Therefore it proves that
$
\hat \nu_0 = \hat \nu_1^{c},
\hat \nu_1 = \hat \nu_0^{c}.
$

Finally, shifting one potential up by a constant and the other down by the same constant preserves
the objective, since both marginals are probability measures. It also preserves $c$-conjugacy,
because for any constant $b$, it is straightforward to see that
\[
(g-b)^{c} = g^{c} + b,
\qquad
(g+b)^{c} = g^{c} - b.
\]
With $a=\hat \nu_1(y^\star)$, the functions
\[
\nu_0 = \hat \nu_0 + a,
\qquad
\nu_1 = \hat \nu_1 - a
\]
satisfy
\[
\nu_1(y^\star)=0,
\qquad
\nu_0 = \nu_1^{c},
\qquad
\nu_1 = \nu_0^{c}.
\]
This proves the claim. 
\end{proof}
Now we are ready to prove dual measurability.

\begin{theorem}\label{thm:dual-meas}
Under Assumption \ref{asm:measureregularity}, let
$$
E=C(\mcY)\times C(\mcY)\times\R_+^L\times\R^K,
$$
where $C(\mcY)$ has the supremum norm. Then $E$ is a fixed Polish
space, and there exists a Borel map
$$
x\longmapsto
(\nu_{0,x}^\star,\nu_{1,x}^\star,\lambda_x^\star,\gamma_x^\star)
\in E
$$
that selects a dual optimizer for every $x$. Consequently,
$$
(x,y_0,y_1)\longmapsto
(\nu_{0,x}^\star(y_0),\nu_{1,x}^\star(y_1))
$$
is Borel on $\mcX\times\mcY\times\mcY$.
\end{theorem}
\begin{proof}
Fix $y_\star\in\mcY$ and write $\eta=(\lambda,\gamma)$, where
\(\lambda=(\lambda_1,\ldots,\lambda_L)\in\mathbb R_+^L\) and
\(\gamma=(\gamma_1,\ldots,\gamma_K)\in\mathbb R^K\). Recall the definition of $c_{x,\lambda,\gamma}$ in \eqref{eq:aug-cost}.
When \(\eta=(\lambda,\gamma)\), we use the shorthand \(c_{x,\eta}=c_{x,\lambda,\gamma}\). Thus \(c_{x,\eta}\) is the cost function obtained by augmenting the original cost with the inequality and equality moment restrictions using multipliers \((\lambda,\gamma)\). The notation \(\nu_1^{c_{x,\eta}}\) and \(\nu_0^{c_{x,\eta}}\) refers to the \(c\)-transform with this augmented cost.
For $n\in\mathbb N$, let
$$
K_n=\left\{(\lambda,\gamma)\in\R_+^L\times\R^K:
\sum_{\ell=1}^L\lambda_\ell+
\sum_{k=1}^K|\gamma_k|\le n\right\}.
$$
For $x\in\mcX$, let $D_n(x)$ be the set of all
$(\nu_0,\nu_1,\eta)\in C(\mcY)\times C(\mcY)\times K_n$ such that
\begin{enumerate}
\item $\nu_1(y_\star)=0$;
\item
$\nu_0=\nu_1^{c_{x,\eta}}$ and
$\nu_1=\nu_0^{c_{x,\eta}}$;
\item
$
\E_{P^\star_{Y(0)\mid X=x}}[\nu_0(Y(0))]
+
\E_{P^\star_{Y(1)\mid X=x}}[\nu_1(Y(1))]
=\theta_L(x).
$
\end{enumerate}

Let
\[
D(x)=\bigcup_{n=1}^\infty D_n(x).
\]
By Theorem \ref{thm:dual-attainment}, there exists triples $(\nu_{0,x}^\star,\nu_{1,x}^\star,\eta_x^\star)$ that attains the optimal value $\theta_L(x)$. Using Lemma \ref{lem:normalization-dual} and the dual optimality of $\nu_{0,x}^\star,\nu_{1,x}^\star$, we can construct a new pair $(\nu_0, \nu_1)$ that satisfies the three conditions above, and thus \(D(x)\neq\varnothing\) for every \(x\). To apply the measurable selection theorem (Lemma \ref{lem:selection}), we now show that each \(D_n(x)\) is compact and that the graph
$$
\Gamma_n=\{(x,u)\in\mcX\times E:u\in D_n(x)\}.
$$
is Borel.
\paragraph{$D_n(x)$ is compact:}
First, fix \(x\) and \(n\). Since \(K_n\) is compact and the functions \(f,w_{x,\ell}, h_{x,k}\) are continuous on the compact space \(\mcY^2\), the family
\[
\{c_{x,\eta}:\eta\in K_n\}
\]
is uniformly bounded and equicontinuous on \(\mcY^2\). Since $\nu_0 = \nu_1^{c_{x,\eta}}, \nu_1 = \nu_0^{c_{x,\eta}}$, then using Lemma \ref{lem:ctrans-bounds}, we know that all such $\nu_0, \nu_1$ are equicontinuous. We now prove the uniform boundedness. For $\nu_1$, Lemma \ref{lem:ctrans-bounds} implies that 
$$
|\nu_1(y_1)|=|\nu_1(y_1)-\nu_1(y_\star)|\le {\sup_{y_0\in\mcY}|c_{x,\eta}(y_0,y_1)-c_{x,\eta}(y_0,y_\star)|}\leq 2\|c_{x,\eta}\|_{\infty}.
$$
For $\nu_0$, using the definition of c-transform
$$
\nu_0(y_0)=\inf_{y_1\in\mcY}\bigl(c_{x,\eta}(y_0,y_1)-\nu_1(y_1)\bigr) \leq c_{x,\eta}(y_0,y_\star)\leq \|c_{x,\eta}\|_{\infty}, 
$$
$$
-\nu_0(y_0)=\sup_{y_1\in\mcY}\bigl(-c_{x,\eta}(y_0,y_1)+\nu_1(y_1)\bigr) \leq3 \|c_{x,\eta}\|_{\infty}. 
$$
Since $\{c_{x,\eta}:\eta\in K_n\}$ is uniformly bounded, the family $\{\nu_0,\nu_1\}$ is uniformly bounded as well. Using Arzelà–Ascoli theorem, the set $D_n(x)$ is relatively compact. 
To prove the compactness of $D_n(x)$, it remains to show it is closed. Let a convergent sequence (in sup norm) in $D_n(x)$ be
\[
(\nu_0^m,\nu_1^m,\eta^m)\in D_n(x),
\qquad
(\nu_0^m,\nu_1^m,\eta^m)\to (\nu_0,\nu_1,\eta).
\]
Since \(K_n\) is closed, \(\eta\in K_n\). Also, by continuity of evaluation, \(\nu_1(y_\star)=0\).

We now show \(\nu_0=\nu_1^{\,c_{x,\eta}}\). Since \(\nu_0^m=(\nu_1^m)^{c_{x,\eta^m}}\), Lemma \ref{lem:ctrans-bounds} and \ref{lem:ctrans-lip} implies that 
\[
\|(\nu_1^m)^{c_{x,\eta^m}}-\nu_1^{\,c_{x,\eta}}\|_\infty
\le
\|\nu_1^m-\nu_1\|_\infty + \|c_{x,\eta^m}-c_{x,\eta}\|_\infty
\]
The right-hand side tends to \(0\), because \(\nu_1^m\to \nu_1\) and $c_{x,\eta^m}\to c_{x,\eta}$ uniformly. Therefore
$
\nu_0^m\to \nu_1^{\,c_{x,\eta}}$
 uniformly. Since \(\nu_0^m\to \nu_0\) uniformly as well, uniqueness of the limit implies
$
\nu_0=\nu_1^{\,c_{x,\eta}}.
$
The proof that \(\nu_1=\nu_0^{\,c_{x,\eta}}\) is identical.
The objective equality passes to the limit because, for a probability
measure $\mu$,
$$
\left|\int(\nu^m-\nu)\,d\mu\right|
\le\|\nu^m-\nu\|_\infty.
$$
Thus $D_n(x)$ is closed and relatively compact, hence compact.
\paragraph{$\Gamma_n$ is Borel.}
Let $\{q_j\}_{j\ge1}$ be a countable dense subset of $\mcY$.
Continuity implies that $\nu_0=\nu_1^{c_{x,\eta}}$ is equivalent to
$$
\nu_0(q_j)=
\inf_{m\ge1}
\{c_{x,\eta}(q_j,q_m)-\nu_1(q_m)\},
\qquad j\ge1,
$$
and the reverse transform equality is equivalent to
$$
\nu_1(q_j)=
\inf_{m\ge1}
\{c_{x,\eta}(q_m,q_j)-\nu_0(q_m)\},
\qquad j\ge1.
$$
Each right-hand side is a countable infimum of Borel real-valued
functions, so the two transform equalities define Borel subsets of
$\mcX\times E$. Multiplier membership and normalization are closed
conditions. Finally, the objective equality is Borel because
$(g,\mu)\mapsto\int g\,d\mu$ is continuous on
$C(\mcY)\times\mcP(\mcY)$, the conditional marginal maps are Borel,
and $\theta_L(x)$ is Borel by Theorem \ref{thm:value-meas}. Hence
$\Gamma_n$ is Borel.
\paragraph{Measurable selection:}
Define
$
\Gamma=\bigcup_{n=1}^\infty \Gamma_n.
$
Since each \(\Gamma_n\) is Borel, \(\Gamma\) is Borel as a countable union of Borel sets. For each fixed \(x\in\mcX\), the section \(\Gamma_x\) equals
\[
\Gamma_x=D(x)=\bigcup_{n=1}^\infty D_n(x),
\]
which is a countable union of compact sets. Therefore \(\Gamma_x\) is \(\sigma\)-compact. By the measurable selection theorem Lemma \ref{lem:selection}, there exists a Borel map
\[
x\longmapsto (\nu_{0,x},\nu_{1,x},\eta_x)\in \Gamma_x=D(x).
\]
By construction, \((\nu_{0,x},\nu_{1,x},\eta_x)\) is dual optimal for each \(x\).
The evaluation map
\[
\operatorname{ev}:C(\mcY)\times \mcY\to \R,\qquad \operatorname{ev}(g,y)=g(y),
\]
is continuous when \(C(\mcY)\) is equipped with the supremum norm. Therefore,
$
(x,y_0,y_1)\longmapsto \bigl(\nu_{0,x}(y_0),\nu_{1,x}(y_1)\bigr)
$
is Borel on \(\mcX\times \mcY\times \mcY\), which finishes the proof.
\end{proof}
\subsection{Proof of \eqref{cex::ate}}\label{appendix::cex_ate}

For this argument, assume that \(\mcY\) contains its minimum and maximum
and that the model class contains degenerate potential-outcome distributions
at every pair in \(\mcY^2\), degenerate covariate distributions at every
\(x\in\mcX\), and treatment propensities arbitrarily close to zero
and one. Thus overlap is imposed separately for each distribution and not by one
uniform constant over the entire model class.
The proof is in three steps. First, we review the derivation of the Lagrange dual. Second, we show the result in the setting where there are no covariates; then, we generalize to the case with covariates.

First, we review the form of the Lagrange dual, following \cite{boyd2004}. Note that the objective function is the map $o : \mcP \to \R$ where $P \mapsto \E_{P}[Y(1) - Y(0)]$ with domain $\mcP$. The optimization variable is $P \in \mcP$, a distribution over $(X, W, Y(0), Y(1))$ satisfying strong ignorability and strict overlap, which induces a distribution $P_{X,W,Y}$ over $(X,W,Y)$. For every $P \in \mcP$, $\E_{P}[Y(1) - Y(0)]$ is a functional of $P_{X, W, Y}$; thus, $\theta(P\opt)$ is identifiable, and we have the equation
\begin{equation*}
    \theta(P\opt) = \min_{P \in \mcP} \E[Y_i(1) - Y_i(0)] \suchthat P_{X,W,Y} = P\opt_{X,W,Y}.
\end{equation*}
Since our constraint is $P_{X,W,Y} = P_{X,W,Y}\opt$, the Lagrangian is simply the objective function plus an additional linear functional of the difference between $P_{X,W,Y}$ and $P\opt_{X,W,Y}$. In other words, for any $h : \mcX \times \{0,1\} \times \mcY \to \R$, the Lagrangian is defined as
\begin{align*}
    L(P, h) &= \E_P[Y(1) - Y(0)] + \E_{P\opt}[h(X,W,Y)] - \E_{P}[h(X,W,Y)]\\
    & = \E_{P\opt}[h(X,W,Y)] + \E_P[Y(1) - Y(0) - h(X,W,Y)].
\end{align*}
The Lagrange dual function is simply the infimum of $L(P, h)$ over $P \in \mcP$:
\begin{equation*}
    g(h) = \E_{P\opt}[h(X,W,Y)] + \inf_{P \in \mcP} \E_P[Y(1) - Y(0) - h(X,W,Y)] = \E_{P\opt}[h(X,W,Y)] + \kappa(h)
\end{equation*}
for $\kappa(h)$ as defined previously, which shows the result.

Now, we show the main result in the setting where $\mcX$ contains one element and thus there are no covariates. In this case, fix any function $h : \{0,1\} \times \mcY \to \R$. For $w, y \in \{0,1\} \times \mcY$, we can write
\begin{equation*}
    h(w, y) = w h_1(y) + (1 - w) h_0(y)
\end{equation*}
for $h_1, h_0 : \mcY \to \R$. Note that under any $P \in \mcP$, we have by weak duality and unconfoundedness that
\begin{align*}
        \theta(P)
    &=
        \E_P[Y(1) - Y(0)] \\
    &\ge
        \E_P[W h_1(Y(1)) + (1-W) h_0(Y(0)) + \kappa(h)] \\
    &=
        P(W = 1) \E_P[h_1(Y(1))] + P(W =0) \E_P[h_0(Y(0))] + \kappa(h).
\end{align*}
Taking limits as $P(W = 1) \to 1$ and $P(W = 0) \to 0$ (note this does not violate strict overlap for each $P$ as $P(W = 1) \in (0,1)$), we obtain that
\begin{equation*}
    \kappa(h) \le \min_{P \in \mcP} \E_P[Y(1) - Y(0)] - \E_P[h_1(Y(1))].
\end{equation*}
Fix any $y\in\mcY$. Choose a sequence of distributions for which
$P(W=1)\to1$, $Y(0)=\max(\mcY)$ a.s., and $Y(1)=y$ a.s. Then
$$
\kappa(h)\le y-h_1(y)-\max(\mcY),
$$
and hence
$$
h_1(y)+\kappa(h)\le y-\max(\mcY).
$$
Because $y$ was arbitrary, this holds for every $y\in\mcY$.

Similarly, choose distributions for which $P(W=0)\to1$,
$Y(1)=\min(\mcY)$ a.s., and $Y(0)=y$ a.s. This gives
$$
h_0(y)+\kappa(h)\le\min(\mcY)-y,
\qquad y\in\mcY.
$$
Therefore, in the no-covariate case,
$$
h(W,Y)+\kappa(h)
\le
\begin{cases}
Y-\max(\mcY),&W=1,\\
\min(\mcY)-Y,&W=0.
\end{cases}
$$

For the general covariate case, fix $x\in\mcX$ and use closure of the
model class under the degenerate distribution $X=x$ a.s. The preceding
no-covariate argument then applies to $h(x,\cdot,\cdot)$. Since $x$
was arbitrary, the pointwise envelope holds for every $x\in\mcX$.
\section{Additional methodological details}\label{apppendix::method}

\subsection{Inference and model selection for generalized estimands with cross-fitting}\label{appendix::delta_method}

This paper primarily considers partially identifiable estimands of the form $\theta(P\opt) = \E_{P\opt}[f(Y(0), Y(1), X)]$. However, many estimands can be written in the form
\begin{equation}\label{eq::delta_method_general}
\theta(P\opt) = h\Bigg(\E_{P\opt}[f(Y(0), Y(1), X)], \,\, \E_{P\opt}[z_1(Y(1), X)], \,\, \E_{P\opt}[z_0(Y(0), X)]\Bigg),
\end{equation}
for some functions $z_0 : \mcY \times \mcX \to \R^{d_0}$ and
$z_1 : \mcY \times \mcX \to \R^{d_1}$. Let $\mathcal K$ be a compact subset
of $\R^{d_0+d_1+1}$ and let $\mathcal O$ be an open neighborhood of
$\mathcal K$. Assume that $h:\mathcal O\to\R$ is continuously differentiable
and nondecreasing in its first argument. Throughout this section, assume that
the population arguments belong to $\mathcal K$ and that, with probability
approaching one, every estimated or intermediate argument at which $h$ or
$\nabla h$ is evaluated belongs to $\mathcal O$.
In other words, $\theta(P\opt)$ can be written as a (nonlinear) function of a partially identifiable expectation and two identifiable expectations. We give two examples of this below.

\begin{example}[Variance of the ITE] If $\theta(P\opt) = \var(Y(1) - Y(0))$, we can write
\begin{equation}
    \theta(P\opt) = \E_{P\opt}[(Y(1) - Y(0))^2] - (\E_{P\opt}[Y(1)] - \E_{P\opt}[Y(0)])^2
\end{equation}
which satisfies Eq. (\ref{eq::delta_method_general}) if we set $f(y_0, y_1, x) = (y_1 - y_0)^2$, $z_0(y_0, x) = y_0$, $z_1(y_1, x) = y_1$, and $h(a, b, c) = a - (b - c)^2$.
\end{example}

\begin{example}[Lee bounds under monotonicity] In the case of Lee bounds (Example \ref{ex::lee}) under monotonicity, use the observable augmented potential outcome $Z(w)=\bigl(Y(w)S(w),S(w)\bigr)$. The estimand can be written as
\begin{equation*}
    \theta(P\opt) = \frac{\E_{P\opt}\left[\left\{Y(1)S(1)-Y(0)S(0)\right\}S(0)\right]}{\E_{P\opt}[S(0)]}
\end{equation*}
which satisfies Eq. (\ref{eq::delta_method_general}) if we set $f((r_0,s_0),(r_1,s_1),x)=(r_1-r_0)s_0$, $z_0((r_0,s_0),x)=s_0$, $z_1((r_1,s_1),x)=0$, and $h(a, b, c) = a / c$.
\end{example}

We now show how to perform inference on estimands in the general case of Eq. (\ref{eq::delta_method_general}). First, we give the main idea without discussing cross-fitting or model selection. Then, we introduce a multiplier bootstrap-like method to select the tightest bounds among $K$  cross-fit estimators of $\theta(P\opt)$. Following Appendix \ref{appendix::proofs}, we assume $n_2 = |\mcD_2| \ge cn$ for some constant $c > 0$.

\subsubsection{Main idea}

As notation, let $\beta = \E_{P\opt}[f(Y(0), Y(1), X)] \in \R$, $\kappa_1 = \E_{P\opt}[z_1(Y(1), X)] \in \R^{d_1}$, and $\kappa_0 = \E_{P\opt}[z_0(Y(0), X)] \in \R^{d_0}$ so that $\theta(P\opt) = h(\beta, \kappa_1, \kappa_0)$. If $\beta_L$ is the sharp lower bound on $\beta$, then $\theta_L \defeq h(\beta_L, \kappa_1, \kappa_0)$ is the sharp lower bound on $\theta(P\opt)$ since $h$ is monotone in its first coordinate and $\kappa_1, \kappa_0$ are identified.

The main idea is as follows. For the partially identified term $\E_{P\opt}[f(Y(0), Y(1), X)]$, estimate dual variables $\hat \nu_0, \hat \nu_1 : \mcY \times \mcX \to \R$ from $\mcD_1$ using techniques from the rest of the paper such that weak duality holds, that is, $\E_P[\hat\nu_{0,X}(Y(0)) + \hat\nu_{1,X}(Y(1)) \mid \mcD_1] \le \E_P[f(Y(0), Y(1), X)]$ for all $P \in \mcP$. Then define the IPW estimators
\begin{equation}
    \hat \beta = \frac{1}{n_2 } \sum_{i \in \mcD_2}  S_i^{(\beta)} \text{ for } S_i^{(\beta)} \defeq \frac{W_i \hat \nu_{1,X_i}(Y_i)}{\pi(X_i)} + \frac{(1- W_i) \hat \nu_{0,X_i}(Y_i)}{1 - \pi(X_i)}
\end{equation}
and for $w \in \{0,1\}$,
\begin{equation}
    \hat \kappa_w = \frac{1}{n_2 } \sum_{i \in \mcD_2}  S_i^{(\kappa_w)} \text{ for } S_i^{(\kappa_w)} \defeq \frac{\I(W_i = w) z_w(Y_i, X_i)}{w \pi(X_i) + (1-w) (1-\pi(X_i))}.
\end{equation}
\begin{equation}
    \sqrt{n_2} \left( \begin{bmatrix} \hat{\beta} \\ \hat\kappa_1 \\ \hat\kappa_0 \end{bmatrix}  - \begin{bmatrix} \tilde{\beta} \\ \kappa_1 \\ \kappa_0 \end{bmatrix} \right) \tod \mcN\left(0, \Sigma \right)
\end{equation}
where $\tilde{\beta} = \E[\hat\beta \mid \mcD_1] \le \beta$ by weak duality and $\Sigma \defeq \cov\left(\left( S_i^{(\beta)}, S_i^{(\kappa_1)}, S_i^{(\kappa_0)} \right) \mid \mcD_1 \right)$. The delta method yields that \begin{equation*}
        \sqrt{n_2} \left( h\left(\hat{\beta}, \hat\kappa_1, \hat\kappa_0 \right) -  h\left(\tilde{\beta}, \kappa_1, \kappa_0 \right) \right) \tod \mcN\left(0, \nabla h\left(\tilde{\beta}, \kappa_1, \kappa_0 \right)^T \Sigma \nabla h\left(\tilde{\beta}, \kappa_1, \kappa_0 \right) \right).
\end{equation*}
By plugging in $\hat \beta, \hat \kappa_1, \hat \kappa_0$, we can get a consistent estimator of the gradient $\nabla h\left(\tilde{\beta}, \kappa_1, \kappa_0 \right)$. Furthermore, we can get a consistent estimator of $\Sigma$ by letting $\hat \Sigma$ denote the empirical covariance matrix of the conditionally i.i.d. vectors $\{(S_i^{(\beta)}, S_i^{(\kappa_1)}, S_i^{(\kappa_0)})\}_{i \in \mcD_2}$. If we set $\hat\theta_L =  h\left(\hat{\beta}, \hat\kappa_1, \hat\kappa_0 \right)$ and $\tilde{\theta}_L = h\left(\tilde{\beta}, \kappa_1, \kappa_0 \right)$, Slutsky's theorem yields
\begin{equation*}
    \sqrt{\frac{n_2}{\nabla h\left(\hat{\beta}, \hat\kappa_1, \hat\kappa_0 \right)^T \hat \Sigma \nabla h\left(\hat{\beta}, \hat\kappa_1, \hat\kappa_0 \right) }} \left(\hat \theta_L - \tilde{\theta}_L \right) \tod \mcN\left(0, 1\right).
\end{equation*}
Using this equation, we note that 
\begin{equation*}
    \hat\theta\lcb = \hat\theta_L - \Phi^{-1}(1-\alpha) \sqrt{\frac{\nabla h\left(\hat{\beta}, \hat\kappa_1, \hat\kappa_0 \right)^T \hat \Sigma \nabla h\left(\hat{\beta}, \hat\kappa_1, \hat\kappa_0 \right)}{n_2}}
\end{equation*}
is an asymptotic $1-\alpha$ lower confidence bound on $\tilde{\theta}_L$. Note that by weak duality, $\tilde{\beta} \le \beta$, and therefore since $h$ is nondecreasing in its first argument, we have that
\begin{equation*}
    \tilde{\theta}_L = h(\tilde{\beta}, \kappa_1, \kappa_0) \le h(\beta, \kappa_1, \kappa_0) = \theta(P\opt)
\end{equation*}
and therefore $\hat \theta\lcb$ is a valid lower confidence bound on $\theta(P\opt)$ as well.

\begin{remark} This calculation requires that the dimensions $d_0, d_1$ are fixed constants that do not grow with $n$. 
\end{remark}

\subsubsection{Cross-fitting and model-selection}

In Section \ref{subsec::multbootstrap}, we introduced a multiplier bootstrap method that selects the tightest possible dual bounds across $K$ dual variable estimates (e.g., fit using different subsets of the covariates), where $K$ may grow exponentially with $n$. We now generalize this method in two ways. First, we now permit the use of cross-fitting. Second, we consider the generalized class of estimands defined in Eq. (\ref{eq::delta_method_general}). However, this generality comes at a cost: unlike Corollary \ref{cor::multiboot_validity}, we require that the number of dual variable estimates $K$ is fixed and does not grow with $n$. 

We first define the method; then we prove its validity. Suppose given the first fold of data $\mcD_1$, we produce $K$ candidate dual variables $\hat\nu^{(1)}, \dots, \hat\nu^{(K)} \in \mcV$, and symmetrically using the second fold $\mcD_2$ we produce $\hat\nu^{(1,\mrswap)}, \dots, \hat\nu^{(K,\mrswap)} \in \mcV$. For ease of exposition, we assume $n$ is even and $n_1 = n_2 = n/2$. Again, the results in this section can be easily extended to $M$-fold cross-fitting for $M > 2$. For each $k \in [K]$, the cross-fit dual lower estimate of $\theta(P\opt)$ is defined by plugging in an IPW-mean estimator of $\kappa_1, \kappa_0$ and a dual cross-fit lower estimator of $\beta \defeq \E_{P\opt}[f(Y(0), Y(1), X)]$ into the definition $\theta(P\opt) = h(\beta, \kappa_1, \kappa_0)$. Precisely: 
\begin{equation}
    \hat\theta_L^{(k)} \defeq h(\hat\beta^{(k)}, \hat\kappa_1, \hat\kappa_0),
\end{equation}
where
\begin{equation}
    \hat\beta^{(k)} = \frac{1}{n} \sum_{i=1}^n S_i^{(\beta, k)} \text{ for } S_i^{(\beta, k)} \defeq \begin{cases} \frac{W_i \hat \nu_{1,X_i}^{(k)}(Y_i)}{\pi(X_i)} + \frac{(1- W_i) \hat \nu_{0,X_i}^{(k)}(Y_i)}{1 - \pi(X_i)} & i \in \mcD_2 \\
    \frac{W_i \hat \nu_{1,X_i}^{(k,\mrswap)}(Y_i)}{\pi(X_i)} + \frac{(1- W_i) \hat \nu_{0,X_i}^{(k,\mrswap)}(Y_i)}{1 - \pi(X_i)} & i \in \mcD_1, \\
    \end{cases}
\end{equation}
and for $w \in \{0,1\}$,
\begin{equation}
    \hat \kappa_w = \frac{1}{n} \sum_{i=1}^n S_i^{(\kappa_w)} \text{ for } S_i^{(\kappa_w)} \defeq \frac{\I(W_i = w) z_w(Y_i, X_i)}{w \pi(X_i) + (1-w) (1-\pi(X_i))}.
\end{equation}
The standard error $\hat\sigma^{(k)}$ of $\sqrt{n} \hat\theta_L^{(k)}$ is defined as:
\begin{equation}
    \hat\sigma^{(k)} = \sqrt{\nabla h\left(\hat{\beta}^{(k)}, \hat\kappa_1, \hat\kappa_0 \right)^T \hat \Sigma^{(k)} \nabla h\left(\hat{\beta}^{(k)}, \hat\kappa_1, \hat\kappa_0 \right)},
\end{equation}
where $\hat\Sigma^{(k)} \in \R^{(1 + d_1 + d_0) \times (1 + d_1 + d_0)}$ is the empirical covariance matrix of the vectors $(S_i^{(\beta, k)}, S_i^{(\kappa_1)}, S_i^{(\kappa_0)})$ for $i \in [n]$. To aggregate evidence across all $K$ lower confidence bounds, we require the following notation. Let $\hat\Sigma_{\mathrm{full}} \in \R^{(K + d_0 + d_1) \times (K + d_0 + d_1)}$ denote the empirical covariance matrix of $\vec{S}_i = (S_i^{(\beta, 1)}, \dots, S_i^{(\beta, K)}, S_i^{(\kappa_1)}, S_i^{(\kappa_0)}) \in \R^{K + d_0 + d_1}$ and let $H : \R^{K + d_0 + d_1} \to \R^K$ be the function defined by $H_k(x)=h\left(x_k,\,x_{K+1:K+d_1},\,x_{K+d_1+1:K+d_1+d_0}\right)$. Here $$\nabla H(x) = [\nabla H_1(x),\ldots,\nabla H_K(x)] \in\R^{(K+d_1+d_0)\times K},$$ so the $k$th column is the gradient of $H_k$. In particular, this definition ensures that if $\bar S$ is the sample average of $\{\vec{S}_i\}_{i \in [n]}$, then $H(\bar S) = (\hat\theta_L^{(1)}, \dots, \hat\theta_L^{(K)})$. Define
\begin{equation}
    \hat \Sigma_H = \nabla H(\bar S)^T \hat\Sigma_{\mathrm{full}} \nabla H(\bar S) 
    \text{ and } \hat C_H = \diag{\hat\Sigma_H}^{-1/2} \hat\Sigma_{H} \diag{\hat\Sigma_H}^{-1/2}.
\end{equation}

Then the final combined lower bound is defined as
\begin{equation}\label{eq::thetaLCB_crossfit_combined}
\hat\theta\lcb\crossfit=\max_{k\in[K]}\left\{\hat\theta_L^{(k)}-\hat q_{1-\alpha}\frac{\hat\sigma^{(k)}}{\sqrt n}\right\},
\end{equation}
where we define $\hat q_{1-\alpha}$ as the $1-\alpha$ quantile of the maximum of a $\mcN(0, \hat C_H)$ vector:
\begin{equation}
    \hat q_{1-\alpha} \defeq Q_{1-\alpha}\left(\max_{k=1}^K Z_k\right) \text{ for } Z \sim \mcN(0, \hat C_H).
\end{equation}
We now show that Eq. (\ref{eq::thetaLCB_crossfit_combined}) defines a valid lower confidence bound under essentially the same assumptions as Theorem \ref{thm::crossfit_validity} as long as the number of models $K$ does not grow with $n$. We implicitly assume that the functions defining the estimand---namely $h$, $f$, $z_0$, $z_1$---do not change with $n$. Below, note that for dual variables $\nu \in \mcV$, $g(\nu) = \E_{P\opt}[\nu_{1,X_i}(Y(1)) + \nu_{0,X_i}(Y(0))]$ is the Lagrange dual function from Section \ref{subsec::setting}.

\begin{corollary} Suppose that $h$ is continuously differentiable and nondecreasing in its first argument. Assume that, for some \(\delta>0\), the full stacked summands
\[
\vec S_i
=
\bigl(S_i^{(\beta,1)},\ldots,S_i^{(\beta,K)},
S_i^{(\kappa_1)},S_i^{(\kappa_0)}\bigr)
\]
and, under Condition~\ref{cond:delta-method-convergence}, their deterministic-limit counterparts have uniformly bounded conditional \((2+\delta)\)-moments on both evaluation folds. Assume also that the corresponding centered stacked means satisfy the joint multivariate CLTs used below and that their empirical first and second moments satisfy the corresponding LLNs. Finally, assume joint nondegeneracy: for some constant \(\underline\sigma^2>0\), all transformed population variances appearing below are at least \(\underline\sigma^2\), and
\[
\P\!\left\{
\min_{k\in[K]}[\widehat\Sigma_H]_{kk}
\ge \underline\sigma^2
\right\}\longrightarrow1.
\]
Under Assumption \ref{assump::rand_experiment}, for $\alpha \le 0.5$,
\begin{equation*}
    \liminf_{n \to \infty} \P(\hat\theta\lcb\crossfit \le \theta_L) \ge 1 - \alpha,
\end{equation*}
holds as long as for each $k \in [K]$, $\hat\nu^{(k)}$ satisfies Assumption \ref{assumption::nu_moment} and one of the two following conditions:
\begin{enumerate}[label=\textnormal{Condition \arabic*:}, ref=\arabic*, topsep=0pt, itemsep=0.5pt, leftmargin=*]
    \item\label{cond:delta-method-convergence} There exist arbitrary deterministic dual variables $\nu\kdagger \in \mcV$ satisfying Assumption \ref{assumption::nu_moment} such that $\E\left[\left(\hat \nu_{w,X}^{(k)}(Y(w)) - \nu\kdagger_{w,X}(Y(w))\right)^2\right] \to 0$ holds at any rate for $w \in \{0,1\}$. The same convergence holds with $\hat\nu^{(k,\mrswap)}$ replacing $\hat\nu^{(k)}$, toward the same deterministic limit $\nu\kdagger$. Note that we do not allow $\{\nu\kdagger_w\}_{w\in \{0,1\}}$ to change with $n$. Furthermore, if $S_i^{(\beta, k, \dagger)}$ is defined analogously to $S_i^{(\beta, k)}$ but with $\nu\kdagger$ replacing $\hat\nu^{(k)}$ and $\hat\nu^{(k,\mrswap)}$, then we require that
    \begin{equation}\label{eq::delta_method_is_allowed}
        \nabla h\left(g(\nu\kdagger), \kappa_1, \kappa_0 \right)^T \cov\!\left(S_i^{(\beta, k, \dagger)}, S_i^{(\kappa_1)}, S_i^{(\kappa_0)}\right)  \nabla h\left(g(\nu\kdagger), \kappa_1, \kappa_0 \right) > 0.
    \end{equation}
    \item\label{cond:delta-method-misspecification} The outcome model is sufficiently misspecified such that the first-stage bias is larger than $n^{-1/2}$, i.e., $\sqrt n\left(\beta_L - \frac{g(\hat\nu^{(k)}) + g(\hat\nu^{(k,\mrswap)})}{2} \right) \toprob \infty.$ Furthermore, the partial derivative $\partial_b h(b, \kappa_1, \kappa_0)$ is uniformly bounded away from zero over the relevant section $\{b\in\R:(b,\kappa_1,\kappa_0)\in\mathcal K\}$.
\end{enumerate}
\end{corollary}

\begin{remark} We recommend that the reader read the proofs of Theorem \ref{thm::alwaysvalid} and Theorem \ref{thm::crossfit_validity} before reading this proof.
\end{remark}

\begin{remark} Conditions~\ref{cond:delta-method-convergence} and~\ref{cond:delta-method-misspecification} are the same conditions required in Theorem \ref{thm::crossfit_validity}, with four changes. First, for simplicity, we do not allow $\nu\kdagger$ to change with $n$. Second, we impose joint moment, CLT, and LLN conditions on the full stacked vector, including the identifiable summands $S_i^{(\kappa_1)}$ and $S_i^{(\kappa_0)}$. Third, we require joint nondegeneracy of all transformed coordinates; Eq. (\ref{eq::delta_method_is_allowed}) states the corresponding population requirement under Condition~\ref{cond:delta-method-convergence}. These conditions ensure that the Gaussian correlation matrices and studentized statistics are well-defined. Fourth, in Condition~\ref{cond:delta-method-misspecification}, we require a lower bound on the partial derivative of $h$ with respect to its first coordinate. This is necessary to guarantee that if $\hat\beta^{(k)}$ is asymptotically conservative for $\beta_L$, then $\hat\theta^{(k)}_L$ will be conservative for $\theta_L$. 
\end{remark}

\begin{proof} We handle the two conditions separately. 

\paragraph{Condition~\ref{cond:delta-method-convergence}.} We first prove the result in the special case where $\hat\nu^{(k)}$ satisfies Condition~\ref{cond:delta-method-convergence} for every $k \in [K]$. As notation, let $S_i^{(k,\beta,\dagger)}, \hat\beta\kdagger, \hat\sigma\kdagger, \hat\Sigma_{\mathrm{full}}\conv, \vec{S}_i\conv$ be defined analogously to $S_i^{(k,\beta)}, \hat\beta^{(k)}, \hat\sigma^{(k)}, \hat\Sigma_{\mathrm{full}}, \vec{S}_i$ but replacing $\hat\nu^{(k)}$ with $\nu\kdagger$, for each $k \in [K]$.
The proof of Theorem \ref{thm::crossfit_validity} in Appendix \ref{appendix::crossfit_proofs} shows the following relationships between these quantities:
\begin{enumerate}[noitemsep, topsep=0pt, series=conditionproof]
    \item\label{res:delta-beta-bound} $\hat\beta^{(k)} \le \hat\beta\kdagger + \Delta_k + o_p(n^{-1/2})$, where $\Delta_k \le \beta_L - \E[\hat\beta\kdagger]$ and $\Delta_k = o_p(1)$. \item\label{res:delta-variance-convergence} $\hat\sigma^{(k)} - \hat\sigma\kdagger = o_p(1)$, and a similar argument shows $\hat\Sigma_{\mathrm{full}} - \hat\Sigma_{\mathrm{full}}\conv = o_p(1)$ holds elementwise (this follows from a uniform LLN as reviewed in Appendix \ref{appendix::crossfit_proofs}). Note that $\hat\Sigma_{\mathrm{full}}\conv$ is simply an empirical covariance matrix of the i.i.d. vectors $\vec{S}_i\conv \defeq (S_i^{(\beta,1,\dagger)}, \dots, S_i^{(\beta, K, \dagger)}, S_i^{(\kappa_1)}, S_i^{(\kappa_0)})$, for $i \in [n]$. Thus, this also implies that $\hat\Sigma_{\mathrm{full}}, \hat\Sigma_{\mathrm{full}}\conv \toprob \Sigma_{\mathrm{full}}\conv \defeq \cov(\vec{S}_i\conv)$. 
\end{enumerate}
These results imply the following results:
\begin{enumerate}[noitemsep, topsep=0pt, resume=conditionproof]
    \item\label{res:delta-transformed-bound} Result~\ref{res:delta-beta-bound} implies that 
    $$\hat\theta_L^{(k)} \defeq h(\hat\beta^{(k)}, \hat\kappa_1, \hat\kappa_0) \le h(\hat\beta\kdagger + \Delta_k + o_p(n^{-1/2}), \hat\kappa_1, \hat\kappa_0) \le h(\hat\beta\kdagger + \Delta_k, \hat\kappa_1, \hat\kappa_0) + o_p(n^{-1/2}).$$
    The first inequality follows because $h$ is nondecreasing in its first argument. The second inequality follows because $h$ is continuously differentiable and $\hat\beta\kdagger + \Delta_k, \hat\kappa_1, \hat\kappa_0$ converges uniformly to $(\E[\hat\beta\kdagger], \kappa_1, \kappa_0)$ by the LLN (remember that all quantities involved have bounded $2+\delta$ moments by assumption). Thus, $h$ is locally Lipschitz at $(\E[\hat\beta\kdagger], \kappa_1, \kappa_0)$ and the result holds.
    \item\label{res:delta-quantile-convergence} Define $C_H\conv \defeq \diag{\Sigma_H\conv}^{-1/2} \Sigma_H\conv \diag{\Sigma_H\conv}^{-1/2}$ where $\Sigma_H\conv \defeq \nabla H(\E[\vec{S}_i\conv])^T \Sigma_{\mathrm{full}}\conv \nabla H(\E[\vec{S}_i\conv])$. In words, $C_H\conv$ is essentially the population variant of $\hat C_H$. Since $\bar S \toprob \E[S_i\conv]$ and $\hat \Sigma_{\mathrm{full}} \toprob \Sigma_{\mathrm{full}}\conv$ and $\nabla H$ is continuous by assumption, we know that $\hat C_H \toprob C_H\conv$. Thus, the continuous mapping theorem yields
    \begin{equation}
        \hat q_{1-\alpha} \toprob q_{1-\alpha}\conv \defeq Q_{1-\alpha}\left(\max_{k=1}^K Z_k\right) \text{ for } Z \sim \mcN(0, C_H\conv).
    \end{equation}\end{enumerate}
Since $K$ does not grow with $n$, we can combine Results~\ref{res:delta-variance-convergence}, \ref{res:delta-transformed-bound}, and~\ref{res:delta-quantile-convergence} to obtain:
\begin{equation}\label{eq::crossfit_mbs_daggerapprx}
    \hat\theta\lcb\crossfit 
    \le \max_{k=1}^K \left( h(\hat\beta\kdagger + \Delta_k, \hat\kappa_1, \hat\kappa_0) - q_{1-\alpha}\conv \frac{\hat\sigma\kdagger}{\sqrt{n}}\right)
    + o_p(n^{-1/2}).
\end{equation}
As notation, let $\hat\kappa = (\hat\kappa_1, \hat\kappa_0) \in \R^{d_1 + d_0}$ and $\kappa = [\kappa_1, \kappa_0] \in \R^{d_1 + d_0}$. Then observe
\begin{align*}
    \star & \defeq \P(\hat\theta\lcb\crossfit\le\theta_L) \\
    &\geq 
    \P\left(\max_{k=1}^K \left(h(\hat\beta\kdagger+\Delta_k, \hat\kappa) - q_{1-\alpha}\conv \frac{\hat\sigma\kdagger}{\sqrt{n}}\right) + o_p(n^{-1/2}) \le \theta_L\right) & \text{ by Eq. (\ref{eq::crossfit_mbs_daggerapprx})} \\
    &=
    \P\left(\max_{k=1}^K \left(\frac{\sqrt{n}\left(h(\hat\beta\kdagger+\Delta_k, \hat\kappa) - \theta_L\right)}{\hat\sigma\kdagger}\right) + o_p(1) \le q_{1-\alpha}\conv \right) & \text{ by rearrangement.}
\end{align*}
Now, we have essentially replaced $\hat\nu^{(k)}$ with $\hat\nu\kdagger$ for each $k$---the next step is to eliminate the random (and non-negligible) $\Delta_k$. We will do this by replacing each $\Delta_k$ with a constant $a_k$; later, we will let $a_k \to 0$.

To be precise, recall that for each $k$, $\Delta_k \le \beta_L - \E[\hat\beta\kdagger]$ \textit{and} $\Delta_k = o_p(1)$. Thus, for each $k$, we may pick a constant $a_k \ge 0$ such that $\Delta_k \le a_k$ with probability approaching one asymptotically\footnote{That is, $\I(\Delta_k \le a_k) \toprob 1$, although the convergence does not necessarily hold a.s.} and $a_k \le \beta_L - \E[\hat\beta\kdagger]$. Since $h$ is nondecreasing in its first argument, this implies both that $h(\hat\beta\kdagger + \Delta_k, \hat\kappa) \le h(\hat\beta\kdagger + a_k, \hat\kappa)$ holds asymptotically with probability approaching one and that $\theta_L = h(\beta_L, \kappa) \ge h(\E[\hat\beta\kdagger] + a_k, \kappa)$. Thus, since $K$ is finite, asymptotically we have that:
\begin{align*}
    \liminf_{n \to \infty} \star 
    &\ge 
    \liminf_{n \to \infty} \P\left(\max_{k=1}^K \frac{\sqrt{n}\left(h(\hat\beta\kdagger+a_k, \hat\kappa) - h(\E[\hat\beta\kdagger] + a_k, \kappa)\right)}{\hat\sigma\kdagger} + o_p(1) \le q_{1-\alpha}\conv \right).
\end{align*}
Now, we have successfully replaced the $\Delta_k$'s with constants $a_k$'s. Our next step is to modify the denominator $\hat\sigma\kdagger$ and replace it with one that accounts for the influence of $a_k$, and then bound the error from this approximation. As notation, let $\hat\sigma\kadagger$ be defined analogously to $\sigma\kdagger$ but replacing $\hat\beta\kdagger$ with $\hat\beta\kdagger + a_k$, that is,
$$
\hat\sigma\kadagger = \sqrt{\nabla h\left(\hat\beta\kdagger + a_k, \hat\kappa \right)^T \hat \Sigma\kdagger \nabla h\left(\hat\beta\kdagger+a_k, \hat\kappa \right)}
$$
and let $\sigma\kadagger$ be the population variant:
$$
\sigma\kadagger = \sqrt{\nabla h\left(\E[\hat\beta\kdagger] + a_k, \kappa \right)^T \Sigma\kdagger \nabla h\left(\E[\hat\beta\kdagger]+a_k, \kappa \right)}.
$$
Lastly, let $\hat Z_k = \frac{\sqrt{n}}{\sigma\kadagger}(h(\hat\beta\kdagger+a_k, \hat\kappa) - h(\E[\hat\beta\kdagger] + a_k, \kappa))$. Rearranging, we obtain
\begin{align}
    \liminf_{n \to \infty} \star 
    &\ge 
    \liminf_{n \to \infty} \P\left(\max_{k=1}^K \frac{\sigma\kadagger}{\hat\sigma\kdagger} \hat Z_k + o_p(1) \le q_{1-\alpha}\conv \right) \\ 
    &\ge
    \liminf_{n \to \infty} \P\left(\max_{k=1}^K \hat Z_k + \max_{k=1}^K \left(\frac{\sigma\kadagger}{\hat\sigma\kdagger}  - 1\right) \hat Z_k + o_p(1) \le q_{1-\alpha}\conv\right).
\end{align}
Now, we observe that $\hat Z \defeq (\hat Z_1, \dots, \hat Z_K)$ is asymptotically multivariate Gaussian by the multivariate delta method. In particular, define the vector of summands
$$V_i \defeq (S_i^{(\beta, 1, \dagger)} + a_1, \dots, S_i^{(\beta, K, \dagger)} + a_K, S_i^{(\kappa_1)}, S_i^{(\kappa_0)}) \in \R^{K + d_0 + d_1}$$
and let $\bar V = \frac{1}{n} \sum_{i=1}^n V_i$. If we define $\tilde{\beta}_L = (\E[\hat\beta^{(1,\dagger)}], \dots, \E[\hat\beta^{(K,\dagger)}])$ and $\vec{a} = (a_1, \dots, a_K)$, the multivariate CLT yields that
\begin{equation}
    \sqrt{n}(\bar V - (\tilde{\beta}_L + \vec{a}, \kappa)) \tod \mcN(0, \Sigma_{\mathrm{full}}),
\end{equation}
where notably $\Sigma_{\mathrm{full}}$ does not depend on $\vec{a}$. For the continuously differentiable function $H : \R^{K + d_0 + d_1} \to \R^K$ defined by $H_k(\bar V)\defeq h\left(\bar V_k,\,\bar V_{K+1:K+d_1},\,\bar V_{K+d_1+1:K+d_1+d_0}\right)=h\left(\hat\beta\kdagger+a_k,\,\hat\kappa_1,\,\hat\kappa_0\right)$, the multivariate delta method yields
\begin{equation*}
    \hat Z \tod \mcN(0, C_{H, \vec{a}}) \text{ where } C_{H, \vec{a}} \text{ is the correlation matrix of } \Sigma_{H, \vec{a}} \defeq \nabla H(\tilde{\beta}_L + \vec{a}, \kappa)^T \Sigma_{\mathrm{full}} \nabla H(\tilde{\beta}_L + \vec{a}, \kappa).
\end{equation*}
(Note that $\Sigma_{H,\vec{a}}$ has nonzero diagonal entries for all $\vec{a}$ sufficiently close to zero because $\Sigma_{H,\vec{0}} = \Sigma_H$ has nonzero diagonal entries by assumption and $\nabla H$ is assumed to be continuous.) Thus, by the continuous mapping theorem, we conclude that as $n \to \infty$,
$$
\max_{k=1}^K \hat Z_k + \max_{k=1}^K \left(\frac{\sigma\kadagger}{\hat\sigma\kdagger}  - 1\right) \hat Z_k \tod \max_{k=1}^K Z_k + \max_{k=1}^K \left(\frac{\sigma\kadagger}{\sigma\kdagger} - 1\right) Z_k \text{ for } Z \sim \mcN(0, C_{H, \vec{a}}).
$$
This implies
\begin{equation*}
    \liminf_{n \to \infty} \star \ge \P_{Z \sim \mcN(0, C_{H, \vec{a}})}\left(\max_{k=1}^K Z_k + \max_{k=1}^K \left(\frac{\sigma\kadagger}{\sigma\kdagger} - 1\right) Z_k \le q_{1-\alpha}\conv \right),
\end{equation*}
where this holds for all $\vec{a}$ sufficiently close to zero. Note that by assumption, the gradient of $h$ is continuous; thus $\nabla H$ is continuous as well. Thus, taking the limit as $\vec{a} \to 0$, we obtain
\begin{equation*}
    \liminf_{n \to \infty} \star \ge \P_{Z \sim \mcN(0, C_{H, \vec{0}})}\left(\max_{k=1}^K Z_k \le q_{1-\alpha}\conv \right) = 1 - \alpha,
\end{equation*}
where the right-hand equality holds by the definition of $q_{1-\alpha}\conv$, since $C_{H,\vec{0}}=C_H\conv$.

\paragraph{Condition~\ref{cond:delta-method-misspecification}.} We now consider the general case. Without loss of generality, suppose $k=1, \dots, K_0$ satisfy Condition~\ref{cond:delta-method-convergence}, and $k=K_0+1, \dots, K$ satisfy Condition~\ref{cond:delta-method-misspecification}. Note that the proof for Condition~\ref{cond:delta-method-convergence} shows that
\begin{equation*}
     \liminf_{n \to \infty} \P\left(\max_{k=1}^{K_0}\left( \hat\theta_L^{(k)} - \hat q_{1-\alpha} \frac{\hat\sigma^{(k)}}{\sqrt{n}} \right)\le \theta_L\right) \ge 1-\alpha,
\end{equation*}
where in particular this holds because the addition of $\hat\nu^{(K_0+1)}, \dots, \hat\nu^{(K)}$ does not affect the values of $\hat\theta_L^{(k)}, \hat\sigma^{(k)}$ and can only increase the value of $\hat q_{1-\alpha}$. Thus, it suffices to show that
\begin{equation*}
    \liminf_{n \to \infty} \P\left(\max_{k > K_0} \left(\hat\theta_L^{(k)} - \hat q_{1-\alpha} \frac{\hat\sigma^{(k)}}{\sqrt{n}} \right)\le \theta_L \right) = 1.
\end{equation*}
To do this, note that whenever $\alpha \le 0.5$, $\hat q_{1-\alpha} \hat\sigma^{(k)} \ge 0$. Thus, it suffices to show that $\hat \theta_L^{(k)} \le \theta_L$ with probability $1$ asymptotically. Yet Condition~\ref{cond:delta-method-misspecification} guarantees both that $\hat\beta^{(k)} \le \beta_L$ with probability one asymptotically and that $ \beta_L - \hat\beta^{(k)} \gg n^{-1/2}$ (see the proof of Theorem \ref{thm::crossfit_validity} in Appendix \ref{appendix::crossfit_proofs}) for $k > K_0$. Furthermore, we assume that $\partial_b h(b, \kappa)\geq\gamma >0$; since $h$ is continuously differentiable, this means that $\partial_b h(b, x)$ is uniformly bounded below by a positive constant $\gamma$ for all $x$ in a neighborhood of $\kappa$. Since $\hat\kappa \toprob \kappa$ asymptotically, we have that with probability one asymptotically,
\begin{align*}
    \hat\theta_L^{(k)} - \theta_L 
    &=
    h(\hat\beta^{(k)}, \hat\kappa) - h(\beta_L, \kappa) \\
    &\le 
    \underbrace{h(\beta_L, \hat\kappa) - h(\beta_L, \kappa)}_{O_p(n^{-1/2})} + \underbrace{\gamma (\hat\beta^{(k)} - \beta_L)}_{\text{nonpositive and }  \gg n^{-1/2}}.
\end{align*}
In particular, the left term is $O_p(n^{-1/2})$ (or smaller) by the delta method, and the right term is $ \gg n^{-1/2}$ by the previous remarks. Since the right-hand term dominates the left-term and is asymptotically less than zero, this implies that $\hat\theta_L^{(k)} - \theta_L \le 0$ with probability one for all $k > K_0$. This completes the proof.
\end{proof}

\section{More about computation}
\subsection{Computational complexity of the dual bounds method}
\label{sec: complexity}
Here we discuss the computational complexity of Algorithm~\ref{alg:crossfit-dual-lcb} and \ref{alg:conditional-dual}. We exclude the complexity of fitting the conditional distribution, as it depends heavily on the concrete model choice. 
Let \(J=\nvals\) be the number of discretization points for each fitted
conditional outcome distribution and let \(L=|\mathcal W_x|\) be the number of additional
moment restrictions. Conditional on the fitted first-stage distributions, Algorithm \ref{alg:crossfit-dual-lcb} solves
the conditional dual problem only at the observed evaluation points \(x=X_i\). Thus, under
cross-fitting, one solves \(n\) conditional problems in total. When the problem can be reduced to a one-dimensional optimization indexed by a parameter $a$, we can simply evaluate the dual value over a discretization of $a$ with $J$ points, which has $O(nJ)$ time complexity. When the reduction is not available, for each fixed \(x\), the
discretized conditional problem is a linear programming problem with
$
d_J = 2J+L
$
variables and
$
r_J = J^2+L
$
constraints. Using a dense interior-point LP solver to accuracy \(\varepsilon\),
one LP solve has time complexity on the order of
\[
O\left(
\sqrt{r_J}\log(1/\varepsilon)
\left\{
r_J d_J^2+d_J^3
\right\}
\right).
\]
If we use the same discretization grids to enforce
dual feasibility, it will add $J^2L$ complexity to each conditional problem. Therefore the total complexity is
\[
O\left(
n\left[
\sqrt{r_J}\log(1/\varepsilon)
\left\{
r_J d_J^2+d_J^3
\right\}
+
J^2L
\right]
\right).
\]
For fixed \(L\), this simplifies to
$O\left(nJ^5\log(1/\varepsilon)\right)$
for a generic interior-point LP implementation.

In the unrestricted case
\(\mathcal W_x=\emptyset\), the same conditional problem is the dual of a standard
discrete optimal transport problem on \(J\) grids. In that case, one can use an optimal transport solver, such as the classical Sinkhorn's algorithm \citet{peyre2019computational} (one run takes $\tilde O(J^2/\epsilon^2)$\footnote{$\tilde O(\cdot)$ means standard $O(\cdot)$ that ignores logarithmic factors}) or the recent primal-dual extragradient algorithm \citet{jambulapati2019direct} (one run takes $\tilde O(J^2/\epsilon)$), which leads to a total complexity of $\tilde O(nJ^2/\epsilon^2)$ or $\tilde O(nJ^2/\epsilon)$. It significantly reduces from the generic interior-point LP implementation when the grid size $J$ is large. It is also worthwhile to notice that several papers \cite{benamou2015iterative, tang2024sinkhorn} have studied efficient algorithms for optimal transport with constraints, which can improve upon the generic interior method and achieve similar performance to Sinkhorn's algorithm in certain scenarios.

\subsection{Computational comparison with \cite{semenova2020better}}
\label{sec: computational-comparison-semenova}
We next use the Lee-bound setting from Section \ref{sec::sims} to illustrate the computational role of our formulation.
We use the same data-generating process as in Section \ref{sec::sims} with $n=100$ data points. To isolate the cost of the sharp-bound computation, we do not include the cost of first-stage nuisance estimation. Instead, for each realized covariate value, we supply the algorithms with the true conditional distributions generated by the data-generating process. Thus all methods start from the same conditional distributions and support grid, and the reported runtimes measure only the computational cost of mapping these conditional distributions into the sharp Lee lower bound. After discretizing each augmented outcome marginal on a common support of total size \(\nvals\), including the atom \((0,0)\), the generic formulation treats the Lee problem as an optimal transport / linear programming problem over the augmented potential outcomes
\[
    Z(0)=(Y(0)S(0),S(0)),
    \qquad
    Z(1)=(Y(1)S(1),S(1)).
\]

\begin{figure}[h!]
    \centering
    \includegraphics[width=0.95\linewidth]{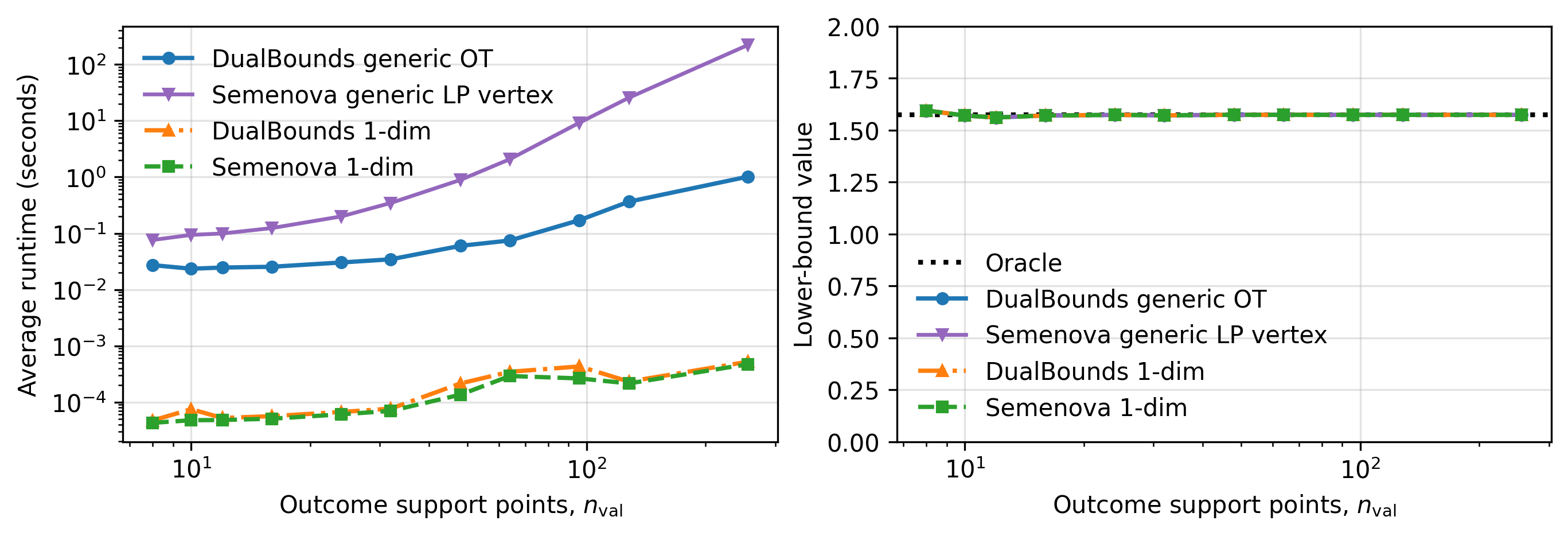}
    \caption{Computational comparison for Lee bounds. Runtime points are averages over three runs.}
    \label{fig::lee_comp}
\end{figure}

Figure \ref{fig::lee_comp} reports the computational comparison. ``dual bounds generic OT'' applies our generic formulation to the augmented potential outcomes and imposes the monotonicity moment $\E[\I\{S(0)>S(1)\}\mid X=x]\le0$. In the primal implementation, this nonnegative moment restriction is equivalently enforced by fixing the coupling masses with $s_0>s_1$ at zero. ``Semenova generic LP vertex'' solves the corresponding discretized dual LP, including the multiplier on $\I\{s_0>s_1\}$ based on the method in \citet{semenova2023aggregated}, with a generic LP solver; equivalently, the LP solver finds an optimal vertex without explicitly constructing the full finite index set. These two methods solve the same discretized sharp-bound problem, and their lower-bound values are numerically identical up to solver tolerance. We also include two Lee-bound-specific one-dimensional reductions. ``Semenova 1-dim'' implements this one-dimensional reduction, and ``dual bounds 1-dim'' applies our dual framework using the same reduction. Section \ref{sec: lee reduction} shows that this one-dimensional reduction is identical to the moment-constrained dual problem on the augmented outcome space.

The computational costs differ sharply. At \(\nvals=256\), our generic OT implementation takes about \(1.03\) seconds on average, while the generic LP vertex method takes about \(226.2\) seconds. The two one-dimensional Lee reductions are much faster: at the same \(\nvals\), both implementations take about \(5\times 10^{-4}\) seconds. The values of the computed dual bounds are numerically identical, which is also explicitly acknowledged in \cite{khandamiryan2026adaptive} (the newest version of \cite{semenova2023aggregated}).

The exact finite-index characterization of
\cite{semenova2023aggregated} is not shown in the main figure because explicitly
enumerating its vertices becomes infeasible even at a moderate support size.
For reference, after one normalization, an unrestricted transport dual on
\(J\) support points per marginal has \(2J-1\) free coordinates and \(J^2\)
transport inequalities. A crude candidate-basis bound is therefore
\[
\binom{J^2}{2J-1},
\]
because a vertex is generically pinned down by \(2J-1\) independent binding
constraints. This is an unrestricted benchmark, not an exact Lee-specific
vertex count: the Lee multiplier and the structural restrictions must be
accounted for separately. Explicit vertex enumeration is distinct from solving
the LP, since a generic LP solver can find an optimum without enumerating every
vertex. In our implementation, direct candidate enumeration takes about
\(0.02\) seconds at \(J=3\) and \(2.71\) seconds at \(J=4\). At \(J=10\), even
the unrestricted benchmark is
\[
\binom{100}{19},
\]
so explicit construction of the finite index set is not computationally
practical.

The comparison in Figure \ref{fig::lee_comp} also highlights the benefit of exploiting target-specific structure. If one uses target-specific structure to reduce the optimization problem to a simpler problem, both frameworks can be significantly improved. If the reduction is not available, then our generic optimal-transport implementation is faster than solving the same full-support dual problem with a generic LP solver, and explicit finite-index vertex construction would be infeasible even with a small support.

\subsection{Alternative approaches for computation}
\subsubsection{Series estimator-based approach to approximate dual variables}\label{appendix::series}
An alternative option to the discretization-based approach discussed in Section \ref{subsec::condcomp} is to choose an integer $M\in\N$, a collection of basis functions $\phi_m : \mcY \to \R$ indexed by $m\in[M]$, and approximate 
$$\nu_{k,x}(y) \approx \sum_{m=1}^M \alpha_{k,m,x} \phi_m(y) \text{ for } \alpha_{k,m,x} \in \R, k \in \{0,1\}, m \in [M].$$
This reduces the problem to fitting the values of $\{\alpha_{k,m,x}\}_{m \in [M], k \in \{0,1\}} \in \R^{2M}$, which is a concave problem with finitely many parameters. Of course, approximating the dual variables with a finite collection of basis functions introduces approximation errors, but we emphasize that our approach still yields \textit{valid} bounds even if our initial estimates $\hat\nu\init$ based on the basis functions are arbitrarily poor. Furthermore, selecting a collection of universal basis functions (e.g., splines, Fourier expansion, Gaussian kernel) can ensure that the approximation errors are not too large.

However, fitting $\{\alpha_{k,m,x}\}_{k \in \{0,1\}, m \in [M]}$ is still challenging because the conditional validity constraint $(\nu_{0,x}, \nu_{1,x}) \in \mcV_x$ is still infinite-dimensional. To overcome this, we use ideas from the optimal transport literature. Indeed, for this particular problem, we can eliminate the effect of the constraints by adding the maximum deviation from the constraints as the penalty in an objective. In particular, for the product measure $\hat P\subprod \defeq \hat P_{Y(1) \mid X = x} \times \hat P_{Y(0) \mid X = x}$, consider the objective 
\begin{align}
    &O(\nu_{0,x}, \nu_{1,x}, \{\lambda_{x,\ell}\}_{\ell=1}^L) \nonumber \\
\defeq&\E_{\hat P\subprod}\bigg[\nu_{0,x}(Y(0)) + \nu_{1,x}(Y(1)) - \underbrace{\max_{y_1, y_0 \in \mcY} \nu_{0,x}(y_0) + \nu_{1,x}(y_1) - \sum_{\ell=1}^{L} \lambda_{x,\ell} w_{x,\ell}(y_0, y_1) - f(y_0, y_1, x)}_{\text{max penalty}}\bigg] \label{eq::odef}  
\end{align}
We can maximize this unconstrained objective to find conditionally optimal dual variables, as stated below. 
\begin{proposition} 
\label{prop: grid search}
Suppose $\hat\nu\init_{0,x}, \hat\nu\init_{1,x} : \mcY \to \R$ and $\hat\lambda_{x,1}, \dots, \hat\lambda_{x,L} \ge 0$ maximize the objective $O(\nu_{0,x}, \nu_{1,x}, \{\lambda_{x,\ell}\}_{\ell=1}^L)$ among all functions $\nu_{0,x}, \nu_{1,x} : \mcY \to \R$ and constants $\lambda_{x,1}, \dots, \lambda_{x,L} \ge 0$. Assume that the conditional dual problem in Eq. (\ref{eq::hatnu_def_comp_sec}) attains its supremum. Let $c_x$ be the minimum constant such that $(\hat\nu\init_{0,x}-c_x,\hat\nu\init_{1,x}-c_x)\in\mcV_x$. Then the pair $(\hat\nu\init_{0,x}-c_x,\hat\nu\init_{1,x}-c_x)$ solves the conditional dual problem Eq. (\ref{eq::hatnu_def_comp_sec}).
\end{proposition}

In other words, if we can find initial solutions $\hat\nu\init_{0,x}, \hat\nu\init_{1,x}, \hat\lambda_{x,1}, \dots, \hat\lambda_{x,L} \in \argmax O(\nu_{0,x}, \nu_{1,x}, \{\lambda_{x,\ell}\}_{\ell=1}^L)$, we can simply apply the grid-search from Section \ref{subsec::compstrat} to find an optimal solution to the conditional dual problem. In practice, we recommend optimizing a sample version of this objective. In particular, let $\{\tilde{Y}_b(0), \tilde{Y}_b(1)\}_{b=1}^B$ denote samples from $\hat P\subprod$ for some large $B$. Then \eqref{eq::odef} can be approximated by 
\begin{align*}
    \hat{O}(\nu_{0,x}, \nu_{1,x}, \{\lambda_{x,\ell}\}_{\ell=1}^L) &= \frac{1}{B} \sum_{b=1}^B \Bigg\{\nu_{0,x}(\tilde{Y}_b(0)) + \nu_{1,x}(\tilde{Y}_b(1))\\
    & - \max_{b\in [B]} \left[\nu_{0,x}(\tilde{Y}_b(0)) + \nu_{1,x}(\tilde{Y}_b(1)) - \sum_{\ell=1}^{L} \lambda_{x,\ell} w_{x,\ell}(\tilde{Y}_b(0), \tilde{Y}_b(1)) - f(\tilde{Y}_b(0), \tilde{Y}_b(1), x)\right]\Bigg\}.
\end{align*}
The subgradient with respect to $\alpha_{k, m, x}$ can be easily computed, and hence the objective can be optimized via gradient-based methods.

One shortcoming of the above approach is that the objective function is non-smooth. An alternative strategy is to use a smooth approximation of the exact objective \eqref{eq::odef}:
\begin{align}
    O_\epsilon(\nu_{0,x}, \nu_{1,x}, \{\lambda_{x,\ell}\}_{\ell=1}^L) \nonumber\defeq\E_{\hat P\subprod}[\nu_{0,x}(Y(0)) + \nu_{1,x}(Y(1))] - \epsilon \log \E_{\hat P\subprod}[R_{\epsilon, x}(Y(1), Y(0))]
\end{align}
where the random variable $R_{\epsilon}(Y(1), Y(0))$ is the following smoothed penalty function:
\begin{equation*}
    R_{\epsilon, x}(Y(1), Y(0)) = \exp\left(\frac{\nu_{0,x}(Y(0)) + \nu_{1,x}(Y(1)) - \sum_{\ell=1}^{L} \lambda_{x,\ell} w_{x,\ell}(Y(0), Y(1)) - f(Y(0), Y(1), x)}{\epsilon}\right).
\end{equation*}

This smooth penalty is typically known as an entropy regularizer in optimal transport theory \citep{villani2009optimal, peyre2019computational}. Note for each $\epsilon$, using the basis approximation $\nu_{k,x}(y) \approx \sum_{m=1}^M \alpha_{k,m,x} \phi_m(y)$, maximizing $O_{\epsilon}(\nu_{0,x}, \nu_{1,x}, \{\lambda_{x,\ell}\}_{\ell=1}^L)$ is now a finite-dimensional concave problem with positive constraints on $\lambda_{x,\ell}$,  which we can solve using gradient based methods. Thus, as a heuristic algorithm (which is commonly used in the optimal transport literature), we suggest using stochastic gradient descent to maximize the smoothed objective and sending $\epsilon \to 0$ along some schedule as we take more gradient steps. This algorithm is closely related to the Sinkhorn algorithm, and indeed, this optimization strategy is widely used in optimal transport literature \citep[e.g.][]{sinkhorn1964relationship, villani2009optimal, cuturi2013sinkhorn, altschuler2017near, peyre2019computational}.  

The main message is as follows: since the conditional problem Eq. (\ref{eq::hatnu_def_comp_sec}) optimizes over only two univariate functions, the literature contains many strategies to solve it approximately, including methods beyond the two mentioned in this paper. When combined with the general strategy outlined in Section \ref{subsec::compstrat}, any of these methods can be used to compute the estimated dual variables $\hat\nu$. Crucially, as long as we effectively perform the two-dimensional grid search in Section \ref{subsec::compstrat}, we will get valid bounds on $\theta_L$ no matter how poorly we solve Eq. (\ref{eq::hatnu_def_comp_sec}).

\paragraph{Proof of Proposition \ref{prop: grid search}.}
\label{sec: comp proof_sec4}
Let
\[
J_x(\nu)
=
\E_{\hat P_{Y(0)\mid X=x}}[\nu_{0,x}(Y(0))]
+
\E_{\hat P_{Y(1)\mid X=x}}[\nu_{1,x}(Y(1))]
\]
and define the maximum violation
\[
V_x(\nu,\lambda)
=
\sup_{y_0,y_1\in\mcY}
\left\{
\nu_{0,x}(y_0)+\nu_{1,x}(y_1)
-\sum_{\ell=1}^L\lambda_{x,\ell}w_{x,\ell}(y_0,y_1)
-f(y_0,y_1,x)
\right\}.
\]
Then
\[
O_x(\nu,\lambda)=J_x(\nu)-V_x(\nu,\lambda).
\]
For any tuple \((\nu,\lambda)\), define
\begin{equation}\label{eq:dual-constant-correction}
\nu_{0,x}^c
=\nu_{0,x}-\frac{V_x(\nu,\lambda)}2,
\qquad
\nu_{1,x}^c
=\nu_{1,x}-\frac{V_x(\nu,\lambda)}2.
\end{equation}
Then \(V_x(\nu^c,\lambda)=0\), so \((\nu^c,\lambda)\) is feasible, and
\[
J_x(\nu^c)
=J_x(\nu)-V_x(\nu,\lambda)
=O_x(\nu,\lambda).
\]
By attainability, let $(\nu^\star,\lambda^\star)$ be a dual-optimal feasible
tuple. Applying \eqref{eq:dual-constant-correction} to this tuple gives another
feasible tuple with zero maximum violation. Since feasibility implies
$V_x(\nu^\star,\lambda^\star)\le0$, the identity above shows that its objective
is weakly larger; optimality therefore forces equality. Relabel this corrected
tuple as $(\nu^\star,\lambda^\star)$. Then
\[
V_x(\nu^\star,\lambda^\star)=0,
\qquad
O_x(\nu^\star,\lambda^\star)=J_x(\nu^\star).
\]
For any $(\nu,\lambda)$, the corrected tuple $(\nu^c,\lambda)$ is feasible, so
\[
O_x(\nu,\lambda)
=J_x(\nu^c)
\le J_x(\nu^\star)
=O_x(\nu^\star,\lambda^\star).
\]
Thus every maximizer
\((\hat\nu^{\rm init},\hat\lambda)\) of \(O_x\) produces, after the
correction in \eqref{eq:dual-constant-correction}, a feasible tuple attaining the optimal dual value.
Consequently,
\(((\hat\nu^{\rm init})^c,\hat\lambda)\) is dual optimal.

\subsubsection{Deep Dual Bounds: an alternative approach for computation}
\label{Appendix: deep dual}

In the main text, we focus on the two-step approach in Section~\ref{subsec::dualbnds}, which first estimates the conditional distributions $\hatPZC$ and $\hatPOC$ and then solves the dual problem \eqref{eq::hatnu_def} for each $x\in \{X_i: i\in\mathcal{D}_2\}$. However, this two-step approach may be infeasible when the covariates are complex and modeling their conditional distributions is challenging. For instance, when $X$ includes unstructured data such as images and text, standard regression-based methods tend to be imprecise because they lack representation learning, while modern machine-learning methods may not be designed to estimate conditional distributions or may require excessive computation. Inspired by the recent success of deep learning with complex data and its application to optimal transport \citep{makkuva2020optimal}, we develop Deep Dual Bounds, an alternative approach that parameterizes the dual variables $\hat{\nu}_{0,x}(y), \hat{\nu}_{1,x}(y)$ with neural networks and computes them by end-to-end training. 

Recall Theorem \ref{thm::kantorovich}, which states that, 
\begin{equation*}
\label{eqn:strong dual appendix}
    \theta_L =  \sup_{\nu_{0}, \nu_1\in\mcV}\E_{P\opt}[\nu_{0, X}(Y(0)) + \nu_{1, X}(Y(1))].
\end{equation*}
First, we transform the above constrained optimization problem into an unconstrained optimization problem by adding a proper penalty onto the objective function. Following \eqref{eq::odef}, we consider the following optimization problem:
\begin{align*}
    &\max\E\bigg[\nu_{0,X}(Y(0)) + \nu_{1,X}(Y(1)) - \max_{y_1, y_0 \in \mcY} \nu_{0,X}(y_0) + \nu_{1,X}(y_1) - \sum_{\ell=1}^{L} \lambda_{X,\ell} w_{X,\ell}(y_0, y_1) - f(y_0, y_1, X)\bigg].
\end{align*}
Unlike the standard supervised learning problems, one cannot directly apply the stochastic gradient-based methods to optimize the above objective because $Y(0)$ and $Y(1)$ cannot be simultaneously observed. 

To address this issue, we construct a pseudo-sample by matching on covariates \citep{abadie2006large, stuart2010matching}. In particular,  for each unit $(X_i, W_i, Y_i)$, we match it to the nearest neighbor $j(i)$ from the other group, i.e., 
$$
j(i) = \argmin_{k: W_k=1-W_i}\|X_i-X_k\|,
$$
For units with $W_i=0$, we impute $Y_i(1)$ by $Y_{j(i)}$; for units with
$W_i=1$, we impute $Y_i(0)$ by $Y_{j(i)}$. This yields a pseudo-sample with triplets
\begin{equation}
\label{eqn:sample matching}
    (\tilde X_i, \tilde Y_i(0), \tilde Y_i(1)) = \begin{cases}
    (X_i, Y_i, Y_{j(i)}) & W_i = 0\\
    (X_i, Y_{j(i)}, Y_i) &W_i = 1
    \end{cases}.
\end{equation}
Then we consider the following proxy objective function, 
\begin{align}\label{eqn:deepdual objective}
    &\hat O(\nu_{0,x}, \nu_{1,x}, \{\lambda_{x,\ell}\}_{\ell=1}^L) \nonumber\\
    \defeq&\frac{1}{n}\sum_{i=1}^{n}\bigg[\nu_{0, \tilde{X}_i}(\tilde{Y}_i(0)) + \nu_{1, \tilde{X}_i}(\tilde{Y}_i(1))\bigg]\nonumber\\
    &- \max_{i=1}^n \bigg[\nu_{0,\tilde X_i}(\tilde Y_i(0)) + \nu_{1,\tilde X_i}(\tilde Y_i(1)) - \sum_{\ell=1}^{L} \lambda_{\ell}(X_i) w_{X_i,\ell}(\tilde Y_i(0), \tilde Y_i(1)) - f(\tilde Y_i(0), \tilde Y_i(1), \tilde X_i)\bigg].
\end{align}
With this sub-differentiable loss function, we parametrize $\nu_{0,x}, \nu_{1,x}, \{\lambda_{x,\ell}\}_{\ell=1}^{L}$ by neural networks and apply stochastic gradient-based methods to learn the dual variables. Finally, we apply the same procedure described in Section \ref{subsec::compstrat} to the solutions to guarantee the dual feasibility and hence the validity of the estimated bounds.

\paragraph{Experimental Results.}
We apply the Deep Dual Bounds to the applications described in Section \ref{sec::applications}. We parametrize the dual variables by 5-layer fully connected ReLU neural nets, and apply the full batch Adam optimizer with a learning rate of 0.05, weight decay 1e-4, to optimize the deep dual objective \eqref{eqn:deepdual objective} for 400 epochs in total. Similar to the two-stage method described in Section \ref{sec::applications}, we use cross-fitting with 10 folds. The experimental results are shown in Table \ref{table::deepdual}.

\begin{table}[h!]
\centering
\begin{tabular}{l|cc|cc}
\toprule
 & Deep Dual LB & Deep Dual UB & Two Stage LB & Two Stage UB \\
Dataset  &  &  &  &  \\
\midrule
Persuasion Effect & 0.0 & 0.6416 & 0.038 & 0.365 \\
 (Section \ref{subsec::persuasion})& (0.000) & (0.097) & (0.027) & (0.019) \\
\midrule
401k Eligibility & 12626 & 69597 & 5564 & 47286 \\
 (Section \ref{subsec::401k})& (1609) & (6945) & (1201) & (1258) \\
\bottomrule
\end{tabular}
\caption{Comparison of Deep Dual Bounds method and the two-stage Dual Bounds on the applications described in Section \ref{sec::applications}. For the two-stage method, we only report the tightest bound from Table \ref{table::persuasion} and \ref{table::401k}. Standard errors are shown in parentheses.}\label{table::deepdual}
\end{table}

For both of the above applications, the Deep Dual Bounds method provides looser bounds than the two-stage method, except for the lower bound in the 401K eligibility example. We could not get a meaningful result for the application in \ref{subsec::carranza_chenroth} due to convergence issues. Therefore, we recommend the two-stage method when the conditional distributions can be estimated and treat the Deep Dual Bounds as a rescue when the two-stage method cannot be implemented.

\end{document}